\documentclass[twocolumn]{aastex631}

\usepackage{bm}
\usepackage{color}
\usepackage{xspace}
\usepackage{multirow}

\def\beq{\begin{eqnarray}}
\def\eeq{\end{eqnarray}}

\usepackage[normalem]{ulem}
\usepackage{physics}
\usepackage{ifthen}
\usepackage{aas_macros}

\graphicspath{{figures/}}

\newcommand{\av}[1]{\left\langle#1\right\rangle}

\newcommand{\hun}{\,\mathrm{km}\,\mathrm{s}^{-1}\mathrm{Mpc}^{-1}}

\newcommand{\arcsecond}{^{\prime \prime}}

\newcommand{\editF}[1]{#1}
\newcommand{\resub}[1]{#1}

\newcommand*{\multilinecell}[2]{\begin{tabular}{@{}#1@{}}#2\end{tabular}}

\begin{document}

\title{The Atacama Cosmology Telescope: Cosmology from cross-correlations of unWISE galaxies and ACT DR6 CMB lensing}
\shorttitle{unWISE x ACT DR6 cosmology}
  \shortauthors{Farren, Krolewski, MacCrann, Ferraro et al.}

\author[0000-0001-5704-1127]{Gerrit~S.~Farren}\affiliation{DAMTP, Centre for Mathematical Sciences, University of Cambridge, Wilberforce Road, Cambridge CB3 OWA, UK}\affiliation{Kavli Institute for Cosmology Cambridge, Madingley Road, Cambridge CB3 0HA, UK}

\author[0000-0003-2183-7021]{Alex~Krolewski}
\affiliation{Perimeter Institute for Theoretical Physics, 31 Caroline St. North, Waterloo, ON NL2 2Y5, Canada}
\affiliation{Waterloo Centre for Astrophysics, University of Waterloo, Waterloo, ON N2L 3G1, Canada}

\author[0000-0002-8998-3909]{Niall~MacCrann}\affiliation{DAMTP, Centre for Mathematical Sciences, University of Cambridge, Wilberforce Road, Cambridge CB3 OWA, UK}

\author[0000-0003-4992-7854]{Simone~Ferraro}\affiliation{Physics Division, Lawrence Berkeley National Laboratory, Berkeley, CA, USA}\affiliation{Berkeley Center for Cosmological Physics, University of California, Berkeley, CA 94720}


\author[0000-0003-3230-4589]{Irene~Abril-Cabezas}\affiliation{DAMTP, Centre for Mathematical Sciences, University of Cambridge, Wilberforce Road, Cambridge CB3 OWA, UK}

\author{Rui An}
\affiliation{Department of Physics and Astronomy, University of Southern California, Los Angeles, CA 90089, USA}

\author[0000-0002-2287-1603]{Zachary~Atkins}
\affiliation{Joseph Henry Laboratories of Physics, Jadwin Hall, Princeton University, Princeton, NJ, USA 08544}

\author[0000-0001-5846-0411]{Nicholas~Battaglia}
\affiliation{Department of Astronomy, Cornell University, Ithaca, NY 14853, USA}

\author[0000-0003-2358-9949]{J.~Richard~Bond}\affiliation{Canadian Institute for Theoretical Astrophysics, University of Toronto, Toronto, ON, Canada M5S 3H8}

\author[0000-0003-0837-0068]{Erminia~Calabrese}\affiliation{School of Physics and Astronomy, Cardiff University, The Parade, Cardiff, Wales CF24 3AA, UK}

\author[0000-0002-9113-7058]{Steve~K.~Choi}\affiliation{Department of Physics, Cornell University, Ithaca, NY 14853, USA}\affiliation{Department of Astronomy, Cornell University, Ithaca, NY 14853, USA}

\author[0000-0003-2946-1866]{Omar~Darwish}\affiliation{Universit\'{e} de Gen\`{e}ve, D\'{e}partement de Physique Th\'{e}orique et CAP, 24 quai Ernest-Ansermet, CH-1211 Gen\`{e}ve 4, Switzerland}

\author[0000-0002-3169-9761]{Mark~J.~Devlin}\affiliation{Department of Physics and Astronomy, University of
Pennsylvania, 209 South 33rd Street, Philadelphia, PA, USA 19104}

\author[0000-0003-2856-2382]{Adriaan~J.~Duivenvoorden}
\affiliation{Center for Computational Astrophysics, Flatiron Institute, New York, NY 10010, USA}
\affiliation{Joseph Henry Laboratories of Physics, Jadwin Hall, Princeton University, Princeton, NJ, USA 08544}

\author[0000-0002-7450-2586]{Jo~Dunkley}\affiliation{Joseph Henry Laboratories of Physics, Jadwin Hall, Princeton University, Princeton, NJ, USA 08544}\affiliation{Department of Astrophysical Sciences, Peyton Hall, Princeton University, Princeton, NJ USA 08544}

\author[0000-0002-9539-0835]{J.~Colin Hill}\affiliation{Department of Physics, Columbia University, 538 West 120th Street, New York, NY, USA 10027}

\author[0000-0002-8490-8117]{Matt~Hilton}
\affiliation{Wits Centre for Astrophysics, School of Physics, University of the Witwatersrand, Private Bag 3, 2050, Johannesburg, South Africa}
\affiliation{Astrophysics Research Centre, School of Mathematics, Statistics, and Computer Science, University of KwaZulu-Natal, Westville Campus, Durban 4041, South Africa}

\author[0000-0001-7109-0099]{Kevin~M.~Huffenberger}\affiliation{Department of Physics, Florida State University, Tallahassee FL, USA 32306}

\author[0000-0002-0935-3270]{Joshua~Kim}\affiliation{Department of Physics and Astronomy, University of Pennsylvania, Philadelphia, PA 19104, USA}

\author[0000-0002-6849-4217]{Thibaut~Louis}\affiliation{Universit\'e Paris-Saclay, CNRS/IN2P3, IJCLab, 91405 Orsay, France}

\author[0000-0001-6740-5350]{Mathew~S.~Madhavacheril}\affiliation{Department of Physics and Astronomy, University of Pennsylvania, 209 South 33rd Street, Philadelphia, PA, USA 19104}

\author[0000-0002-8571-8876]{Gabriela~A.~Marques}\affiliation{Fermi National Accelerator Laboratory, P. O. Box 500, Batavia, IL 60510, USA}\affiliation{Kavli Institute for Cosmological Physics, University of Chicago, 5640 S. Ellis Ave., Chicago, IL 60637, USA}

\author[0000-0002-7245-4541]{Jeff~McMahon}
\affiliation{Department of Astronomy and Astrophysics, University of Chicago, 5640 S. Ellis Ave., Chicago, IL 60637, USA}
\affiliation{Kavli Institute for Cosmological Physics, University of Chicago, 5640 S. Ellis Ave., Chicago, IL 60637, USA}
\affiliation{Department of Physics, University of Chicago, Chicago, IL 60637, USA}
\affiliation{Enrico Fermi Institute, University of Chicago, Chicago, IL 60637, USA}

\author[0000-0001-6606-7142]{Kavilan~Moodley}\affiliation{Astrophysics Research Centre, School of Mathematics, Statistics and Computer Science, University of KwaZulu-Natal, Durban 4001, South Africa}

\author[0000-0002-9828-3525]{Lyman~A.~Page}\affiliation{Joseph Henry Laboratories of Physics, Jadwin Hall,
Princeton University, Princeton, NJ, USA 08544}

\author[0000-0001-6541-9265]{Bruce Partridge}\affiliation{Department of Physics and Astronomy, Haverford College, Haverford PA, USA 19041}

\author[0000-0001-7805-1068]{Frank~J.~Qu}\affiliation{DAMTP, Centre for Mathematical Sciences, University of Cambridge, Wilberforce Road, Cambridge CB3 OWA, UK}

\author[0000-0002-4619-8927]{Emmanuel Schaan}\affiliation{SLAC National Accelerator Laboratory, Menlo Park, CA 94025, USA}
\affiliation{Kavli Institute for Particle Astrophysics and Cosmology and Department of Physics, Stanford University, Stanford, CA 94305, USA}

\author[0000-0002-9674-4527]{Neelima~Sehgal}\affiliation{Physics and Astronomy Department, Stony Brook University, Stony Brook, NY USA 11794}

\author[0000-0002-4495-1356]{Blake~D.~Sherwin}\affiliation{DAMTP, Centre for Mathematical Sciences, University of Cambridge, Wilberforce Road, Cambridge CB3 OWA, UK}\affiliation{Kavli Institute for Cosmology Cambridge, Madingley Road, Cambridge CB3 0HA, UK}

\author[0000-0002-8149-1352]{Crist\'obal Sif\'on}\affiliation{Instituto de F\'isica, Pontificia Universidad Cat\'olica de Valpara\'iso, Casilla 4059, Valpara\'iso, Chile}

\author[0000-0002-7020-7301]{Suzanne~T.~Staggs}\affiliation{Joseph Henry Laboratories of Physics, Jadwin Hall, Princeton University, Princeton, NJ, USA 08544}

\author[0000-0002-3495-158X]{Alexander~Van~Engelen}\affiliation{School of Earth and Space Exploration, Arizona State University, Tempe, AZ 85287, USA}

\author[0000-0001-5327-1400]{Cristian~Vargas}
\affiliation{Instituto de Astrof\'isica and Centro de Astro-Ingenie\'ia, Facultad de F\'isica, Pontificia Universidad Cat\'olica de Chile, Av. Vicu\~na Mackenna 4860, 7820436 Macul, Santiago, Chile}

\author[0000-0001-5245-2058]{Lukas~Wenzl}\affiliation{Department of Astronomy, Cornell University, Ithaca, NY, 14853, USA}

\author[0000-0001-9912-5070]{Martin~White}\affiliation{Department of Physics, University of California, Berkeley, 366 LeConte Hall MC 7300, Berkeley, CA 94720-7300, USA}

\author[0000-0002-7567-4451]{Edward~J.~Wollack}\affiliation{NASA/Goddard Space Flight Center, Greenbelt, MD, USA 20771}


\correspondingauthor{Gerrit~S.~Farren}
\email{gsf29@cam.ac.uk}



\begin{abstract} 
We present tomographic measurements of structure growth using cross-correlations of Atacama Cosmology Telescope (ACT) DR6 and \textit{Planck} CMB lensing maps with the unWISE Blue and Green galaxy samples, which span the redshift ranges $0.2 \lesssim z \lesssim 1.1$ and $0.3 \lesssim z \lesssim 1.8$, respectively. We improve on prior unWISE cross-correlations not just by making use of the new, high-precision ACT DR6 lensing maps, but also by including additional spectroscopic data for redshift calibration and by analysing our measurements with a more flexible theoretical model. We determine the amplitude of matter fluctuations at low redshifts ($z\simeq 0.2-1.6$), finding $S_8 \equiv \sigma_8 (\Omega_m / 0.3)^{0.5} = 0.813 \pm 0.021$ using the ACT cross-correlation alone and $S_8 = 0.810 \pm 0.015$ with a combination of \textit{Planck} and ACT cross-correlations; these measurements are fully consistent with the predictions from primary CMB measurements assuming standard structure growth. The addition of Baryon Acoustic Oscillation data breaks the degeneracy between $\sigma_8$ and $\Omega_m$, allowing us to measure $\sigma_8 = 0.813 \pm 0.020$ from the cross-correlation of unWISE with ACT and $\sigma_8 = 0.813\pm 0.015$ from the combination of cross-correlations with ACT and \textit{Planck}. These results also agree with the expectations from primary CMB extrapolations in $\Lambda$CDM cosmology; the consistency of $\sigma_8$ derived from our two redshift samples at $z \sim 0.6$ and $1.1$ provides a further check of our cosmological model. Our results suggest that structure formation on linear scales is well described by $\Lambda$CDM even down to low redshifts $z\lesssim 1$.

\end{abstract}


\section{Introduction}\label{sec: intro}
Measuring the amplitude of low redshift matter fluctuations can probe the growth of cosmic structure over time, reveal the properties of dark matter and dark energy, constrain the masses of neutrinos, and provide important tests of general relativity. A number of lensing-related techniques have been developed for this purpose, including both galaxy lensing and CMB lensing.

Recent measurements of galaxy weak lensing from the Dark Energy Survey \citep[DES;][]{2015AJ....150..150F,2022PhRvD.105b3520A}, the Kilo-Degree Survey \cite[KiDS;][]{2015MNRAS.454.3500K,2021A+A...646A.140H}, Hyper Suprime-Cam \cite[HSC;][]{2018PASJ...70S...4A,2023PhRvD.108l3520M,2023PhRvD.108l3517M,2023PhRvD.108l3521S}, among others, have found a $2-3\sigma$ lower amplitude of fluctuations, compared to the prediction from \textit{Planck} primary CMB assuming standard structure growth. In particular, parameterising the low-redshift amplitude by $S_8 \equiv \sigma_8 (\Omega_m / 0.3)^{0.5}$ (which is the quantity best measured by galaxy weak lensing surveys), DES, KiDS, and HSC obtain $S_8 =0.782\pm 0.019$ \footnote{This value is obtained from the public DES parameter chain with fixed neutrino mass \href{http://desdr-server.ncsa.illinois.edu/despublic/y3a2_files/chains/chain_3x2pt_fixednu_lcdm.txt}{here}. We adopt this result throughout for better comparability with our analysis which fixes the neutrino mass to the minimum value allowed in the normal hierarchy.},  $S_8 = 0.765^{+0.017}_{-0.016}$ \footnote{This differs from the results reported by KiDS, $S_8=0.766^{+0.020}_{-0.014}$. KiDS by default reports the maximum a posteriori (MAP) value along with the projected joint highest posterior density region (PJ-HPD). For better comparability, however, we adopt throughout the marginalised mean and credible interval instead.}, $S_8=0.775^{+0.043}_{-0.038}$ \footnote{We note that the HSC results we compare to in Sec.\,\ref{subsec:comp_lss_surveys} differ slightly from these values as they are derived from a reanalysis of the HSC data with prior choices consistent with those adopted in this work.\label{footnote:HSC}} respectively from a combination of cosmic shear and galaxy clustering \citep{2022PhRvD.105b3520A,2021A+A...646A.140H,2023PhRvD.108l3521S}. HSC also presents an alternative analysis using smaller scales and a halo model based approach using a cosmological emulator (rather than the linear bias approximation used on large scales) finding $S_8=0.763^{+0.040}_{-0.036}$ \footnote{See footnote \ref{footnote:HSC}.} \citep{2023PhRvD.108l3517M}. A recent joint reanalysis of the DES and KiDS cosmic shear data hints at a slightly higher value of $S_8 = 0.790_{-0.014}^{+0.018}$ \citep{2023OJAp....6E..36D}. Comparing to the value $S_8= 0.834 \pm 0.016$ from the \textit{Planck} primary CMB \citep[or $S_8= 0.832 \pm 0.013$ when including CMB lensing;][]{2020A+A...641A...6P}, we see that most recent galaxy weak lensing surveys have a preference for slightly lower $S_8$, even though no single experiment is in significant tension with \textit{Planck}. The recent DES + KiDS reanalysis of cosmic shear is consistent at the 1.7$\sigma$ level. 

CMB lensing probes the same physics as galaxy weak lensing, but has different potential systematics; a known redshift of the ``source'' (the primary CMB) and well-understood statistical properties make it a particularly reliable and independent cosmological probe. The CMB lensing auto-correlation is in excellent agreement with the primary CMB predictions: recent measurements from the Atacama Cosmology Telescope \citep[ACT][]{2024ApJ...962..112Q, 2024ApJ...962..113M} combined with Baryon Acoustic Oscillation (BAO) data yield $S_8 = 0.840 \pm 0.028$ from ACT alone and $S_8 = 0.831 \pm 0.023$ when combined with \textit{Planck} lensing \citep{2020A+A...641A...8P, 2022JCAP...09..039C}. 

However, the cross-correlation between CMB lensing and galaxies has at times shown a preference for a lower $S_8$. For example, cross-correlations of DESI Luminous Red Galaxy (LRG) targets \citep{2022JCAP...02..007W, 2021MNRAS.501.6181K, 2021MNRAS.501.1481H} or galaxies from the Baryon Oscillation Spectroscopic Survey (BOSS) \citep{2020MNRAS.491...51S, 2022JCAP...07..041C} with \textit{Planck} lensing show a preference for a lower $S_8$ at up to $3\sigma$. An analysis of the cross-correlation between CMB lensing from the South Pole Telescope (SPT) and \textit{Planck} and various data sets from DES finds varying levels of tension between $2.2$ and $3\sigma$ \citep[see Sec.\,\ref{subsec:comp_lss_surveys} for details;][]{2023PhRvD.107b3530C,2023PhRvD.107b3531A}. Cross-correlations with previous CMB lensing reconstructions from ACT have also found this preference at the $\sim 2\sigma$ level, albeit with substantial uncertainties. \editF{Using CMB lensing data from ACT DR4 and \textit{Planck} together with galaxy shear from KiDS-1000, \cite{2021A+A...649A.146R} found $S_8=0.64\pm 0.08$; similarly an analysis of the cross-correlation between ACT DR4 CMB lensing and galaxy clustering from DES-Y3 yields $S_8=0.75^{+0.04}_{-0.05}$ \citep{2024JCAP...01..033M}.}

Moreover, \cite{2021JCAP...12..028K} used the unWISE full-sky galaxy catalogue split into three tomographic bins spanning the redshift range $0 \lesssim z \lesssim 2$ \citep[with the power spectra and redshift distributions measured in][]{2020JCAP...05..047K}, together with the \textit{Planck} 2018 CMB lensing maps \citep{2020A+A...641A...8P}, to set a constraint $S_8 = 0.784 \pm 0.015$, about 2.4$\sigma$ lower than estimated from the primary CMB. 
Motivated by the availability of additional data for the unWISE redshift distribution calibration, improved understanding of the calibration of \textit{Planck} CMB lensing maps and the availability of lower-noise CMB lensing maps from \textit{Planck} PR4 and ACT DR6, we perform a new and improved analysis of the cross-correlation between unWISE galaxies and CMB lensing. An expanded set of null-tests, enhanced foreground control and improved theoretical modelling add to the robustness of the measurement.

This paper is structured as follows, in Sec.\,\ref{sec:key_results} we briefly summarise the key results of this work before introducing the data sets used in this work in Sec.\,\ref{sec:data}. In Sec.\,\ref{sec:tomography} we present the measurements of the galaxy auto-correlation power spectra and the galaxy-CMB lensing cross-power spectra. In Sections \ref{sec:sims} and \ref{sec:covmat} we introduce the simulations we use in this work and describe how we estimate the data covariance. Sec.\,\ref{sec:sys_test} focuses on various test for systematic errors. In Sec.\,\ref{sec:model} we describe the theoretical model that we use for parameter inference. The parameter inference pipeline and the results of our analysis are presented in Sec.\,\ref{sec:cosmo}. Finally, we compare our results to other results from the literature in Sec.\,\ref{sec:discussion} and discuss the implications of our findings for the ``$S_8$ tension". 

\section{Summary of key results}\label{sec:key_results}

The key results of this work are constraints on structure formation at $z \simeq 0.2$--$1.6$. Within a $\Lambda$CDM cosmology the best constrained parameter combination in the analysis presented here is approximately $\sigma_8 \Omega_m^{0.45}$. This differs slightly from the best constrained parameter combination in galaxy weak lensing analyses, $\sigma_8 \Omega_m^{0.5}$. To facilitate comparisons with other probes we mainly present constraints on the commonly used parameter $S_8 \equiv \sigma_8 (\Omega_m/0.3)^{0.5}$ \citep{2022PhRvD.105b3520A,2021A+A...646A.140H,2023PhRvD.108l3518L,2023PhRvD.108l3519D} in this paper; however, we also define a cross-correlation equivalent $S_8^\times \equiv \sigma_8 (\Omega_m/0.3)^{0.45}$. CMB lensing auto-spectrum analyses, in contrast, primarily probe the combination $S_8^{\rm {CMBL}} \equiv \sigma_8 (\Omega_m/0.3)^{0.25}$ \citep{2024ApJ...962..112Q,2024ApJ...962..113M,2022JCAP...09..039C,2020A+A...641A...8P}.  In this paper, we also combine with BAO measurements that probe $\Omega_m$ independently of $\sigma_8$ to obtain constraints on $\sigma_8$ that are directly comparable to other probes.

\begin{figure*}
    \includegraphics[width=\linewidth]{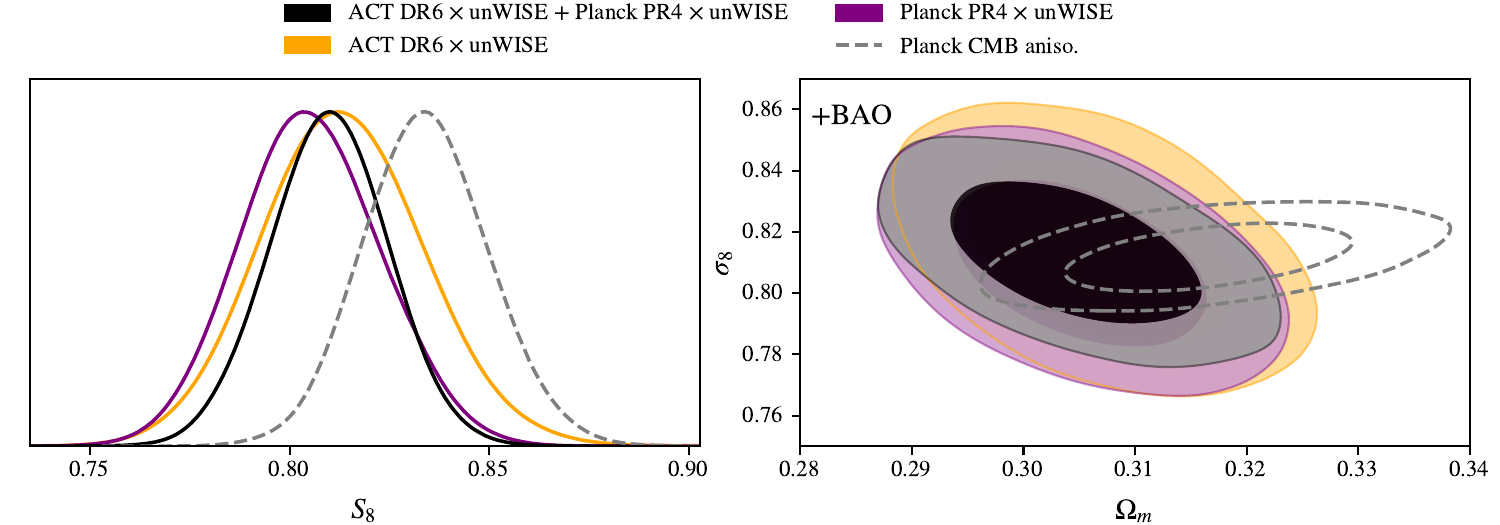}%
    \caption{The main result from this work is a constraint on the amplitude of low redshift structure captured by the parameter $S_8\equiv \sigma_8 (\Omega_m/0.3)^{0.5}$ (\textbf{left}). We obtain $S_8=0.810\pm0.015$ using the combination of cross-correlation measurements of the unWISE galaxies with ACT DR6 and \textit{Planck} PR4 lensing reconstructions. Combining with baryon acoustic oscillations (BAO), which constrain $\Omega_m$, we obtain constraints on $\sigma_8$ of $\sigma_8 = 0.813 \pm 0.015$ (\textbf{right}). Our results show no significant tension with values inferred from the primary CMB from \textit{Planck}.\label{fig:key_results}}
\end{figure*}

When using only the cross-correlation between the ACT DR6 CMB lensing reconstruction and the two unWISE galaxy samples, Blue and Green \citep[previously defined in][]{2020JCAP...05..047K}, as well as the auto-correlation of the galaxy samples measured on the overlapping area, we find $S_8=0.813\pm 0.021$ ($S^\times_8=0.817\pm 0.019$). \resub{We combine this data with an equivalent cross-correlation analysis using the same unWISE galaxy samples and a lensing reconstruction from \textit{Planck} PR4 \citep{2022JCAP...09..039C} while taking into account the relevant covariance matrix and obtain improved constraints of $S_8=0.810\pm0.015$ ($S^\times_8=0.814\pm 0.014$).} The posterior distributions for $S_8$ are shown in the left panel of Fig.~\ref{fig:key_results}. With the addition of BAO we find $\sigma_8 = 0.813 \pm 0.020$ from unWISE and ACT alone and $\sigma_8 = 0.813 \pm 0.015$ when combining with \textit{Planck} lensing (see the right panel of Fig.\,\ref{fig:key_results}). 

\begin{figure*}
    \includegraphics[width=\linewidth,trim=1.2cm 1.5cm 1.2cm 0.2cm, clip]{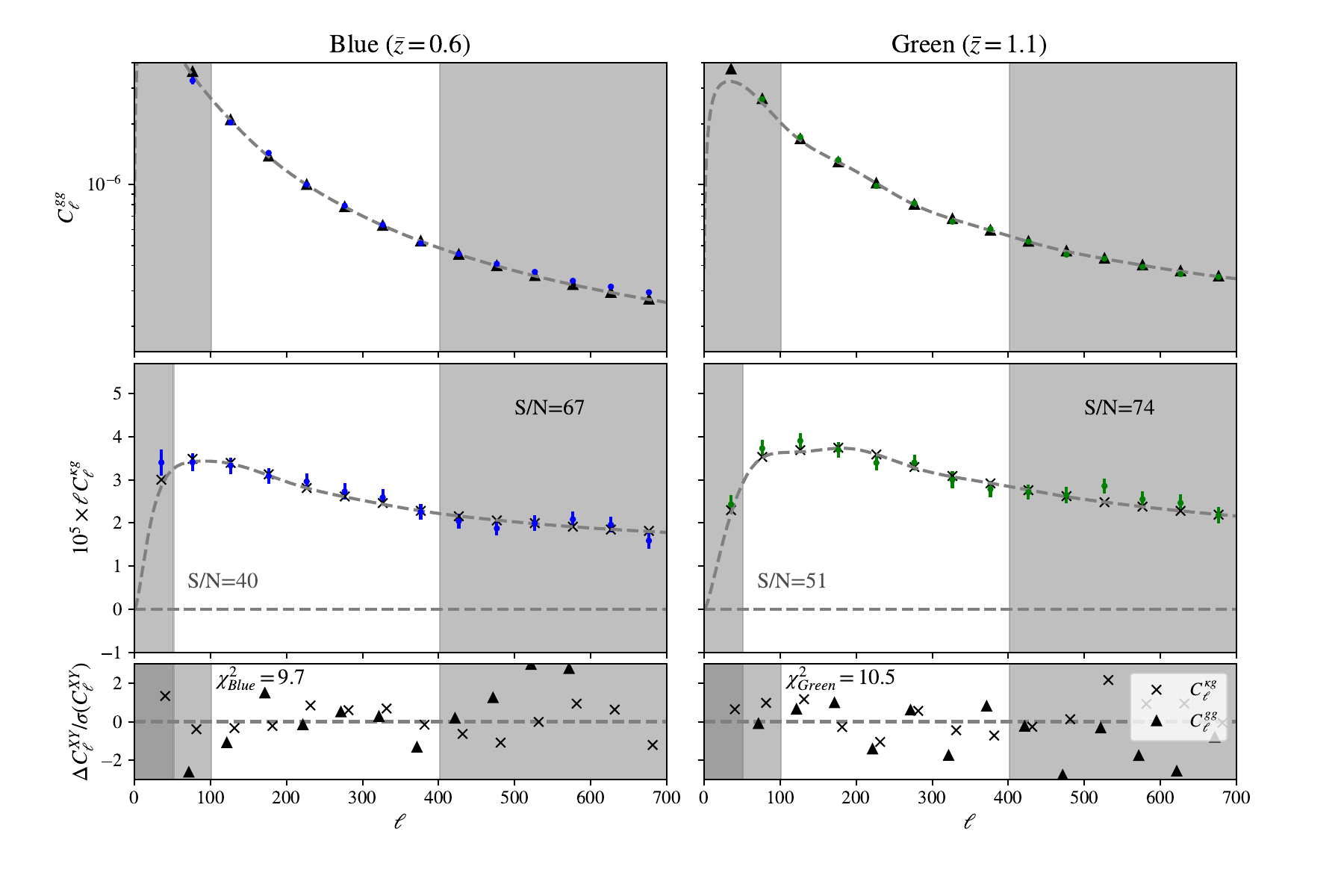}
    \caption{We measure $C_\ell^{gg}$ (\textbf{top} row) and $C_\ell^{\kappa g}$ (\textbf{middle} row) for the Blue and Green samples of unWISE galaxies. The total signal to noise in $C_\ell^{\kappa g}$ on all measured scales ($20\leq\ell\leq3000$) is approximately 67 and 74 respectively. Within the analysis range ($50\leq \ell \leq 400$) we obtain SNR of 40 and 51 for Blue and Green respectively. The grey line shows the best fit from the joint fit to both samples, with the model residuals shown in the \textbf{bottom} row. The total model $\chi^2$ for the joint fit is $21$, slightly larger than the sum of the $\chi^2$ for each of the two samples (9.7 and 10.5 for Blue and Green respectively) due to the non-zero off-diagonal covariance between them. We estimate the model PTE at $0.5$.}
    \label{fig:spectra}
\end{figure*}

The model best fitting the measured $C_\ell^{gg}$ and $C_\ell^{\kappa g}$ is shown in Fig.\,\ref{fig:spectra}. We find a minimum $\chi^2=20.9$. Our model, described in detail in Sec.\,\ref{sec:model}, includes 20 model parameters. Of those 15 are largely prior dominated yielding an approximate number of degrees of freedom of $21$ (there are a total of 26 bandpowers for the Blue and Green samples combined). Therefore, we estimate the probability to exceed (PTE) for the model at $0.5$.

These conclusions differ from previous results from the cross-correlation of unWISE with CMB lensing reconstruction from \textit{Planck} PR3 \citep{2020A+A...641A...8P} presented in \cite{2021JCAP...12..028K}. In Appendix~\ref{app:planck_reanalysis} we present a reanalysis of the \textit{Planck} cross-correlation, finding results consistent with those presented here. We discuss the corrections and improvements that lead to the discrepancy. Other cross-correlation measurements still exhibit a preference for lower $S_8$ \citep{2020MNRAS.491...51S,2021MNRAS.501.6181K,2022JCAP...02..007W,2022JCAP...07..041C,2023PhRvD.107b3530C,2023PhRvD.107b3531A,2021A+A...649A.146R,2024JCAP...01..033M} and while some of the improvements made here (e.g., the inclusion of a Monte Carlo correction for the lensing normalisation) are also relevant to those works, preliminary results show that they are likely insufficient to alleviate the discrepancy completely \citep{Kim2024,Sailer2024}.

Our results are in good agreement with the predictions for the amplitude of low redshift structure obtained from the primary CMB observed by the \textit{Planck} satellite \citep[within $1.1\sigma$ for $S_8$ and $0.1\sigma$ for $\sigma_8$;][]{2020A+A...641A...6P} as well as measurements from CMB lensing using either ACT \citep[$0.3\sigma$ for $\sigma_8$;][]{2024ApJ...962..112Q,2024ApJ...962..113M} or \textit{Planck} \citep[$0.04\sigma$ for $\sigma_8$;][]{2022JCAP...09..039C,2020A+A...641A...8P}. On the other hand, our results are also not in significant tension with recent results from any one galaxy weak lensing survey such as DES, KiDS, HSC \citep[$1.1\sigma$, $2.0\sigma$, $0.8\sigma$ respectively for $S_8$;][]{2022PhRvD.105b3520A,2021A+A...646A.140H,2023PhRvD.108l3521S}, although our results favour a higher amplitude of low redshift structure. Overall, our results do not strengthen the case for a real discrepancy in structure growth or for problems with the $\Lambda$CDM model, at least in the low-redshift, linear-scale regime to which we are sensitive.

\section{The data}
\label{sec:data}
We use the unWISE galaxy catalogue \citep{2020JCAP...05..047K,2019ApJS..240...30S} and lensing reconstructions from the anticipated Data Release 6 (DR6) of ACT \citep{2024ApJ...962..112Q, 2024ApJ...962..113M, 2024ApJ...966..138M}. We also combine our cross-correlation analysis (results given in Sec.~\ref{subsec:cosmo}) with the cross-correlation obtained using the latest CMB lensing reconstruction based on data from the \textit{Planck} satellite \citep{2022JCAP...09..039C}.

In Sec.\,\ref{subsec:unwise_gals} we briefly describe the unWISE samples, their selection, the measurement of their redshift distribution, and the mitigation of observational systematics. Subsequently, we describe the ACT DR6 lensing reconstruction in Sec.\,\ref{subsec:act_lensing} and the \textit{Planck} PR4 lensing reconstruction in Sec.\,\ref{subsec:planck_lensing}. 

\subsection{unWISE galaxies}\label{subsec:unwise_gals}
The unWISE galaxy catalogue is constructed from the Wide-Field Infrared Survey Explorer (WISE) survey \citep{2010AJ....140.1868W}, including four years of the post-hibernation NEOWISE phase \citep{2011ApJ...731...53M,2014ApJ...792...30M}. The WISE satellite mapped the entire sky at 3.4 (W1), 4.6 (W2), 12 (W3) and 22 (W4) $\mu$m, although NEOWISE only measured in bands W1 and W2 due to a lack of cryogen necessary for the longer-wavelength bands. As a result, unWISE \citep{2014AJ....147..108L,2017AJ....153...38M} is constructed only from the much deeper W1 and W2 bands.

\subsubsection{Galaxy Selection} 
\label{subsubsec:unwise_sel}
We select three galaxy samples from unWISE using W1-W2 colour cuts, called the Blue, Green, and Red samples, at $z \sim 0.6$, 1.1, and 1.5, respectively. These samples are extensively described in \cite{2020JCAP...05..047K} and \cite{2019ApJS..240...30S}. The Red sample of galaxies has a significantly lower number density than the other two samples and was shown to contribute negligible constraining power \citep{2021JCAP...12..028K}. \cite{2021JCAP...12..028K} also showed that it was difficult to reproduce the red sample in $N$-body simulations and to recover unbiased cosmology from simulations tuned to reproduce the properties of the red sample. Thus we do not use the Red sample. 

In addition to the unWISE colour cuts, we additionally remove any unWISE source within 2.75$\arcsecond$ of a Gaia \citep{2018A+A...616A...1G} point source,\footnote{2.75$\arcsecond$ is the size of a WISE pixel, and we use it as the match radius when cross-matching unWISE and Gaia. Relative to the 6$\arcsecond$ WISE PSF, this is similar to the 0.5$\arcsecond$ match radius often used between optical catalogs with 1$\arcsecond$ seeing.} and remove potentially spurious sources in unWISE imaging, as described in \cite{2020JCAP...05..047K}. Residual stellar contamination is $<2\%$.
As pointed out in \cite{2020JCAP...05..047K}, the power spectrum of stars
drops very rapidly with $\ell$, such that even at $\ell \sim 50$, 
the stellar power spectrum multiplied by the contamination fraction 
is $< 0.5\%$ of the galaxy auto power spectrum, compared to the 2\% statistical errors of the galaxy auto-spectrum at that bandpower. In the limit that the stellar power spectrum contributes negligible power, stellar
contamination simply modifies the number density in a way that is completely
degenerate with the linear galaxy bias.

The unWISE mask is based on the \textit{Planck} 2018 CMB lensing mask \citep{2020A+A...641A...8P} with a Gaussian apodisation of 1$^\circ$, and additional cuts around bright stars, nearby galaxies, planetary nebulae, and optical diffraction spikes. Because we remove any source within 2.75$\arcsecond$ of a Gaia point source and mask out diffraction spikes and other narrow regions around stars, the effective area within each HEALPix pixel is less than the full area of the pixel. We correct the counts in each pixel for the effective area (and mask out any pixel with $<80\%$ coverage), before applying the unWISE mask. The large-scale component of the unWISE mask, derived from the \textit{Planck} lensing mask, is apodised by smoothing with a 1$^{\circ}$ FWHM Gaussian; the other components of the mask are treated as binary. The effective sky fraction after masking is $f_{\rm sky} = 0.586$.

\subsubsection{Galaxy Redshift Distributions}\label{subsubsec:dndz}

Since the unWISE galaxies are selected from two-band photometry, it is not possible to determine photometric redshifts for individual galaxies. Instead, we use cross-correlations with Sloan Digital Sky Survey (SDSS) spectroscopy to measure the redshift distributions of the samples \citep[e.g.,][]{2008ApJ...684...88N,2013MNRAS.433.2857M,2013arXiv1303.4722M}.  
\resub{The cross-correlation between a photometric sample with unknown redshift distribution and a spectroscopic sample is proportional to the biases of the two galaxy samples and the overlap in redshift distributions. Therefore, by repeating the cross-correlation measurement over a large number of narrowly-spaced spectroscopic bins and measuring spectroscopic sample's bias in each bin, we can measure the product of the photometric sample's bias evolution and its redshift distribution. The formalism is described in Appendix~\ref{app:fid_cosmo_correction} and Section 5.2 of \cite{2020JCAP...05..047K}.}

The unWISE cross-correlation redshifts were originally presented in \cite{2020JCAP...05..047K} using spectroscopic samples from LOWZ and CMASS galaxies \citep{2016MNRAS.455.1553R}, BOSS quasars \citep{2020MNRAS.498.2354R}, and eBOSS (extended BOSS) DR14 quasars \citep{2018MNRAS.473.4773A}. \cite{2020JCAP...05..047K} only considered objects in the Northern Galactic Cap (NGC) where most of the spectroscopy lies. These tracers span the full redshift range of the unWISE samples. 

We update the unWISE cross-correlation redshifts first measured in \cite{2020JCAP...05..047K} to include additional data. We use the Southern Galactic Cap (SGC) footprints for all tracers, due to the better overlap with the mostly-southern ACT footprint. We replace the eBOSS DR14 quasars with the final eBOSS DR16 quasars \citep{2020MNRAS.498.2354R}, leading to a significant increase in the area (1178 to 4752 deg$^2$) and number of quasars used (54708 to 343708). Finally, we also use the eBOSS DR16 LRGs \citep{2020MNRAS.498.2354R}. We do not use the eBOSS DR16 emission line galaxies (ELGs), due to the significantly smaller area compared to the LRGs or quasars (1120 deg$^2$ vs. 4202 deg$^2$ for LRG and 4808 deg$^2$ for quasars) and potential systematics in the ELG autocorrelation needed to measure the ELGs' spectroscopic bias. These additions significantly improve the cross-correlation redshifts at $z \sim 0.7$--2, the redshift range where most of the unWISE galaxies lie. In Appendix~\ref{app:spec_samples} we describe the spectroscopic samples used and their relevant properties (linear and magnification biases), then discuss the impact of the additional data used.

Conveniently, cross-correlation redshifts are sensitive to the product of the galaxy bias and the number density, $b(z) dN/dz$, that appears in the dominant terms in our model (Section~\ref{subsec:pk_model}). The conversion from the measured correlation function to $b(z) dN/dz$ leads to a dependence on a fiducial cosmology, although we note that there is no dependence on the amplitude of the power spectrum because we require that $b(z) dN/dz$ is normalised to integrate to unity. The residual cosmology dependence is quite minor, and the correction is described in detail in Appendix~\ref{app:fid_cosmo_correction} \citep[updating the heuristic correction presented in][]{2021JCAP...12..028K}. Although the spectra we measure primarily depend on $b(z) dN/dz$, the lensing magnification and some higher-order galaxy bias terms also depend on $dN/dz$, which we measure by cross-matching unWISE galaxies to deep photometric redshifts from the COSMOS2015 catalogue \citep{2016ApJS..224...24L}. This redshift distribution is consistent with the cross-correlation $b(z) dN/dz$ assuming a simple halo occupation distribution (HOD), consistent with the unWISE number density,  to determine the bias evolution $b(z)$. From the COSMOS cross-match, we find no evidence for unWISE galaxies to lie at $z > 1.7$ (2.5) for Blue (Green) and therefore set the smoothed redshift distribution to zero above these high redshift thresholds.

\begin{figure}
    \centering
    \includegraphics[width=\linewidth]{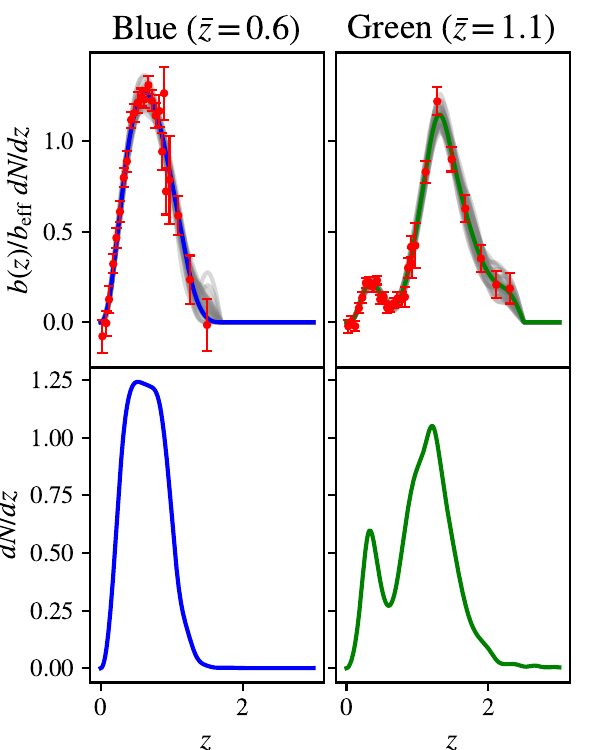}
    \caption{The Blue and Green samples of unWISE galaxies span the redshift ranges $z\simeq0.2-1.1$ and $z\simeq0.3-1.8$ with mean redshifts of approximately $0.6$ and $1.1$ respectively. In the \textbf{top} panel we show the normalised estimate of $b(z) dN/dz$ obtained from cross-correlating with spectroscopic tracers from SDSS, BOSS, and eBOSS. The Blue and Green curves are the spline interpolations of the best fitting estimates of $b(z) dN/dz$ for the two galaxy samples. We also show several noise realisations in grey and the clustering redshift measurements in red. The \textbf{lower} panel shows estimates of $dN/dz$ obtained by cross-matching with photometric data from COSMOS2015.}
    \label{fig:dndz}
\end{figure}

The cross-correlation estimates of $b(z)dN/dz$ are shown in the top panel in Fig.\,\ref{fig:dndz}. We normalise the cross-correlation redshift estimates (for more details see Sec.\,\ref{subsec:pk_model}). We also show in the lower panel the cross-match $dN/dz$ from COSMOS2015.

\resub{The mean bias-weighted redshift is $0.697 \pm 0.023$ (Blue) and $1.355 \pm 0.022$ (Green), with uncertainties estimated from the samples drawn according to the cross-correlation redshift error estimates. The median and uncertainty in the cross-match redshifts were estimated in Table 2 of \cite{2020JCAP...05..047K}, $0.63 \pm 0.022$ (Blue) and $1.09 \pm 0.019$ (Green); the significant difference between the bias weighted and unweighted redshift distributions for green is due to the strong bias evolution consistent with the simple HOD described above. The uncertainties were estimated by cross-matching to wider-area HSC DR1 photometric redshifts \citep{2018PASJ...70S...8A,2018PASJ...70S...9T} at $z_{\textrm{phot}} < 1$ and extrapolating the additional variance relative to Poisson to $z_{\textrm{phot}} > 1$ (where the HSC redshifts become unreliable).}

\subsubsection{Removing unWISE correlations with stellar density and WISE depth}\label{subsec:sys_weights}

In addition to cosmological fluctuations, the observed galaxy number densities are determined by foregrounds such as Galactic dust and stars, as well as survey depth and other imaging properties. Both foregrounds and imaging properties affect the galaxy selection. The power spectra of these contaminants are generally red; if uncorrected they therefore add significant power to the galaxy auto-correlation at low $\ell$. The standard approach for galaxy surveys is to create a set of weights that are designed to remove any relationship between the galaxy overdensity and various imaging property maps  \citep{2012MNRAS.424..564R,2017MNRAS.464.1168R,2018MNRAS.473.4773A,2018ApJ...863..110B,2018PhRvD..98d2006E,2020MNRAS.498.2354R,2022MNRAS.511.2665R}. \cite{2021JCAP...12..028K} did not follow this approach; instead, the authors applied a high-pass filter to the galaxy data, removing all modes of the galaxy survey at $\ell < 20$, and found that this led to better agreement between the $\ell < 100$ auto-correlation and a theory model. Additionally, \cite{2021JCAP...12..028K} did not use the galaxy auto-correlation at $\ell < 100$, where changing the Galactic mask significantly changed the shape of the unWISE auto-correlation\footnote{At $\ell \gtrsim 100$, changing the Galactic (or ecliptic latitude) mask produced a scale-independent change in the unWISE auto-correlation \citep{2020JCAP...05..047K}. This is due to the fact that the selection properties of the galaxy catalogue vary with Galactic or ecliptic latitude due to variations in the WISE coverage depth. This induces differences in the galaxy bias. We return to this in Sec.\,\ref{subsec:galaxy_sys}.}.

In this work, we update the method used in \cite{2021JCAP...12..028K} to apply weights that explicitly remove correlations between the galaxy density and maps of stellar density and WISE depth. This is similar to the approach taken by other galaxy surveys and ensures that our unWISE galaxy maps are uncorrelated with known foreground survey systematics that may affect the autocorrelation at $\ell > 100$. These weights were originally created in \cite{2022JCAP...04..033K} to use the low-$\ell$ unWISE data in cross-correlation with CMB temperature to measure the integrated Sachs-Wolfe effect. We also no longer filter out the low-$\ell$ ($\ell < 20$) modes in the unWISE map. The large-scale filtering has a similar effect to weighting, also reducing large-scale power by removing correlations between systematics and the true galaxy density. However, removing large scales in harmonic space complicates the use of the \textsc{MASTER} algorithm \cite{2002ApJ...567....2H} to obtain unbiased bandpowers through mode decoupling. Hence, we no longer adopt this method. This change has only a small impact on the power spectra ($<0.5\%$ on the auto- and cross-correlation \resub{corresponding to $<0.3\sigma$ in terms of the bandpower uncertainties}; see Fig.\,\ref{fig:weights_impact}), but we consider the updated results more robust because the galaxy density has a significantly reduced dependence on both Galactic stellar density and WISE depth. 

\editF{In the remainder of this section we provide more detail on the construction of these weights.} We measure the correlation between unWISE galaxy density and several templates: 1) Gaia stellar density; 2\&3) W1 and W2 limiting magnitude; 4) dust extinction $E(B-V)$ from the Schlegel-Finkbeiner-Davis map corrected for cosmic infrared background (CIB) contamination \citep[cSFD,][]{1998ApJ...500..525S,2023ApJ...958..118C}\footnote{The relationship between unWISE galaxy density and cSFD is nearly identical if we use the uncorrected SFD instead.}; 5) neutral hydrogen column density NHI from the H14PI survey \citep{2016A+A...594A.116H} as an alternative dust map that is noisier than SFD but has much reduced extragalactic contamination \citep{2019ApJ...870..120C}; 6\&7) a 3.5 and 4.9 $\mu$m sky brightness from the DIRBE Zodi-Subtracted Mission Average (ZSMA)\footnote{\url{https://lambda.gsfc.nasa.gov/product/cobe/dirbe_zsma_data_get.cfm}}; 8) a DIRBE measurement of the total 4.9 $\mu$m background light  at solar elongation of 90$^{\circ}$\footnote{\url{https://lambda.gsfc.nasa.gov/product/cobe/c_90deg_skymap.html}} \citep[WISE always observed at 90$^{\circ}$ solar elongation;][]{2011ApJ...731...53M}; and 9) a separate model of the zodiacal background light at 4.9$\mu$mu from the DIRBE Sky and Zodi Atlas (DSZA)\footnote{\url{https://lambda.gsfc.nasa.gov/product/cobe/dirbe_dsza_data_get.cfm}} \citep{1998ApJ...508...44K}.

We find that the strongest correlations are with Gaia stellar density and W2 limiting magnitude. To conservatively guard against the possibility of overfitting \citep{2021MNRAS.503.5061W}, we only regress against these two maps to define the systematics weights. Several other correlations are also quite significant (particularly with W1 limiting magnitude), but are significantly mitigated after weighting against stellar density and W2 depth, due to correlations between these contaminant templates. For example, both W1 and W2 depths are set by the WISE scan strategy, so the two maps exhibit the same structure.

For the Blue sample, we fit a linear trend between unWISE galaxy density and Gaia stellar density, and a three parameter piecewise linear trend to the W2 5$\sigma$ limiting magnitude. We determine uncertainties on the measured correlations using the variance of density values from 100 isotropic Gaussian mocks (no correlation with the maps of imaging systematics). For the Green sample, we use a piecewise linear fit with three free parameters for both stellar density and W2 5$\sigma$ magnitude. We show the relationship between unWISE density and imaging systematics in Appendix~\ref{app:sys_weights} (Figs.\,\ref{fig:imaging_systematics_blue} and~\ref{fig:imaging_systematics_green}), both before and after applying weights. Before applying weights, the correlations between the galaxy density and all nine templates listed above are highly significant; after weighting, induced variations in the galaxy number density are reduced to $\sim$2\%. While the $\chi^2$ for all correlations is significantly reduced, it is still indicative of statistically significant correlations in some cases. However, the covariance estimate is approximate (for instance, the errors are assumed to be diagonal) and potentially is an underestimate. Even the dramatic reduction in $\chi^2$ after applying weights to correct the correlations with stellar density and W2 limiting magnitude leads to only a minor impact on the auto-spectrum ($<3\%$ on scales $\ell>100$ \resub{corresponding to 0.5-1$\sigma$ in terms of the bandpower uncertainties within our analysis range}) and does not impact the cross-correlation spectrum significantly. We conclude that any residual correction will be negligible given the size of our statistical errors. The impact of applying these weights on our parameter inference is $\Delta S_8=0.006$ ($\sim0.3\sigma$)\footnote{This includes the small change in the measured clustering redshifts due to the application of the weights (see Sec.\,\ref{app:spec_samples}). The mean redshift of the Blue and Green samples changes by $\Delta \bar{z}=0.007$ and $\Delta \bar{z}=-0.002$ respectively.}.


\subsection{CMB Lensing from the Atacama Cosmology Telescope} \label{subsec:act_lensing}
We use the CMB lensing convergence map reconstructed from the CMB temperature and polarisation anisotropy data from the upcoming DR6 of ACT \citep{2024ApJ...962..112Q,2024ApJ...962..113M,2024ApJ...966..138M}. This release is based on CMB measurements made between 2017 and 2021 (relying only on the night-time data) at $\sim$90 and $\sim$150 GHz. It uses an early version of the ACT DR6 maps, labelled \texttt{dr6.01}.

The lensing maps cover $9400\,\rm{deg}^2$ of the sky and are signal dominated on scales of $\ell<150$. These maps are reconstructed using a cross correlation based estimator \citep{2020arXiv201102475M}. Rather than using a single CMB map, the cross correlation estimator uses several time-interleaved splits ensuring that the instrumental noise of each map is independent. The resulting CMB lensing map is therefore insensitive to modelling of the instrumental noise.

Lensing is reconstructed with CMB scales from $600<\ell<3000$. The large scales of the input CMB map, $\ell<600$, are excluded due to significant atmospheric noise, Galactic foregrounds and a $>10\%$ correction for the large scale transfer functions. The small scales are excluded to minimise contamination from astrophysical foregrounds like the thermal Sunyaev-Zeldovich (tSZ) effect, the cosmic infrared background (CIB), and radio sources \cite{2024ApJ...966..138M}. To further mitigate extragalactic foregrounds, the lensing reconstruction uses a profile-hardened lensing estimator \citep{2020PhRvD.102f3517S}. This involves constructing a quadratic estimator that is immune to the contribution to the CMB mode coupling arising from objects with radial profiles similar to those expected from tSZ clusters.

\editF{The baseline ACT DR6 lensing mask is constructed from a Galactic mask which selects the 60\% of the sky with the lowest dust contamination \citep[see][for a detailed discussion of the ACT lensing mask]{2024ApJ...962..112Q}. As a consistency test we also use a ACT lensing reconstruction using only the area within the 40\% of the sky with the lowest dust contamination (see Sec.\,\ref{sec:sys_test}). We will subsequently refer to these masks as the 60\% and 40\% Galactic masks respectively.}

\subsection{CMB Lensing from \textit{Planck}} \label{subsec:planck_lensing}

The \textit{Planck} PR4 lensing analysis \citep{2022JCAP...09..039C} reconstructs lensing with CMB angular scales from $100\leq\ell\leq2048$ using the quadratic estimator. This analysis is based on the reprocessed PR4 \texttt{NPIPE} maps that incorporated around $8\%$ more data compared to the 2018 \textit{Planck} PR3 release. It also includes pipeline improvements such as optimal (anisotropic) filtering of the input CMB fields resulting in an increase of the overall signal-to-noise ratio by around $20\%$ compared to  \textit{Planck} PR3 \citep{2020A+A...641A...8P} and a detection of the lensing power spectrum  at $42\sigma$.

\cite{2024ApJ...962..112Q} demonstrated good consistency between the ACT and \textit{Planck} PR4 lensing bandpowers justifying combining the measurements at the likelihood level to obtain tighter constraints. Similar to the analysis in \cite{2024ApJ...962..112Q} and \cite{2024ApJ...962..113M}, we combine the cross correlation of ACT DR6 CMB lensing and unWISE with the cross correlation measurement between \textit{Planck} PR4 CMB lensing and unWISE at the likelihood level \resub{taking into account the full relevant covariance matrix. The simulated ACT-\textit{Planck} covariance matrix (see Sec.~\ref{sec:covmat} for details) accounts for the use of identical galaxy samples, the partially overlapping areas, and the covariance between the ACT and \textit{Planck} lensing reconstructions.}

\section{CMB Lensing Tomography}\label{sec:tomography}

We measure the auto-spectra of the two galaxy samples described in Sec.\,\ref{subsubsec:unwise_sel} and their cross-correlation spectra with the lensing map from ACT described in \ref{subsec:act_lensing} on the cut sky using a pseudo-$C_\ell$ estimator \citep{2002ApJ...567....2H}. 

Given two fields in the sky $a(\bm{\theta})$ and $b(\bm{\theta})$ a simple estimator of the pseudo-$C_\ell$ is
\beq
\tilde{C}_\ell^{ab} = \frac{1}{2\ell+1} \sum_{m=-\ell}^\ell a_{\ell m} (b_{\ell m})^\dagger,
\eeq
where $a_{\ell m}$  and $b_{\ell m}$ are the spherical harmonic transforms (SHT) of the fields $a$ and $b$. These pseudo-$C_\ell$ differ from the true $C_\ell$ due to mask induced mode-coupling. Their expectation value $\av{\tilde{C}_\ell}$ can be related to the true $C_\ell$ as
\beq
\av{\tilde{C}_\ell} = \sum_{\ell'} C_{\ell'} M_{\ell \ell'} \label{eq:mode-coupling}
\eeq
where $M_{\ell \ell'}$ is a mode coupling matrix which can be computed from the power spectrum of the mask alone. This relation can be approximately inverted if the power spectrum is assumed to be piecewise constant across a number of discrete bins \citep{2002ApJ...567....2H,2019MNRAS.484.4127A}. To perform the mode-decoupling of the binned $C_\ell$ we use the implementation in the \texttt{NaMaster} code\footnote{\url{https://github.com/LSSTDESC/NaMaster}} \citep{2019MNRAS.484.4127A}. It should be noted that lensing reconstruction with a quadratic estimator, when performed on the masked sky, effectively convolves the mask with the signal in a non-trivial manner that is not captured exactly by the \texttt{NaMaster} algorithm. We approximate this added complexity by taking the mask used to compute the mode-coupling matrix to be the square of the ACT mask\footnote{This is a good approximation (at least when the mask varies on much larger scales than the CMB and lensing scales of interest) since the lensing signal is reconstructed using a quadratic estimator which reconstructs the lensing from the off-diagonal correlations in the temperature and polarisation maps which each carry one power of the mask.}. Our tests on simulations (see Sec.~\ref{sec:sims}) suggest that this approximation performs to within better than 1\% on the scales of interest. Small residuals are corrected using a transfer function as described in section Sec.\,\ref{sec:sims}. We use HEALPix maps with $\texttt{nside}=2048$ and run \texttt{NaMaster} with $\ell_{\rm{max}}^{\texttt{NaMaster}}=3000$ even though we use only multipoles $\ell \leq 400$ in our analysis to avoid bias from the pseudo-$C_\ell$ method.

We measure the cross-correlation between the unWISE galaxies and the ACT DR6 lensing reconstruction (see Fig.\,\ref{fig:spectra}) with a signal-to-noise ratio of 40 and 51 within our cosmological analysis range of $50\leq\ell\leq400$ for the two galaxy samples, Blue and Green, respectively. This range was set prior to unblinding. If we were able to reliably model smaller scales we would be able to leverage even larger signal-to-noise, increasing to 67 and 74 respectively for $20\leq \ell \leq 3000$. 

The minimum multipole $\ell_{\rm min}$ is chosen based on our systematics tests presented in Sec.\,\ref{sec:sys_test} to guard against contamination from large scale observational systematics that may be correlated between the galaxy sample and the lensing reconstruction (as well as against potential mis-estimation of the lensing reconstruction mean field, which could lead to an underestimate of errors). The maximum multipole scale cut, $\ell_{\rm max}$, is predominantly set by the requirement for unbiased recovery of cosmological parameters, which becomes challenging with our chosen model on smaller scales (see Sec.\,\ref{subsec:model_test}). For the galaxy auto-correlation, $C_\ell^{gg}$, we adopt a more conservative minimum scale cut, $\ell_{\rm min}=100$, because the auto-correlation is more susceptible to observational systematics than the cross-correlation. On larger scales we observe significant fluctuations of the auto-correlation signal, for example for different choices of the mask \citep[see Sec.\,\ref{subsec:galaxy_sys} and also Sec.\,7.3 in][]{2020JCAP...05..047K}.

\section{Simulations for power spectrum recovery and covariance estimation} \label{sec:sims}
To test recovery of unbiased power and to compute covariances we use $400$ CMB lensing reconstruction simulations and appropriately correlated Gaussian simulations of the galaxy number density fields. The lensing simulations use a Gaussian lensing convergence to lens a randomly drawn CMB realisation. We add realistic survey noise and masking \citep[for details on the noise simulations see][]{2023JCAP...11..073A}, and reconstruct the lensing convergence with the pipeline used on the data \citep{2024ApJ...962..112Q}. 

\begin{figure*}
    \centering
    \includegraphics[width=\linewidth, trim=0cm 0.2cm 0cm 0cm, clip]{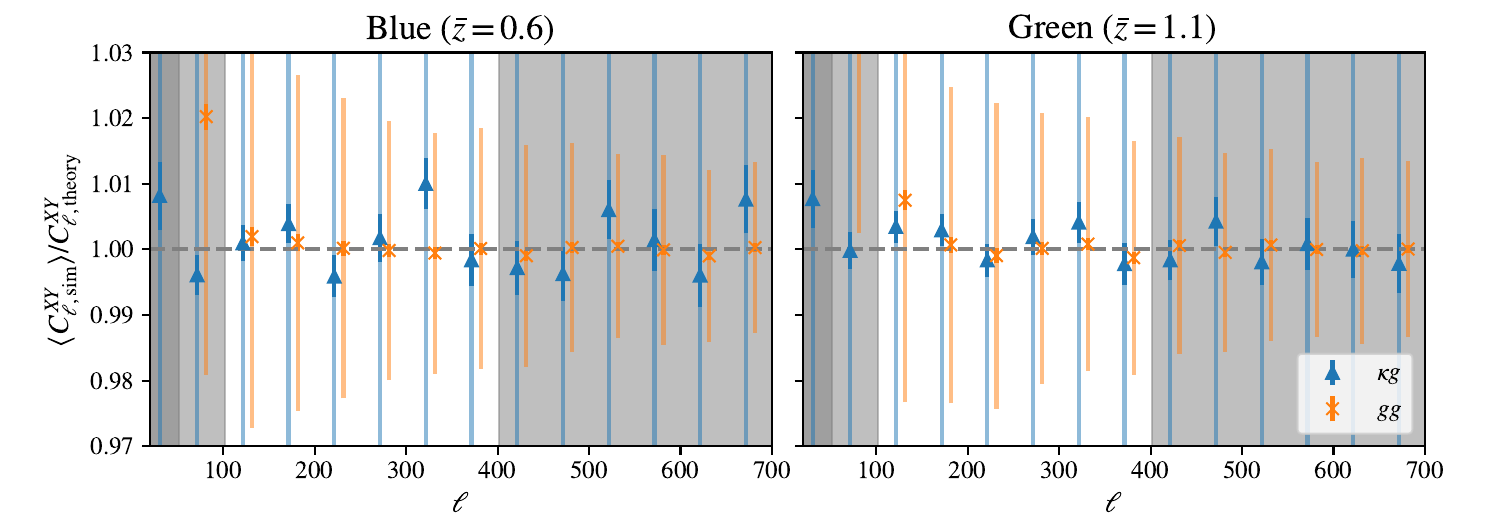}
    \caption{The recovery of $C_\ell^{gg}$ and $C_\ell^{\kappa g}$ on simulations when correctly accounting for the effect of mode coupling and the approximate inversion of the mode coupling matrix is better than 1\% on all scales. The lightly coloured errorbars indicate our measurement errors, while the dark errorbars show the error on the mean of 400 Gaussian simulations.}
    \label{fig:transfer_function}
\end{figure*}

To generate the galaxy simulations we measure $\hat{C}_\ell^{gg}$ and $\hat{C}_\ell^{\kappa g}$ on the two unWISE samples in narrow $\ell$-bins. We fit a model with fixed cosmology and free nuisance parameters\footnote{The model and its parameters are described in Sec.~\ref{sec:model}. For this fit we vary all nuisance parameters, including the galaxy bias, the shot-noise, the higher order bias parameters and the principal component expansion coefficients for the cross-correlation redshifts. In total there are 8 parameters for the Blue sample and 10 parameters for Green}. We use the range $20 \leq \ell \leq 3000$ for $C_\ell^{\kappa g}$ and $100\leq \ell\leq 3000$ for $C_\ell^{gg}$. We do not fit $\hat{C}_\ell^{gg}$ and $\hat{C}_\ell^{\kappa g}$ jointly but rather individually. When fitting $\hat{C}_\ell^{gg}$ and $\hat{C}_\ell^{\kappa g}$ individually the amplitude of the spectrum is completely degenerate with the galaxy bias. We do not assess the consistency of the galaxy bias found for the auto and cross-correlation in order not to compromise our blinding procedure. Using the model fit we compute fiducial $C_{\ell, \rm{fid}}^{gg}$ and $C_{\ell, \rm{fid}}^{\kappa g}$ and convolve them with the appropriate pixel window-function for our chosen map resolution of $\texttt{nside}=2048$. To capture the excess noise present on large scales in $\hat{C}_\ell^{gg}$, we additionally fit a smoothly broken power-law with an exponential cut-off at $\ell_{\rm{noise, max}} \leq 100$ to $\hat{C}_\ell^{gg} - C_{\ell, \rm{fid}}^{gg}$ on large scales. Using the Gaussian realisations of the lensing convergence field employed in generating the lensing simulations, we compute Gaussian realisations of the galaxy field that match the observed cross-correlation between CMB lensing and galaxy number density as
\beq
a_{\ell m}^{g} = \frac{C_{\ell, \rm{fid}}^{\kappa g}}{C_{\ell, \rm{fid}}^{\kappa \kappa}} a_{\ell m}^{\kappa} + a_{\ell m}^{g, \rm{uncorrelated}} + a_{\ell m} ^ {g, \rm{noise}}.
\eeq
The contributions $a_{\ell m}^{g, \rm{uncorrelated}}$ and $a_{\ell m} ^ {g, \rm{noise}}$ are drawn such that
\begin{eqnarray}
    \av{a_{\ell m}^{g, \rm{uncorr.}} (a_{\ell' m'}^{g, \rm{uncorr.}})^*} &=& \delta_{\ell \ell'} \delta_{m  m'} \left(C_{\ell, \rm{fid}}^{gg} - \frac{(C_{\ell, \rm{fid}}^{\kappa g})^2}{C_{\ell, \rm{fid}}^{\kappa \kappa}} \right)\nonumber\\
    &\\
    \av{a_{\ell m}^{g, \rm{noise}} (a_{\ell' m'}^{g, \rm{noise}})^*} &=& \delta_{\ell \ell'} \delta_{m  m'} C_{\ell, \rm{noise}}^{gg}.
\end{eqnarray} 
This yields the correct auto- and cross-spectra. We do not include correlations between the different galaxy samples (for the purposes of the covariance computation in Sec.~\ref{sec:covmat}, we include an analytic approximation for the covariance between different samples). We measure the auto-correlations of these simulations as well as their correlation with lensing reconstruction simulations in exactly the same way as we treat the data. We then compare the measured and decoupled power spectra to the input power spectra. Note that, since the mode decoupling is only approximate, in order to make a fair comparison we have to first convolve the input $C_{\ell, fid}$ with the appropriate mode-coupling matrix (see Eq. \ref{eq:mode-coupling}) before binning them and applying the approximate decoupling applied to the measured $\hat{C}_\ell$. We find that the recovered $\hat{C}_\ell^{gg}$ and $\hat{C}_\ell^{\kappa g}$ are consistent with the inputs to within $\lesssim1\%$ in our cosmology range (see Fig.\,\ref{fig:transfer_function}). Our simulations mimic the large scale excess of power present in the galaxy data and we use them to compute a multiplicative transfer function correcting for this effect and small inaccuracies in the mode decoupling. This excess ($<0.5\%$ for Blue and $<1\%$ for Green), which is present only in the lowest-$\ell$ bandpower in $C_\ell^{gg}$, is much smaller than the uncertainty on our data bandpowers ($\sim 3\%$ in the lowest-$\ell$ bin), but is detected significantly with 400 simulations.

\section{Covariance Matrices}\label{sec:covmat}

\begin{figure}
    \centering
    \includegraphics[width=\linewidth,trim=0.6cm 0.6cm 0cm 0cm,clip]{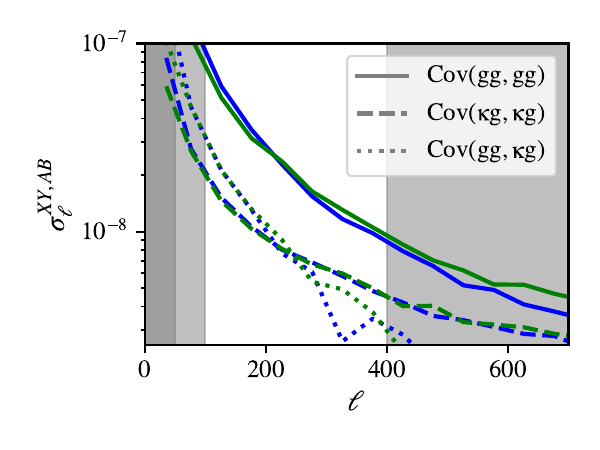}
    \caption{This figure shows the size of the diagonal elements of the covariance matrix as well as the off-diagonal covariance between $C_\ell^{gg}$ and $C_\ell^{\kappa g}$ which is substantial due to the contribution from auto-correlation of the galaxy sample.}
    \label{fig:covmat_diag}
\end{figure}

\begin{figure}
    \centering
    \includegraphics[width=\linewidth, trim=0cm 0.2cm 0cm 0cm, clip]{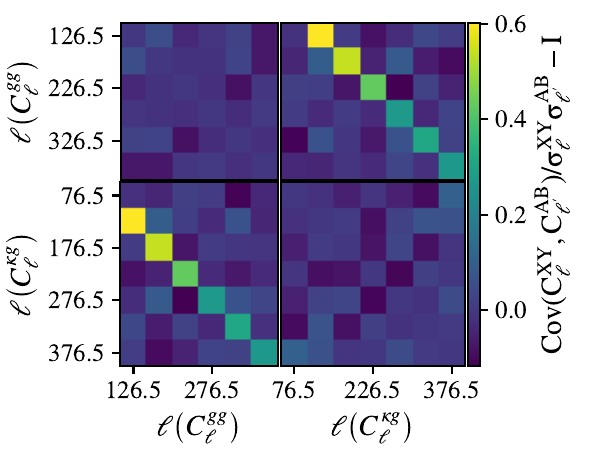}
    \caption{Correlations between $C_\ell^{gg}$ and $C_\ell^{\kappa g}$ are up to 60\% on large scales (here shown for the Green sample of unWISE galaxies). Correlations between different scales are small; less than 10\% in magnitude. We have subtracted the diagonal elements of the correlation matrix for better legibility.}
    \label{fig:corrmat_green}
\end{figure}

We use the suite of $400$ Gaussian simulations of the galaxy and lensing fields discussed in Sec.\,\ref{sec:sims} to estimate the covariance $\rm{Cov}(C_\ell^{XY}, C_{\ell'}^{AB})$ ($XY,AB \in \{gg, \kappa g\}$). The diagonal elements of the covariance matrix, as well as the diagonals of the $\rm{Cov}(C_\ell^{gg}, C_{\ell'}^{\kappa g})$ part of the covariance are shown in Figure \ref{fig:covmat_diag}, and Figure \ref{fig:corrmat_green} shows the correlation matrix for one of the two samples of galaxies.

We find large correlations (up to 60\%) between $C_\ell^{gg}$ and $C_\ell^{\kappa g}$ at the same $\ell$. Off-diagonal correlations (between unequal $\ell$), on the other hand, are small ($\leq 10\%$). To account for the fact that the inverse of the above covariance matrix is not an unbiased estimate of the inverse covariance matrix, we rescale the inverse covariance matrix by the Hartlap factor \citep{2007A+A...464..399H}:
\begin{equation}
    \alpha_\mathrm{cov}=\frac{N_s-N_{\text{bins}}-2}{N_s-1}.
\end{equation}
Given our 400 simulations and the 13 combined data points for $C_\ell^{gg}$ and $C_\ell^{\kappa g}$ the Hartlap factor is approximately $\alpha_{\rm{cov}} = 0.96$ for the analysis of a single sample and $\alpha_{\rm{cov}} = 0.93$ for the joint analysis of the two galaxy samples.

Since our Gaussian simulations do not include correlations between the two galaxy samples we are unable to use these simulations to estimate the off-diagonal blocks of the covariance used in the joint analysis of both galaxy samples. Instead we approximate those blocks analytically using the fiducial input spectra used for our simulations, a theory curve for the $C_\ell^{g_{\rm{Blue}} g_{\rm{Green}}}$-power spectrum obtained from the fixed cosmology fits we employed to obtain the input auto-spectra for our Gaussian simulations (described in Sec.~\ref{sec:sims}), and a measurement of $C_\ell^{\kappa \kappa}$ including the reconstruction noise from our simulations. Using the Gaussian covariance module implemented in \texttt{NaMaster} \citep{2019MNRAS.484.4127A,2019JCAP...11..043G} we compute a theory expression for the off-diagonal covariance. Correlations between the $C_\ell^{gg}$ for different samples are small ($<20\%$ on all scales of interest). The $C_\ell^{\kappa g}$ on the other hand are significantly correlated ($40$--$60\%$) because this part of the covariance has a contribution proportional to the product of the cross-correlations, $C_\ell^{\kappa g_{\rm Blue}} C_\ell^{\kappa g_{\rm Green}}$.

In Sec.~\ref{subsec:combination_planck} we also present a combined analysis of the ACT DR6 cross-correlation with the unWISE galaxies and the corresponding cross-correlations with \textit{Planck} PR4 lensing reconstructions. To estimate the covariance for this joint analysis we proceed as follows: We use a set of 480 FFP10 CMB simulations used by \cite{2022JCAP...09..039C}. These are Gaussian simulations very similar to those used in the ACT analysis, and for which there exist corresponding lensing reconstructions obtained using the \textit{Planck} PR4 pipeline and the corresponding noise model. In an identical manner to the process described in Sec.~\ref{sec:sims} we obtain a set of correlated Gaussian galaxy realisations allowing us to estimate the covariance for the \textit{Planck} cross-correlation analysis.

As described in \cite{2024ApJ...962..112Q} we also obtained corresponding ACT DR6 reconstructions by analysing the FFP10 simulations using the ACT DR6 pipeline which uses a different, but overlapping, area of the sky and a different range of CMB scales to reconstruct the lensing convergence. These ACT simulations do not include the ACT noise model, but since instrumental and atmospheric noise in the ACT data are not correlated with \textit{Planck} \editF{and ACT uses a lensing estimator based on cross-correlations of independent data splits}, the noise does not enter the ACT-\textit{Planck} covariance. The omission of the ACT noise in fact improves the convergence of the cross-covariance. The partially correlated lensing reconstruction noise arising from CMB fluctuations, on the other hand, is captured by these simulations.

The correlation between the ACT and \textit{Planck} lensing reconstructions is relatively small ($\lesssim\!\!30\%$), as demonstrated in \cite{2024ApJ...962..112Q}. However, the cross-correlation covariance is substantial even though the survey areas only partially overlap, since the identical galaxy sample is used in both cross-correlations. The correlation between the galaxy auto-correlations measured on the ACT and \textit{Planck} footprints is up to $60\%$, while the cross-correlations are up to $50\%$ correlated. As with the ACT-only analysis we analytically estimate the covariance between different galaxy samples, which is small in the case of $C_\ell^{gg}$ ($<20\%$) but again substantial for $C_\ell^{\kappa g}$ (up to 50\%). We show a full summary of the level of off-diagonal correlations in Appendix~\ref{app:off-diag_correlations}.

To assess the convergence of our covariance we produce an independent set of 400 further simulations. We find that the signal to noise is changed by $\lesssim5\%$ when computing the covariance using this set of simulations. In addition we perform a consistency test with this set of simulations finding that our results are stable (see Sec.\,\ref{subsec:param_consistency_tests}). We thus conclude that our covariance matrix is sufficiently converged.

\subsection{Marginalisation over uncertainty in the lensing normalisation}
As discussed in more detail in \cite{2024ApJ...962..112Q} (see in particular Appendix B there) the unnormalised lensing reconstruction is sensitive to the product of the lensing deflection field, $\phi_{\ell m}$, and the lensing response function, $\mathcal{R}_\ell$. The lensing normalisation, $\mathcal{R}^{-1}_\ell$, is in principle a function of the true underlying CMB two-point spectrum, but is in practice computed assuming a fiducial spectrum, $C^{\rm{CMB}, \rm{fid}}_\ell$\footnote{The ACT lensing reconstruction adopts as the fiducial spectrum the $\Lambda$CDM model from \textit{Planck}~2015 TTTEEE cosmology with an updated $\tau$ prior as in~\cite{2017PhRvD..95f3525C}.}. While the CMB power spectrum is well constrained by \textit{Planck}, some residual uncertainty remains that must be propagated to the lensing normalisation. Any deviation from the fiducial spectrum is expected to be small and we thus Taylor expand the normalisation around the fiducial value
\beq \label{eq:response_expansion}
\begin{aligned}
    \mathcal{R}^{-1}_{L}\rvert_{C_\ell^{\textrm{CMB}}} \approx &  \mathcal{R}^{-1}_{L}\rvert_{C_\ell^{\textrm{CMB,fid}}} \\ &\times \left[1 + M_{L}^{\ell} (C_\ell^{\textrm{CMB}}-C_\ell^{\textrm{CMB,fid}}) \right] ,
\end{aligned}
\eeq
where $M_{L}^{\ell} = \partial \ln \mathcal{R}^{-1}_L / \partial C_\ell^{\text{CMB}} \rvert_{C_\ell^{\text{CMB, fid}}}$ is the linearised normalisation-correction matrix. As in \cite{2024ApJ...962..112Q} we sample 1000 $\Lambda$CDM CMB power spectra from the ACT DR4 + \textit{Planck} parameter chains presented in~\citet{2020JCAP...12..047A} to generate an ensemble of smooth power spectrum curves consistent with the ACT DR4 + \textit{Planck} power spectrum measurements. Using Eq.~\ref{eq:response_expansion}, we then propagate the scatter in these power spectra to additional uncertainties in the cross-correlation bandpowers. \cite{2024ApJ...962..112Q} find this approach to be robust despite the use of smooth $\Lambda$CDM power spectra and the use of spectra drawn from the ACT DR4 + \textit{Planck} chains. We note that at the time of writing the ACT DR6 CMB power spectra are not yet available so that a direct marginalisation over the underlying spectra is not possible. In addition, the effect is significantly smaller for the cross-correlation than for the auto-spectrum so that we do not expect to incur any bias due to our treatment of the normalisation marginalisation.

\begin{figure}
    \centering
    \includegraphics[width=\linewidth,trim=0cm 0.6cm 0cm 0cm, clip]{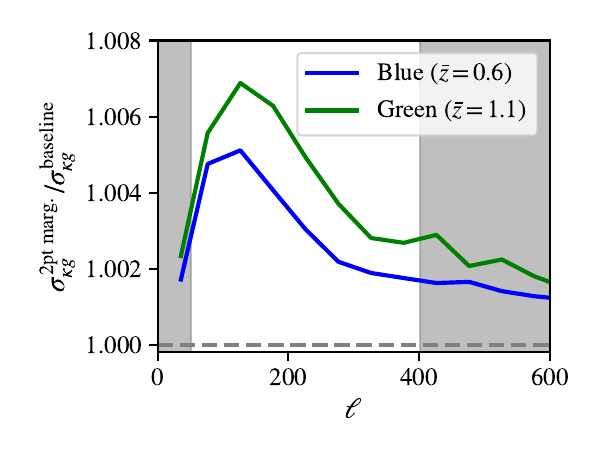}
    \caption{The contribution to the bandpower errors from marginalisation over the lensing normalisation is negligible. Nevertheless, because the contributions to the off-diagonal elements of the covariance are more significant, we include this effect in our analysis.}
    \label{fig:2pt_marg_errors}
\end{figure}

In Figure~\ref{fig:2pt_marg_errors} we show the increase in the bandpower errorbars; the change is less than 1\%. For consistency with the analysis in \cite{2024ApJ...962..112Q}, and because the contributions to the off-diagonal elements of the covariance are slightly more significant, we still include these contributions in the covariance. We also implement the normalisation marginalisation self consistently for the covariance between different galaxy samples and the covariance between the ACT and \textit{Planck} cross-correlations. However, accounting for the uncertainty in the normalisation has no impact on our parameter constraints.

\section{Tests for Systematic Errors}\label{sec:sys_test}

To test that our data is free from significant contamination that would affect our cosmological constraints, we perform a series of consistency checks. The tests are aimed at systematic errors in the lensing reconstruction that may correlate with the galaxy distribution; they are summarised in Sec.\,\ref{subsec:lensing_sys}, with more details provided in Appendix \ref{app:null-tests}. Sec.\,\ref{subsec:galaxy_sys} summarises the consistency tests performed that target, in particular, spatial inhomogeneities in the galaxy samples. In addition, we estimate biases in the lensing cross-spectra due to extragalactic foreground contamination of the lensing reconstruction (see Sec.\,\ref{subsec:sim_foregrounds}). 

\subsection{Testing for Contamination of the Lensing Reconstruction}\label{subsec:lensing_sys}

We perform a series of null-tests to demonstrate that our lensing reconstruction is free from systematic effects that could correlate with the galaxy samples. Such contamination could arise, for example, from the thermal Sunyaev-Zeldovich (tSZ) effect which originates from inverse Compton scattering of CMB photons off thermal electrons in galaxy clusters; it can, if not appropriately mitigated, bias the lensing reconstruction \citep{2024ApJ...966..138M}. The tSZ signal also correlates with the large scale matter distribution and hence the galaxy density. Similarly, CIB contamination, which originates from unresolved dusty galaxies, also correlates with the galaxy densities and can bias the lensing reconstruction. The ACT DR6 lensing reconstruction used in this work adopts bias hardening methods to mitigate these biases \citep[for details see][]{2024ApJ...966..138M,2024ApJ...962..112Q}. We verify here that contaminants are mitigated sufficiently and that potential residual contamination does not introduce significant biases in our analysis. In addition, there is, in principle, the potential of bias due to Galactic contamination arising from a correlation of galaxy density with Galactic structure that also impacts the lensing reconstruction; this effect is easily tested and is mitigated by both masking of dust-contaminated regions and by the systematics weighting described in Sec.\,\ref{subsec:sys_weights}.

All our null-tests are summarised within Appendix \ref{app:null-tests}, where we show $\chi^2$ and PTEs in Table \ref{table:clkg_nulls} and several figures showing the null-test bandpowers (see Figures~\ref{fig:null_test_curl} through \ref{fig:null_test_stellar_mask}). We define a null-test to be `passing' if it returns a $\rm{PTE}>0.05$. Overall we consider our tests to be passing if the number of failures ($\rm{PTE}<0.05$) is consistent with expectations due to random fluctuations. 

We observe a total of four failures\footnote{But no catastrophic failures with $\rm{PTE}<0.01$.}; given that we perform a total of 28 null-tests for $C_\ell^{\kappa g}$ across the two samples\footnote{This excludes all tests for which passing is not necessarily expected due to spatial variations in the galaxy selection which affects the galaxy bias. See Sec.\,\ref{subsec:galaxy_sys}.}, Blue and Green, this is slightly more than would be expected due to random fluctuations when neglecting correlations between the tests. We note, however, that all failing tests concern the Blue sample of galaxies and at least three of them are highly correlated. To estimate the number of expected failures given the correlations between the null-test we performed these null-tests on our 400 Gaussian simulations discussed in Sec.~\ref{sec:sims}. We find that in 11\% of cases we observe more than the 4 failures found on the data. We conclude that the number of observed failures is not inconsistent with what could be expected due to random fluctuations. Nevertheless, we investigate below whether there is additional evidence that would suggest that the observed failures indicate systematic contamination of our data.

\begin{figure*}
    \centering
    \includegraphics[width=0.5\linewidth, trim=0cm 0.7cm 0cm 0cm, clip]{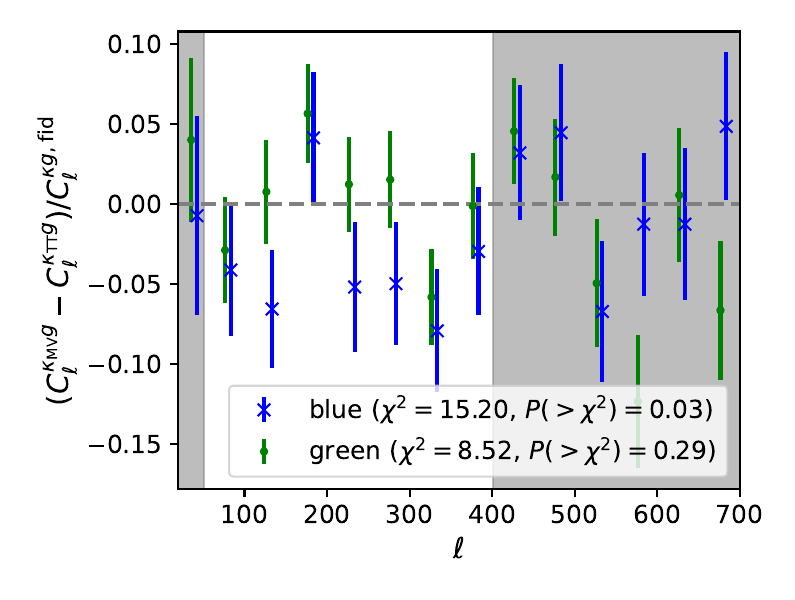}%
    \includegraphics[width=0.5\linewidth, trim=0cm 0.7cm 0cm 0cm, clip]{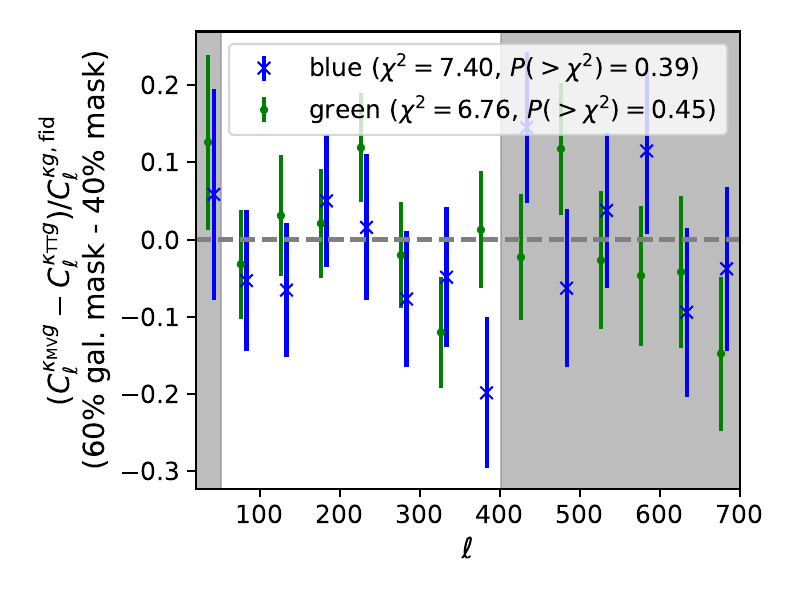}
    \caption{We show here the bandpowers for two of our null-tests comparing different lensing reconstructions (further tests are shown in Appendix~\ref{app:null-tests}). When comparing bandpowers between the cross-correlations of the unWISE galaxies with the minimum variance lensing reconstruction and the temperature only reconstruction, we observe a null-test failure for the Blue sample of galaxies (\textbf{left}). As discussed in the body of the text, the one plausible systematic contamination which could cause this behaviour is contamination from polarised Galactic dust emission. However, the failure does not show the scale dependence that is expected for such a failure, with the discrepancy increasing on larger scales, nor is the discrepancy observed in regions  near the Galactic plane where this effect is expected to be most significant (\textbf{right}). As we discuss in the text the number of observed failures is consistent with random fluctuations.}
    \label{fig:null_test_MV-TT}
\end{figure*}

The failing tests are the following: (1) bandpower difference between the cross-correlations of the unWISE galaxies with the minimum variance lensing reconstruction and the temperature only reconstruction (henceforth MV-TT, see left panel in Fig.~\ref{fig:null_test_MV-TT}), (2) the difference between the cross-correlation bandpowers between the temperature only and polarisation only reconstruction (henceforth TT-MVPOL), and (3) bandpower difference between the cross-correlations of the unWISE galaxies with the minimum variance lensing reconstruction and a reconstruction that explicitly deprojects the CIB (henceforth MV-CIBD). Additionally, we also observe $\rm{PTE}<0.05$ for (4) the null-test between minimum variance lensing reconstruction and polarisation only reconstruction when performed using the 40\% Galactic mask (henceforth MV\_GAL040-MVPOL\_GAL040)\footnote{The test of MV-MVPOL on the full footprint yields $\rm{PTE}=0.06$}. We find the tests TT-MVPOL and MV\_GAL040-MVPOL\_GAL040 to be $\sim 77\%$ and $\sim32\%$ correlated with MV-TT leading us to conclude that the joint failure of these three tests is not unexpected.

Regarding the failure of MV-CIBD we note that the CIB deprojected analysis uses a slightly different lensing mask, removing a small additional area near the Galactic plane due to problems with Galactic dust. As we discuss in more detail in Sec.~\ref{subsec:galaxy_sys} the galaxy selection is not expected to be uniform on large scales so that variations in the mask can lead to a differences in the galaxy bias, leading null-tests to fail without significant impact on the inferred cosmology. When comparing the galaxy auto-correlation on the mask used in the CIB deprojected analysis with the baseline mask we also obtain a low PTE of 0.09. It is thus likely that the observed failure in this case is at least in part due to varying galaxy selection properties which should not affect our cosmological inference.

One plausible systematic contamination that could induce the observed failures, given that all other null-tests are passing, is contamination by polarised emission from Galactic dust. Dust emission correlates with dust extinction which may also affect the galaxy samples. Weighting the unWISE galaxy number density to correct for a correlation with stellar density also significantly reduces the correlation with dust extinction (see Appendix \ref{app:sys_weights}), but we nevertheless consider the possibility of residual contamination. Such contamination is expected to be significantly stronger near the Galactic plane. We hence perform the same tests of comparing minimum variance, temperature-only and polarisation-only reconstructions on the areas of the sky included in our 60\% Galactic mask but not in the more restrictive 40\% Galactic mask; most of which are at low Galactic latitude. We find all these tests are passing comfortably and hence, given their high sensitivity to dust contamination, conclude that there is no evidence for any contamination from Galactic dust emission.

In Sec.\,\ref{subsec:param_consistency_tests} we furthermore discuss a parameter consistency test between our baseline analysis and an analysis using the temperature only reconstruction. We find the inferred parameters to be consistent within the expected error on the difference.

\begin{figure*}
    \centering
    \includegraphics[width=0.5\linewidth, trim=0cm 0.7cm 0cm 0cm, clip]{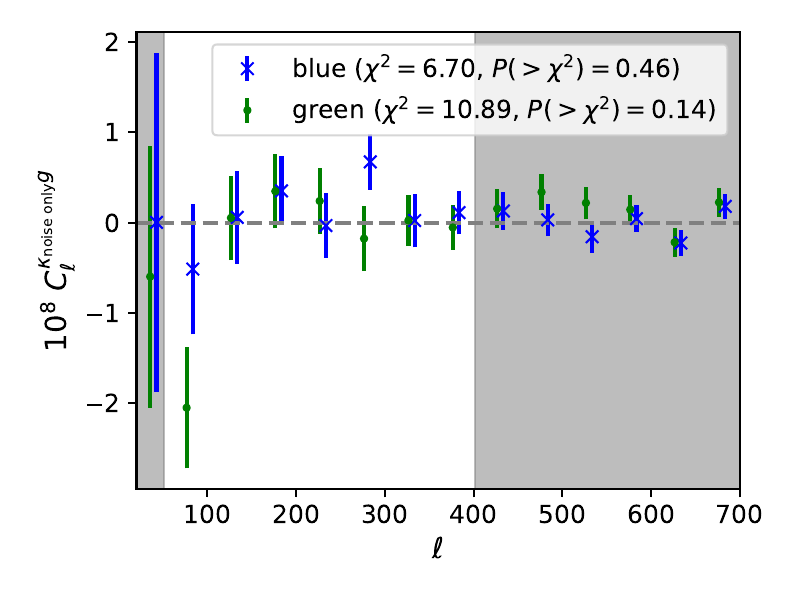}%
    \includegraphics[width=0.5\linewidth, trim=0cm 0.7cm 0cm 0cm, clip]{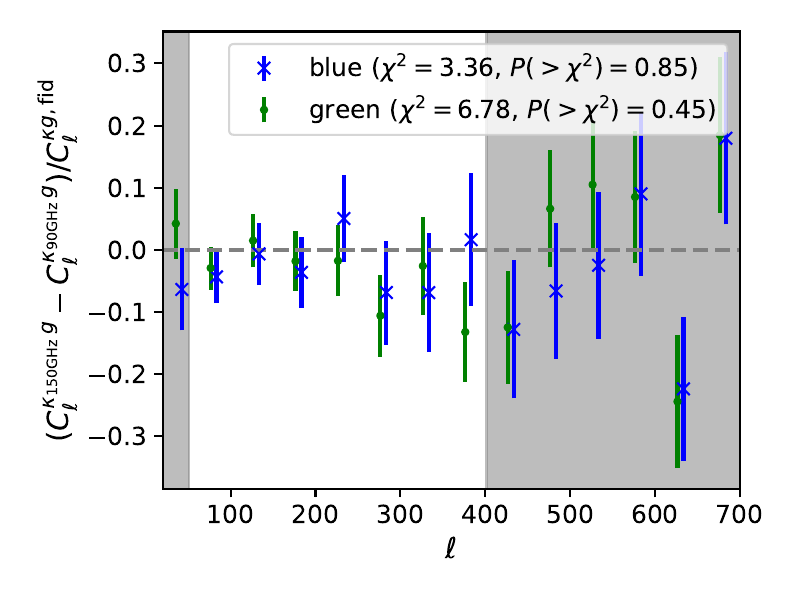}
    \caption{To test for contamination by extragalactic foregrounds we conduct two kinds of frequency level null-tests. First, we take the difference of the observed temperature and polarisation maps at 150 and 90 GHz to obtain a map containing only noise (and foregrounds) on which we perform lensing reconstruction. We then cross-correlate this map with our two galaxy samples (\textbf{left}). Secondly, we investigate the bandpower difference between cross-correlations measured using the reconstructions performed only on the 90 and 150 GHz data respectively. In both cases our null-tests are passed.}
    \label{fig:null_test_f150-f090}
\end{figure*}

To target extragalactic contamination we perform a series of tests that leverage the distinctive frequency dependence of extragalactic foregrounds. We perform two kinds of frequency level null-tests. First, we take the difference of the observed temperature and polarisation maps at 150 and 90 GHz to obtain a map containing only noise (and foregrounds) on which we perform lensing reconstruction. We then cross-correlate this map with our two galaxy samples. Secondly, we investigate the bandpower difference between cross-correlations measured using the reconstructions performed only on the 90 and 150 GHz data respectively (see Fig.~\ref{fig:null_test_f150-f090} for both versions of this test). We find no failures for those tests.

\subsection{Testing for Contamination and Homogeneity of the Galaxy Samples}\label{subsec:galaxy_sys}

To test the homogeneity of our observations we perform null-tests using different masks. We perform a test on the difference between cross-correlations using our 60\% and 40\% Galactic masks. Additionally, we construct a null-test that compares our baseline footprint with a footprint restricted to ecliptic latitude larger than 30 degrees. Due to the unWISE survey's scan strategy, the survey depth varies with ecliptic latitude which may affect our inference. Furthermore, zodiacal light, sun light scattering of interplanetary dust, may contaminate the galaxy selection. Finally, we also split our sample into the northern and the southern Galactic cap to perform a null-test with two completely independent samples. 

\begin{figure*}
    \centering
    \includegraphics[width=0.5\linewidth, trim=0cm 0.7cm 0cm 0cm, clip]{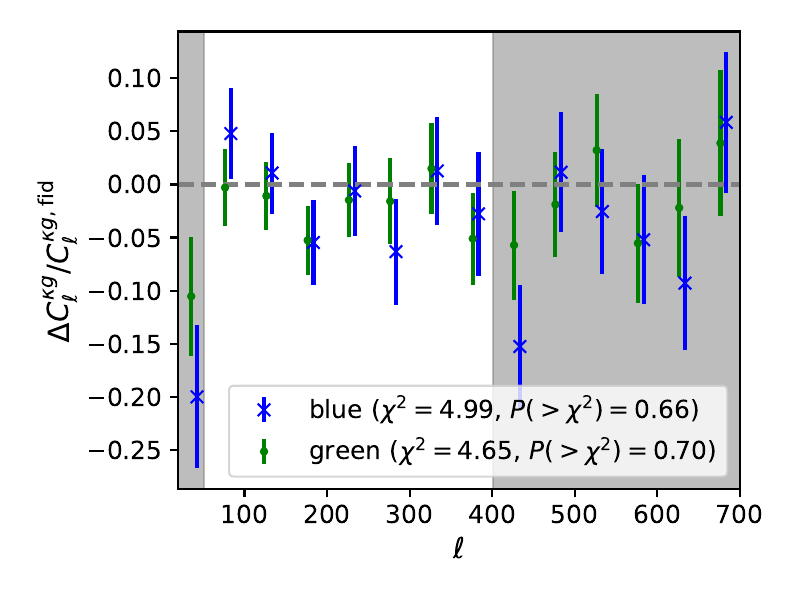}%
    \includegraphics[width=0.5\linewidth, trim=0cm 0.7cm 0cm 0cm, clip]{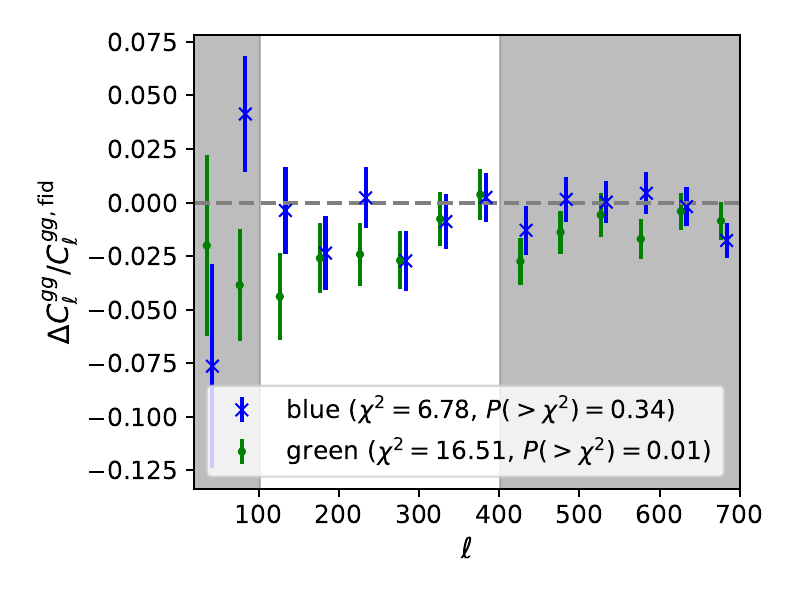}
    \caption{One of the tests targeting the spatial homogeneity of the sample is the comparison of our baseline footprint with a more conservative Galactic mask. While this null-test passes for $C_\ell^{\kappa g}$ (\textbf{left}) we observe a failure in $C_\ell^{gg}$ for the Green sample (\textbf{right}). This is not unexpected due to varying galaxy bias arising from large scale inhomogeneity of the galaxy selection. We thus investigate the approximately bias independent combination $(C_\ell^{\kappa g})^2/C_\ell^{gg}$ which we observe to be passing (see Fig.~\ref{fig:null_test_GAL040_kg^2_gg}).}
    \label{fig:null_test_GAL040}
\end{figure*}

\begin{figure}
    \centering
    \includegraphics[width=\linewidth, trim=0cm 0.7cm 0cm 0cm, clip]{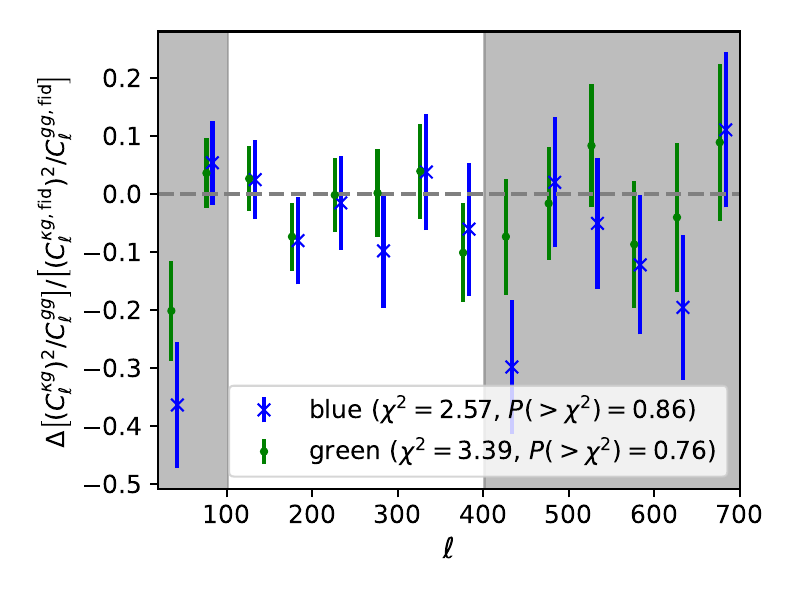}
    \caption{The combination $(C_\ell^{\kappa g})^2/C_\ell^{gg}$ is approximately independent of the linear order bias and thus less sensitive to spatial variations in the galaxy selection. Comparing this combination between our baseline mask and a more conservative Galactic mask shows consistency within the errors.}
    \label{fig:null_test_GAL040_kg^2_gg}
\end{figure}

All tests employing the differences between different regions of the sky are complicated by the fact that the galaxy selection is not expected to be uniform on very large scales, due to the varying WISE depth of coverage. Thus, the galaxy bias, the shot noise and the redshift distribution of our samples are expected to vary across the sky \cite[see for example Figs.~10 and 11 in][]{2020JCAP...05..047K}. In particular when comparing $C_\ell^{gg}$, which is measured to extremely high precision, we expect null-tests to fail due to the differences in the galaxy bias and shot noise (see e.g., Fig.~\ref{fig:null_test_GAL040} for the comparison of different Galactic masks). However, since we marginalise over the galaxy bias, shot noise and uncertainties in $b(z) dN/dz$ in our analysis this does not bias our cosmological inference. We thus construct the quantity $(C_\ell^{\kappa g})^2/(C_\ell^{gg} - \hat{n}_{\rm{shot}})$\footnote{For brevity we will usually neglect the shot noise subtraction in our notation.} which at linear order is independent of the bias and directly proportional to $S_8$, the main parameter of interest in our analysis. The shot noise is here simply estimated as the inverse of the galaxy number density in the respective footprint. None of the scales used in our analysis are shot noise dominated and hence our tests are insensitive to small misestimations in the shot noise. Fig.~\ref{fig:null_test_GAL040_kg^2_gg} shows the null-test for two different Galactic masks using this bias-independent quantity. When considering $(C_\ell^{\kappa g})^2/C_\ell^{gg}$, we find $\rm{PTE}>0.05$ for all tests, but for $C_\ell^{gg}$ alone we find several tests with $\rm{PTE}\ll0.05$ as expected. This implies that, up to differences in the galaxy bias which does not affect our cosmological inference, the samples are homogeneous, yielding consistent amplitudes for different areas on the sky.


\subsection{Simulation Based Test for Extragalactic Foregrounds}\label{subsec:sim_foregrounds}

As a further test for systematic contamination, in particular from extragalactic foregrounds which may contaminate the lensing reconstruction, we perform a series of tests on realistic foreground maps from the \textsc{WebSky} simulations \citep{2019MNRAS.483.2236S,2020JCAP...10..012S}. Following \cite{2024ApJ...966..138M} we make the assumption that any extragalactic contamination affects exclusively the observed CMB temperature so that $T = T_{\rm{CMB}} + T_{\rm{fg}}$, but leaves the polarisation unaffected. We do not expect contaminants such as tSZ and CIB to be polarised at a significant level for our observations, and any bright polarised sources in the DR6 data are masked and inpainted \citep{2024ApJ...962..112Q}. Denoting the quadratic estimator used to estimate the lensing convergence field, $\hat{\kappa}$, from the two temperature fields $T_A$ and $T_B$ as $Q(T_A, T_B)$, and the cross-correlation with the galaxy field as $C_\ell^{Xg} = \av{Xg}$, we obtain the bias on the cross-correlation of the temperature only lensing reconstruction with the galaxy sample due to foreground contamination as \citep[compare][]{2024ApJ...966..138M}
\beq
\Delta C_\ell^{\hat{\kappa} g} = 2 \av{Q(T_{\rm{CMB}}, T_{\rm{fg}}) g} + \av{Q(T_{\rm{fg}}, T_{\rm{fg}}) g}.
\eeq

Assuming that the foregrounds are uncorrelated with the CMB temperature, which will be true to high accuracy, we can see that the only relevant bias arises from the correlation of a lensing-like signal due to foreground contamination with the galaxy sample, $\av{Q(T_{\rm{fg}}, T_{\rm{fg}}) g}$.

We quantify the bias due to foreground contamination in terms of the bias on the cross-correlation amplitude, $A_{\times}$. Since we mostly aim to constrain, and are sensitive to, the amplitude of structure growth this captures the relevant bias for our analysis well. The bias on $A_{\times}$ in terms of the uncertainty on this amplitude is given by
\beq
\frac{\Delta A_{\times}}{\sigma(A_{\times})} = \frac{\sum_{\ell, \ell'} \Delta C_\ell^{\kappa g} \mathbb{C}_{\ell \ell'} C_{\ell'}^{\kappa g, \rm{fid}}}{\sqrt{\sum_{\ell, \ell'} C_\ell^{\kappa g, \rm{fid}} \mathbb{C}_{\ell \ell'} C_{\ell'}^{\kappa g, \rm{fid}}}}.
\eeq
Here $\Delta C_\ell^{\kappa g}$ is the foreground induced bias, i.e. the cross-correlation spectrum between the galaxy sample and the lensing estimated from the foreground only maps. Furthermore, $C_{\ell}^{\kappa g, \rm{fid}}$ is the true cross-correlation signal and $\mathbb{C}_{\ell \ell'}$ is the appropriate covariance matrix.

In addition to estimating the lensing signal from the foreground-only maps as discussed in \cite{2024ApJ...966..138M} we populate the \textsc{WebSky} halo catalogue with galaxies according to an HOD similar to the one described in \cite{2020JCAP...05..047K}, with the minimum halo mass increased by 3-5\% to better account for the different mass resolution of the \textsc{WebSky} simulations. We then sample the resulting galaxy catalogue to match the redshift distribution of our samples. We find a match between the spectra obtained from \textsc{WebSky} and from \textsc{CrowCanyon} described in \cite{2020JCAP...05..047K} to within about 5\% on the scales used in our cosmology analysis\footnote{Since we are interested mainly in the amplitude of the fractional bias, an exact match to the data, in particular to the galaxy bias, is not required.}.

\begin{figure*}
    \centering
    \includegraphics[width=\linewidth]{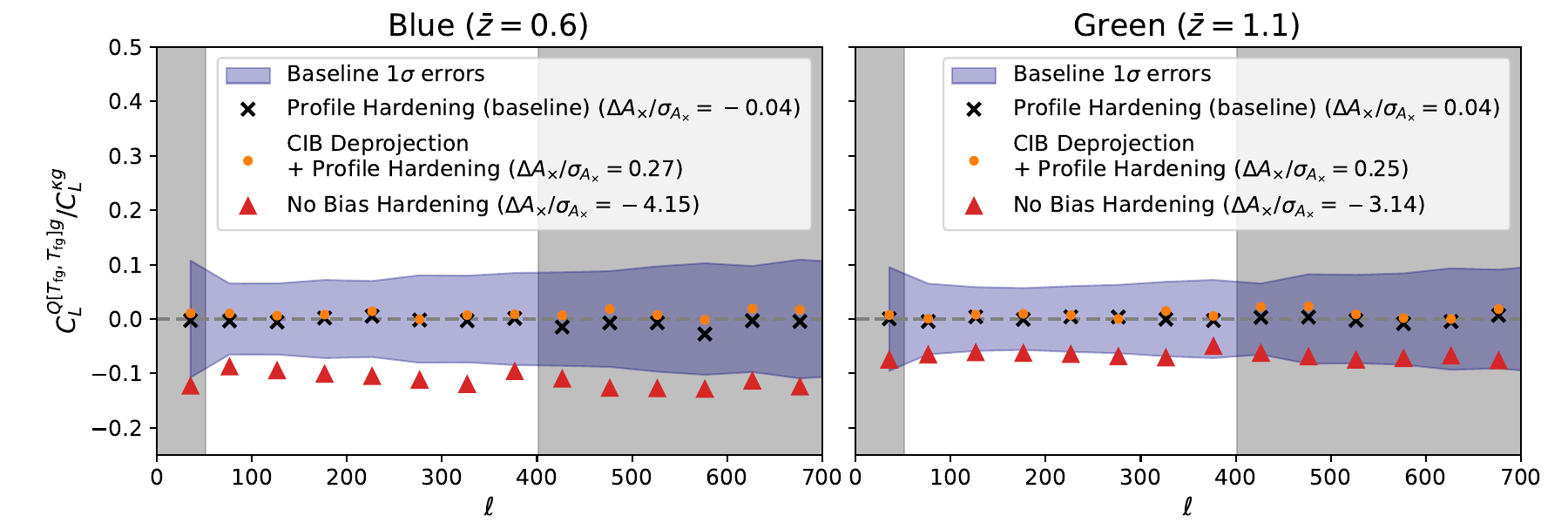}
    \caption{We estimate biases due to extragalactic foregrounds using realistic foreground simulations from \textsc{WebSky} \citep{2019MNRAS.483.2236S,2020JCAP...10..012S}. We perform lensing reconstruction on foreground only maps using different foreground mitigation strategies and cross-correlate them with a galaxy number density maps that we obtain by populating the \textsc{WebSky} halo catalogue using a HOD. We find that our baseline analysis reduces all biases to $<0.05\sigma$ while the analysis without any mitigation yields significant biases (up to $\sim 4.2\sigma$).}
    \label{fig:websky_biases}
\end{figure*}

We find that in our baseline analysis, which employs frequency coadds and geometric profile hardening \citep{2020PhRvD.102f3517S}, the residual bias is highly subdominant and can be safely neglected; $\left|\Delta A_{\times}/\sigma(A_{\times})\right| \simeq 0.04$ for both our galaxy samples\footnote{Note that this is a upper limit on the bias since the minimum variance reconstruction we employ in our analysis also receives contributions from polarisation which contributes with a weight of about 30\%.} (see Fig.\,\ref{fig:websky_biases}). Additionally, explicitly deprojecting the CIB does not yield any further gains. This is consistent with findings from \cite{2024ApJ...966..138M} and indicates that profile hardening, despite adopting a profile appropriate for tSZ clusters, also partially accounts for contamination from CIB, as was also found in \cite{2023PhRvD.107b3504S}. (For comparison, when performing no bias hardening we find significant biases of $A_{\times}/\sigma(A_{\times}) \simeq -4.2$ and $-3.1$ for the Blue and Green samples respectively.) We conclude that biases due to extragalactic foregrounds are small in comparison to the statistical errors.

\section{The model}\label{sec:model}
In Sec.\,\ref{sec:tomography} we discussed the measurement of the angular power spectra of the unWISE galaxies, $C_\ell^{gg}$ and their cross-spectra with CMB lensing reconstruction from the ACT. In this section we introduce the model to fit these data when measuring cosmological parameters. We present the 3D power spectrum model and its projection in Sec.\,\ref{subsec:pk_model} and detail our procedure for marginalising over redshift uncertainties in Sec.\,\ref{subsec:dndz_marg}. Finally, in Sec.\,\ref{subsec:model_test} we test the model on $N$-body simulations to verify its accuracy.

\subsection{Hybrid Power Spectrum Model} \label{subsec:pk_model}
In the following four subsections we detail the different components of our model. First, in Sec.\,\ref{subsubsec:3d_power}, we describe our model for the three dimensional clustering of galaxies and matter. Because we observe the galaxy number density and CMB lensing only in projection, we describe in Sec.\,\ref{subsubsec:projected_spectra} how we obtain the power spectra for the projected fields. 

The projection kernel requires knowledge of the redshift distribution of the galaxy samples. Previously, we discussed how we measure these redshift distributions (see Sec.\,\ref{subsubsec:dndz} with more details provided in Appendix~\ref{app:spec_samples}). Sec.\,\ref{subsubsec:dndz_bias_evol} introduces the way in which we implement these redshift distribution measurements in the projected power spectra. Subsequently, in Sec.\,\ref{subsubsec:higher_order_biases}, we describe our model for the redshift evolution of all higher order biases and the marginalisation over their contribution to the observed power spectra. Finally, we show the relative contributions of various model components (Sec.\,\ref{subsubsec:model_contributions}).

\subsubsection{3D Power Spectra and Bias Expansion}\label{subsubsec:3d_power}
We adopt an empirical hybrid model for the three dimensional power spectra of the clustering of galaxies, matter, and their cross-clustering ($P_{gg}$, $P_{mm}$, and $P_{mg}$ respectively). This model combines fits to numerical simulations with beyond linear order terms from Lagrangian perturbation theory \citep[LPT;][]{2015JCAP...09..014V,2016JCAP...12..007V,2017JCAP...08..009M,2021JCAP...03..100C,2020JCAP...07..062C}. We note that these additional terms beyond linear bias are necessary to ensure unbiased results, as discussed in Sec.\,\ref{subsec:model_test}. In particular, we adopt the \texttt{HMCode} model from \cite{2016MNRAS.459.1468M} as implemented in \textsc{CAMB}\footnote{\url{https://github.com/cmbant/CAMB}} \citep{2000ApJ...538..473L,2012JCAP...04..027H} and the formalism for $P_{gg}$ and $P_{gm}$ described in \cite{2021MNRAS.501.6181K} with some substitutions described below \cite[similar to][]{2020PhRvD.102l3522P}.


In the LPT model the bias expansion for the clustering of galaxies and their cross-clustering with matter is given by \citep{2017JCAP...08..009M,2021JCAP...03..100C,2020JCAP...07..062C,2021MNRAS.501.6181K}
\begin{subequations}\label{eq:full_lpt_model}
\begin{align} 
    P_{gg}&=\left(1 - \frac{\alpha_{\rm{auto}} k^2}{2} \right)P_Z + P_{1-\rm{loop}}\label{eq:full_lpt_model_gg}\\ 
    \nonumber&\phantom{{}={}} \negmedspace+ b_{1,L}P_{b_1} + b_{2,L}P_{b_2} + b_{s,L}P_{b_s}\\
    \nonumber&\phantom{{}={}} \negmedspace+ b_{1,L}b_{2,L}P_{b_1 b_2} + b_{1,L}b_{s,L}P_{b_1 b_s} + b_{2,L}b_{s,L} P_{b_2 b_s}\\
    \nonumber&\phantom{{}={}} \negmedspace+ b_{1,L}^2P_{b_1^2} + b_{2,L}^2P_{b_2^2} + b_{s,L}^2P_{b_s^2}
    + P_{\rm{shot\ noise}} \hspace{3cm} \rm{and}\\
    P_{mg}&=\left(1 - \frac{\alpha_{\rm{cross}} k^2}{2} \right)P_Z + P_{1-\rm{loop}}\label{eq:full_lpt_model_mg}\\ 
    \nonumber&\phantom{{}={}} \negmedspace+ \frac{b_{1,L}}{2}P_{b_1}+ \frac{b_{2,L}}{2}P_{b_2} + \frac{b_{s,L}}{2}P_{b_s}
\end{align}
\end{subequations}
where the $b_{X,L}$ are free bias coefficients and $P_Z$ and $P_{1-\rm{loop}}$ are the Zeldovich and 1-loop power spectra, the lowest order contributions to the theory. The $P_Z$, $P_{1-\rm{loop}}$, and higher order contributions ($P_{b_2}$, $P_{b_s}$ etc.) can be computed analytically from the linear power spectrum $P_{mm,\rm{lin}}$.

The term
\beq
P_{mm} = \left(1 - \frac{\alpha k^2}{2} \right)P_Z + P_{1-\rm{loop}}
\eeq
captures the dark matter contribution to the clustering \citep{2015JCAP...09..014V}. The parameter $\alpha$ allows marginalisation over the impact from small scale physics beyond the cut-off of the theory, and is generically not assumed to be identical in the auto- and cross-clustering. This contribution is what the \texttt{HMCode} model emulates numerically, which motivates the following substitution
\begin{eqnarray}\label{eq:matter_power_sub}
    P_{mm} = \left(1 - \frac{\alpha k^2}{2} \right)P_Z + P_{1-\rm{loop}} \to P_{mm,\rm{HM}}.
\end{eqnarray}
This approach incorporates all higher order effects in the dark matter clustering.

The remaining terms in the LPT bias expansion encode the response of the galaxy clustering to the large scale matter distribution. Assuming that at lowest order galaxies are linearly biased tracers of the matter density yields  
\begin{subequations}
\begin{align}
    P_{gg} &=b_{1,E}^2 P_{mm, \rm{HM}} + \rm{higher\ order\ terms} \hspace{0.4cm}\rm{and}\\
    P_{mg} &=b_{1,E} P_{mm, \rm{HM}} + \rm{higher\ order\ terms},
\end{align}
\end{subequations}
where $b_{1,E}$ is the lowest order Eulerian bias. It is related to the lowest order bias in the Lagrangian formalism, $b_{1,L}$, as $b_{1,E} = b_{1,L} + 1$ (we use subscripts $E$ and $L$ to distinguish Eulerian and Lagrangian biases). Assuming furthermore that the impact of small scale physics on the cross- and auto-clustering is identical (i.e. applying the substitution from Eq.\,\ref{eq:matter_power_sub} to both Eqs.\,\ref{eq:full_lpt_model_gg} and \ref{eq:full_lpt_model_mg}, setting $\alpha = \alpha_{\textrm{auto}} = \alpha_{\textrm{cross}}$) then requires the further substitutions 
\begin{subequations}
\begin{align}
    P_{b_1} &\to 2 P_{mm, \rm{HM}}\hspace{1cm}\rm{and}\\
    P_{b_1^2} &\to P_{mm, \rm{HM}},
\end{align}
\end{subequations}
to recover the linear bias model at lowest order.

With these substitution the galaxy-galaxy, galaxy-matter and matter-matter power spectra are given in detail by
\begin{widetext}
    \begin{subequations}
    \begin{align}
        P_{gg}(k,z) &= b^2_{1,E}(z)P_{mm, \rm{HM}} + b_{2,L}(z)P_{b_2}(k, z) + b_{s,L}(z)P_{b_s}(k, z)\\
        \nonumber&\phantom{{}={}} \negmedspace+ b_{1,L}(z)b_{2,L}(z)P_{b_1 b_2}(k, z) + b_{1,L}(z)b_{s,L}(z)P_{b_1 b_s}(k, z) + b_{2,L}(z)b_{s,L}(z) P_{b_2 b_s}(k, z)\\
        \nonumber&\phantom{{}={}} \negmedspace+ b_{2,L}^2(z)P_{b_2^2}(k, z) + b_{s,L}^2(z)P_{b_s^2}(k, z)
        + P_{\rm{shot\ noise}},\\
        P_{gm}(k,z) &= b_{1,E}(z)P_{mm, \rm{HM}} + \frac{b_{2,L}(z)}{2}P_{b_2}(k, z) + \frac{b_{s,L}(z)}{2}P_{b_s}(k, z), \hspace{1cm} \rm{and}\\
        P_{mm}(k,z) &= P_{mm, \rm{HM}}(k,z).
    \end{align}
    \end{subequations}
\end{widetext}
Approaches similar to this have been shown to work well on simulations \citep[e.g.][]{2020PhRvD.102l3522P}.

To compute the effective field theory contributions, $P_{b_2}$, $P_{b_s}$ etc., we rely on the \texttt{velocileptors} code\footnote{\url{https://github.com/sfschen/velocileptors}}. Following \cite{2022JCAP...07..041C} we adopt the baryon and dark matter only power spectrum for all contributions to $P_{gg}$ (i.e. we exclude the contribution from neutrinos which, to a good approximation, do not cluster and hence do not contribute to the galaxy clustering). For $P_{mg}$ we use the cross power spectrum between the baryon and dark matter density on one hand and the total matter density (including neutrinos) on the other.

\subsubsection{Power Spectra of Projected Fields}\label{subsubsec:projected_spectra}
Since we observe the galaxy overdensity and CMB lensing only projected along the line of sight we require a model for the projected galaxy-galaxy and galaxy-CMB lensing power spectra. To obtain those spectra we employ the Limber approximation\footnote{\editF{Using the implementation in the \texttt{Core Cosmology Library} (\texttt{CCL}) package \cite[\url{https://github.com/LSSTDESC/CCL};][]{2019ApJS..242....2C} we checked that the leading beyond-limber contribution, arising from redshift-space distortions, is less than 0.1\% on all scales used in our analysis and thus safely negligible.}} \citep{1953ApJ...117..134L,2008PhRvD..78l3506L}
\begin{eqnarray}
    C_\ell^{gg} &=& \int d\chi \frac{W_g^2(z)}{\chi^2} P_{gg}(k\chi = l + 1/2, z)\\
    C_\ell^{\kappa g} &=& \int d\chi \frac{W_g(z) W_\kappa(z)}{\chi^2} P_{gm}(k\chi = l + 1/2, z).
\end{eqnarray}
Here $\chi$ is the comoving distance to redshift $z$ along the line of sight. The galaxy and lensing projection kernels, $W_g$ and $W_\kappa$, are given by
\begin{eqnarray}
    W_g(z) &=& H(z) \frac{dN}{dz}\label{eq:g_projection_kernel}\ \rm{and}\\
    W_\kappa(z) &=& \frac{3}{2}\Omega_m H_0^2(1+z) \frac{\chi(\chi_\star-\chi)}{\chi_\star} \label{eq:kappa_projection_kernel}.
\end{eqnarray}
The total matter density, $\Omega_m$, includes the density of neutrinos which are non-relativistic at low redshifts. Additionally, $\chi_\star$ is the comoving distance to the last scattering surface, $H(z)$ is the Hubble rate, and $H_0=H(z=0)$.

In addition to the contribution from the correlation of galaxy clustering with the matter field that gives rise to the CMB lensing, the measured angular power spectra also contain relevant contributions from the lensing-magnification bias \citep{2020JCAP...05..047K}. This effect arises because individual galaxies may be gravitationally lensed and magnified (or demagnified) by foreground structure, affecting our sample selection by artificially increasing (or decreasing) the magnitude of a galaxy in a way that is correlated with the large scale structure. We use the index $\mu$ to denote quantities related to the magnification bias. We model the additional contributions as 
\begin{eqnarray}
    C_\ell^{g \mu} &=& \int d\chi \frac{W_g(z) W_\mu(z)}{\chi^2} P_{gm}(k\chi = l + 1/2, z),\\
    C_\ell^{\kappa \mu} &=& \int d\chi \frac{W_\kappa(z) W_\mu(z)}{\chi^2} P_{mm}(k\chi = l + 1/2, z),\\\nonumber \rm{and}\\
    C_\ell^{\mu \mu} &=& \int d\chi \frac{W_\mu(z) W_\mu(z)}{\chi^2} P_{mm}(k\chi = l + 1/2, z)
\end{eqnarray}
with the lensing-magnification kernel given by \citep{1995astro.ph.12001V,1998MNRAS.294..291M,2001PhR...340..291B}
\begin{equation}
\begin{split}
W_\mu(z) = (5 s_\mu -2)&\frac{3}{2}\Omega_m H_0^2(1+z)\\ &\times \int_\chi^{\chi_\star} d\chi' \frac{\chi(\chi' -\chi)}{\chi'} H(z') \frac{dN}{dz'}\label{eq:mu_projection_kernel}.
\end{split}
\end{equation}
The parameter $s_\mu \equiv d \log_{10} N/dm$ is the response of the galaxy number density to a change in magnitude. This parameter is measured from the data by perturbing the photometry of the unWISE galaxies and reapplying the selection criteria \citep[see Appendix D of][]{2020JCAP...05..047K}. However, in principle the magnification bias parameter is the derivative of the number density of the true underlying galaxy population (rather than the observed one) which is unknown. In addition, \cite{2020JCAP...05..047K} showed some variations of $s_\mu$ with survey depth \resub{(by about 5\%; see Fig. 20 in \citealt{2020JCAP...05..047K})}. Therefore, we marginalise over this parameter in our analysis with a conservative 10\% Gaussian prior. We note that $C_\ell^{\kappa \mu}$ and  $C_\ell^{\mu \mu}$ depend on the matter-matter power spectrum, $P_{mm}$. The total, observed, galaxy-galaxy and galaxy-CMB lensing spectra are then given by $C_\ell^{gg} + 2C_\ell^{g\mu} + C_\ell^{\mu \mu}$ and $C_\ell^{\kappa g} + C_\ell^{\kappa \mu}$.

\subsubsection{Redshift Distribution and Bias Evolution} \label{subsubsec:dndz_bias_evol}

While the lensing projection kernel (Eq.~\ref{eq:kappa_projection_kernel}) is determined by the cosmological model alone, the galaxy and magnification bias kernels (Eqs.~\ref{eq:g_projection_kernel} and \ref{eq:mu_projection_kernel}) depend on measurements of the galaxy redshift distributions, $dN/dz$. As discussed in Sec.~\ref{subsubsec:dndz} we have two ways of measuring the redshift distribution of the unWISE samples, one relying on cross-correlations with spectroscopic tracers and one relying on cross-matching with a deep photometric sample from COSMOS.

Conveniently, cross-correlation redshifts are sensitive to the product $b_{1,E}(z) dN/dz$. We normalise those redshift distributions so that 
\beq
W^{\rm{xc}}(z) = \frac{\left(\widehat{b_{1,E}(z) \frac{dN}{dz}}\right)_{\rm{cross-correlation}}}{b^{\rm{eff}}_{1,E}},
\eeq
where $b_{\rm{eff}}$ is chosen such that $\int dz W^{\rm{xc}}(z) =1$. In this context $\widehat{X}$ shall denote our observational estimate of the quantity $X$. Wherever the product $b_{1,E}(z) W_g(z)$ appears in our model we thus use $b_g W^{\rm{xc}}(z)H(z)$\footnote{Note the factor of the Hubble rate $H(z)$ in the definition of $W_g(z)$ (Eq.~\ref{eq:g_projection_kernel}).} where $b_g$ is a free parameter in our analysis over which we marginalise. In this manner we encode the correct bias evolution for all the dominant terms in $C_\ell^{gg}$ and $C_\ell^{\kappa g}$ while allowing the amplitude of the bias to vary. This comes at the cost of introducing a dependence on the fiducial cosmology assumed when computing the cross-correlation redshifts. In Appendix~\ref{app:fid_cosmo_correction} we discuss how we correct for this effect. Where $W_g(z)$ appears in a term that is not also proportional to $b_{1,E}$ we rely on the cross-match derived redshifts, defining a normalised kernel
\beq
W_g(z) \to W^{\rm{xm}}(z)H(z) = \frac{\left(\widehat{\frac{dN}{dz}}\right)_{\rm{cross-match}}}{N} H(z)
\eeq
with $N$ chosen again such that $\int dz W^{\rm{xm}}(z) = 1$. A special case arises for terms proportional to $b_{1,L}(z)W_g(z)$. Since $b_{1,L} = b_{1,E} -1$ we model these terms as $b_{1,L}(z)W_g(z) \to b_g W^{\rm{xc}}(z)H(z) - W^{\rm{xm}}(z)H(z)$.

\subsubsection{Redshift Evolution of Higher Order Biases and Marginalisation}\label{subsubsec:higher_order_biases}

\begin{figure}
    \centering
    \includegraphics[width=\linewidth]{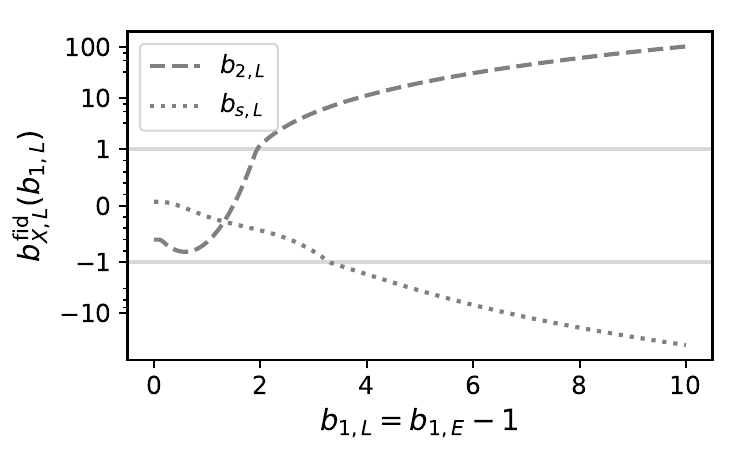}
    \caption{We show here the coevolution relations between the higher order Lagrangian bias parameters and the lowest order bias used to obtain the redshift evolution of higher order biases. The higher order Lagrangian biases, $b_{2,L}$ and $b_{s,L}$, are given as functions of the lowest order bias, $b_{1,L}$. These relations are derived from fits to numerical simulations presented in \cite{2018JCAP...07..029A}.}
    \label{fig:bias_coevol}
\end{figure}

\begin{figure}
    \centering
    \includegraphics[width=\linewidth]{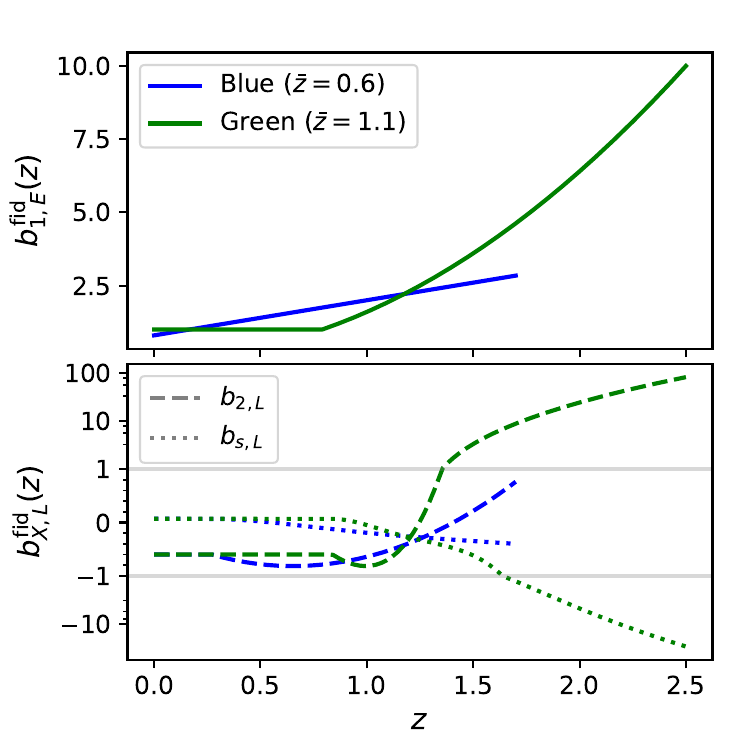}
    \caption{Here we show the fiducial redshift evolution of lowest order linear bias parameters (taken from \cite{2020JCAP...05..047K, 2021JCAP...12..028K}; \textbf{upper} panel) and the resulting Lagrangian bias evolution for all two unWISE samples (\textbf{lower} panel). The redshift evolution of the higher order biases is obtained by applying the coevolution relation shown in Fig.\,\ref{fig:bias_coevol} to the fiducial lowest order bias.}
    \label{fig:bias_redshift_evol}
\end{figure}

In the preceeding section we discussed the inclusion of the bias evolution for the lowest order bias. The higher order biases, relevant to the LPT terms in our model, also exhibit significant redshift evolution. Keeping with the model in \cite{2021JCAP...12..028K}, we capture the redshift dependence of $b_{2,L}$ and $b_{s,L}$ (in part) by defining them through co-evolution relations as functions of the lowest order bias $b_{1,L} = b_{1,E} -1$. To obtain $b^{\rm{co-evol.}}_{2,L}(b_{1,L})$ and $b^{\rm{co-evol.}}_{s,L}(b_{1,L})$ we use the relations presented in \cite{2018JCAP...07..029A} (see Fig.\,8 in \cite{2018JCAP...07..029A} and Fig.\,\ref{fig:bias_coevol} for our fit to those measurements) which are measured from clustering of protohalos in N-body simulations. Since we do not know $b_{1,L}$ this requires assuming a fiducial redshift evolution, $b_{1,L}^{\rm{fid}}$ which we adopt from \cite{2020JCAP...05..047K,2021JCAP...12..028K},
\begin{subequations}
\begin{align}
   b_{1,L}^{{\rm fid}}(z) &= 0.8 + 1.2 z - 1 &&\text{Blue} \\
   b_{1,L}^{{\rm fid}}(z) &= \max{(1.6z^{2},1)} - 1 &&\text{Green} \\
   b_{1,L}^{{\rm fid}}(z) &= \max{(2z^{1.5},1)} - 1 &&\text{Red}
\end{align}
\label{eqn:bias_evolution}
\end{subequations}
This evolution is consistent with the observed clustering and with the expected bias evolution from a simple HOD of the unWISE samples \citep{2020JCAP...05..047K}. We show these fiducial bias evolutions as well as the resulting evolution of $b_{2,L}$ and $b_{s,L}$ in Fig.~\ref{fig:bias_redshift_evol}.

In going beyond the model described in \cite{2021JCAP...12..028K}, we allow some freedom in the co-evolution by adding a free offset over which we marginalise in our analysis. The higher order biases are thus given effectively as $b_{X, L}(z) = b^{\rm{co-evol.}}_{X, L}(b_{1,L}^{\rm{fid}}(z)) + c_{X, L}$. To set priors on each $c_{X, L}$ we run our pipeline on the $N$-body simulations we use for model verification while holding the cosmology fixed. We then take the prior on $c_{X, L}$ to be a Gaussian approximation of the resulting posterior but increase the width (if necessary) such that the standard co-evolution is allowed at one sigma, i.e. $\mathcal{N}(\mu_{c_{X, L}}, \max(|\mu_{c_{X, L}}|, \sigma_{c_{X, L}}))$. 

\editF{Since we observe only projected spectra, redshift dependent deviations from the fiducial bias evolution can largely be absorbed in an effective amplitude. To demonstrate this we tested a rescaling of the fiducial bias evolution (in addition to the free offset), letting $b_{X, L}(z) = a_{X,L} b^{\rm{co-evol.}}_{X, L}(b_{1,L}^{\rm{fid}}(z)) + c_{X, L}$ where both $a_{X,L}$ and $c_{X, L}$ are free parameters. From runs on simulations we found these parameters to be nearly completely degenerate. Furthermore, we tested including an offset in the form of a third degree polynomial ($b_{X, L}(z) = b^{\rm{co-evol.}}_{X, L}(b_{1,L}^{\rm{fid}}(z)) + c_{X,L} + \sum_{n=1}^{3} c^{(n)}_{X, L} (1+z)^n$) after minimising over the constant offset within the prior range used in our model (while holding all other parameters fixed to their best-fit values) we find no significant deviations in the goodness of fit ($\Delta \chi^2<1$) even for a bias evolution which differs significantly ($\mathcal{O}(1)$) from the fiducial one. This indicates that the parameterisation through a single offset allows for sufficient freedom in the higher order contributions to the spectrum.}

In simulations we also find the cosmology dependence of all higher order contributions to the final projected spectra (arising from $P_{b_2}$, $P_{b_s}$, $P_{b_1b_2}$, etc) can effectively be absorbed into a change in the amplitude of those model components. We thus choose to compute them at a fixed fiducial cosmology taken to be the mean cosmology found in \cite{2020A+A...641A...6P}\footnote{Adopting the anlysis including $TT$, $TE$, and $EE$ spectra for multipoles $\ell \geq 30$, as well as the $EE$ spectrum for multipoles $2 \leq \ell \leq 29$, \textit{Planck} lensing, and BAO.}.

\subsubsection{Relative Contributions of Model Components} \label{subsubsec:model_contributions}

\begin{figure*}
    \centering
    \includegraphics[width=\linewidth]{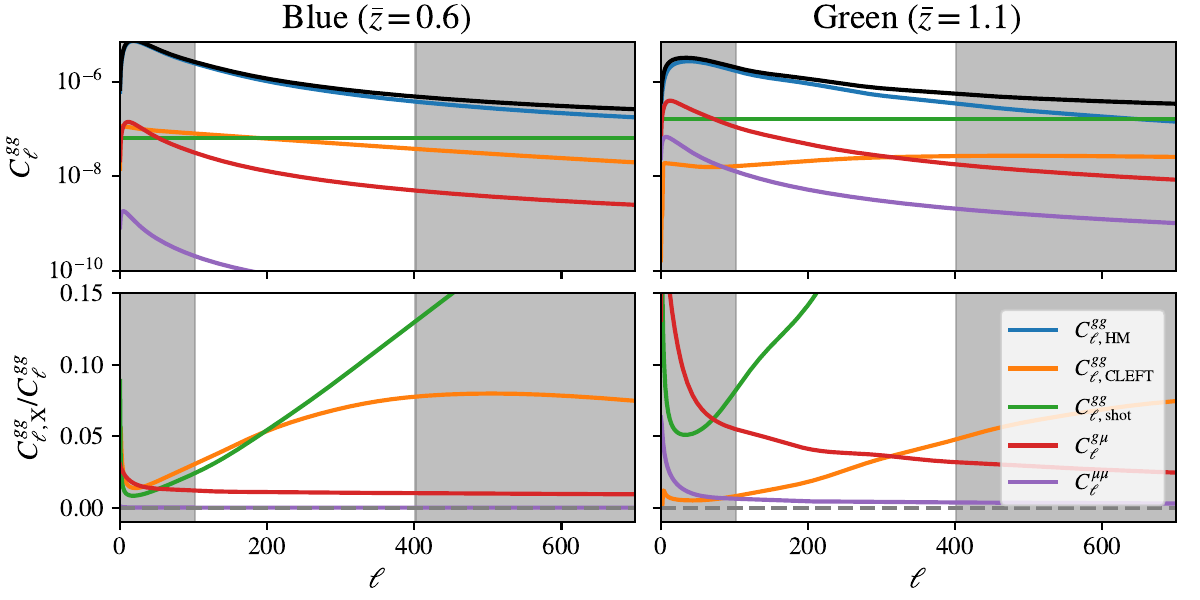}
    \caption{Contribution of various model components to $C_\ell^{gg}$ ($C_{\ell,\rm{HM}}^{gg}$: \texttt{HMCode} matter power spectrum; $C_{\ell,\rm{CLEFT}}^{gg}$: higher order terms; $C_{\ell,\rm{shot}}^{gg}$: shot noise; $C_{\ell}^{g\mu}$ and $C_{\ell}^{\mu\mu}$: magnification bias). The most important contributions to the galaxy power spectrum arise from the \texttt{HMCode} contribution at lowest order and the shot noise. The higher order contributions contribute at most 7\% and 5\% of the signal for the Blue and Green samples respectively. While lensing magnification plays a negligible role for the Blue sample it contributes up to 5\% of the signal on large scales for the Green sample. This figure assumes the best fitting cosmology for the joint fit to the cross-correlation of the Blue and Green samples with ACT DR6.\label{fig:gg_cobtributions}}
\end{figure*}

\begin{figure*}
    \centering
    \includegraphics[width=\linewidth]{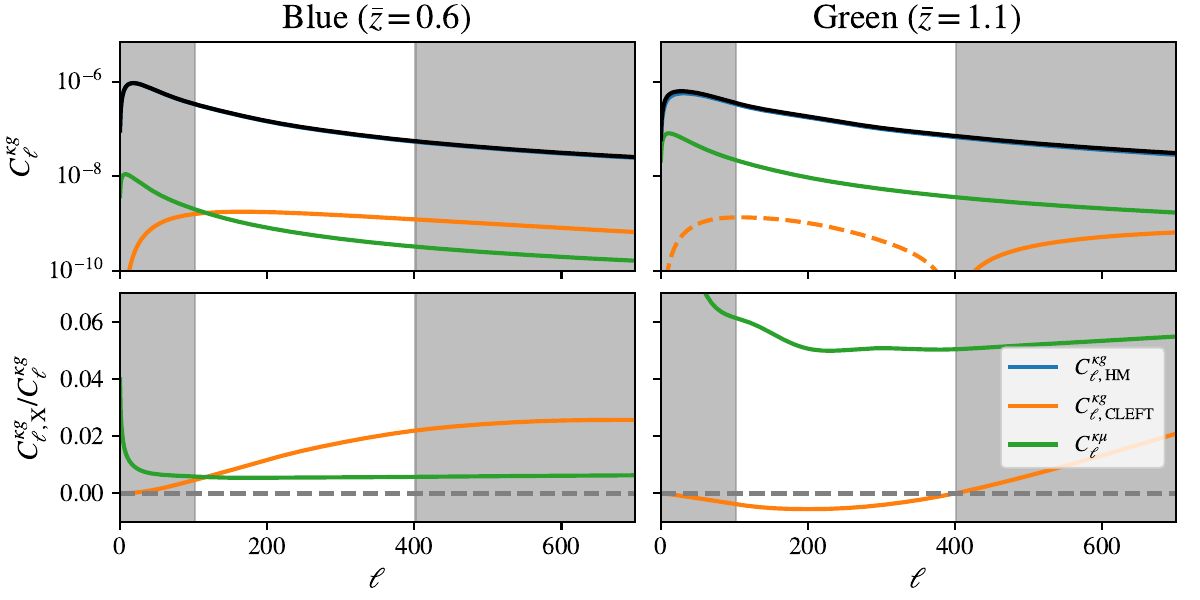}
    \caption{Contribution of various model components to $C_\ell^{\kappa g}$ ($C_{\ell,\rm{HM}}^{\kappa g}$: \texttt{HMCode} matter power spectrum; $C_{\ell,\rm{CLEFT}}^{\kappa g}$: higher order terms; $C_{\ell}^{\kappa \mu}$: magnification bias). As in the case of $C_\ell^{g g}$ the signal is dominated by the contribution from the \texttt{HMCode} power spectrum and has small but non-negligible contributions from lensing magnification ($<1\%$ for Blue and about 6\% for Green) and higher order contributions (up to about 2\% for Blue and $<1\%$ for Green).\label{fig:kg_cobtributions}}
\end{figure*}

In Fig.~\ref{fig:gg_cobtributions} we show the contributions of various model components assuming the best fitting cosmology for the joint fit to the cross-correlation of the Blue and Green samples with ACT DR6 including BAO information on the matter density (as described in Sec.~\ref{subsec:cosmo}). The contribution of higher order terms to the total $C_\ell^{gg}$ is at most 7\% for the Blue sample and 5\% for the Green sample. While lensing magnification does not exceed 1\% for the Blue sample it contributes up to $5\%$ to the signal on large scales for the Green sample. For $C_\ell^{\kappa g}$ the contributions from higher order terms are even more subdominant (at most 2\% for the Blue sample) (see Fig.~\ref{fig:kg_cobtributions}). Lensing magnification contributes about 6\% to the observed $C_\ell^{\kappa g}$ for the Green sample. Since these contributions to our model are subdominant we do not expect our results to be sensitive to the detailed modelling choices (e.g. the exact higher order bias evolution or the estimate of the magnification bias parameter).

\resub{In summary our power spectrum model has four free bias parameters per galaxy sample. This includes the amplitude of the linear order bias, free offsets to the fiducial redshift evolution for the second order and shear biases, and the magnification bias parameter. In addition we will also marginalise over the amplitude of the shot noise contribution.} In Sec.\,\ref{subsec:param_consistency_tests} we return to the impact of the different model components on our cosmological inference. We find that neglecting all higher order corrections only yields a small shift in $S_8$ ($\Delta S_8 \simeq 0.1\sigma$). However, \resub{it is know that the observed clustering statistics are in principle sensitive to the higher moments of the matter distribution and tidal effects which are encoded in the LPT terms. The impact the inclusion of these terms has on the mean inferred cosmological parameters} is sensitive to the value of the higher order biases in our data which is a priori unknown and so we conservatively maintain those free parameters in our model. \resub{This approach propagates the modelling uncertainty arising from our lack of knowledge of the higher order biases which our data does not constrain well to the inferred cosmological parameters of interest}. Neglecting the magnification bias terms leads to a shift of $\Delta S_8\simeq -0.2\sigma$\resub{, but again since such an effect is in principle expected we include it in our model to appropriately propagate the resulting uncertainties}.




\subsection{Marginalisation over Redshift Uncertainties} \label{subsec:dndz_marg}
As discussed in Sec.\,\ref{subsubsec:dndz} and \cite{2020JCAP...05..047K, 2021JCAP...12..028K} the measured redshift distribution of galaxies is subject to significant statistical errors. Here we describe a method based on principle component analysis for propagating these uncertainties to our cosmological results. 

In principle both our measurements of the redshift distribution, those based on cross matching with spectroscopic samples, $W^{\rm{xm}}(z)$, as well as those based on cross-correlations with spectroscopic surveys, $W^{\rm{xc}}(z)$, are affected by statistical uncertainty. However, the contribution to the uncertainty in our final spectra from any uncertainty in the cross-match redshift distributions is expected to be subdominant since those measurements only affect terms that do not depend on the galaxy bias, and these terms contribute at most at the few percent level to the total signal (see Figures~\ref{fig:gg_cobtributions} and \ref{fig:kg_cobtributions}). Hence we choose to neglect this source of uncertainty and focus instead on the uncertainty in the cross-correlation based redshift distribution measurements.

\begin{figure*}
    \centering
    \includegraphics[width=\linewidth]{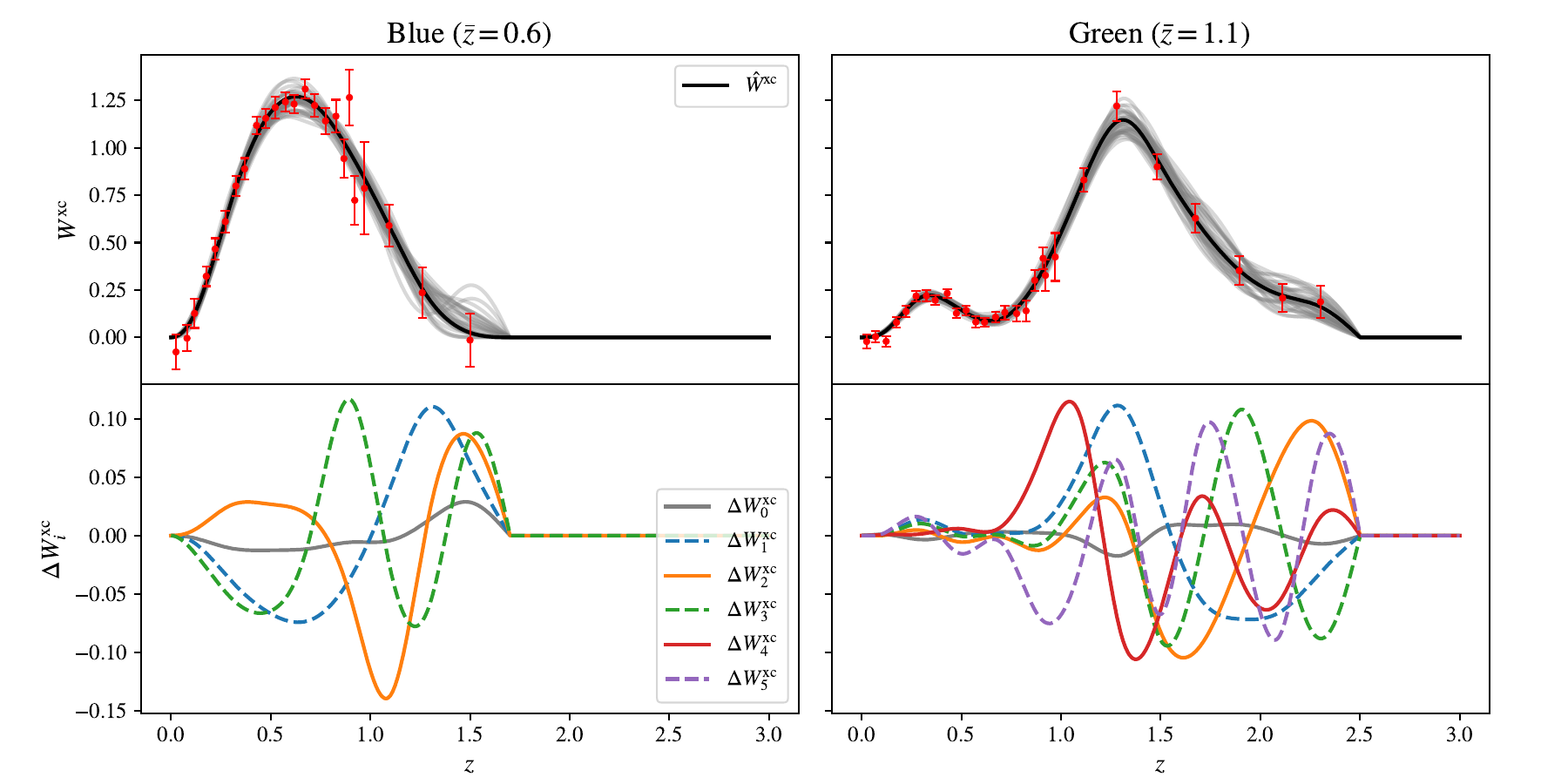}
    \caption{To propagate uncertainties in the measurement of the cross-correlation redshifts to our cosmological analysis we draw 1000 Gaussian realisations of $W^{\rm{xc}}$ in a manner consistent with the noise covariance (the best fitting $\hat{W}^{\rm{xc}}$ and a subset of the noisy realisations are are reproduced here from Fig.\,\ref{fig:dndz} to facilitate comparison with the PCA noise components). We then perform a principal component analysis on the difference between the realisations and the best fit. The lower panel shows the principal components ($i=1, \dots, 4$) along with the mean difference between the best fitting $\hat{W}^{\rm{xc}}$ and the noise realisations ($i=0$). In our cosmological analysis we marginalise over the principal component coefficients.}
    \label{fig:dndz_pca}
\end{figure*}

We generate 1000 realisations of $W^{\rm{xc}} \propto b(z) dN/dz$ as described in \cite{2021JCAP...12..028K}. We draw 1000 Gaussian realisations of the cross-correlation redshift estimate using the noise covariance obtained for those measurements and interpolate the resulting samples using a B-Spline with positivity constraint before normalising them. We then perform a principal component analysis on the difference of these realisations and the normalised spline interpolation of the best fit, $\hat{W}^{\rm{xc}}(z)$. We show the resulting principal components in Fig.\,\ref{fig:dndz_pca}. In our cosmology analysis we then use
\beq
W^{\rm{xc}}(z) = \hat{W}^{\rm{xc}}(z) + \Delta W_0^{\rm{xc}}(z) + \sum_{i=1}^n c_i \Delta W_i^{\rm{xc}}(z)
\eeq
where $\Delta W_0^{\rm{xc}}(z)$ is the mean difference between the 1000 random realisations and the best fit and $\Delta W_i^{\rm{xc}}(z)$ are the $i$-th principal components. The coefficients $c_i$ are marginalised over. We keep principal components up to $n=3$ and $5$ for the Blue and Green samples respectively. This $n$ is chosen to account for at least 90\% of the variance observed in the realisations. \resub{Due to the projection over redshift, oscillatory features in the redshift distribution of galaxies as expressed by higher principle components (which have an increasing number of nodes) will contribute less to the projected power spectrum. Therefore, it is expected that while we capture only 90\% of the variance in the bias weighted redshift distribution, a larger fraction of the induced power spectrum variance is due to the first few principal components. Using a larger number of principal components (capturing 99\% of the $W^{\rm xc}$ variance; $n=6$ and 9 for the Blue and Green samples respectively) we checked that the residual $C_\ell^{gg}$ variance not captured by the first few principle components is at most $\sim$10\% of the uncertainty on galaxy power spectrum, while the variance captured by this marginalisation is comparable to the $1\sigma$ errorbars. Nevertheless, we also perform a consistency test by running the full parameter inference pipeline with this larger set of principal components and find no significant shift in the mean cosmological parameters. The uncertainty on $S_8^{\times}$ is inflated by 2\% and 14\% when analysing the Blue and Green samples respectively and remains the same within the convergence limits of our chains when combining both samples (see also Sec.\,\ref{subsec:param_consistency_tests}).}

\begin{figure*}
    \centering
    \includegraphics[width=\linewidth]{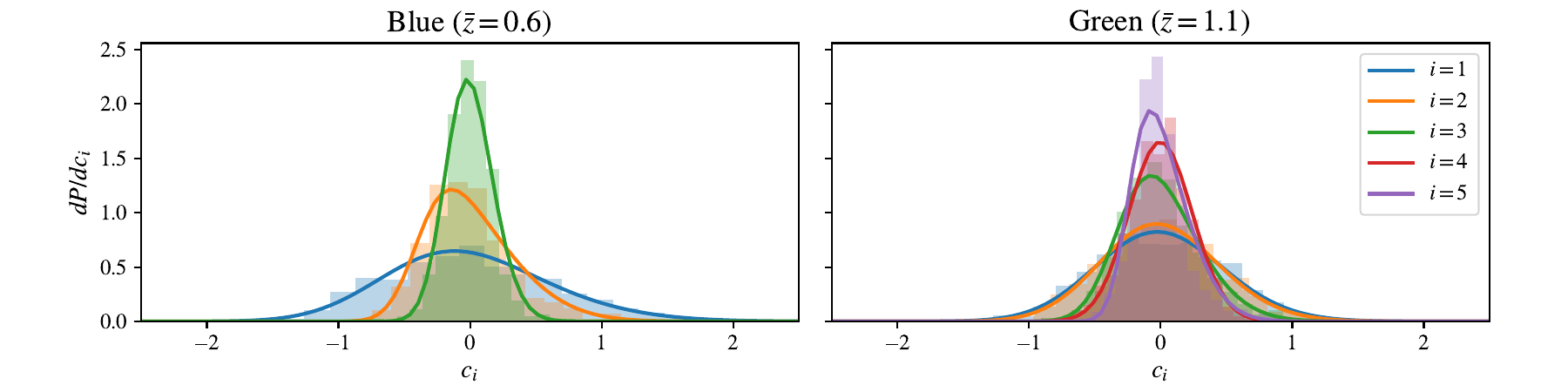}
    \caption{We obtain priors on the principal component coefficients, $c_i$, by transforming the 1000 realisations of $W^{\rm{xc}}$ into the basis of principal components and fit the distribution of coefficients observed using skewnormal distributions.}
    \label{fig:dndz_pca_coeff}
\end{figure*}
The uncertainty on our measurement of the clustering redshifts is then expressed in terms of priors on the $c_i$. We transform the 1000 realisations above into the basis of principal components and fit the distribution of coefficients observed using skew-normal distributions (shown in Fig.~\ref{fig:dndz_pca_coeff}) which we use as priors on each $c_i$ in our analysis.

We note, that the marginalisation over the uncertainties in the redshift distribution measurements introduces a noise bias in our model prediction. This is because we enforce a positivity constraint on the spline interpolation of the cross-correlation redshift estimates. Additionally, $C_\ell^{gg}$ depends on the square of $W^{xc}(z)$ and so the final spectra are a non-linear function of this noisy quantity. The resulting noise bias can be as large as 5\% for some of the components contributing to $C_\ell^{gg}$. Our method for mitigating this bias is discussed in Appendix~\ref{app:noise_bias}.

\subsection{Testing the Model on Mocks}\label{subsec:model_test}

For model verification we adopt the CrowCanyon simulations discussed in \cite{2021JCAP...12..028K}. These simulations are generated using the FastPM code \citep{2016MNRAS.463.2273F} using $8192^3$ equal mass particles in a $4096 h^{-1} \rm{Mpc}$ a side cubic volume with periodic boundary conditions. The resulting mass resolution is $1.1\times 10^{10} h^{-1} M_\odot$. The simualtion uses a fiducial cosmology close to the mean cosmology found in \cite{2020A+A...641A...6P} with $\Omega_m = 0.3092$, $\Omega_b = 0.0496$, $h = 0.677$, $n_s =0.968$, and $\sigma_8=0.822$.

The lensing convergence and magnification fields are computed by integrating the matter density on the lightcone and weighting by the appropriate kernels \citep[for details see][]{2021JCAP...12..028K}. We note that since no lensing reconstruction is performed the lensing observable does not include any reconstruction noise. To obtain realistic galaxy samples a HOD tuned to reproduce the unWISE samples is applied to the halos identified in the CrowCanyon simulation. This HOD is available as part of the \textsc{SIMPLEHOD} package\footnote{\url{https://github.com/bccp/simplehod/blob/master/scripts/wlen.py}}. Finally, $C_\ell^{gg}$ and $C_\ell^{\kappa g}$ are measured from these simulations on the full sky; they hence differ from data in that no mask has been applied and consequently no mode coupling is introduced. 

The total simulated volume is $\sim 69 h^{-3} \rm{Gpc}^3$ and thus not overwhelmingly larger than the volume probed by the data (approximately by a factor of 4). Since we restrict ourselves to relatively large scales in our analysis we expect some fluctuations due to cosmic variance and sample variance. Therefore, we consider our model unbiased if we are able to recover the input cosmology to within $0.5\sigma$ when using the covariance appropriate for our observations with real data. We also perform a test in which we adopt a theory covariance appropriate for the simulation volume which additionally neglects the lensing reconstruction noise as this is not present in the mocks. For this test we still recover the input cosmology to within $0.7\sigma$ across all cosmological parameters (see Table~\ref{table:param_recovery}).

\cite{2021JCAP...12..028K} measured $b(z) dN/dz$ for these mock samples and found a good match to the observed clustering redshift estimates. Random realisations of $W^{xc}(z)$ were generated in a similar way as for the data assuming the same noise covariance found for the data.

We find that modelling biases on all cosmological parameters considered in our analysis are well under control, not exceeding $0.3\sigma$ in all cases (see Table \ref{table:param_recovery}). This is a slight improvement over the $\sim$0.5$\sigma$ modelling biases in \cite{2021JCAP...12..028K}.

\begin{deluxetable*}{lcDcDcDcD}
    \tablehead{
    \nocolhead{}    & \colhead{$\Omega_m$} & \multicolumn{2}{c}{$\Delta\Omega_m/\sigma_{\Omega_m}$} & \colhead{$\sigma_8$} & \multicolumn{2}{c}{$\Delta\sigma_8/\sigma_{\sigma_8}$} & \colhead{$S_8$} & \multicolumn{2}{c}{$\Delta S_8/\sigma_{S_8}$} & \colhead{$S_8^{\times}$} & \multicolumn{2}{c}{$\Delta S_8^{\times}/\sigma_{S_8^{\times}}$} }
    \startdata
    \decimals
    Simulation input value& 0.3092 & \multicolumn{2}{c}{$-$} & 0.8222 & \multicolumn{2}{c}{$-$} & 0.8347 & \multicolumn{2}{c}{$-$} & 0.8335 & \multicolumn{2}{c}{$-$}\\
    &\multicolumn{12}{c}{Simulation errors}\\
    Blue & $0.319^{+0.034}_{-0.043}$ & 0.19 & $0.825\pm 0.043$ & 0.05 & $0.8466^{+0.0091}_{-0.021}$ & 0.60 & $0.8443^{+0.0076}_{-0.018}$ & 0.67 \\
    Green & $0.304^{+0.016}_{-0.018}$ & -0.31 & $0.830\pm 0.018$ & 0.40 & $0.8350^{+0.0095}_{-0.013}$ & -0.05 & $0.8344^{+0.0080}_{-0.012}$ & 0.01 \\
    Joint & $0.306^{+0.013}_{-0.015}$ & -0.26 & $0.829\pm 0.016$ & 0.39 & $0.8360^{+0.0068}_{-0.0089}$ & 0.10  & $0.8353^{+0.0059}_{-0.0078}$ & 0.20 \\
    &\multicolumn{12}{c}{ACT DR6 $\times$ unWISE-like errors}\\
    Blue   & $0.335^{+0.053}_{-0.100}$ & 0.18 & $0.819^{+0.083}_{-0.098}$ & -0.07  & $0.848^{+0.026}_{-0.041}$ & 0.28 & $0.844^{+0.024}_{-0.036}$ & 0.26  \\
    Green  & $0.318^{+0.035}_{-0.043}$ & 0.18 & $0.817^{+0.043}_{-0.048}$ & -0.14 & $0.837\pm 0.025$ & 0.07 & $0.835\pm 0.023$ & 0.05 \\
    Joint & $0.318^{+0.033}_{-0.041}$ & 0.19 & $0.816^{+0.042}_{-0.047}$ & -0.16 & $0.836\pm 0.020$ & 0.08 & $0.834\pm 0.019$ & 0.03
    \enddata
    \caption{When analysing synthetic galaxy clustering and galaxy-lensing cross-correlation observations from $N$-body simulations we recover the input cosmology to within better than $0.3\sigma$ when adopting the appropriate data covariance for our ACT DR6 cross-correlation analysis. With the smaller theory covariance appropriate for the (somewhat larger) volume of the simulations we still recover the input cosmology to within about $0.6\sigma$ in terms of the now substantially tighter constraints. \label{table:param_recovery}}
\end{deluxetable*}

To establish that an alternative, much simpler choice of model, would not suffice to model the data we analyse the $N$-body simulations described above with alternative models. We consider two such simpler models: using only the matter power spectrum from \texttt{HMCode} or adopting a fixed fiducial evolution of higher order bias parameters without a free offset \citep[as in][]{2021JCAP...12..028K}. For this test we adopt the covariance appropriate for the simulation volume to perform a maximally stringent test of the input parameter recovery. We find that both models are significantly biased, exhibiting shifts much larger than the expected $\sim$$1\sigma$. A \texttt{HMCode}-only model underpredicts $S_8$ by $\sim$$2.2\sigma$, while the model from \citep{2021JCAP...12..028K} yields a value for $S_8$ which is about $4\sigma$ larger than the true value in the simulations. The shifts are less significant when considering a realistic data covariance ($\sim$$0.3\sigma$ for \texttt{HMCode}-only model and $\sim$$1.2\sigma$ for the model with fixed higher order biases), but because the value of the true higher order biases is unknown and could induce larger biases in our data analysis we adopt the more conservative model.



\section{Cosmological Analysis}\label{sec:cosmo}

\subsection{Blinding Policy} \label{subsec:blinding}
In the process of preparing the analysis presented in this paper, constraints on cosmological parameters were blinded until we could demonstrate passing a sequence of tests detailed below. We were, however, not blind to the measured spectra which, at least in the case of $C_\ell^{gg}$ had already been presented in previous work \citep{2020JCAP...05..047K}. Our process towards unblinding cosmological constraints involved the following steps:
\begin{itemize}
    \item We tested parameter recovery with our model on N-body simulations as discussed in Sec.\,\ref{subsec:model_test}. Since the volume of these simulations is comparable to that of the observations, deviations from the input cosmology are expected. Hence, we consider our model to be unbiased if we are able to recover the input cosmology to within $0.5\sigma$ when using the covariance appropriate for our observations.
    \item To verify that our data are not contaminated by foregrounds, including Galactic and extragalactic foregrounds, we run a series of tests described in Sec.\,\ref{sec:sys_test}. We do not perform such extensive testing for the galaxy-galaxy auto spectrum since these have been presented in previous work \citep{2020JCAP...05..047K,2021JCAP...12..028K}. However, since we added systematics weighting for this work, we perform a smaller number of tests to re-verify the stability of the galaxy-galaxy autospectrum to different choices of the mask. 
    
    We qualify a test to be passing if it yields a PTE greater than 0.05. We are satisfied that our mitigation strategies are sufficient if the number of failures ($\rm{PTE}<0.05$) is approximately consistent with what would be expected from random fluctuations when accounting for the correlations between different tests\footnote{We chose to not consider tests yielding $\rm{PTE}>0.95$ as failures, to minimise the potential for spurious failures and to focus our null-tests on systematic contamination of the data in contrast to a miss-modelling of the covariance matrix. Furthermore, the lensing noise modelling has already been tested in detail in \cite{2024ApJ...962..112Q}. We also note, after the fact, that we do not observe any $\rm{PTE}>0.95$.}.
    \item As a final test we investigate the consistency of parameters recovered with different analysis choices, including different sky masks, different lensing reconstruction and bias mitigation methods, varying choices for the measurement of the cross-correlation redshifts, and for the higher order bias priors. For this purpose we run parameter inference on the data, but add random offsets to all cosmological parameters that we explicitly sample in the analysis or vary as a function of other parameters ($A_s$, $\Omega_m$, $\omega_{\rm{cdm}}$ $H_0$, $\sigma_8$ and $S_8$).
    \item Before unblinding we freeze all baseline analysis choices. This includes the range of angular scales used in $C_\ell^{gg}$ and $C_\ell^{\kappa g}$ and all priors on cosmological as well as nuisance parameters.
\end{itemize}

\subsection{Post-unblinding Changes}\label{subsec:post_unblinding_changes}

A few minor aspects of the analysis were corrected post-unblinding, but the impact of these change is small and they partially offset.  Consequently, the inferred value of $S_8$ for our baseline analysis including only the galaxy auto-correlation of both redshift samples and the cross-correlations with CMB lensing from ACT is practically unaffected, changing by $-0.1\sigma$ from our initial unblinded results ($\Delta S_8=-0.002$, $-0.25\%$). The post-unblinding changes included a modification to the treatment of mask induced mode coupling of the shot noise component and a small correction to the transfer function obtained from simulations ($+0.2\sigma$). Additionally, the cross-covariance between the two redshift samples had erroneously been neglected ($-0.3\sigma$). Finally, we noticed a bug in the model code which meant that the fiducial bias evolution for $b_{2,L}$ was erroneously adopted for $b_{s,L}$ as well. This correction required redetermining the priors on the higher order biases from the $N$-body simulations and re-verifying the model. Fortunately, the model remains unbiased and we infer the same value of $S_8$ for the joint analysis of the cross-correlations of the Blue and Green sample with ACT with only a 5\% increase in uncertainty. More detail on these post-unblinding changes is provided in Appendix~\ref{app:post-unblinding_changes}.

\subsection{Parameter Inference and Priors} \label{subsec:inference_priors}
We obtain cosmological constraints by constructing a Gaussian likelihood
\beq
-2 \ln \mathcal{L} \propto \sum_{bb'}\begin{bmatrix}\Delta \hat{C}_b^{gg}(\bm{\theta}) \\ \Delta \hat{C}_b^{\kappa g}(\bm{\theta}) \end{bmatrix} \mathbb{C}^{-1} \begin{bmatrix}\Delta \hat{C}_{b'}^{gg}(\bm{\theta}) \\ \Delta \hat{C}_{b'}^{\kappa g}(\bm{\theta})\end{bmatrix} 
\eeq
where the $\Delta \hat{C}_b^{gg}$ and  $\Delta \hat{C}_b^{\kappa g}$ are the residuals between our observed galaxy-galaxy and galaxy-lensing spectra, $\hat{C}_b^{gg}$ and $\hat{C}_b^{\kappa g}$, and the respective band window convolved theory spectra, $C_b^{gg}$ and $C_b^{\kappa g}$. The covariance $\mathbb{C}$ has the form
\beq
\mathbb{C} = \begin{bmatrix}\mathbb{C}_{b b'}^{gg-gg} & \mathbb{C}_{b b'}^{gg-\kappa g} \\ \left(\mathbb{C}_{b b'}^{gg-\kappa g}\right)^T & \mathbb{C}_{b b'}^{\kappa g-\kappa g}\end{bmatrix}
\eeq
where $\mathbb{C}_{b b'}^{gg-gg}$, $\mathbb{C}_{b b'}^{\kappa g-\kappa g}$, and $\mathbb{C}_{b b'}^{gg-\kappa g}$ are the galaxy auto-spectrum covariance, the galaxy-lensing cross-spectrum covariance, and the cross-covariance between them. These are estimated from simulations as described above in Sec.\,\ref{sec:covmat}.

We use the Markov chain Monte Carlo code \texttt{cobaya}\footnote{\url{https://github.com/CobayaSampler/cobaya}}\citep{2021JCAP...05..057T} to infer parameters from our galaxy-galaxy and galaxy-CMB lensing data using the model described in Sec.\,\ref{sec:model}. We consider chains to be converged if the Gelman-Rubin statistic \citep{1992StaSc...7..457G,Gelman1998} satisfies $R-1 \leq 0.01$.

Our data is insensitive to the optical depth to reionisation and we thus fix it to the best fit value from \textit{Planck}\footnote{We adopt the results obtained in \cite{2020A+A...641A...6P} by including the TT,TE, and EE spectra for multipoles $\ell \geq 30$, the EE spectrum for lower multiples $2 \leq \ell \leq 29$ ("lowE" likelihood), and additionally \textit{Planck} lensing and BAO \cite[see Table 2 therein]{2020A+A...641A...6P}.}, $\tau = 0.0561$ \citep{2020A+A...641A...6P}. Since we are using low redshift projected tracers alone, the information from the BAO feature in the power spectrum is largely erased and thus our data is mostly sensitive to the total matter density, $\Omega_m$. We therefore again choose to fix $\Omega_b h^2$ to the central value from \textit{Planck}, $\Omega_b h^2 = 0.2242$ \citep{2020A+A...641A...6P}, and vary only $\Omega_m$. Additionally, our measurement of the lensing amplitude is largely degenerate with the distance to the unWISE redshifts which is primarily set by the Hubble parameter, $h$. In keeping with \cite{2021JCAP...12..028K} we break this degeneracy by fixing the projected angular size of the sound horizon to the value measured from the CMB by the Planck Collaboration, $\theta_{\rm{MC}}$. It can be shown that $\theta_{\rm{MC}}$ is predominantly sensitive to the product $\Omega_m h^3$ \citep{2002MNRAS.337.1068P}. Hence, we fix this combination to $\Omega_m h^3=0.09635$, the mean value obtained by the Planck Collaboration \citep{2020A+A...641A...6P}. It should be noted that this combination is largely independent of the complex physics and possible observational systematics that affect the broadband shape of the CMB. It is determined to approximately 0.3\% by \textit{Planck} so that any uncertainty on this combination would be subdominant to our measurement uncertainties. Finally, we also fix the tilt of the primordial power spectrum to $n_s=0.9665$ \citep{2020A+A...641A...6P} and assume the minimum neutrino mass allowed in the normal hierarchy ($\sum m_\nu = 0.06$ eV). \resub{Low redshift data alone is largely insensitive to the sum of the neutrino masses; we will explore the implications of our data in combination with high redshift information from the primary CMB for neutrino mass constraints in future work.} We explore an alternative set of priors on cosmological parameters adopted from the ACT DR6 lensing power spectrum analysis \citep{2024ApJ...962..112Q} in Appendix~\ref{app:alt_priors}. We show that these alternative priors do not significantly affect any of our conclusions.

Our procedure for setting priors on the higher order bias parameters, $c_{X, L}$, and the redshift marginalisation parameters, $c_{dN/dz \rm{\ PCA}, i}$, are described in detail in Sections \ref{subsec:pk_model} and \ref{subsec:dndz_marg}, respectively, and the resulting priors are summarised in Table \ref{tab:priors}. We note that the priors on the redshift marginalisation parameters shown here are the ones adopted for our baseline analysis which includes both the full spectroscopic data set from BOSS and eBOSS and the appropriate weighting of the unWISE data to determine the cross-correlation redshifts (see Sec.\,\ref{subsubsec:dndz}). For some of our consistency tests we adopt different subsets of the spectroscopic data or neglect the systematics weighting when measuring the cross-correlation redshifts; in each case we adopt the priors appropriate for the specific cross-correlation measurement used which can differ somewhat from the ones shown here. As in \cite{2021JCAP...12..028K} we adopt 10\% Gaussian priors on the measured lensing magnification bias and Gaussian priors with standard deviation $0.2$ on the logarithm of the shot noise, $\log_{10} P_{\rm{shot}}$ (translating to a prior of approximately $60\%$ on the amplitude of the shot noise).

\begin{table}
    \centering
    \begin{tabular*}{\columnwidth}{ l  @{\extracolsep{\fill}} c  c}
    \hline
    \hline
    Parameter   & Prior  \\ \hline
    \multicolumn{2}{l}{\bf Sampled cosmological parameters} \\
    $\Omega_m$ & $[0.01, 0.95]$\\ 
    $\ln(10^{10} A_s)$ & $[1.0, 4.0]$\\\hline
    \multicolumn{2}{l}{\bf Fixed and derived cosmological parameters} \\
    $n_s$& fixed (0.9665)\\
    $\tau$& fixed (0.0561)\\
    $\Omega_m h^3$ & fixed (0.09635)\\
    $\Omega_b h^2$ & fixed (0.02242)\\
    $\sum m_\nu$ & fixed (0.06 eV)\\
    $h$ & derived ($\left[\Omega_m h^3 /\Omega_m\right]^{1/3}$)\\
    $\Omega_c h^2$ & derived ($\Omega_m h^2 - \Omega_b h^2 - \sum m_\nu / 93.14 \rm{\ eV}$)\\ \hline \hline
    \multicolumn{2}{l}{\bf Galaxy model parameters} \\ \hline
    \multicolumn{2}{l}{\bf Blue sample ($\bar{z}=0.6$)}\\
    $b^{\rm{Blue}}_{1,E}$&$[0.5,3.9]$\\
    $\log_{10} P_{\rm{shot}}^{\rm{Blue}}$& $\mathcal{N}(-7.05, 0.2)$\\
    $s^{\rm{Blue}}_{\mu}$&$\mathcal{N}(0.455, 0.0455)$\\
    $c^{\rm{Blue}}_{2,L}$& $\mathcal{N}(0.55, 0.55)$\\
    $c^{\rm{Blue}}_{s,L}$& $\mathcal{N}(0.17, 0.29)$\\
    $c_{dN/dz \rm{\ PCA}, 1}^{\rm{Blue}}$& $\mathcal{N}_s(-0.58, 0.86, 1.61)$\\
    $c_{dN/dz \rm{\ PCA}, 2}^{\rm{Blue}}$& $\mathcal{N}_s(-0.40, 0.54, 2.94)$\\
    $c_{dN/dz \rm{\ PCA}, 3}^{\rm{Blue}}$& $\mathcal{N}_s(-0.14, 0.23, 1.27)$\\ \hline
    \multicolumn{2}{l}{\bf Green sample ($\bar{z}=1.1$)}\\
    $b^{\rm{Green}}_{1,E}$&$[0.7,4.2]$\\
    $\log_{10} P_{\rm{shot}}^{\rm{Green}}$& $\mathcal{N}(-6.79, 0.2)$\\
    $s^{\rm{Green}}_{\mu}$&$\mathcal{N}(0.653, 0.0653)$\\
    $c^{\rm{Green}}_{2,L}$& $\mathcal{N}(0.42, 0.42)$\\
    $c^{\rm{Green}}_{s,L}$& $\mathcal{N}(0.22, 0.50)$\\
    $c_{dN/dz \rm{\ PCA}, 1}^{\rm{Green}}$& $\mathcal{N}_s(-0.30, 0.57, 0.88)$\\
    $c_{dN/dz \rm{\ PCA}, 2}^{\rm{Green}}$& $\mathcal{N}_s(-0.30, 0.54, 0.99)$\\
    $c_{dN/dz \rm{\ PCA}, 3}^{\rm{Green}}$& $\mathcal{N}_s(-0.31, 0.44, 1.90)$\\
    $c_{dN/dz \rm{\ PCA}, 4}^{\rm{Green}}$& $\mathcal{N}_s(-0.09, 0.26, 0.51)$\\
    $c_{dN/dz \rm{\ PCA}, 5}^{\rm{Green}}$& $\mathcal{N}_s(-0.24, 0.33, 2.65)$\\ \hline

    \end{tabular*}
    \caption{Parameters and priors used in this work. $\mathcal{N}( \mu, \sigma)$ indicates a Gaussian prior with mean $\mu$ and variance $\sigma^2$, while $\mathcal{N}_s(\xi, \omega, \alpha)$ indicates a skew-normal prior with location $\xi$, scale $\omega$ and shape $\alpha$. Uniform priors are indicated by square brackets. The priors on the magnification bias parameters are adopted from \cite{2020JCAP...05..047K}, the procedure for obtaining priors on the free parameters in the higher order bias evolution is described in Sec.\,\ref{subsubsec:higher_order_biases}, and the priors on the redshift marginalisation parameters are described in Sec.\,\ref{subsec:dndz_marg}.}
    \label{tab:priors}
\end{table}

\subsection{BAO Likelihoods}\label{subsec:bao_lklh}

Weak lensing measurements depend primarily on the amplitude of matter fluctuations $\sigma_8$ and the matter density $\Omega_m$. In order to break the degeneracies of our $\sigma_8$ constraint with the latter parameter and allow for more powerful comparisons with other probes (including the primary CMB, optical weak lensing measurements, and constraints from CMB lensing auto-spectra), we include information on Baryon Acoustic Oscillations from the 6dF and SDSS surveys. We adopt the same likelihoods used in \cite{2024ApJ...962..113M}. The data we include measure the BAO signature in the clustering of galaxies with samples spanning redshifts up to $z\simeq 1$, including 6dFGS \citep{2011MNRAS.416.3017B}, SDSS DR7 Main Galaxy Sample (MGS;~\citealt{2015MNRAS.449..835R}), BOSS DR12 LRGs \citep{2017MNRAS.470.2617A}, and eBOSS DR16 LRGs \citep{2021PhRvD.103h3533A}. We do not use the higher-redshift ELGs \citep{2016A+A...592A.121C}, Lyman-$\alpha$ \citep{2020ApJ...901..153D}, and quasar samples \citep{2021PhRvD.103h3533A}, though we hope to include these in future analyses. We only include the BAO information from these surveys (which for our purposes primarily constrains $\Omega_{m}$) and do not include the structure growth information in the redshift-space distortion (RSD) component of galaxy clustering. We make this choice so as to isolate information on structure formation purely from the cross-correlation alone.

\subsection{Constraints on Cosmological Parameters from the cross-correlation of ACT DR6 Lensing and unWISE} \label{subsec:cosmo}

\begin{figure*}
    \centering
    \includegraphics[width=0.5\linewidth,trim=0.5cm 0.5cm 0.5cm 0.5cm, clip]{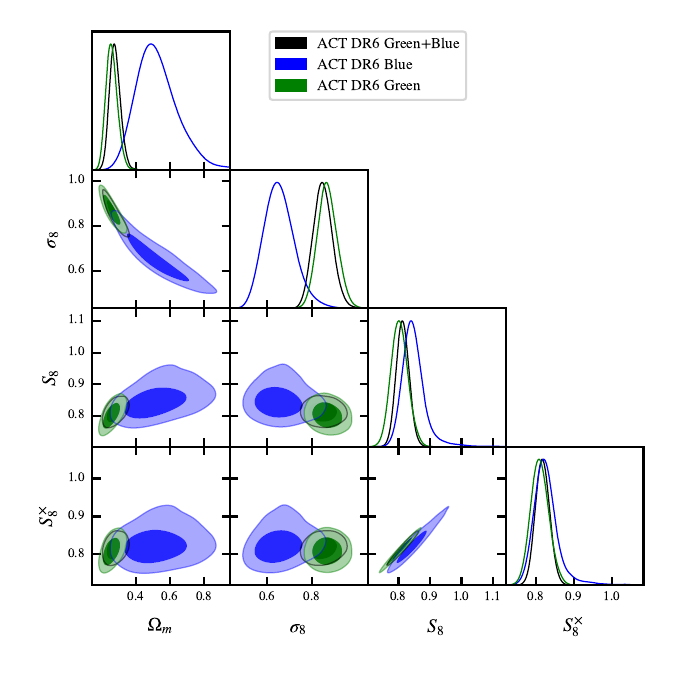}%
    \includegraphics[width=0.5\linewidth,trim=0.5cm 0.5cm 0.5cm 0.5cm, clip]{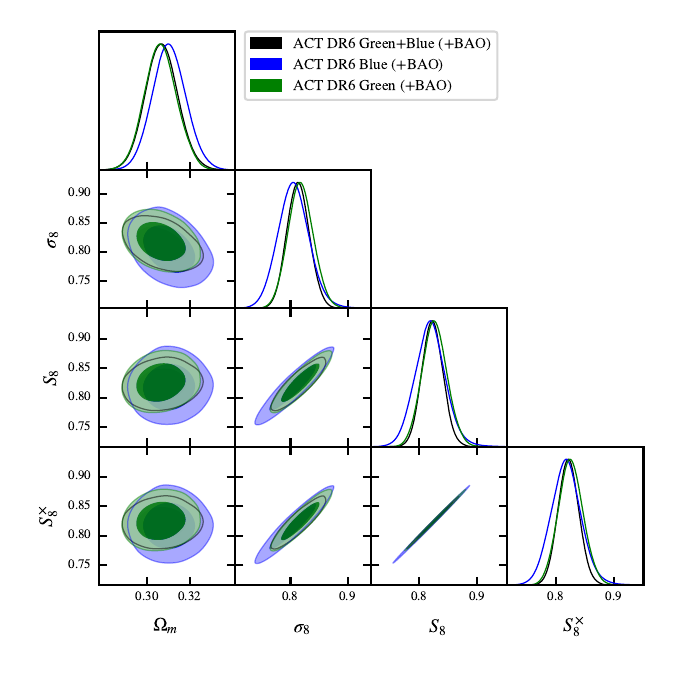}
    \caption{Parameter constraints from the cross-correlation of ACT DR6 lensing and unWISE galaxies. From the combination of $C_\ell^{gg}$ and $C_\ell^{\kappa g}$ only for both samples of unWISE galaxies (\textbf{left}) we find $S_8 = 0.813\pm 0.021$. As described in the test the Blue and Green sample are consistent to within $1.7\sigma$. With additional information on $\Omega_m$ from BAO (\textbf{right}) the degeneracy between $\Omega_m$ and $\sigma_8$ is broken and we place competitive constraints on $\sigma_8=0.813\pm 0.020$.}
    \label{fig:actXunWISE_corner_plots}
\end{figure*}

Analysing jointly the auto-correlation of the two samples of unWISE galaxies and the cross-correlation of each with the ACT DR6 lensing reconstruction we obtain a $2.6\%$ constraint on $S_8\equiv\sigma_8 (\Omega_m/0.3)^{0.5}$ of 
\beq
S_8 = 0.813\pm 0.021.
\eeq 
The best-constrained parameter in our analysis actually differs slightly from $S_8$ and we empirically determined it to be closer to $S_8^{\times} \equiv \sigma_8 (\Omega_m/0.3)^{0.45}$ which we constrain to $2.3\%$ ($S_8^{\times} = 0.817 \pm 0.019$). All parameter constraints are summarised in Table \ref{table:results}, with the posteriors on cosmological parameters shown in Fig.~\ref{fig:actXunWISE_corner_plots}. 

We note that when analysing only a single tomographic sample without additional information to break the degeneracy between $\sigma_8$ and $\Omega_m$ these parameters are individually only poorly constrained. While the marginalised one dimensional posteriors on $\Omega_m$ and $\sigma_8$ obtained from the Blue and Green sample may visually appear to be in some tension, the two samples are consistent within $1.7\sigma$ in the full $\sigma_8$-$\Omega_m$ parameter space\footnote{We used the \textsc{tensiometer} package to estimate the significance (\url{https://github.com/mraveri/tensiometer})}. We discuss the consistency of the two redshift samples in more detail in Appendix~\ref{app:sample_consistency}.

We can break the degeneracy between $\sigma_8$ and $\Omega_m$ through the addition of external information. In particular, we combine our analysis with publicly available BAO likelihoods described in Sec.~\ref{subsec:bao_lklh} which primarily constrain the matter density, $\Omega_m$. In the combination with BAO we are able to place competitive constraints on $\sigma_8$ alone. With the degeneracy broken the amplitude of the spectra becomes the dominant source of information on $\sigma_8$. We find that the combination of the cross- and auto-correlations of the Blue and Green samples with BAO yields 
\beq
\sigma_8=0.813\pm 0.020.
\eeq

The change in $S_8$ inferred from the joint analysis of the Blue and Green cross-correlations using the alternative priors from Appendix~\ref{app:alt_priors} is small (0.3\% or $0.14\sigma$); we find $S_8=0.810 \pm 0.021$ indicating that the choice of alternative priors does not significantly affect our constraining power on $S_8$. When including BAO we find $\sigma_8=0.813 \pm 0.022$ for this alternative set of priors, showing no shift from the mean value obtain with our baseline analysis, but 10\% wider errors.

\begin{deluxetable*}{lCCCCCCCC}
    \tablehead{
    \nocolhead{}    & \multicolumn{2}{c}{$\Omega_m$} & \multicolumn{2}{c}{$\sigma_8$} & \multicolumn{2}{c}{$S_8$} & \multicolumn{2}{c}{$S_8^{\times}$} }
    \startdata
    &\multicolumn{8}{c}{ACT DR6 $\times$ unWISE only}\\
    Blue & 0.523^{+0.093}_{-0.14} & (0.476) & 0.654^{+0.058}_{-0.074} & (0.659) & 0.849^{+0.024}_{-0.040} & (0.831) & 0.827^{+0.021}_{-0.035} & (0.812)\\
    Green & 0.259^{+0.029}_{-0.037}& (0.245) & 0.868^{+0.041}_{-0.046}& (0.878) & 0.803^{+0.025}_{-0.028} & (0.793) & 0.809^{+0.022}_{-0.025} & (0.801)\\
    \textbf{Joint} & 0.279^{+0.028}_{-0.035} & (0.276) & 0.843^{+0.038}_{-0.044} & (0.848) & 0.813\pm 0.021 & (0.813) & 0.817\pm 0.019 & (0.816)
    \\ \hline
    &\multicolumn{8}{c}{ACT DR6 $\times$ unWISE + BAO}\\
    Blue & 0.3102\pm 0.0079 & (0.310) & 0.806\pm 0.028 & (0.808) & 0.819\pm 0.027& (0.821) & 0.818\pm 0.027 & (0.820)\\
    Green & 0.3063\pm 0.0076& (0.306) & 0.818\pm 0.022& (0.816) & 0.826\pm 0.022& (0.824) & 0.825\pm 0.021& (0.823)\\
    \textbf{Joint} & 0.3068\pm 0.0077& (0.306) & 0.813\pm 0.020& (0.811) & 0.822\pm 0.018& (0.819) & 0.822\pm 0.018 & (0.818)\\ \hline
    &\multicolumn{8}{c}{ACT DR6 $\times$ unWISE + \textit{Planck} PR4 $\times$ unWISE}\\ 
    Blue & 0.440^{+0.056}_{-0.082} & (0.424) & 0.692\pm 0.051 & (0.690) & 0.830\pm 0.021 & (0.820) & 0.815\pm 0.019 & (0.806)\\
    Green & 0.263^{+0.020}_{-0.023} & (0.275) & 0.861\pm 0.029 & (0.857) & 0.804\pm 0.019 & (0.793) & 0.809\pm 0.017 & (0.799)\\
    \textbf{Joint} & 0.274^{+0.019}_{-0.022} & (0.264) & 0.849\pm 0.029 & (0.860) & 0.810\pm 0.015 & (0.808) & 0.814\pm 0.014 & (0.813)\\ \hline
    &\multicolumn{8}{c}{ACT DR6 $\times$ unWISE + \textit{Planck} PR4 $\times$ unWISE + BAO}\\ 
    Blue & 0.3109\pm 0.0079 & (0.309) & 0.801\pm 0.021 & (0.805) & 0.816\pm 0.019 & (0.817) & 0.814\pm 0.019 & (0.816)\\
    Green & 0.3039\pm 0.0074 & (0.303) & 0.818\pm 0.017 & (0.818) & 0.823\pm 0.016 & (0.821) & 0.823\pm 0.016 & (0.821)\\
    \textbf{Joint} & 0.3046\pm 0.0074 & (0.305) & 0.813\pm 0.015 & (0.809) & 0.819\pm 0.014 & (0.815) & 0.819\pm 0.014 & (0.815)\\ \hline
    \enddata
    \caption{Summary of the constraints on cosmological parameters obtained from the cross-correlation of unWISE galaxies with ACT DR6 lensing reconstruction. Best-fit values (maximum a posteriori) are shown in parentheses following the one dimensional marginalised constraints. The combination of the galaxy-lensing cross-correlation and the galaxy clustering auto-spectrum is primarily sensitive to the parameter $S_8$. To break the degeneracy between the matter density, $\Omega_m$ and the amplitude of fluctuations $\sigma_8$ we analyse our data jointly with BAO. We also present constraints from the joint analysis of the \textit{Planck} and ACT cross-correlations.\label{table:results}}
\end{deluxetable*}

\subsection{Combination with the cross-correlation of \textit{Planck} Lensing and unWISE} \label{subsec:combination_planck}

In Appendix~\ref{app:planck_reanalysis} we present a reanalysis of the cross-correlation between \textit{Planck} CMB lensing and the two unWISE galaxy samples studied in this work using our improved model $dN/dz$ and systematics weights, implementing the previously neglected Monte Carlo lensing norm correction, and updating to \textit{Planck} PR4. We find
\beq
S_8=0.805\pm 0.018
\eeq 
from the cross-correlation alone and 
\beq
\sigma_8=0.810\pm 0.018
\eeq 
when combining with BAO. Despite the significant improvement in lensing noise for ACT these results from the cross-correlation with \textit{Planck} constrain $S_8$ and $\sigma_8$ more tightly than the results presented in Sec.~\ref{subsec:cosmo}. This is due to the significantly smaller sky area in ACT ($\sim 60\%$ vs. $\sim 20\%$). Since the results from ACT and \textit{Planck} are consistent we also present a joint analysis of the cross-correlation of unWISE galaxies with ACT DR6 and \textit{Planck} PR4 lensing. 

In Sec.~\ref{sec:covmat} we described how we estimate the cross-covariance between the ACT and \textit{Planck} cross-correlations. We established in Sec.~\ref{subsec:galaxy_sys} that the galaxy nuisance parameters, principally the linear galaxy bias, are expected to vary across the sky. Hence we do not assume these to be identical in the ACT and \textit{Planck} footprints. We therefore include $C_\ell^{gg}$ measured with both the ACT and \textit{Planck} masks in our analysis and independently marginalise over the nuisance parameters for both sets of observations. We include the significant cross-covariance between $C_\ell^{gg, \rm{ACT}}$ and $C_\ell^{gg, \rm{Planck}}$ which is also estimated from our Gaussian simulations and correctly captures the overlap in area of both observations.

\begin{figure*}
    \centering
    \includegraphics[width=0.5\linewidth,trim=0.5cm 0.5cm 0.5cm 0.5cm, clip]{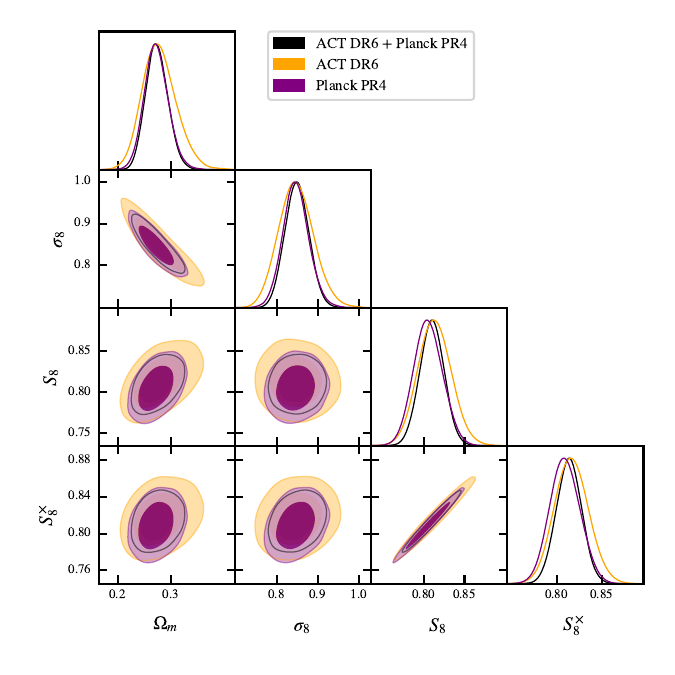}%
    \includegraphics[width=0.5\linewidth,trim=0.5cm 0.5cm 0.5cm 0.5cm, clip]{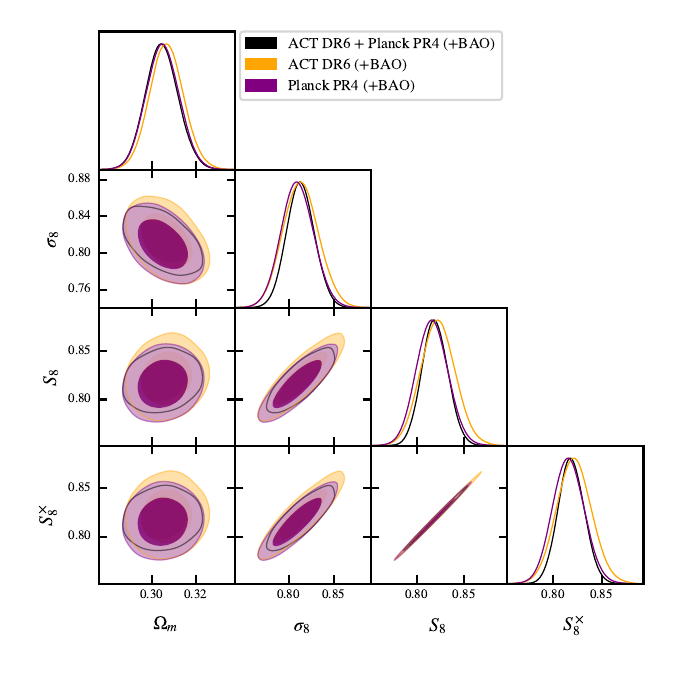}
    \caption{We combine the cross-correlations of unWISE galaxies with lensing reconstructions from ACT DR6 and \textit{Planck} PR4 at the likelihood level to obtain further tightened constraints on cosmological parameters. Here we show a comparison of the joint analysis of both redshift samples from ACT only, \textit{Planck} only and the combination of ACT and \textit{Planck}. As above, the \textbf{left} panel shows constraints using $C_\ell^{gg}$ and $C_\ell^{\kappa g}$ only, for which we find $S_8 = 0.810\pm 0.015$. Including BAO information (\textbf{right}) yields $\sigma_8=0.813\pm 0.015$.}
    \label{fig:act+planckXunWISE_corner_plots}
\end{figure*}

The combination of ACT DR6 and \textit{Planck} PR4 cross-correlations with unWISE yields a joint constraint 
\beq S_8=0.810\pm 0.015,
\eeq 
a $29\%$ and $17\%$ improvement over the constraints from ACT DR6 and \textit{Planck} PR4 alone respectively. In combination with BAO we find 
\beq
\sigma_8=0.813\pm 0.015,
\eeq 
again improving the individual constraints from ACT(+BAO) and \textit{Planck}(+BAO) by $25\%$ and $17\%$ respectively. The joint constraints are shown alongside the constraints from ACT and \textit{Planck} alone in Fig.\,\ref{fig:act+planckXunWISE_corner_plots}.

\subsection{Parameter Based Consistency Test} \label{subsec:param_consistency_tests}

\begin{figure*}
    \includegraphics[width=\linewidth]{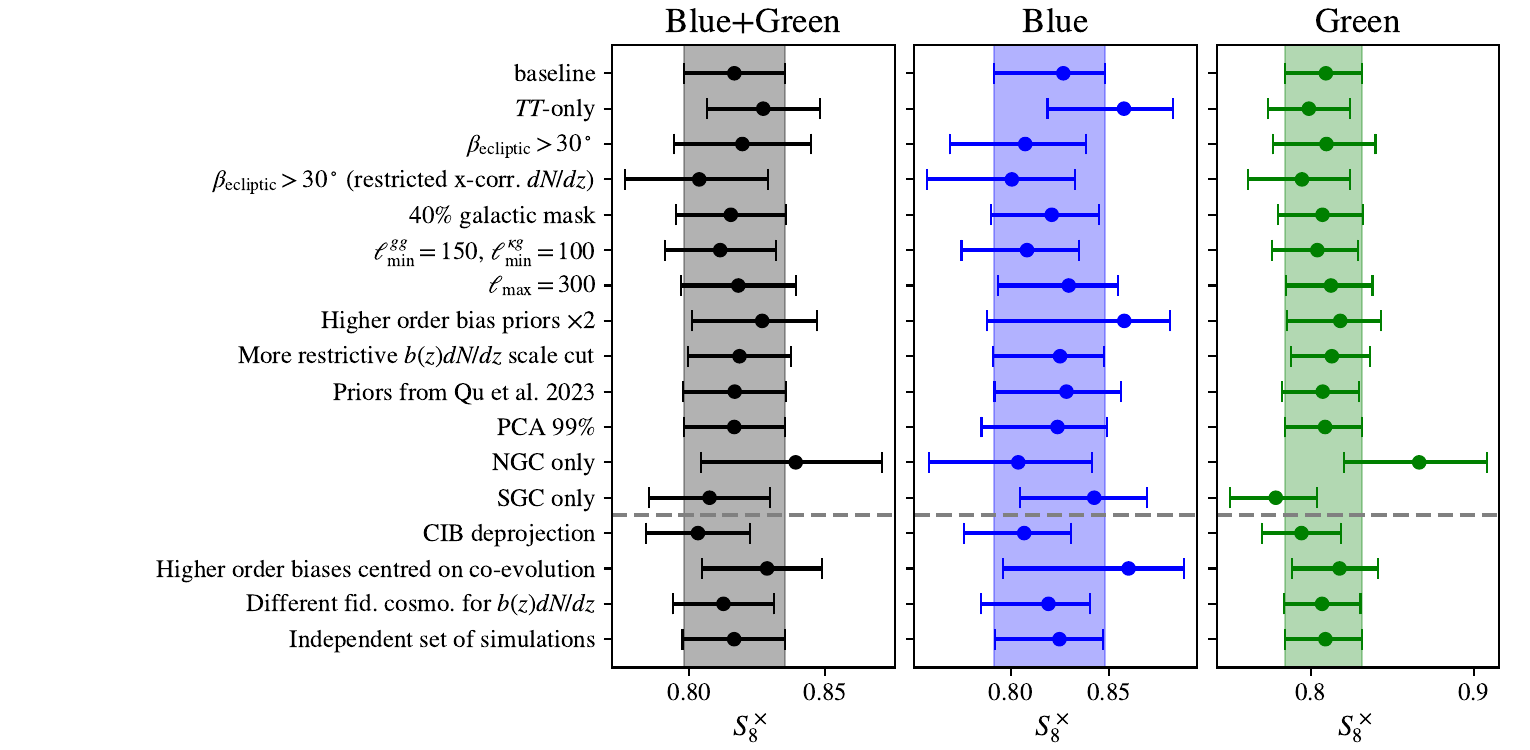}
    \caption{We tested various analyses of data subsets and alternative choices for higher order bias parameters and hyperparameters relating to the measurement of the cross-correlation redshifts. Here we show the impact on the best constrained parameter, $S_8^\times$, of these alternative analysis choices. We note that these results are not independent and where possible we estimate the likelihood of the shifts observed in Table~\ref{table:param_consistency_subset}. More consistency tests are also summarised in Table~\ref{table:param_consistency_alt_analysis}. We find consistency across all our tests, indicating that our results are unlikely to be significantly affected by residual systematic contamination.}
    \label{fig:param_consistency}
\end{figure*}

As discussed in Sec.\,\ref{subsec:blinding} we performed a series of parameter level consistency tests on the ACT data by running our full analysis pipeline using different lensing reconstruction methods and data subsets. These tests were originally performed blind by consistently adding random offsets to the cosmological parameters. We present here the unblinded versions of these tests. The alternative analyses using simplified models, discussed towards the end of this section, were not performed prior to unblinding. The results of these tests are summarised visually in Fig.~\ref{fig:param_consistency}.

We compare the following data subsets to our baseline analysis: 1) cross-correlations measured using the temperature only lensing reconstruction, 2) an analysis restricted to ecliptic latitudes greater than $30^\circ$ \resub{(additionally, 2b) also restricting the cross-correlation redshift estimation to this region)}, 3) an analysis using a more restrictive Galactic mask which includes only the 40\% of the sky with the lowest Galactic contamination, 4) two analyses further restricting the range of scales used in our analysis, once by increasing $\ell_{\rm{min}}$ to $150$ and $100$ for $C_\ell^{gg}$ and $C_\ell^{\kappa g}$ respectively, and once by reducing $\ell_{\rm{max}}$ to $300$ for both $C_\ell^{gg}$ and $C_\ell^{\kappa g}$, 5) an analysis using twice wider priors on the higher order bias parameters, 6) an analysis using a more restrictive scale cut for the cross-correlation redshift estimation (minimum scale of 4 $h^{-1}$Mpc rather than 2.5 $h^{-1}$Mpc)\resub{, and 7) an analysis using more principal components for redshift marginalisation (capturing 99\% of the variance in the bias weighted redshifts; $n=6$ and 9 for the Blue and Green samples respectively)}.

\resub{Furthermore, we compare our results to the constraints obtained using a different set of priors adopted from the analysis in \cite{2024ApJ...962..112Q} as discussed above and in Appendix~\ref{app:alt_priors}.} Finally, we also produce a comparison between measurements conducted exclusively on the northern and southern Galactic regions (NGC and SGC). These are fully independent as they also rely on measuring the cross-correlation redshifts on the respective regions.

We estimate the statistical likelihood of observing the shifts obtained by considering the shift in the best constrained parameter $S_8^{\times}$, effectively assuming that we are only measuring a single parameter related to the amplitude of the spectrum. For the comparison of nested datasets, i.e. such datasets where one is a proper subset of the other, we estimate the statistical uncertainty on the expected shift in parameters as the difference between the marginalised posterior width of each of the samples approximating them as Gaussian \citep{2020MNRAS.499.3410G}. In the case where both datasets are independent we add the variance of both measurements. Detailed results of all consistency tests are provided in Table~\ref{table:param_consistency_subset} in Appendix~\ref{app:param_consistency_tests}. 

\resub{For the other parameters of interest (in particular $\sigma_8$ and $\Omega_m$) such comparisons are less straightforward, as these parameters are individually only poorly constraint by the cross-correlations alone and the posteriors are far from Gaussian. However, we find that the parameter means fall within the 68\% confidence region of our baseline analysis in all but three cases for which such shifts are not unexpected due to a significant reduction in data volume (when restricting the ecliptic latitude to be greater than 30$^\circ$ for the Blue sample and when considering only the NGC for the Blue or Green samples). We show the constraints on $\sigma_8$ and $\Omega_m$ from the various subset analyses in Figs.\,\ref{fig:param_consistency_sigma8} and \ref{fig:param_consistency_OmegaM} in Appendix~\ref{app:param_consistency_tests}.}

We also conduct three variations of our baseline analysis that do not simply consider subsets of the full data set (also shown in Fig.~\ref{fig:param_consistency}). First, we adopt a lensing reconstruction in which the contamination from CIB has explicitly been deprojected using high frequency data from \textit{Planck}. Secondly, we centre the higher order bias parameters on the fiducial coevolution relations rather than the best fit from simulations. Finally, we use a slightly different fiducial cosmology to measure the cross-correlation redshifts, with $h = 0.702$, $\Omega_m = 0.278$ and $\sigma_8 = 0.805$. For these tests an estimate of the significance of the shifts observed cannot easily be obtained due to the non-trivial covariance between these measurements and our baseline analysis. However, the shifts in the parameters $S_8$ and $S_8^\times$ are small compared to our measurement uncertainties ($<0.35\sigma$; see Table~\ref{table:param_consistency_alt_analysis} within Appendix~\ref{app:param_consistency_tests}). As discussed in Sec.\,\ref{sec:covmat} we also conduct a stability test using an independent set of simulations to estimate our covariance. This test is also shown in Fig.\,\ref{fig:param_consistency}. We find that our results are unchanged with this independent covariance estimate.

All consistency tests pass and the shifts in the parameters of interest are small compared to our uncertainties. This indicates that our baseline cosmology results are unlikely to be significantly affected by remaining systematic contamination. 

We note that the null-test failures discussed in Sec.\,\ref{subsec:lensing_sys} indicate a slightly larger lensing amplitude preferred by the temperature-only data. We showed that the difference is consistent with random fluctuations, nevertheless we assessed the impact on cosmological parameters. Using the temperature-only lensing reconstruction leads to an increase in the inferred $S_8$ compared to our baseline analysis of $\Delta S_8=0.014$ ($+0.7\sigma$; see Table~\ref{table:param_consistency_subset}). The shift is consistent within the expected error on the difference between the baseline and temperature-only analysis, and increases the discrepancy with $S_8$ inferred from galaxy surveys (see comparisons in Sec.\,\ref{subsec:comp_lss_surveys}).

\begin{figure*}
    \includegraphics[width=\linewidth]{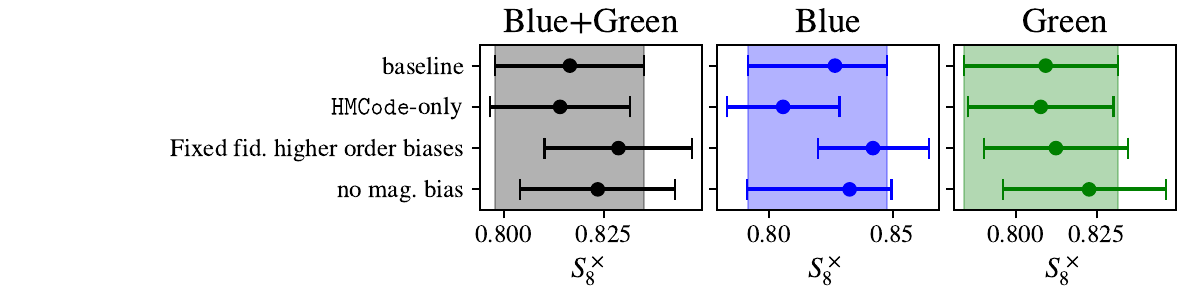}
    \caption{The impact of neglecting various components of our model is very small with all resulting shifts well below $1\sigma$ in terms of our statistical uncertainty. This indicate that our results will only have a small dependence on the detailed modelling choices for those model components.}
    \label{fig:model_comps_data}
\end{figure*}

In Fig.\,\ref{fig:model_comps_data} we also show the impact of neglecting some of the model components in our analysis and find this to have a small impact on the inferred values of $S_8^{\times}$ ($<1\sigma$ in terms of our statistical uncertainties; these analyses are also summarised in Table~\ref{table:param_consistency_alt_analysis}). While analyses with these models do not appear strongly biased, they also do not dramatically improve our constraining power.

\editF{Finally, we compare model predictions for the cross-correlation between the two galaxy samples with the observed $C_\ell^{g_{\rm{Blue}}g_{\rm{Green}}}$. We find that the model prediction, fit only to the two galaxy auto-spectra and the cross-correlations with the ACT DR6 lensing reconstruction, is consistent with the observed cross-spectrum within the \resub{expected uncertainty} (see Appendix~\ref{app:gg_cross_test} for details).}

\section{Discussion and Conclusions} \label{sec:discussion}
Measuring the amplitude of low redshift fluctuations can reveal the nature of dark matter and dark energy, measure the masses of neutrinos and provide crucial tests of gravity. The cross-correlation of CMB lensing maps with matter tracers (such as galaxies in this work) with a known redshift distribution -- or CMB lensing tomography -- can isolate the redshifts of interest, study the time-dependence of the signal, break the degeneracy with galaxy bias and reduce the sensitivity to several sources of systematics. In this paper, we leveraged the low shot noise and large redshift lever arm of the unWISE Blue and Green galaxy samples, which fully cover the dark energy dominated era as well as the transition to matter domination. Our results improve and build upon a previous analysis in \cite{2021JCAP...12..028K}, and the main differences are highlighted in Appendix \ref{app:planck_reanalysis}.

\subsection{Comparison with predictions from the primary CMB}

\begin{figure*}
    \centering
    \includegraphics[width=\linewidth]{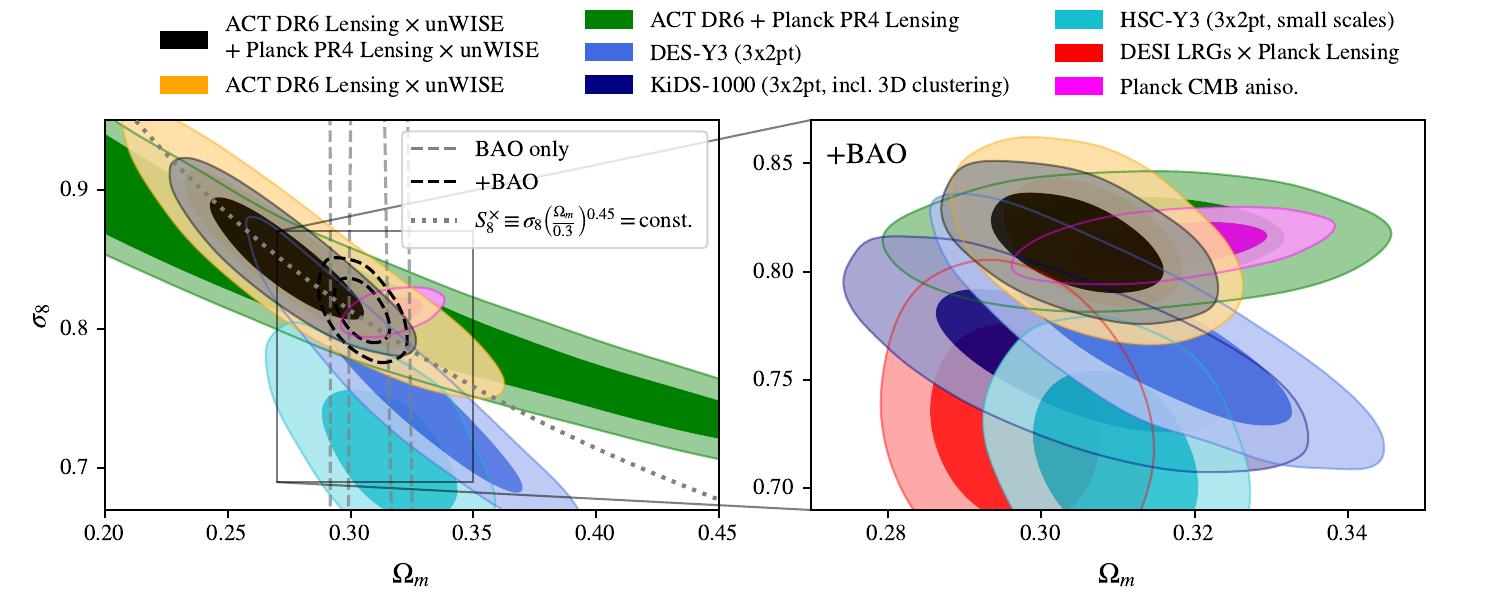}
    \caption{Here we show constraints in the $\Omega_m-\sigma_8$ plane for our baseline results and selected other datasets. The \textbf{left} panel shows results without external information on $\Omega_m$; all large scale structure probes show a significant degeneracy between $\Omega_m$ and $\sigma_8$. For our analysis the best constraint parameter combination is proportional to $\sigma_8 \Omega_m^{0.45}$ shown here as a dotted line. This degeneracy is broken by the addition of BAO information. To illustrate the effect of BAO information \resub{we show the BAO-only constraint as a dashed grey contour and} the joint cross-correlation analysis of ACT and \textit{Planck} with BAO as a dashed black contour (in the \textbf{left} panel). The \textbf{right} panel shows a comparison with other datasets when adding BAO information. \resub{We note that given that we do not reanalyse these other data sets, they do not all use the same BAO likelihoods as in this work.} We can see that our analysis is in no significant tension with any one of the other datasets shown, when considering the full $\Omega_m-\sigma_8$ parameter space, despite favouring a larger value of $S_8$ and $\sigma_8$ compared to galaxy weak lensing surveys. Our results are consistent with predictions from the primary CMB and CMB lensing.}
    \label{fig:sigma8-omegam_2d_comp}
\end{figure*}

By combining galaxy clustering and cross-correlation measurements, we find $S_8 = 0.813 \pm 0.021$ using ACT lensing, and $S_8 = 0.810 \pm 0.015$ with a combination of \textit{Planck} and ACT lensing, consistent with the prediction from the primary CMB from \textit{Planck} at the $0.8\sigma$ and $1.1\sigma$ level respectively \citep[\textit{Planck} obtains $S_8= 0.834 \pm 0.016$ or $S_8= 0.832 \pm 0.013$ with CMB lensing;][]{2020A+A...641A...6P} and with similar statistical uncertainty. BAO data breaks the degeneracy between $\sigma_8$ and $\Omega_m$ and we find $\sigma_8 = 0.813 \pm 0.020$ from ACT alone and $\sigma_8 = 0.813\pm 0.015$ from the combination of \textit{Planck} and ACT. Again, this is in very good agreement with the \textit{Planck} primary CMB results of $\sigma_8 = 0.8120 \pm 0.0073$ ($0.05\sigma$ and $0.07\sigma$ respectively) using primary CMB only \citep[or $ \sigma_8 = 0.8111 \pm 0.0060$ when including CMB lensing;][]{2020A+A...641A...6P}. The comparisons with the \textit{Planck} results is shown in Figs.\,\ref{fig:sigma8-omegam_2d_comp} along with other large scale structure measurements of $S_8$ and $\sigma_8$. Our results are also consistent with predictions from independent primary CMB observations, including from the combination of WMAP and ACT \citep{2020JCAP...12..047A}. We show those constraints in Figs.\,\ref{fig:s8_whisker} and \ref{fig:sigma8_whisker} together with a more comprehensive set of other measurements.

\subsection{Comparison with other measurements of the large scale structure of the universe}\label{subsec:comp_lss_surveys}

In addition to the comparison to predictions for the amplitude of structure formation from the primary CMB (using \textit{Planck} and ACT) we also compare our results to measurements from other large scale structure observables. We include results from CMB lensing auto-spectrum analyses \citep{2020ApJ...888..119B,2022JCAP...09..039C,2024ApJ...962..112Q,2024ApJ...962..113M}, galaxy weak lensing surveys \citep[DES, KiDS and HSC;][]{2022PhRvD.105b3520A,2021A+A...646A.140H,2023PhRvD.108l3520M,2023PhRvD.108l3521S,2023PhRvD.108l3517M}, from other cross-correlations with CMB lensing from ACT, SPT, and \textit{Planck} \citep{2022JCAP...02..007W,2022JCAP...07..041C,2023PhRvD.107b3530C,2023PhRvD.107b3531A,2024JCAP...01..033M}, and from the three dimensional clustering of galaxies \citep{2021PhRvD.103h3533A,2023PhRvD.107h3515I}. We subsequently briefly introduce the datasets we employ:
\begin{itemize}
    \item \textbf{CMBL:} These are constraints from the analysis of the auto-spectrum of CMB lensing reconstructions. These results are mostly sensitive to linear scales at $z=1-2$ and primarily constrain the parameter combination $\sigma_8 \Omega_m^{0.25}$. We include results from \textit{Planck} PR4 \citep{2022JCAP...09..039C}, SPTpol \citep{2020ApJ...888..119B}, ACT DR6, and a joint analysis of ACT DR6 and \textit{Planck} PR4 \citep{2024ApJ...962..113M}. To make fair comparisons we combine the CMB lensing measurements with BAO information which breaks the degeneracy between $\sigma_8$ and $\Omega_m$.

    \item \textbf{WL:} From galaxy weak lensing surveys we include constraints from `3x2pt'-analyses, combining measurements of galaxy clustering, galaxy shear, and their cross-correlation. For DES we adopt the results obtained in \cite{2022PhRvD.105b3520A} when fixing the neutrino mass. For KiDS we show results presented in \cite{2021A+A...646A.140H}. We note, that in contrast to the DES analysis these results are obtained from the combination of projected galaxy shear and galaxy-galaxy lensing measurements with a measurement of the three dimensional clustering of galaxies in the spectroscopic BOSS and 2dfLenS surveys. Therefore, the KiDS analysis already contains the BAO information. For HSC we adopt a set of results from a reanalysis of the HSC galaxy shear, galaxy-galaxy lensing and galaxy clustering measurements \citep{2023PhRvD.108l3520M} using priors on cosmological parameters consistent with those used in this work (see Table~\ref{tab:priors}). We show results for an analysis using a linear bias model \citep{2023PhRvD.108l3521S} and from an analysis using a halo model based emulator which includes non-linear scales \citep{2023PhRvD.108l3517M}.

    \item \textbf{CMBLX}: We compare our results with other cross-correlations between CMB lensing and galaxy surveys. \cite{2022JCAP...02..007W} analysed the cross-correlation between DESI LRG targets and a lensing reconstruction from \textit{Planck} PR3 \citep{2020A+A...641A...8P}. We also include a series of works cross-correlating DES-Y3 galaxy shear and galaxy clustering ($\delta_g + \gamma$) with a joint SPT and \textit{Planck} lensing reconstruction \citep{2023PhRvD.107b3530C}. Building on this work \cite{2023PhRvD.107b3531A} present a `5x2pt' and `6x2pt' analysis combining with all DES internal cross- and auto-correlations and additionally with the lensing auto-spectrum respectively. The joint lensing reconstruction from SPT-SZ and \textit{Planck} is presented in \cite{2023PhRvD.107b3529O}. A recent cross-correlation between DES-Y3 clustering ($\delta_g$) and ACT DR4 lensing \citep{2024JCAP...01..033M} based on the lensing reconstruction from \cite{2021MNRAS.500.2250D} is also included. While all aforementioned analyses use photometric galaxy samples in their cross-correlations, the final cross-correlation study included in our comparisons, \cite{2022JCAP...07..041C}, jointly models the three dimensional clustering of the spectroscopic BOSS galaxies and their cross-correlation with \textit{Planck}.

    \item \textbf{GC}: Finally, we compare our results with constraints from the three dimensional clustering of galaxies as measured by BOSS and eBOSS. We include the analysis of BAO and RSD from \cite{2021PhRvD.103h3533A}. Furthermore, we include a independent analysis based on the effective theory of large scale structure (EFTofLSS) that fits the `full shape' of the power spectrum and bispectrum measured in redshift-space \citep{2023PhRvD.107h3515I}. Similar, previous analyses found compatible results \citep[see e.g.,][]{2020JCAP...05..005D,2020JCAP...05..042I,2022JCAP...02..008C}.
\end{itemize}

\begin{figure}
    \centering
    \includegraphics[width=\linewidth]{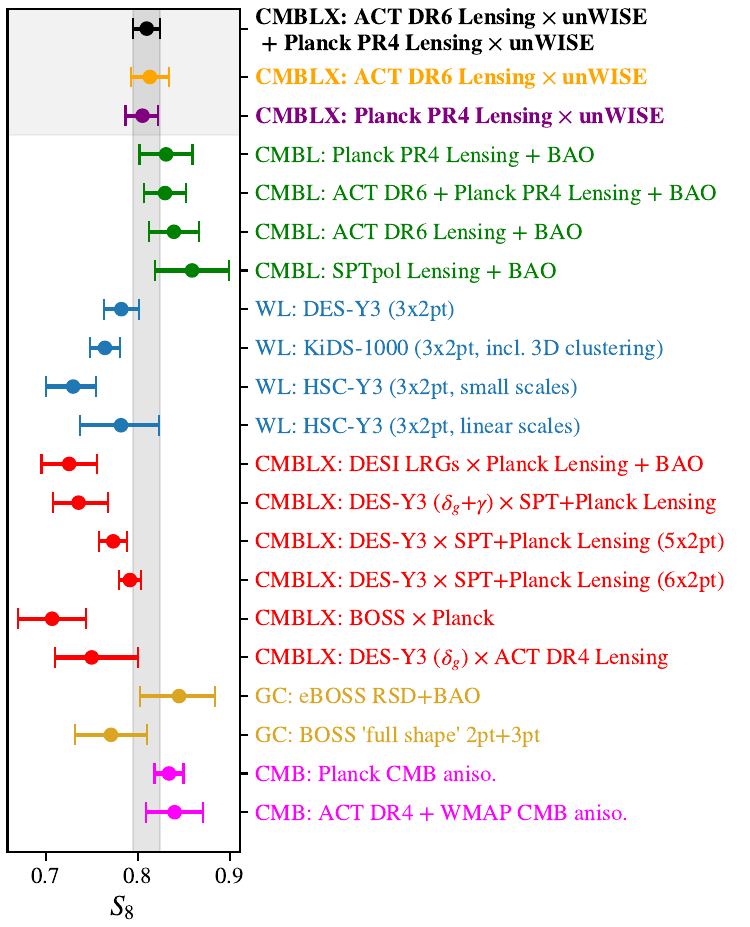}
    \caption{Here we compare with an extensive set of measurements of $S_8$. We include measurements from CMB lensing analyses in combination with BAO (in green), from galaxy weak lensing surveys (in blue), from cross-correlations with CMB lensing (in red), from  the three dimensional clustering for spectroscopic galaxy surveys (in gold), and predictions from the primary CMB (in magenta). Our results are in good agreement with CMB lensing analyses, as well as with the predictions from the primary CMB. They are also in no significant tension with any one galaxy weak lensing survey, or the $S_8$ inferred from the three dimensional clustering of spectroscopic galaxies. We do, however, find some discrepancies at up to $3\sigma$ with previous cross-correlation work.}
    \label{fig:s8_whisker}
\end{figure}

\begin{figure}
    \centering
    \includegraphics[width=\linewidth]{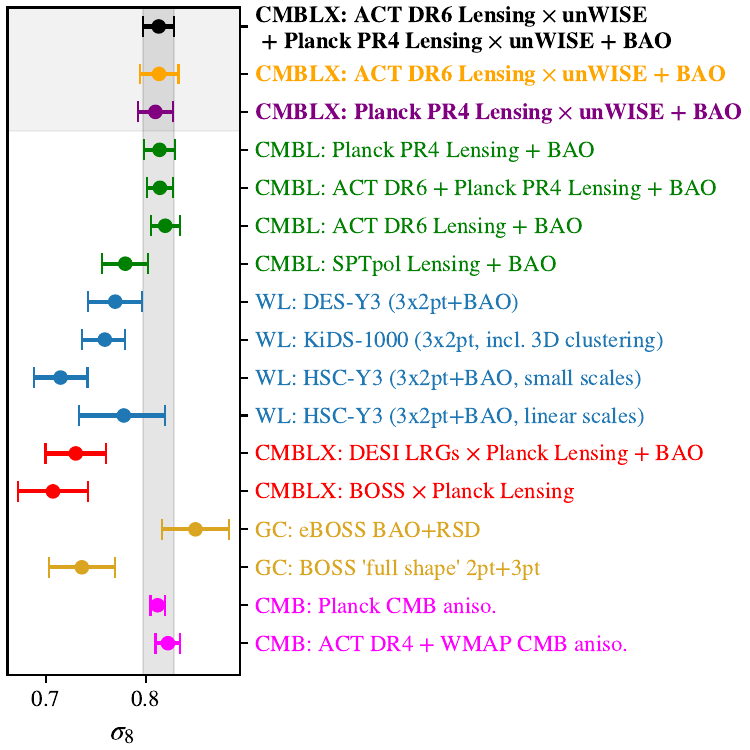}
    \caption{Here we compare with an extensive set of measurements of $\sigma_8$. As above we include measurements from CMB lensing analyses in combination with BAO (in green), from galaxy weak lensing surveys (+BAO) (in blue), from cross-correlations with CMB lensing (+BAO) (in red), from the three dimensional clustering for spectroscopic galaxy surveys (in gold), and predictions from the primary CMB (in magenta). As in the case of $S_8$, our results are in good agreement with most CMB lensing analyses, as well as with the predictions from the primary CMB. Galaxy weak lensing surveys favour smaller $\sigma_8$ at varying levels of significance ($0.8-3.2\sigma$). We also find some discrepancies at up to $3\sigma$ with previous cross-correlation work.}
    \label{fig:sigma8_whisker}
\end{figure}

In the left panel of Fig.\,\ref{fig:sigma8-omegam_2d_comp} we show a comparison of our results with various other probes in the $\Omega_m - \sigma_8$ parameter space. We can see that the large scale structure probes shown exhibit a significant degeneracy between $\sigma_8$ and $\Omega_m$. That degeneracy is broken when adding BAO information as we show in the right panel of Fig.\,\ref{fig:sigma8-omegam_2d_comp}. A more extensive set of comparisons for $S_8$ and $\sigma_8$ is shown in Figs.\,\ref{fig:s8_whisker} and \ref{fig:sigma8_whisker} respectively. Generically, we include BAO information in all comparisons of $\sigma_8$ to break the $\Omega_m - \sigma_8$ degeneracy. An exception are those measurements employing three dimensional galaxy clustering either on its own, as part of a `3x2pt' analysis, or in cross-correlation. Those measurements already implicitly contain the BAO information. 

We note that we do not reanalyse these datasets with our prior choices. An exception are the results from HSC for which the Markov Chain Monte Carlo runs are not publicly available at the time of writing and which were kindly provided by the HSC team with prior choices that match ours. \cite{2024ApJ...962..113M} discussed the impact of reanalysing the galaxy survey datasets with priors matching those in the ACT DR6 lensing auto-spectrum analysis and found only minor changes \cite[see Appendix~C in][]{2024ApJ...962..113M}. As previously discussed we also explore an alternative set of priors in Appendix~\ref{app:alt_priors}. We find that while our constraints on $S_8$ are not very sensitive to the choice of priors, the constraints on $\sigma_8$ in combination with BAO are slightly degraded ($\sim$$10\%$) when the \textit{Planck} prior on $\Omega_m h^3$ is removed. However, we do not observe any significant shifts in the mean $\sigma_8$ or $S_8$ and therefore do not expect our conclusions to differ substantially.

Our results are in good agreement with various CMB lensing auto-spectrum analyses. Given that CMB lensing is primarily sensitive to the parameter combination $\sigma_8 \Omega_m^{0.25}$ we combine our measurements as well as the CMB lensing measurements from \textit{Planck}, SPTpol and ACT DR6 with BAO to be able to directly compare the resulting $\sigma_8$ constraints. The value of $\sigma_8$ inferred from the cross-correlation of unWISE with ACT DR6 and \textit{Planck} together with BAO is consistent within $0.3\sigma$ with the constraints from ACT DR6, within $0.04\sigma$ with the result from \textit{Planck} PR4, and within $0.06 \sigma$ with the results from the joint analysis of ACT DR6 and \textit{Planck} PR4. In the parameter $\sigma_8$ the agreement with SPTpol is less good, though not at any statistical significance ($1.2\sigma$).

These results can also be compared to galaxy weak lensing measurements of $S_8$. Analyses of DES and KiDS obtain $S_8 =0.782\pm0.019$ and $S_8 = 0.765^{+0.017}_{-0.016}$ respectively from a combination of cosmic shear and galaxy clustering. The reanalysis of HSC cosmic shear and galaxy clustering with our prior choices yields $S_8=0.782\pm 0.043$ when using only linear scales and $S_8=0.730^{+0.025}_{-0.029}$ from an analysis using a halo model based emulator\footnote{\editF{These results differ by approximately $0.2$ and $1.2\sigma$ respectively from the fiducial results reported by HSC \citep{2023PhRvD.108l3521S,2023PhRvD.108l3517M}. The difference is largely driven by the fact that we fix $\Omega_m h^3$ to the mean value for this parameter combination found by \textit{Planck}. In terms of the expected uncertainty on the difference between the two analyses the shifts correspond to about $0.3-0.6$ and $1.2-1.5\sigma$ respectively \citep[assuming that the fiducial analysis constitutes a subset of the analysis with our priors;][]{2020MNRAS.499.3410G}. These shifts are thus within what is expected given the difference in prior choices.}}. They are all lower than our results, but not in significant statistical tension (approximately $1.1$, $2.0$, $0.6$ and $2.5\sigma$ for DES, KiDS, and the two results from HSC respectively). 
We note that the DES and KiDS `3x2pt' analyses draw on cosmic shear measurements for which recent reanalyses have found shifts towards higher $S_8$. A recent and improved reanalysis of KiDS-1000 cosmic shear data obtains $S_8 = 0.776^{+0.031}_{-0.030}$ \citep{2023A+A...679A.133L} \citep[increased from the previous $S_8 = 0.759^{+0.024}_{-0.021}$;][]{2021A+A...645A.104A}, while a recent joint reanalysis of the DES and KiDS cosmic shear data also hints at a higher value of $S_8 = 0.790_{-0.014}^{+0.018}$ \citep[also higher than previous DES results from cosmic shear alone which yielded $S_8 = 0.759^{+0.025}_{-0.023}$ and the KiDS results above;][]{2023OJAp....6E..36D,2022PhRvD.105b3514A}. For these two renewed cosmic shear analyses the discrepancy with our results is reduced to $\sim 1.5$ and $\sim0.9\sigma$ respectively. It is possible that the shear recalibration from \cite{2023A+A...679A.133L} and changes to the pipeline and prior choices in \cite{2023OJAp....6E..36D} will also increase $S_8$ inferred from the `3x2pt' analyses to which we compare here.

In addition to evaluating the consistency of our cosmological parameters with various weak lensing surveys, we also test the goodness of fit when fixing the cosmology to one similar to the one preferred by DES. This is to test the possibility that our preference for a higher value of $S_8$ is driven in part by projection effects which have been seen (albeit in the opposite direction) in some other cross-correlation analyses \citep{2022JCAP...02..007W,2022JCAP...07..041C}. We fixed $S_8$ to $0.78$, close to the mean value obtained by DES \citep{2022PhRvD.105b3520A}. We then vary $\Omega_m$ and all nuisance parameters in our analysis to obtain the best fit to the auto-correlations of the Blue and Green samples and their cross-correlations with the ACT lensing reconstruction. We find a minimum $\chi^2=22.8$. When instead fixing the cosmology to $S_8=0.83$, similar to the \textit{Planck} cosmology, we similarly find $\chi^2=23.2$. Our best-fitting baseline model, in which $S_8$ and $\Omega_m$ are free to vary has a minimum $\chi^2$ of $20.9$ but includes one more degree of freedom suggesting that both the DES and \textit{Planck} values of $S_8$ are compatible with our data at $\sim 1\sigma$. However, consistency with the DES cosmology comes at the cost of some tension with the matter density inferred from BAO. When we add BAO information to the minimisation at fixed $S_8$ we find $\Delta \chi^2=3.6$ for the DES-like cosmology relative to our baseline fit and $\Delta \chi^2=0.3$ for a \textit{Planck}-like cosmology (including the $\chi^2$ contributions from the BAO likelihoods). This corresponds to a cosmology with the DES-like value of $S_8$ being slightly disfavoured by $\sim1.6\sigma$.

Our results appear to show a discrepancy with previous CMB lensing cross-correlations. This is true for the previous analysis of the cross-correlation of unWISE and \textit{Planck} lensing \citep{2021JCAP...12..028K}; we discuss this comparison and the origin of the discrepancy in detail in Appendix~\ref{app:planck_reanalysis}. We also find some discrepancy with the results from the cross-correlation of the DESI LRG targets with \textit{Planck}. \cite{2022JCAP...02..007W} measured $S_8=0.73\pm0.03$ corresponding to about a $2.5\sigma$ discrepancy with our analysis using both ACT and \textit{Planck}. A similarly significant difference is also found with the cross-correlation of BOSS and \textit{Planck}, which yields $S_8=0.707\pm 0.037$ ($\sim$$2.6\sigma$) and with the cross-correlation of DES-Y3 with SPT and \textit{Planck} ($S_8=0.736^{+0.032}_{-0.028}$; $\sim$$2.2\sigma$). This discrepancy is reduced when adding the DES-internal cross- and auto-correlations. The `5x2pt' analysis yields $S_8=0.773\pm0.016$ ($\sim$$1.7\sigma$). Adding the CMB lensing auto-correlation from \textit{Planck} further increases $S_8$ to $0.792\pm0.012$ and reduces the discrepancy to $\sim$$1.0\sigma$ (`6x2pt'). Given the consistency of our results with \textit{Planck} lensing this is unsurprising. The analysis of DES-Y3 and ACT DR4 lensing also obtains a lower value of $S_8$, albeit with much larger errors ($S_8=0.75^{+0.04}_{-0.05}$; $\sim$$1.3\sigma$).

Some of the improvements made here to the previous analysis of the cross-correlation between unWISE and \textit{Planck} are also relevant for some of the other cross-correlation measurements discussed above. In particular, the inclusion of the Monte Carlo correction to the normalisation of the \textit{Planck} lensing maps (for details see Appendix~\ref{app:planck_reanalysis}) will also impact the analyses in \cite{2022JCAP...02..007W,2022JCAP...07..041C,2023PhRvD.107b3530C} and \cite{2023PhRvD.107b3531A}. However, at least in some cases preliminary results show that they are unlikely to completely alleviate the discrepancy. \cite{2022JCAP...02..007W} and \cite{2022JCAP...07..041C} also found indications for significant projection effects in their analyses. We also note the importance of the better redshift calibration of the unWISE samples which contributed a shift of $\sim$$0.8\sigma$ to our reanalysis of the cross-correlation with \textit{Planck} (see Appendix~\ref{app:planck_reanalysis}) which may also affect other analyses and should be significantly improved in the near future with the availability of spectroscopic redshifts from DESI.

Our results are consistent with the eBOSS analysis of BAO and RSD which finds $\sigma_8=0.850\pm0.035$ (consistent with our baseline results within $1.0\sigma$). The `full shape' measurements on the other hand prefer a lower value of $\sigma_8$ by about $2\sigma$.


\subsection{Our results in the context of the $S_8$/$\sigma_8$ ``tension"}

\begin{figure}
    \centering
    \includegraphics[width=\linewidth,trim=0cm 0.5cm 0cm 0cm,clip]{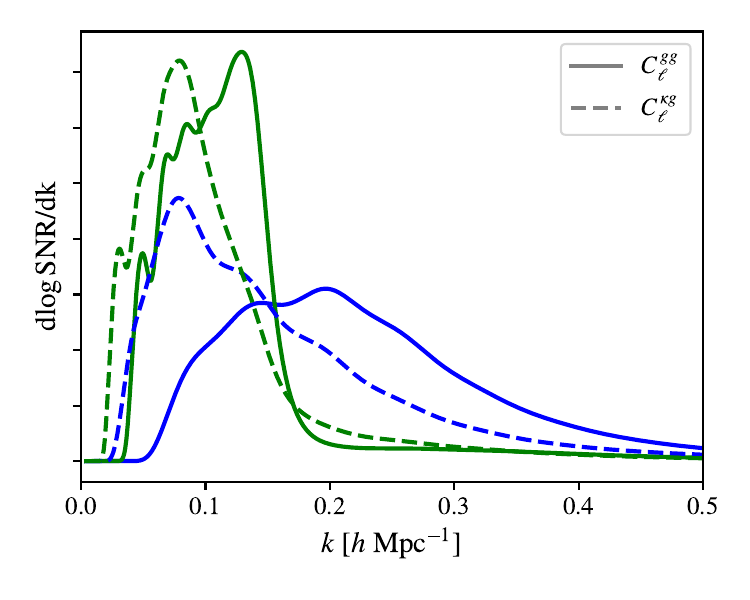}
    \caption{We show the change in the total signal-to-noise as function of the maximum $k$ considered. With our scale cuts ($50\leq\ell\leq400$ for $C_\ell^{\kappa g}$ and $100\leq\ell\leq400$ for $C_\ell^{g g}$) the signal for the galaxy cross- and auto-correlations originates largely on scales $k \lesssim 0.3$ $h$/Mpc. For the Green sample of galaxies and its cross-correlation with CMB lensing almost all the signal comes from $k<0.2$ $h$/Mpc. While the Blue sample has minor contributions from $k\gtrsim0.3$ $h$/Mpc, most of the cosmological constraining power comes from larger scales since we marginalise over higher order contributions to the signal which become important on smaller scales.}
    \label{fig:k_origin_signal}
\end{figure}

We briefly discuss the implications of our results for the claimed $S_8$ tension. It is worth noting that most of the galaxy weak lensing constraints come at least partially from smaller scales, where gravitational non-linearities, as well as the effects of galaxy formation (assembly bias, baryonic feedback, etc.) are more pronounced. Our cosmological constraints derive almost entirely from perturbative scales at the redshifts of interest ($k \lesssim 0.3$ $h$/Mpc), and therefore we expect uncertainties due to galaxy formation to be negligible and fully absorbed in our bias expansion. In Fig.\,\ref{fig:k_origin_signal} we show the scales on which our signal originates. Extensive tests on mocks presented in this paper also show no evidence for ``projection effects'' in the Bayesian parameter estimation. Together with the large number of consistency and null tests, this provides confidence in the results.

The fact that CMB lensing power spectrum measurements \citep{2024ApJ...962..112Q, 2024ApJ...962..113M, 2022JCAP...09..039C, 2020A+A...641A...8P} are consistent with the CMB power spectrum prediction deriving from high redshifts excludes the possibility of substantial effects of new physics on structure growth at high redshifts and large scales; however, the possibility of new physics at low redshifts ($z\lesssim1$) or smaller scales remains when considering only CMB lensing power spectrum data. Although the analyses presented in this paper are sensitive to the growth of structure at $z\simeq0.2-1.6$ (on large scales), our measurements show excellent consistency with CMB power spectrum predictions. In short, using the cross-correlations of CMB lensing and galaxy density we find no evidence for a $S_8$ (or $\sigma_8$) tension on scales $k \lesssim 0.3$ $h$/Mpc and for $z\gtrsim0.2$. Our results together with CMB lensing auto-spectrum analyses, imply that, if new physics is indeed responsible for the $S_8$ tension observed in galaxy weak lensing data sets, any effects of new physics must be mostly limited to small scales ($k \gtrsim 0.3$ $h$/Mpc) to which our unWISE measurements are insensitive. \editF{This is consistent with resolutions to the $S_8$ ``tension" based on modifications to the matter power spectrum on non-linear scales \citep{2022MNRAS.516.5355A,2023MNRAS.525.5554P}}. Our work therefore motivates further efforts to test the $\Lambda$CDM predictions for structure formation on smaller, non-linear scales.

\subsection{Conclusions}

We have presented cosmological results from the cross-correlation of the Blue and Green unWISE galaxy samples and the latest \textit{Planck} and ACT CMB lensing maps. Thanks to its large footprint on the sky, extended redshift lever arm and high number density, the unWISE extragalactic samples are ideal for a number of cross-correlations, including CMB lensing tomographic analyses.

Using conservative choices and a blinded analysis we were able to obtain a 1.7\% and 1.8\% measurement of $S_8$ and $\sigma_8$, respectively, when combining the cross-correlations of the Blue and Green unWISE samples with ACT and \textit{Planck} lensing reconstructions. Both samples and their combination are in full agreement with CMB lensing and the primary CMB results, providing an important test of the standard cosmological model at $z \simeq 0.2-1.6$. \editF{Our results imply that, if new physics is responsible for the $S_8$ tension, any effects of new physics must be mostly limited to small scales ($k \gtrsim 0.3$ $h$/Mpc) to which our unWISE measurements are insensitive.}

The statistical consistency with CMB lensing auto-correlation results, together with the comprehensive suite of systematic and null tests, motivates deriving joint constraints between the cross-correlation presented here and the CMB lensing auto-power spectrum. This analysis will be presented in upcoming work. 

\section*{Acknowledgements}

The authors wish to thank David Alonso, Julien Carron, Benjamin Joachimi, Giulio Fabbian and Noah Sailer for useful discussions and comments. We are particularly grateful to Hironao Miyatake and the HSC team for making a consistent reanalysis of their Y3 results available to us. \resub{We also thank the anonymous reviewer for a very thorough review of this work and their constructive feedback.}

Support for ACT was through the U.S.~National Science Foundation through awards AST-0408698, AST-0965625, and AST-1440226 for the ACT project, as well as awards PHY-0355328, PHY-0855887 and PHY-1214379. Funding was also provided by Princeton University, the University of Pennsylvania, and a Canada Foundation for Innovation (CFI) award to UBC. The development of multichroic detectors and lenses was supported by NASA grants NNX13AE56G and NNX14AB58G. Detector research at NIST was supported by the NIST Innovations in Measurement Science program. 
ACT operated in the Parque Astron\'omico Atacama in northern Chile under the auspices of the Agencia Nacional de Investigaci\'on y Desarrollo (ANID). We thank the Republic of Chile for hosting ACT in the northern Atacama, and the local indigenous Licanantay communities whom we follow in observing and learning from the night sky.

Computing was performed using the Princeton Research Computing resources at Princeton University, the Niagara supercomputer at the SciNet HPC Consortium, and the Symmetry cluster at the Perimeter Institute. This research also used resources provided through the STFC DiRAC Cosmos Consortium and hosted at the Cambridge Service for Data Driven Discovery (CSD3). SciNet is funded by the CFI under the auspices of Compute Canada, the Government of Ontario, the Ontario Research Fund–Research Excellence, and the University of Toronto. Research at Perimeter Institute is supported in part by the Government of Canada through the Department of Innovation, Science and Industry Canada and by the Province of Ontario through the Ministry of Colleges and Universities. This research also used resources of the National Energy Research Scientific Computing Center (NERSC), a U.S. Department of Energy Office of Science User Facility located at Lawrence Berkeley National Laboratory, operated under Contract No. DE-AC02-05CH11231 using NERSC award HEP-ERCAPmp107.

GSF acknowledges support through the Isaac Newton Studentship and the Helen Stone Scholarship at the University of Cambridge. 
GSF, NM, IAC, FJQ, BDS acknowledge support from the European Research Council (ERC) under the European Union’s Horizon 2020 research and innovation programme (Grant agreement No. 851274). BDS further acknowledges support from an STFC Ernest Rutherford Fellowship.
IAC also acknowledges support from Fundaci\'on Mauricio y Carlota Botton.
ZA acknowledges support from NSF grant AST-2108126.
EC acknowledges support from the European Research Council (ERC) under the European Union’s Horizon 2020 research and innovation programme (Grant agreement No. 849169).
OD acknowledges support from SNSF Eccellenza Professorial Fellowship (No. 186879).
AJD acknowledges support from the Flatiron Institute which is supported by the Simons Foundation.
MH acknowledges financial support from the National Research Foundation of South Africa (Grant No. 137975).
KMH acknowledges support from the NSF under awards 1815887, 2009870, and 2206344, and from NASA under award 80NSSC23K0466.
LP acknowledges support from the Misrahi and Wilkinson funds.
MM acknowledges support from NASA grant 21-ATP21-0145.
GAM is part of Fermi Research Alliance, LLC under Contract No. DE-AC02-07CH11359 with the U.S. Department of Energy, Office of Science, Office of High Energy Physics.
MW is funded by the U.S. Department of Energy.
NS acknowledges support from DOE award number DE-SC0020441.
CS acknowledges support from the Agencia Nacional de Investigaci\'on y Desarrollo (ANID) through FONDECYT grant no.\ 11191125 and BASAL project FB210003.

\subsection*{Software}

Some of the results in this paper have been derived using the \texttt{healpy} \citep{2019JOSS....4.1298Z} and \texttt{HEALPix} \citep{2005ApJ...622..759G} packages. This research made use of \texttt{Astropy}\footnote{\url{https://www.astropy.org/}} a community-developed core Python package for Astronomy \citep{2013A+A...558A..33A,2018AJ....156..123A}. We also acknowledge use of the \texttt{matplotlib} \citep{2007CSE.....9...90H} package and the Python Image Library for producing plots in this paper. Furthermore, we acknowledge use of the \texttt{numpy} \citep{2020Natur.585..357H} and \texttt{scipy} \citep{2020NatMe..17..261V} packages. We use the Boltzman code \texttt{CAMB} \citep{2000ApJ...538..473L,2012JCAP...04..027H} for calculating theory spectra, and use \texttt{GetDist} \citep{2019arXiv191013970L} and \texttt{Cobaya} \citep{2021JCAP...05..057T} for likelihood analysis. We acknowledge work done by the Simons Observatory Pipeline and Analysis Working Groups in developing open-source software used in this paper.

\bibliography{joined_ACT_DR6_lensing,software_bib}

\appendix

\section{Off-diagonal correlations in the covariance matrix}\label{app:off-diag_correlations}

\begin{figure}
    \centering
    \includegraphics[width=0.6\linewidth]{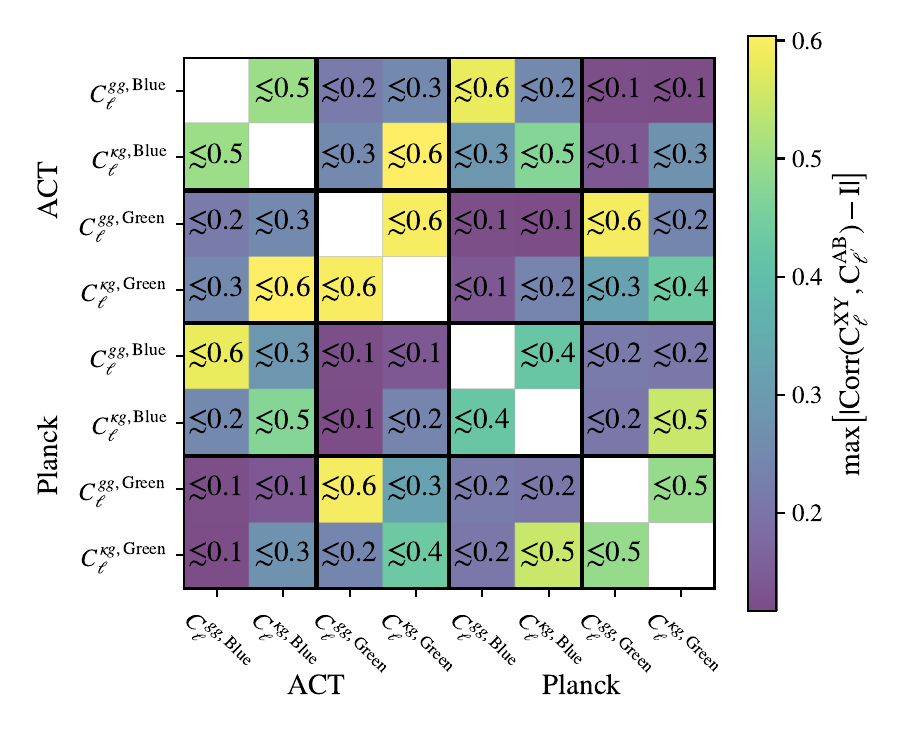}
    \caption{Maximum off-diagonal correlations between different samples and between cross-correlations using the ACT and \textit{Planck} lensing reconstructions. The most significant off-diagonal correlations are present between the $C_\ell^{\kappa g}$ from different samples which is due to the use of identical lensing maps and between $C_\ell^{gg}$ measured on the ACT and \textit{Planck} footprints due to the use of the identical galaxy samples.}
    \label{fig:off-diag_correlations}
\end{figure}

Further to the discussion of the off-diagonal correlations in Sec.\,\ref{sec:covmat} we show here a summary of the level of correlations between different samples and between the cross-correlations using ACT and \textit{Planck} lensing reconstructions. Fig.\,\ref{fig:off-diag_correlations} shows the maximum off-diagonal correlations within our analysis range that uses bandpowers in the range $50\leq\ell\leq400$ in $C_\ell^{\kappa g}$ and $100\leq\ell\leq400$ in $C_\ell^{gg}$.

\section{Detailed summary of null-tests}\label{app:null-tests}

\begin{figure}
    \centering
    \includegraphics[width=0.5\linewidth,trim=0cm 0.7cm 0cm 0cm, clip]{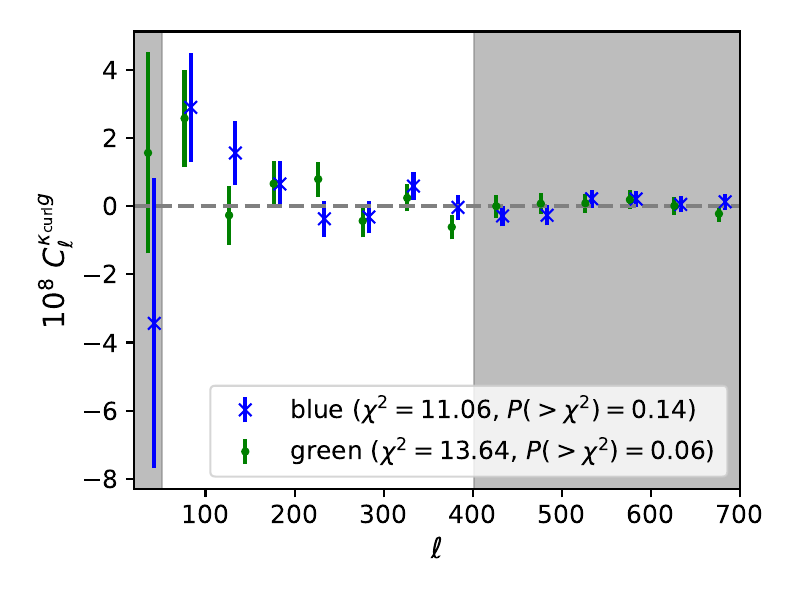}%
    \caption{At current levels of precision we do not expect any signal from the curl mode of the lensing reconstruction. The cross-correlation of the unWISE samples with the curl mode of the lensing reconstruction thus serves as a test of our covariance estimation.}
    \label{fig:null_test_curl}
\end{figure}

\begin{figure}
    \centering
    \begin{tabular}[b]{c c}
         \includegraphics[width=0.5\linewidth]{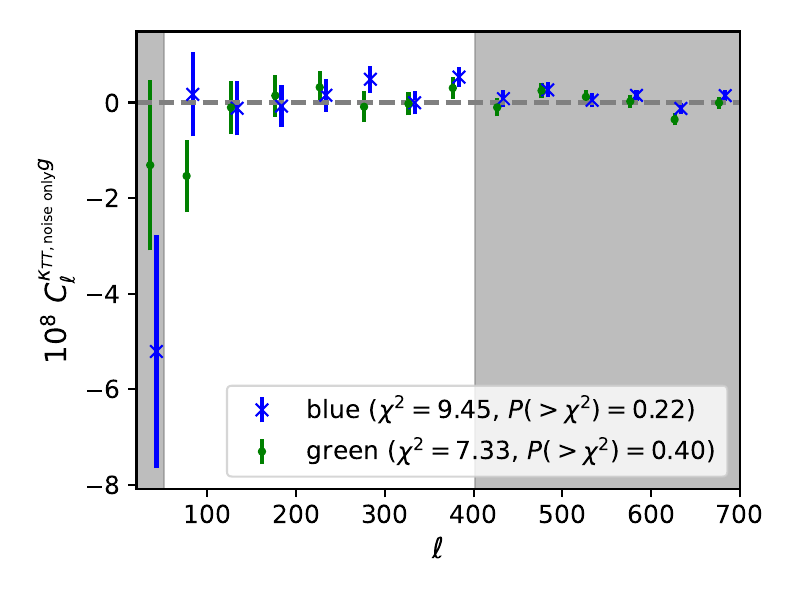} &
         \includegraphics[width=0.5\linewidth]{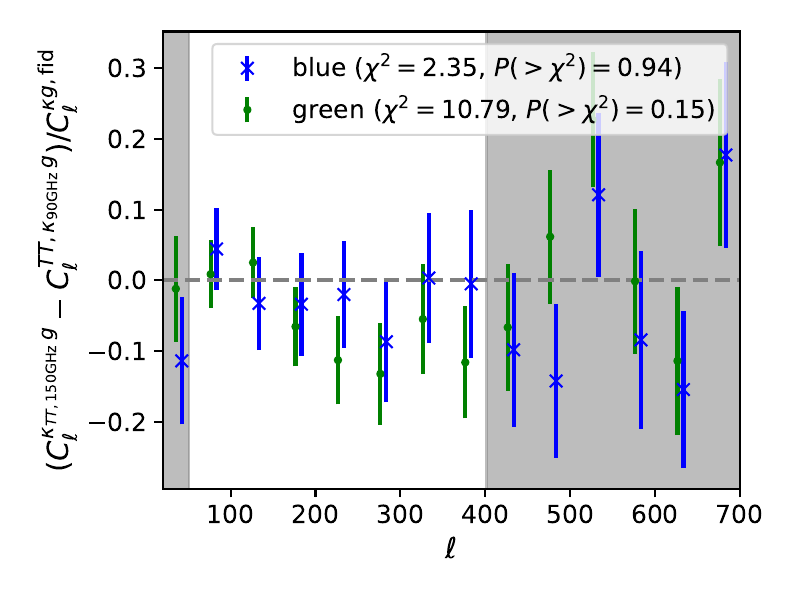} \\[-1.5em]
         \multilinecell{c}{(a) temperature only reconstruction on\\150 and 90GHz (noise only) difference maps} & \multilinecell{c}{(b) difference of cross-correlation bandpowers using\\temperature only reconstructions at 150 and 90GHz }\\[0.5em]

         \includegraphics[width=0.5\linewidth]{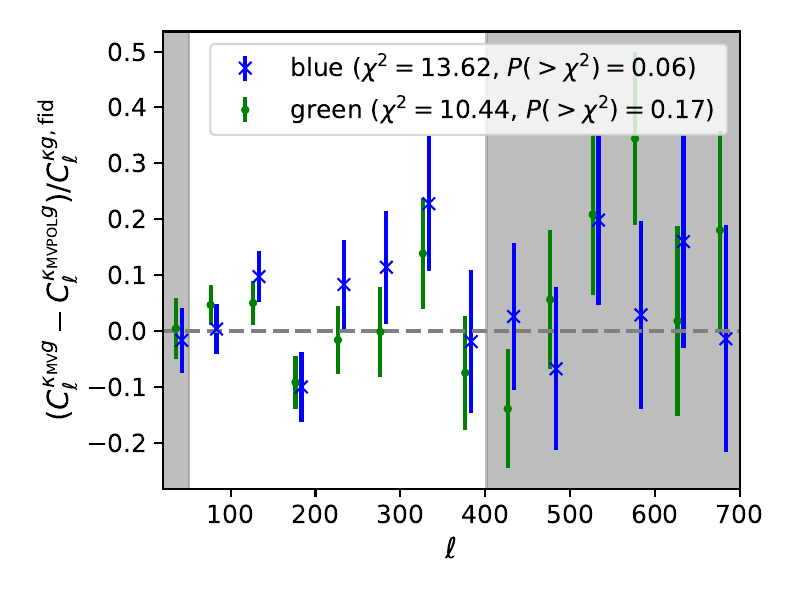} &
         \includegraphics[width=0.5\linewidth]{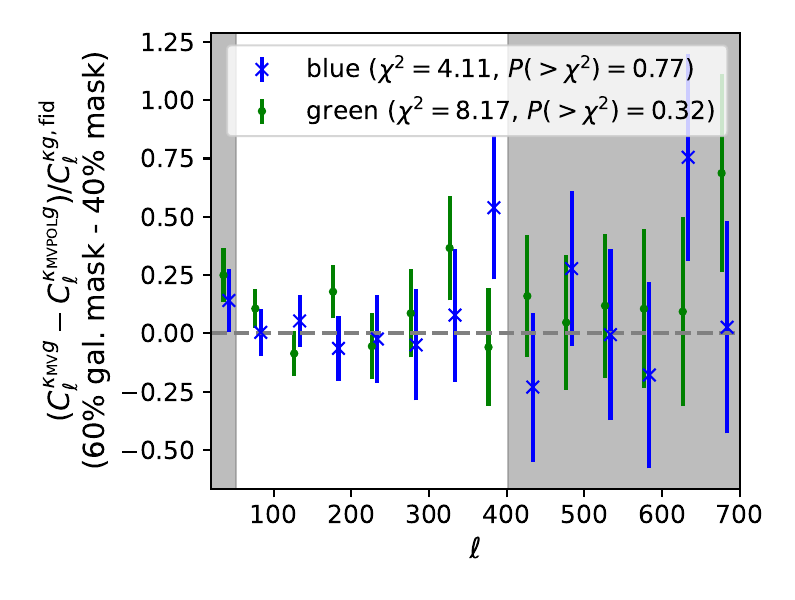}\\[-1.5em]
         \multilinecell{c}{(c) difference of cross-correlation bandpowers using\\minimum variance and polarisation only reconstructions} & \multilinecell{c}{(d) same as (c) but on the difference\\region of the 60\% and 40\% masks}\\[0.5em]

         \includegraphics[width=0.5\linewidth]{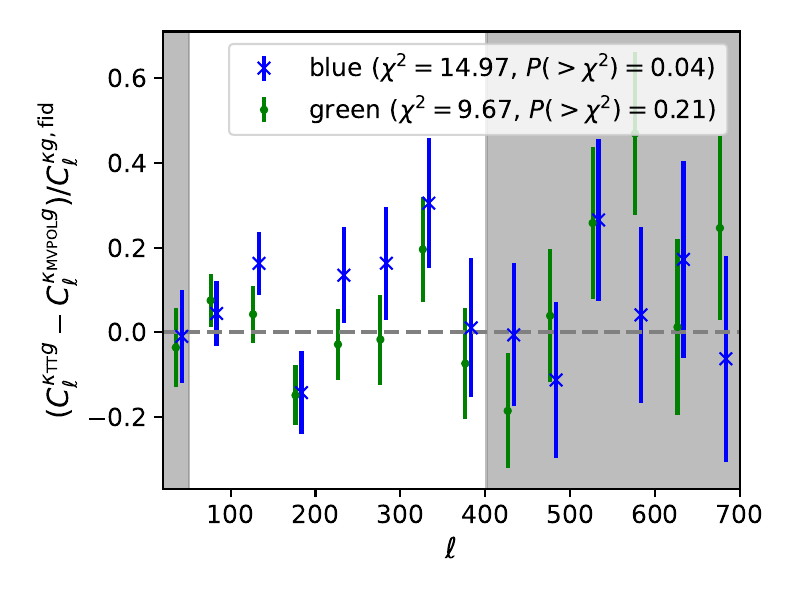} &
         \includegraphics[width=0.5\linewidth]{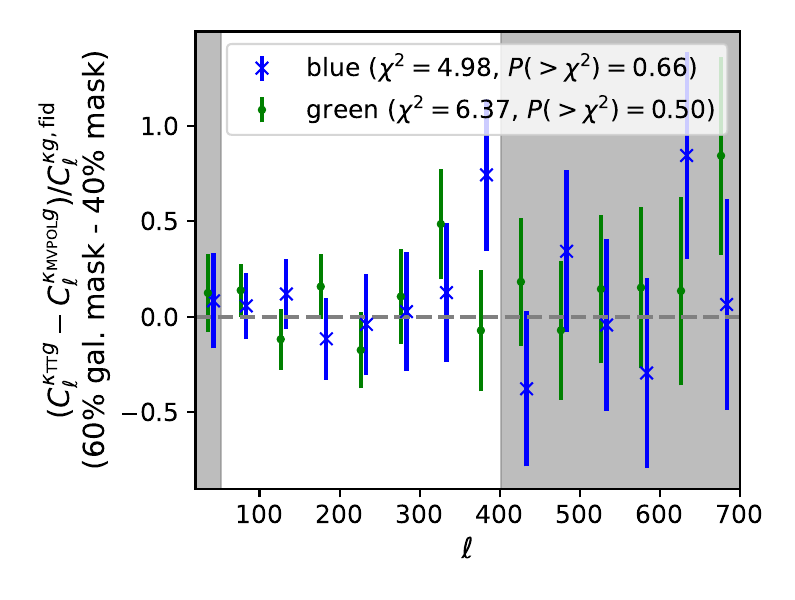}\\[-1.5em]
         \multilinecell{c}{(e) difference of cross-correlation bandpowers using\\temperature only and polarisation only reconstructions} & \multilinecell{c}{(f) same as (e) but on the difference\\region of the 60\% and 40\% masks}\\[0.5em]
    \end{tabular}
    \caption{Summary of additional null-tests related to potential foreground systematics in the lensing reconstruction. We compare cross-correlations with lensing reconstructions using only one of the two ACT frequencies (panels (a) and (b)), cross-correlations using the minimum variance and polarisation only reconstructions (panels (c) and (d)), and cross-correlations using the temperature only and polarisation only reconstructions (panels (e) and (f)). The number of failures is consistent with expectations when taking into account the correlations between different tests.}
    \label{fig:null_test_summary_lensing}
\end{figure}

A detailed summary of all our lensing reconstruction null-test results can be found in Table~\ref{table:clkg_nulls}. We also show the corresponding bandpowers not already shown in the body of this paper in Figures~\ref{fig:null_test_curl}, \ref{fig:null_test_summary_lensing}, and \ref{fig:null_test_CIBD}. The tests targeting the spatial homogeneity of the data are summarised in Table~\ref{table:mask_null_tests} with bandpowers shown in Fig.\,\ref{fig:null_test_summary_homogeneity} and additionally Fig.\,\ref{fig:null_test_stellar_mask}.

We perform twelve null-tests using different lensing reconstructions for each of the two galaxy samples, Blue and Green, and additionally a test on the cross-correlation between our galaxy samples and the curl mode of the baseline lensing reconstruction. The latter is primarily a test of our covariance estimation since we do not expect any signal in the curl mode of the lensing reconstruction at current levels of precision.

\begin{figure}
    \centering
    \includegraphics[width=0.5\linewidth,trim=0cm 0.7cm 0cm 0cm, clip]{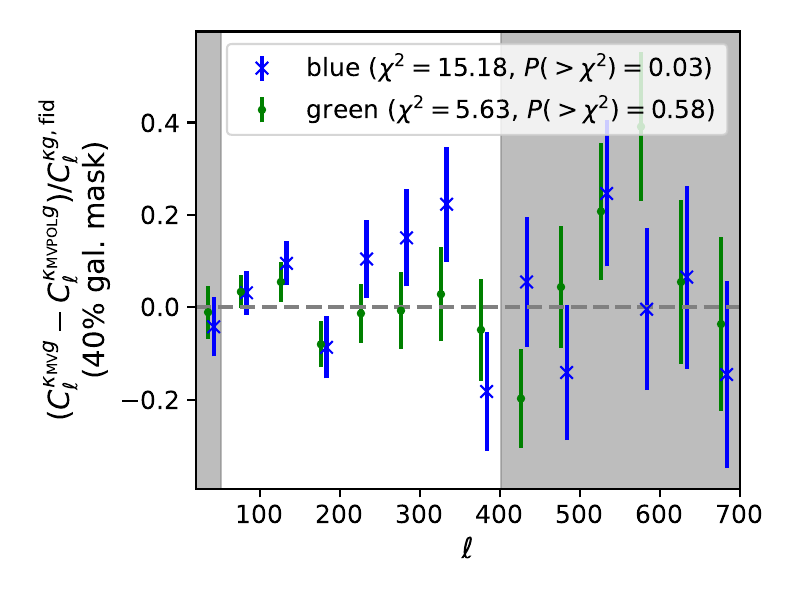}%
    \caption{Null-test showing the difference of cross-correlation bandpowers using minimum variance and polarisation only reconstructions on the more restrictive (40\%) Galactic mask. We observe a failure for the Blue sample which is highly correlated with other observed failures shown above. Overall the number of failures is consistent with expectations taking into account the correlations between our tests.}
    \label{fig:null_test_MV-MVPOL_GAL040}
\end{figure}

\begin{figure}
    \centering
    \includegraphics[width=0.5\linewidth,trim=0cm 0.7cm 0cm 0cm, clip]{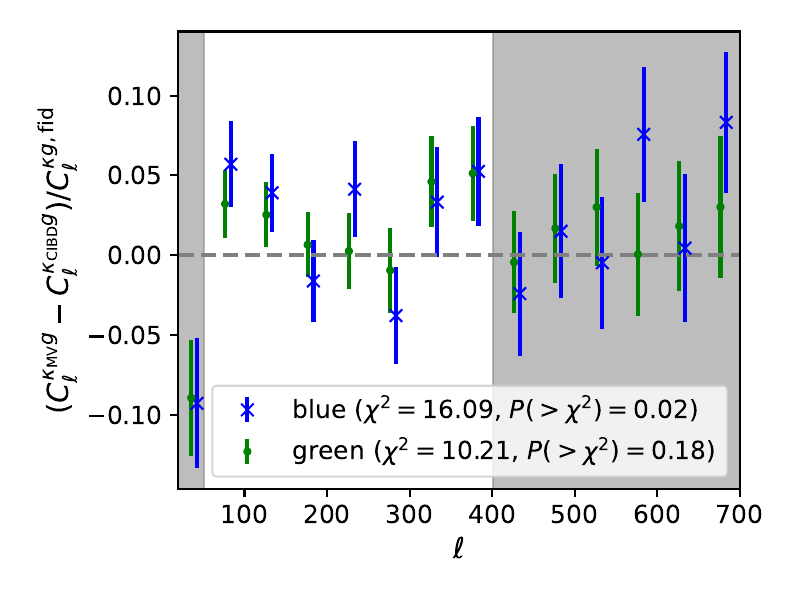}%
    \caption{Null-test showing the difference of cross-correlation bandpowers using our baseline reconstruction and one which explicitly deprojects CIB contamination using higher frequency data from \textit{Planck}. Again we see a failure for the Blue sample.  We note however that we cannot rule out the possibility that this is related to a slightly different mask used for the CIB deprojected reconstruction which removes some additional areas near the Galactic plane. Looking at the null-test for the galaxy auto-correlation on the modified masks also yields a borderline PTE of 0.09.}
    \label{fig:null_test_CIBD}
\end{figure}

\begin{figure}
    \centering
    \begin{tabular}[b]{c c}
        \includegraphics[width=0.5\linewidth]{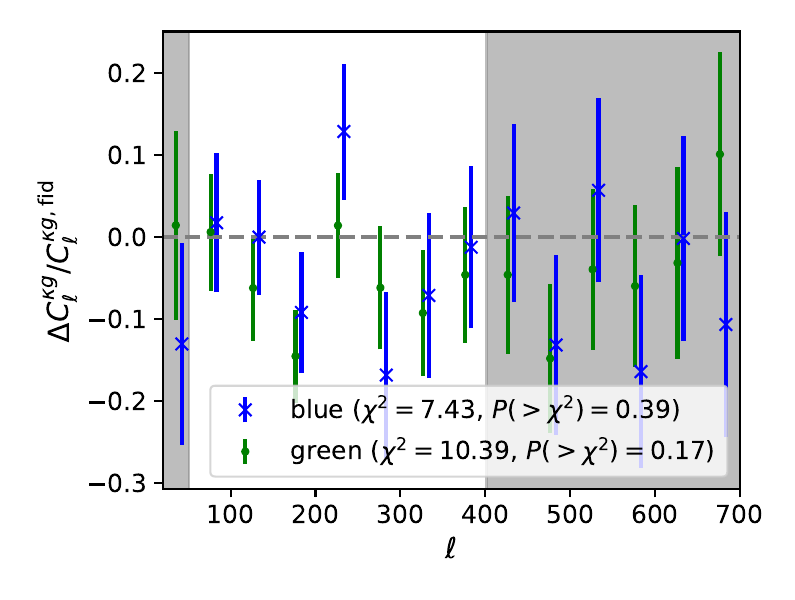} &
        \includegraphics[width=0.5\linewidth]{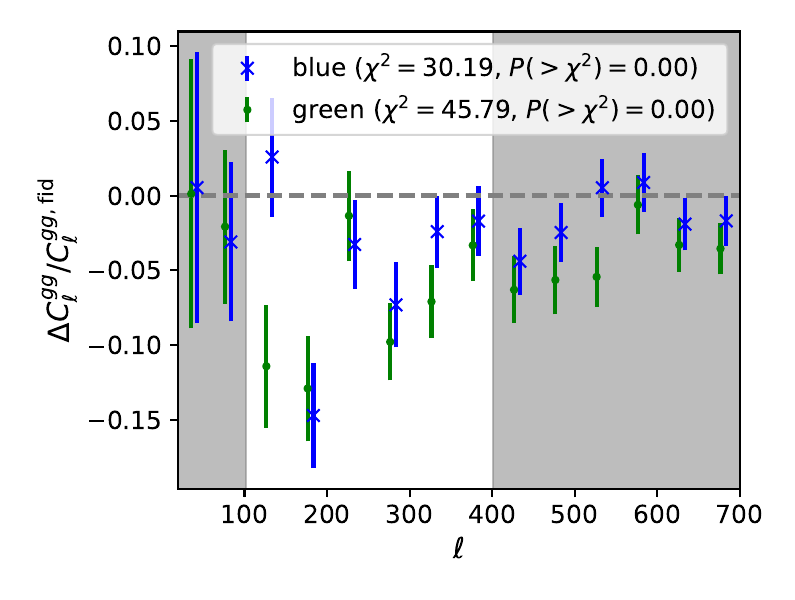}\\[-1.5em]
        \multilinecell{c}{(a) Difference of $C_\ell^{\kappa g}$ measured in the baseline footprint\\and when restricted to ecliptic latitudes $>30^\circ$.} & \multilinecell{c}{(b) Same as (a) but for $C_\ell^{gg}$.}\\[0.5em]

        \includegraphics[width=0.5\linewidth]{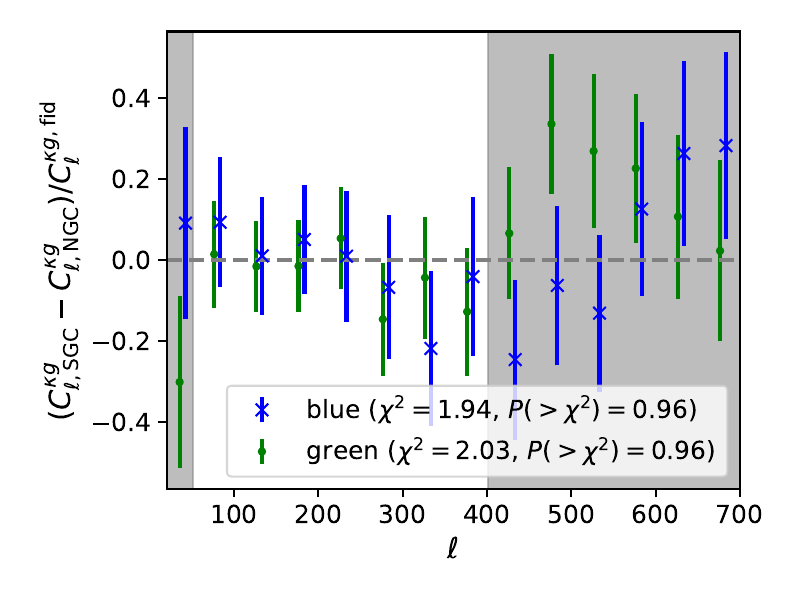} &
        \includegraphics[width=0.5\linewidth]{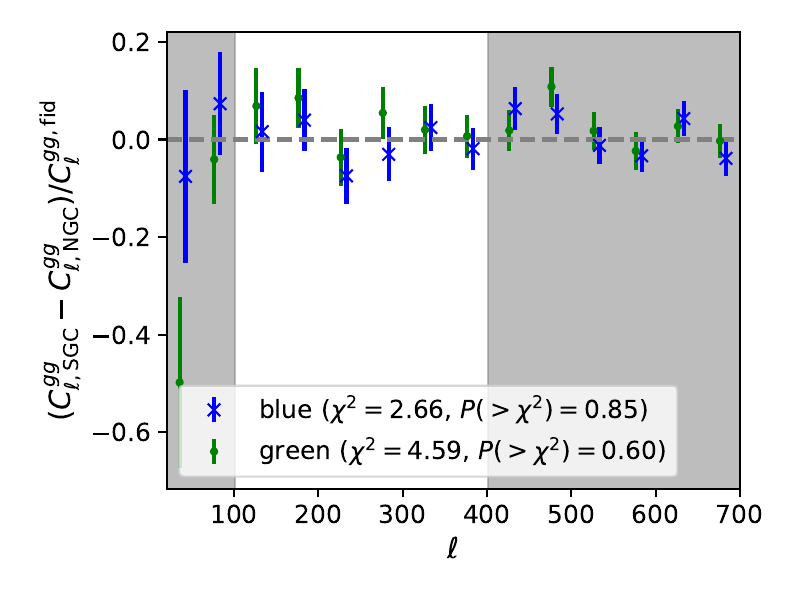}\\[-1.5em]
        \multilinecell{c}{(c) Difference of $C_\ell^{\kappa g}$ measured in the southern\\and northern galactic cap (SGC and NGC)} & \multilinecell{c}{(d) Same as (c) but for $C_\ell^{gg}$.}\\[0.5em]

        \includegraphics[width=0.5\linewidth]{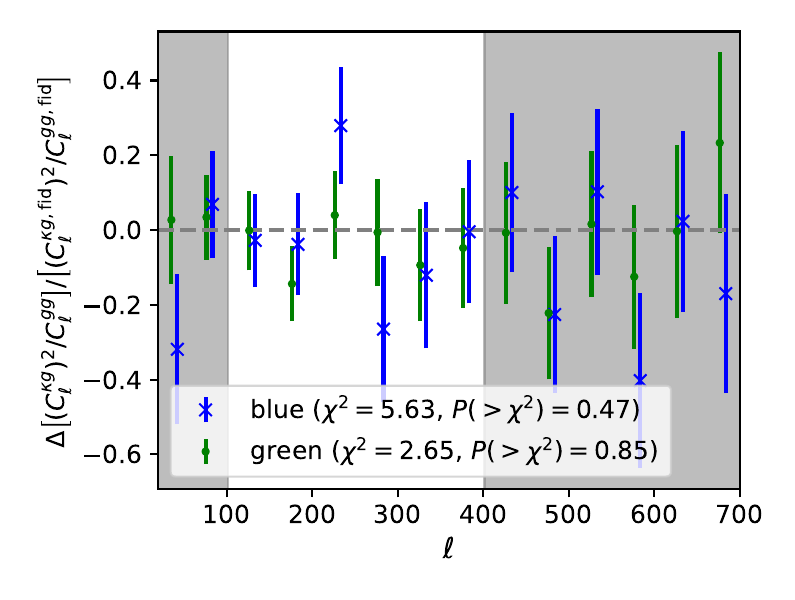} & 
        \includegraphics[width=0.5\linewidth]{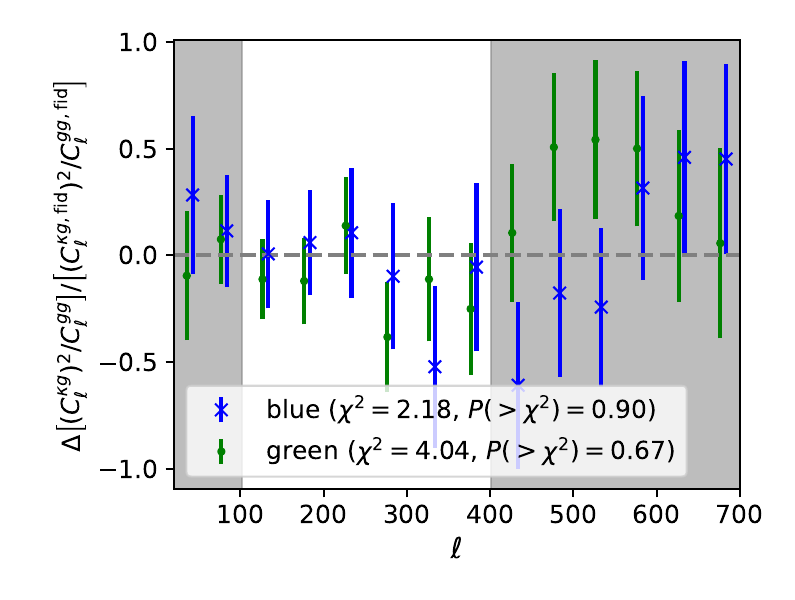}\\[-1.5em]
        \multilinecell{c}{(e) Difference in the bias independent quantity\\$(C_\ell^{\kappa g})^2/C_\ell^{gg}$ measured in the baseline footprint\\and when restricted to ecliptic latitudes $>30^\circ$.} & \multilinecell{c}{(f) Same as (e) but for the comparison\\between SGC and NGC.}\\[0.5em]
         
    \end{tabular}
    \caption{Summary of additional null-tests related to the spatial homogeneity of the observed spectra. We compare cross-correlations and galaxy auto-correlation spectra for different ecpliptic latitudes (panels (a) and (b)) and between the SGC and NGC (panels (c) and (d)). In addition, we show the approximately bias independent quantity $(C_\ell^{\kappa g})^2/C_\ell^{gg}$ for these two comparisons (panels (e) and (f)). While we find several highly significant failures for $C_\ell^{gg}$ caused by  variations in the galaxy bias arising from large scale fluctuations in the galaxy selection properties, the bias independent quantity $(C_\ell^{\kappa g})^2/C_\ell^{gg}$ passes our tests.}
    \label{fig:null_test_summary_homogeneity}
\end{figure}

\begin{figure*}
    \centering
    \includegraphics[width=0.5\linewidth]{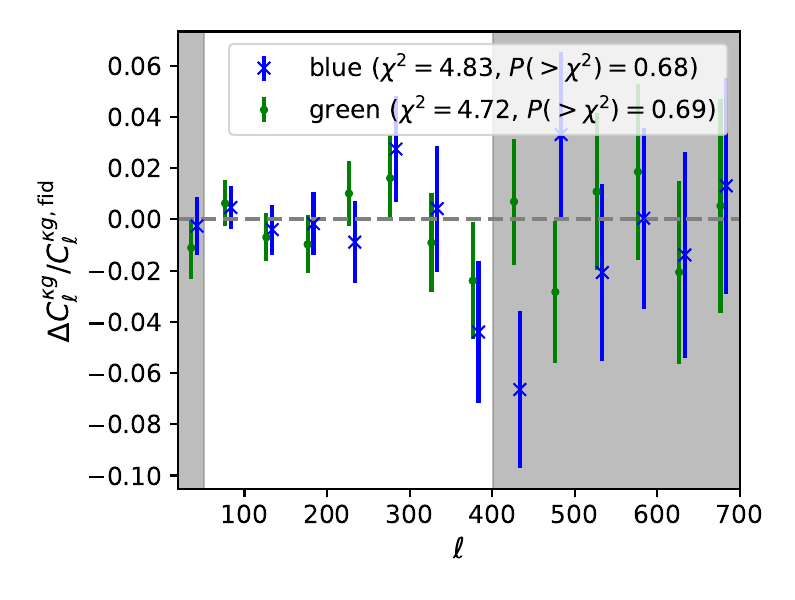}%
    \includegraphics[width=0.5\linewidth]{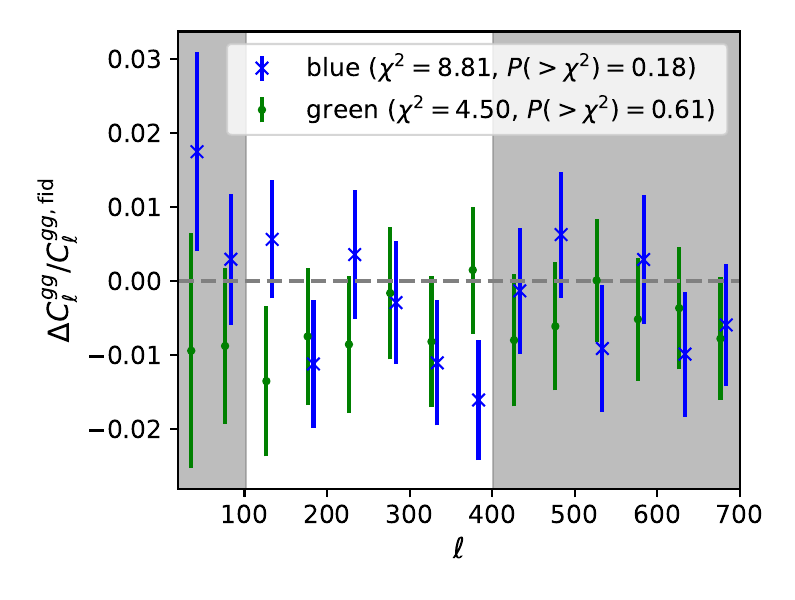}
    \caption{These null-tests compare our baseline mask, which removes any source within $2.75^{\prime \prime}$ from a Gaia point source (with W2 $< 2.5$), with one which additionally masks any source within 0.1$^\circ$ from a further 1.85 million stars brighter than 8th magnitude. We find no evidence for problems with stellar contamination. This is consistent with results from alternative estimates of the stellar contamination in \cite{2020JCAP...05..047K}.}
    \label{fig:null_test_stellar_mask}
\end{figure*}

In Sec.\,\ref{subsec:lensing_sys} we discuss the observed failures and summarise our null-tests targeting contamination of the lensing reconstruction. The number of observed failures is consistent with what would be expected due to random fluctuations given the correlations between the various tests. We nevertheless investigated whether the observed failures were indicative of contamination, for example by considering alternative masks to maximise or minimise the contribution from potential contaminants. We performed the comparison of the minimum variance and polarisation reconstructions on a more restrictive Galactic mask (see Fig.\,\ref{fig:null_test_MV-MVPOL_GAL040}); if the failures observed on the full footprint was due to contamination by polarised Galactic dust, we should have seen an improvement further away from the Galactic plane and we do not.   Alternatively, the effect should be more significant near the Galactic plane (i.e. in the region which is included in our baseline mask but not the more restrictive Galactic mask), but this is not the case (see Fig.\,\ref{fig:null_test_summary_lensing} panels (d) and (f)).

Our tests for the homogeneity of the data sets were discussed in Sec.\,\ref{subsec:galaxy_sys}. As pointed out there we do not expect null-tests to pass due to fluctuations in the selection properties of the galaxy sample on large scales. Those fluctuations affect, among other things, the galaxy bias. We thus construct an approximately bias independent quantity, $(C_\ell^{\kappa g})^2/C_\ell^{gg}$, on which we perform null-tests to confirm the homogeneity of the data. In Fig.\,\ref{fig:null_test_summary_homogeneity} (panel (e)) we show this null-test for the difference between our baseline analysis and an analysis restricted to ecliptic latitude larger than $30^\circ$. This test primarily targets trends with unWISE depth which varies with ecliptic latitude due to the satellites scan strategy. We also compared the bandpowers obtained on the SGC and NGC (panel (f)) and find those to be consistent as well. Finally, in Fig.\,\ref{fig:null_test_stellar_mask}, we show tests for $C_\ell^{\kappa g}$ and $C_\ell^{gg}$ comparing different stellar masks. For this test we mask 0.1$^{\circ}$ around an additional 1.85 million stars with W2, W2 or 2MASS $K_m$ brighter than 8th magnitude, beyond the 12890 very bright stars (W2 $< 2.5$) that we mask by default. This decreases the unmasked sky fraction in the unWISE mask from 0.58 to 0.45. The tests are passing and because we are changing the mask on small scales rather than selecting a different region we expect this test to be unaffected by large scale fluctuations in the galaxy selection.

\begin{deluxetable}{lDDDD}

\tablehead{
\nocolhead{}    & \multicolumn{4}{c}{Blue} & \multicolumn{4}{c}{Green} \\
\multicolumn{1}{l}{Null-test}     & \multicolumn2c{$\chi^2$} & \multicolumn2c{PTE} & \multicolumn2c{$\chi^2$} & \multicolumn2c{PTE}}
\startdata
\decimals
Curl & 11.06                & 0.136                & 13.64                & 0.058               \\ 
\cutinhead{\bf Minimum Variance - Temperature Only - Polarisation Only}\\
$C_\ell^{\kappa_{\rm{MV}}g} - C_\ell^{\kappa_{\rm{TT}} g}$ & 15.20                & 0.033                & 8.52                 & 0.289         \\
$C_\ell^{\kappa_{\rm{MV}}g} - C_\ell^{\kappa_{\rm{MVPOL}} g}$ & 13.62                & 0.058                & 10.44                 & 0.165                \\
$C_\ell^{\kappa_{\rm{TT}}g} - C_\ell^{\kappa_{\rm{MVPOL}} g}$ & 14.97                & 0.036                & 9.67                 & 0.208                \\
$C_\ell^{\kappa_{\rm{MV}}g} - C_\ell^{\kappa_{\rm{MVPOL}} g}$ (40\% gal. mask)  & 15.18                & 0.034                & 5.63                 & 0.584                \\
\cutinhead{\bf Minimum Variance - Temperature Only - Polarisation Only (near Galactic plane)}
$C_\ell^{\kappa_{\rm{MV}}g} - C_\ell^{\kappa_{\rm{TT}} g}$ (near Galactic plane)    & 7.40                 & 0.388                & 6.76                 & 0.454                \\
$C_\ell^{\kappa_{\rm{MV}}g} - C_\ell^{\kappa_{\rm{MVPOL}} g}$ (near Galactic plane) & 4.11                 & 0.767                & 8.17                 & 0.318                \\
$C_\ell^{\kappa_{\rm{TT}}g} - C_\ell^{\kappa_{\rm{MVPOL}} g}$ (near Galactic plane) & 4.98                 & 0.663                & 6.37                 & 0.497                \\
\cutinhead{\bf Frequency difference tests}
$C_\ell^{\kappa_{\rm{MV},\rm{noise\ only}}g}$ (150GHz - 90GHz map level null-test, minimum variance)                         & 6.70                 & 0.460                & 10.89                & 0.143                \\
$C_\ell^{\kappa_{\rm{TT},\rm{noise\ only}}g}$ (150GHz - 90GHz map level null-test, temperature only)                           & 9.45                 & 0.222                & 7.33                 & 0.396                \\
$C_\ell^{\kappa_{\rm{MV}, 150\rm{GHz}}g} - C_\ell^{\kappa_{\rm{MV}, 90\rm{GHz}} g}$ (150GHz - 90GHz bandpower null-test, minimum variance)                     & 3.36                 & 0.850                & 6.78                 & 0.452                \\
$C_\ell^{\kappa_{\rm{TT}, 150\rm{GHz}}g} - C_\ell^{\kappa_{\rm{TT}, 90\rm{GHz}} g}$ (150GHz - 90GHz bandpower null-test, temperature only)                     & 2.35                 & 0.938                & 10.79                 & 0.148                \\
\cutinhead{\bf Other}
baseline vs. CIB deprojection                       & 16.09                & 0.024                & 10.21                 & 0.177                \\
\enddata
\caption{Summary of null-tests for $C_\ell^{\kappa g}$. We compute the $\chi^2$ within our cosmological analysis range of $50\leq\ell\leq400$ and estimate the PTE using a $\chi^2$-distribution with 7 degrees of freedom (given the seven bandpowers in the analysis range). We find 4 failures (PTE$<0.05$) which are highly correlated; all relate to comparing the cross-correlation of the Blue sample of galaxies with different lensing reconstruction options (minimum variance, temperature only, and polarisation only). We investigate in Sec.\,\ref{subsec:lensing_sys} potential systematic contamination, but stress that the number of failures is not inconsistent with random fluctuations and no further evidence of contamination is found. \label{table:clkg_nulls}}
\end{deluxetable}

\begin{deluxetable}{l DDDDDD DDDDDD}

\tablehead{
\nocolhead{}    & \multicolumn{12}{c}{Blue} & \multicolumn{12}{c}{Green} \\
\nocolhead{}    & \multicolumn{4}{c}{$C_\ell^{\kappa g}$} & \multicolumn{4}{c}{$C_\ell^{g g}$} & \multicolumn{4}{c}{$(C_\ell^{\kappa g})^2/C_\ell^{gg}$} & \multicolumn{4}{c}{$C_\ell^{\kappa g}$} & \multicolumn{4}{c}{$C_\ell^{g g}$} & \multicolumn{4}{c}{$(C_\ell^{\kappa g})^2/C_\ell^{gg}$} \\
\multicolumn{1}{l}{Null-test}     & \multicolumn2c{$\chi^2$} & \multicolumn2c{PTE} & \multicolumn2c{$\chi^2$} & \multicolumn2c{PTE} & \multicolumn2c{$\chi^2$} & \multicolumn2c{PTE} & \multicolumn2c{$\chi^2$} & \multicolumn2c{PTE} & \multicolumn2c{$\chi^2$} & \multicolumn2c{PTE} & \multicolumn2c{$\chi^2$} & \multicolumn2c{PTE}}
\startdata
\decimals
40\% Galactic mask & 4.99 & 0.662 & 6.78  & 0.341 & 2.57 & 0.861 & 4.65 & 0.702 & 16.51 & 0.011 & 3.39 & 0.758 \\
$\beta_{\rm ecliptic}>30^{\circ}$ & 7.43 & 0.386 & 30.19 & 0.000 & 5.63 & 0.466 & 10.39 & 0.168 & 45.79 & 0.000 & 2.65 & 0.851 \\
SGC - NGC       & 1.94 & 0.963 & 2.66  & 0.850 & 2.18 & 0.903 & 2.03 & 0.958 & 4.59  & 0.597 & 4.04 & 0.672 \\
stricter stellar mask & 4.83 & 0.681 & 7.65  & 0.265 & \multicolumn{2}{c}{$-$} & \multicolumn{2}{c}{$-$} & 4.72 & 0.694 & 4.50  & 0.609 & \multicolumn{2}{c}{$-$} & \multicolumn{2}{c}{$-$}
\enddata
\caption{Summary of null-tests targeting the spatial homogeneity of the galaxy samples. As expected we find some inhomogeneity in the galaxy sample, in particular with respect to cuts in ecliptic latitude. The $\chi^2$ is computed within our cosmological analysis ranges of $50\leq\ell\leq400$ for $C_\ell^{\kappa g}$ and $100\leq\ell\leq400$ for $C_\ell^{g g}$. We estimate the PTE using a $\chi^2$-distribution with 7(6) degrees of freedom (given the 7(6) bandpowers in the analysis range for $C_\ell^{\kappa g}$ ($C_\ell^{gg}$)). As argued in Sec.~\ref{subsec:galaxy_sys}, the failures are likely due to large scale inhomogeneities in the galaxy selection and thus varying galaxy bias. The bias independent combination $(C_\ell^{\kappa g})^2/C_\ell^{gg}$, however, passes our null-tests.\label{table:mask_null_tests}}
\end{deluxetable}

\section{Details of power spectrum model} \label{app:model_details}
In this appendix we discuss the details of our correction for the dependence of the cross-correlation redshifts on the fiducial cosmology, as well as our correction for the noise bias introduced by marginalisation over redshift uncertainties.

In Sec.\,\ref{subsec:dndz_marg} we raised the need to assume a fiducial cosmology to interpret the cross-correlation between the photometric and spectroscopic samples as estimates of $b(z)dN/dz$. In this appendix we lay out a formalism for correcting for this assumption to avoid biasing our results (Sec.\,\ref{app:fid_cosmo_correction}). 

In Sec.\,\ref{subsec:dndz_marg} we discussed the effect of imposing smoothness and positivity on the normalised cross-correlation redshifts, $W^{\rm{xc}}(z)$.  The result is a noise bias introduced by our marginalisation procedure, since the final $C_\ell$ effectively become a non-linear function of a noisy quantity. This bias can be as large as $\sim 5\%$ for some of the terms contributing to $C_\ell^{gg}$ in some range of $\ell\rm{s}$. The effect is less significant for $C_\ell^{\kappa g}$, but still remains comparable to the data covariance. In Sec.\,\ref{app:noise_bias} we detail how we subtract this noise bias.

\subsection{Correction for the Fiducial Cosmology Assumed in Determining Cross-correlation Redshifts} \label{app:fid_cosmo_correction}
As described in Sec.\,\ref{subsubsec:dndz} (with further details provided in \cite{2020JCAP...05..047K} and Appendix~\ref{app:spec_samples}) we measure the redshift distribution of unWISE galaxies using cross-correlations with spectroscopic surveys. This introduces a dependence on the fiducial cosmology assumed in the process of converting from the measured cross-correlation functions to an estimate of $b(z) dN/dz$ \citep{2021JCAP...12..028K}. To correct for this dependence we introduce a new method which is exact in the limit of narrow redshift bins used to measure the cross-correlations between the photometric and spectroscopic samples.

We measure the cross-correlations between the photometric unWISE sample (subsequently indicated by a subscript $p$) and multiple spectroscopic tracers (subsequently denoted with a subscript $s$) as well as the spectroscopic sample's auto-correlation in $N$ narrow redshift bins between $z_{\rm{min},i}$ and $z_{\rm{max},i}$ ($i=1,\dots, N$). Assuming scale independent bias\footnote{This assumption is tested to some extent in our mocks used for model verification, where we have also measured the cross-correlation redshifts on the same scales as in our analysis. The ability to recover an unbiased cosmology discussed in Sec.\,\ref{subsec:model_test} indicates that this does not bias our constraints. Moreover, we find similar cosmological parameters (shift of $-0.1\sigma$ in $S_8$) when increasing the minimum scale used in the cross-correlation redshifts, which decreases the sensitivity to scale-dependent bias.} \resub{and linear bias} the relevant correlation functions are given in theory by
\begin{eqnarray}
    w_{sp}(\theta, z_i) &=& \int \frac{k dk}{2\pi} \int_{z_{\rm{min},i}}^{z_{\rm{max},i}} dz J_0\left[k \theta \chi(z)\right] b_s \frac{dN_s}{dz} b_p \frac{dN_p}{dz} H(z) P_{mm}(k, z)\\
    w_{ss}(\theta, z_i) &=& \int \frac{k dk}{2\pi} \int_{z_{\rm{min},i}}^{z_{\rm{max},i}} dz J_0\left[k \theta \chi(z)\right] \left(b_s \frac{dN_s}{dz}\right)^2 H(z) P_{mm}(k, z).
\end{eqnarray}
Here $b_s$ and $b_p$ are the bias of the spectroscopic and photometric samples respectively and $dN_s/dz$ and $dN_p/dz$ are the corresponding redshift distributions. The comoving radial distance, $\chi(z)$, the Hubble rate $H(z)$ and the matter-matter power spectrum $P_{mm}(k,z)$ depend on the assumed cosmology.

Furthermore, the correlation functions are binned into $n=3$ log-spaced angular bins between $\theta_{\rm{min}} = s_{\rm{min}}/\chi_{\rm{fid}}(z_{\rm{central}})$ and $\theta_{\rm{max}} = s_{\rm{max}}/\chi_{\rm{fid}}(z_{\rm{central}})$ where $s_{\rm{min}}= 2.5\rm{\ h^{-1} Mpc}$ and $s_{\rm{max}} = 10 \rm{\ h^{-1} Mpc}$. Here $\chi_{\rm{fid}}$ denotes the comoving radial distance within the chosen fiducial cosmology. Finally, the resulting quantity is integrated over the full range of scales weighting by $\theta^{-1}$ to increase the signal to noise \citep{2013arXiv1303.4722M}, yielding
\beq
\bar{w}_{XY}(z_i) = \sum_{j=1}^n \frac{2\Delta \theta_j}{\theta_j (\theta^2_{\rm{max}, j} - \theta^2_{\rm{min}, j})} \int_{\theta_{\rm{min}, j}}^{\theta_{\rm{max}, j}} \theta d\theta \ w_{XY}(\theta, z_i).
\eeq

In the limit of narrow redshift bins we can estimate the spectroscopic bias, $b_s$, from the measured binned and integrated auto-correlation function as
\beq
\hat{b}_s(z_i) = \sqrt{\hat{\bar{w}}_{ss}(z_i)} \left[\Delta z_i \left(\frac{dN_s}{dz}\right)^2 H(z_i) \int k dk  P_{mm}(k, z_i) W(k, z_i)\right]^{-1/2},
\eeq
where $W(k, z)$ is a weighting function which can be computed analytically as
\begin{equation}
    W(k, z) = \sum_{j=1}^n \frac{\Delta \theta_j \left[\theta_{\rm{max},j} J_1\left(k \chi(z) \theta_{\rm{max},j}\right) - \theta_{\rm{min},j} J_1\left(k \chi(z) \theta_{\rm{min},j}\right) \right]}{k \chi(z) \theta_j \pi \left(\theta^2_{\rm{max},j}-\theta^2_{\rm{min},j}\right)}.
\end{equation}
Finally, we can obtain an estimate of $b_p dN_p/dz$ as
\beq
\widehat{b_p \frac{dN_p}{dz}} = \frac{\hat{\bar{w}}_{sp}(z_i)}{\sqrt{\hat{\bar{w}}_{ss}(z_i)}} \left[\Delta z_i H(z_i) \int k dk P_{mm}(k, z_i) W(k, z_i)\right]^{-1/2}.
\eeq
Given $\left(\widehat{b_p dN_p/dz}\right)_{\rm{fid}}$ measured in a chosen fiducial cosmology the dependence on the choice of fiducial cosmology can therefore be corrected as
\beq \label{eq:fid_cosmo_correction}
\widehat{b_p \frac{dN_p}{dz}} = \left(\widehat{b_p \frac{dN_p}{dz}}\right)_{\rm{fid}} \frac{\mathcal{C}}{\mathcal{C}_{\rm{fid}}}.
\eeq
with
\beq
\mathcal{C} = \left[\Delta z_i H(z_i) \int k dk P_{mm}(k, z_i) W(k, z_i)\right]^{-1/2}.
\eeq
\label{eqn:cfactor}

In our analysis we compute $\mathcal{C}$ for every new cosmology to avoid biasing our results through the dependence on the assumed fiducial cosmology. For consistency with the method used to measure the cross-correlation redshifts we use a \texttt{HMCode} model for $P_{gg}(k,z)$.
We show in Table~\ref{table:param_consistency_subset} that our $S_8$ constraints are unchanged even when using a different fiducial cosmology, validating the correction of Eq.~\ref{eqn:cfactor}.

\subsection{Noise Bias Correction}\label{app:noise_bias}
Similarly to \cite{2021JCAP...12..028K} we compute a correction for the noise bias introduced by marginalisation over uncertainties in the redshift distributions within a fiducial cosmology \cite[ again  taken to be the mean cosmology from][]{2020A+A...641A...6P}. First we draw 9000 sets of PCA coefficients for each of the galaxy samples according to the skewnormal priors obtained as described in Sec.\,\ref{subsec:dndz_marg} and summarised in Table~\ref{tab:priors}. This number is chosen such that the resulting uncertainty in the correction is highly subdominant compared to the data covariance. We then evaluate the power spectrum model term by term for each set of coefficients and obtain the noise bias as
\beq
(\Delta C^{XY}_\ell)_{\rm{noise\ bias}} = \langle{C^{XY, \rm{theory}}_\ell\rangle}_{c_1, \dots, c_n} - \left. C^{XY, \rm{theory}}_\ell \right|_{c_1= \dots= c_n=0}
\eeq
where $C_\ell^{XY, \rm{theory}}$ are the various terms contributing to our theory expressions for $C_\ell^{gg}$ and $C_\ell^{\kappa g}$. During cosmology analysis we then correct the theory spectra as
\beq
(C_\ell^{XY, \rm{theory}})_{\rm{corrected}} = C_\ell^{XY, \rm{theory}} - (\Delta C^{XY}_\ell)_{\rm{noise\ bias}}.
\eeq
In this process we scale our noise bias correction appropriately with the value of the nuisance parameters including galaxy bias parameters and lensing magnification parameters.

\section{Posterior distributions for the full parameter space}

\begin{figure}
    \centering
    \includegraphics[width=\linewidth]{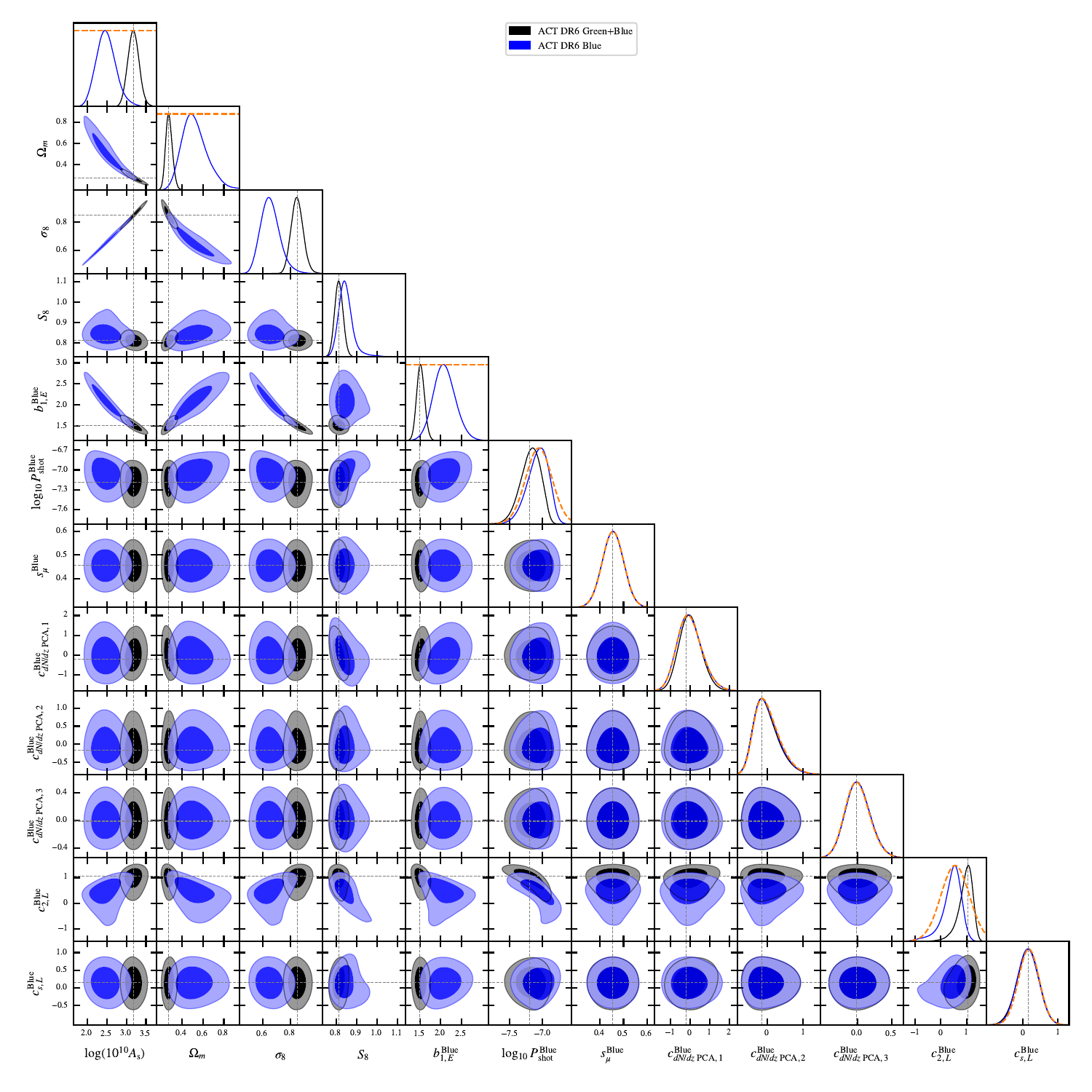}
    \caption{Posterior distributions in the full parameter space including the galaxy nuisance parameters for the Blue sample of unWISE galaxies. The dashed grey lines indicate the maximum a posterior (MAP) and the orange dashed lines show the priors on all samples parameters. We observe that the data does not constrain the redshift distribution beyond the priors provided which are described in Sec.\,\ref{subsec:dndz_marg}. On the other hand the data is informative for the second order Lagrangian bias.}
    \label{fig:full_post_wpriors_blue}
\end{figure}

\begin{figure}
    \centering
    \includegraphics[width=\linewidth]{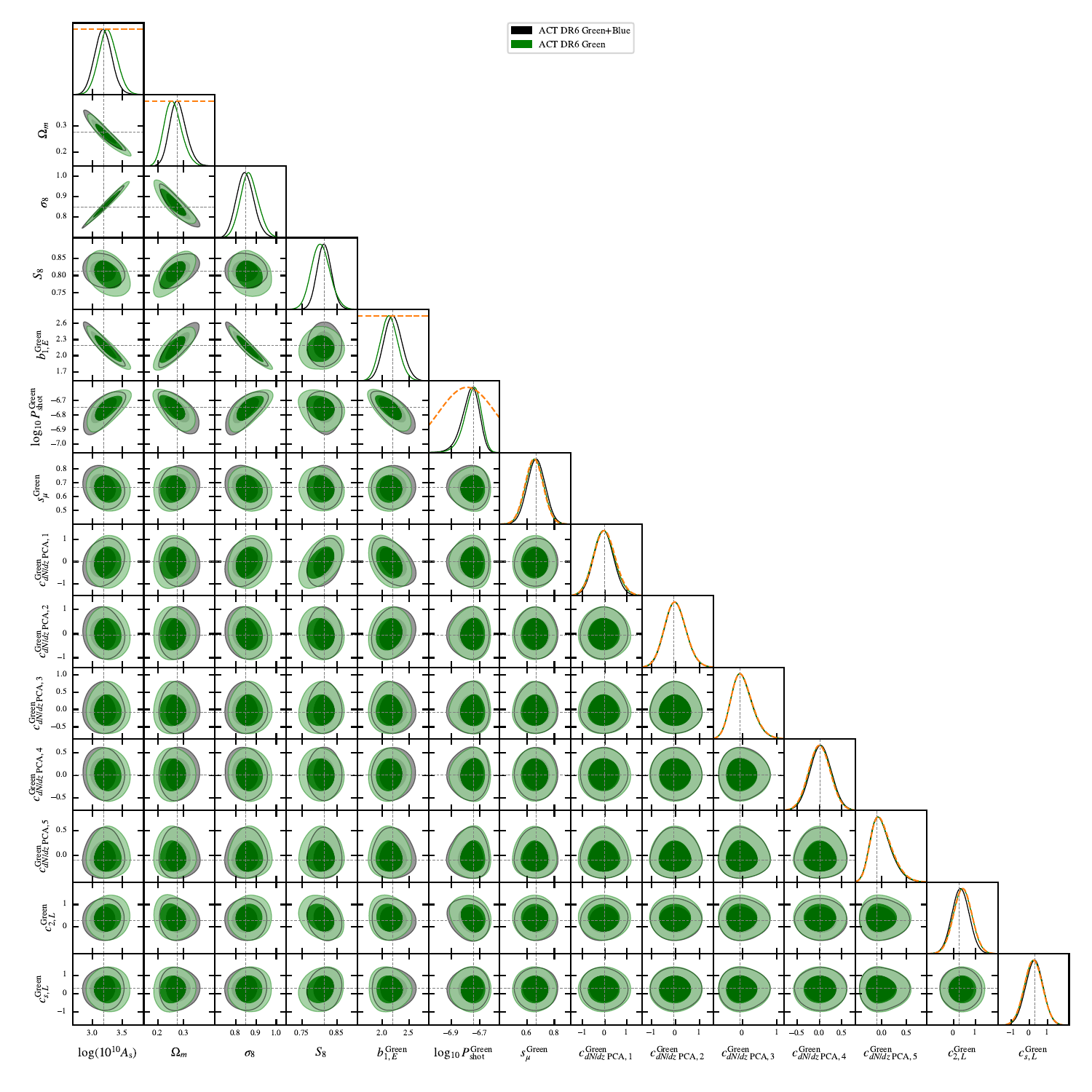}
    \caption{Same as Fig.\,\ref{fig:full_post_wpriors_blue} for the Green sample of unWISE galaxies. As with the Blue sample the redshift distribution parameters are not constrained by the data. In contrast with Blue the higher order biases are also prior dominated since these contributions are significantly smaller for the Green sample.}
    \label{fig:full_post_wpriors_green}
\end{figure}

In Figs.\,\ref{fig:full_post_wpriors_blue} and \ref{fig:full_post_wpriors_green} we show the posterior distributions for the full parameter space including the galaxy nuisance parameters for the Blue and Green sample of unWISE galaxies respectively. The posteriors shown are for our baseline analysis without BAO. Alongside the posterior distributions we also show the maximum a posterior (MAP) and the priors on all sampled parameters. We observe that the data does not constrain the redshift distribution marginalisation parameters beyond the information provided by the priors. These parameters are also only very weakly degenerate with the cosmological parameters of interest. In the case of the free parameters modifying the higher order bias evolution we see that the only parameter which is constrained by the data is the offset in the second order Lagrangian bias for the Blue sample. This parameter has some degeneracy with cosmology (see Appendix~\ref{app:sample_consistency} for some discussion of this). For the Green sample higher order contributions are less significant and the constraints on the higher order bias parameters are prior dominated.

\section{Parameter consistency from Blue and Green}\label{app:sample_consistency}

\begin{figure*}
    \centering
    \includegraphics[width=0.5\linewidth,trim=0.3cm 0.5cm 0.5cm 0.5cm, clip]{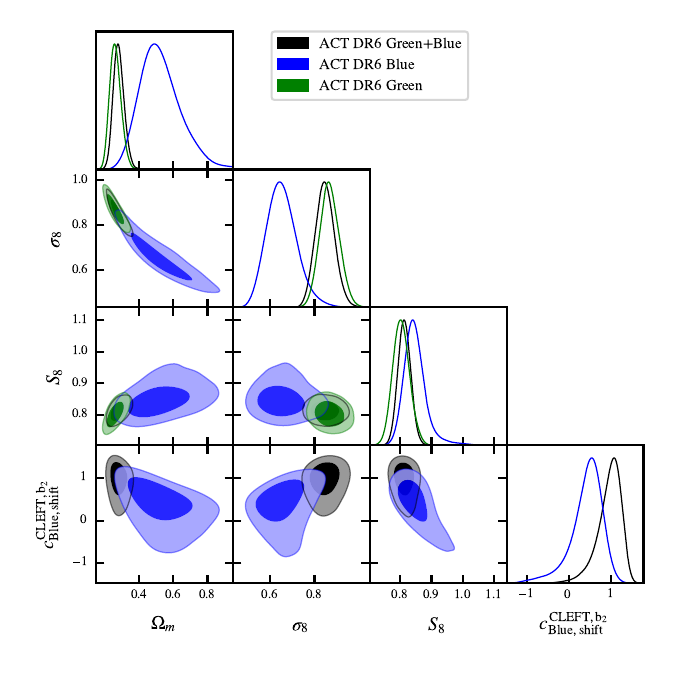}%
    \includegraphics[width=0.5\linewidth,trim=0.3cm 0.5cm 0.5cm 0.5cm, clip]{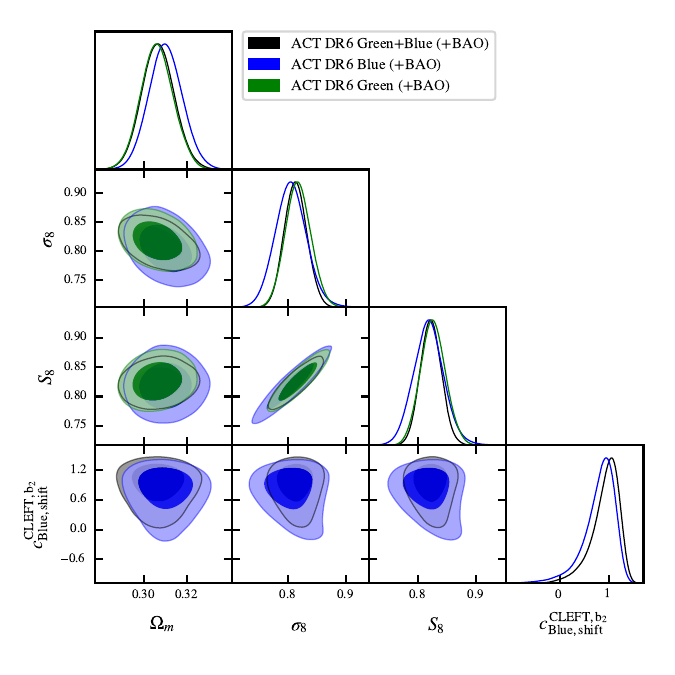}
    \caption{When plotting the posterior for the second order Lagrangian bias parameter, $c_{2,L}^{\rm Blue}$, along with the cosmological parameters we can see that the $\sigma_8$ and $\Omega_m$ are highly degenerate with this parameter when using only the combination of $C_\ell^{gg}$ and $C_\ell^{\kappa g}$ (\textbf{left}). This is in contrast to $S_8$ which is only weakly dependent on the higher order bias. With BAO data (\textbf{right}) this degeneracy is broken and the inferred parameters are less sensitive to the higher order bias priors.}
    \label{fig:actXunWISE_corner_plots_withCLEFTparam}
\end{figure*}

In Sec.\,\ref{subsec:cosmo} we presented posteriors on cosmological parameters obtained from the analysis of the auto-correlation of the two unWISE redshift samples, Blue and Green independently, and their cross-correlation with CMB lensing reconstructions from ACT. Our analysis is primarily sensitive to the amplitude of the signal which depends on the parameter combination $S_8^\times\equiv \sigma_8 (\Omega_m/0.3)^{0.45}$. When using only a single tomographic bin (and neglecting structure growth within the bin) $\sigma_8$ and $\Omega_m$ are exactly degenerate at linear order. Constraints on either parameter individually therefore arise only from beyond linear order contributions to the spectrum. Therefore, the inference of $\sigma_8$ and $\Omega_m$ is very sensitively dependent on the priors on higher order parameters. In Fig.\,\ref{fig:actXunWISE_corner_plots_withCLEFTparam} we can see that $\sigma_8$ and $\Omega_m$ show a strong degeneracy with $c_{2,L}^{\rm Blue}$, the free parameter offsetting the second order Lagrangian bias for the Blue sample. $S_8$ on the other hand shows a much weaker degeneracy with the higher order bias parameters. The degeneracy is broken when both tomographic samples are included or when we additionally include BAO information.

Given the sensitivity to the higher order bias priors we thus do not necessarily expect the two samples to yield the same values for $\sigma_8$ or $\Omega_m$ and the fact that they are consistent at $1.7\sigma$ suggest that our choice of priors is reasonable.

Furthermore, our tests described in Sec.~\ref{subsec:galaxy_sys} specifically ensured that the bias-independent amplitude of our spectra remained consistent across different subsets of the data. We can thus not rule out that the exact scale dependence of the spectrum which gives rise to the weak constraints on $\sigma_8$ and $\Omega_m$ in this type of analysis is affected to some degree by residual systematic contamination. 

In the parameter which is sensitive to the amplitude of the spectrum, $S_8$ (or $S_8^\times$), on the other hand, we find good agreement between the samples and thus consider it justified to combined them.

\section{Summary of all parameter consistency tests} \label{app:param_consistency_tests}

\begin{figure*}
    \includegraphics[width=\linewidth]{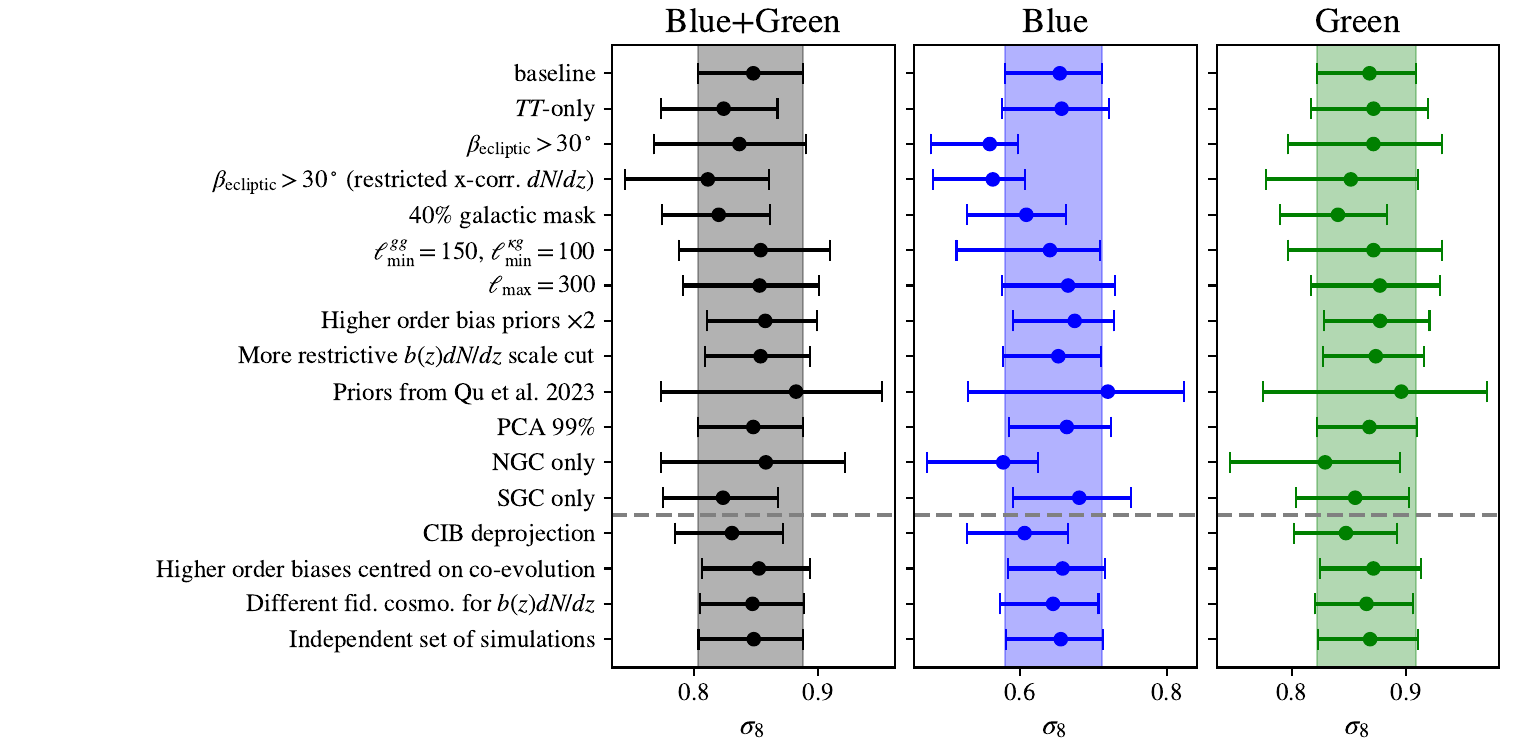}
    \caption{Like Fig.\,\ref{fig:param_consistency} but showing the constraints on $\sigma_8$ for various subset analyses. The posteriors are highly non-Gaussian and because the different analyses are not independent we are unable to reliably estimate the significance of the parameter shifts observed. However, as argued in Sec.\,\ref{subsec:param_consistency_tests} comparisons of the best constraint parameter combination $S_8^{\times}$ show no evidence for any significant bias with all shifts consistent with expectations based on reduced data volume.}
    \label{fig:param_consistency_sigma8}
\end{figure*}

\begin{figure*}
    \includegraphics[width=\linewidth]{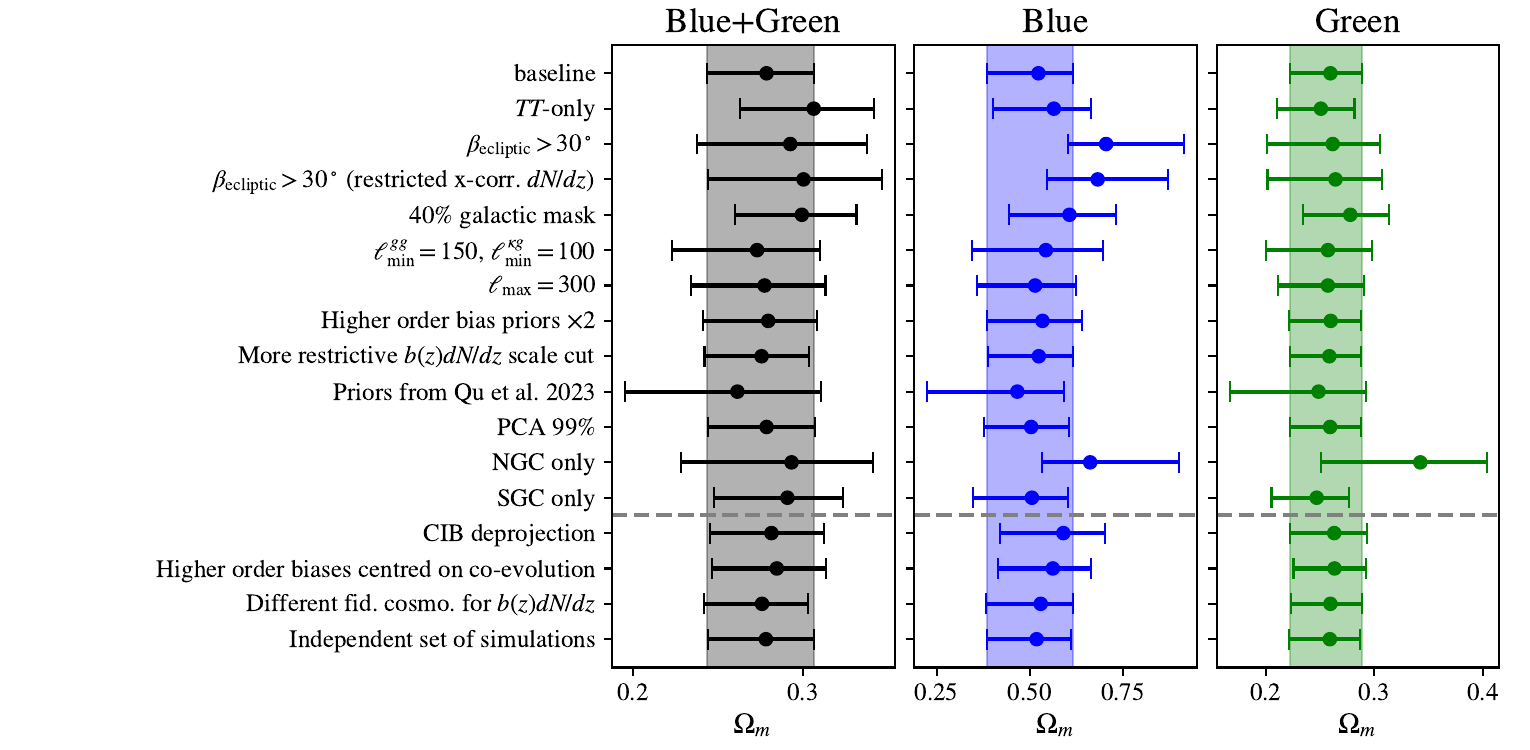}
    \caption{Like Fig.\,\ref{fig:param_consistency} and Fig.\,\ref{fig:param_consistency_sigma8} but showing the constraints on $\Omega_m$. As in Fig.\,\ref{fig:param_consistency_sigma8} we are unable to reliably estimate the significance of the shifts given the highly non-Gaussian posteriors and the correlations between different data subsets.}
    \label{fig:param_consistency_OmegaM}
\end{figure*}

In Table~\ref{table:param_consistency_subset} we show a summary of several parameter consistency tests performed in this work; these employ various subsets of our baseline dataset. These are described in Sec.\,\ref{subsec:param_consistency_tests}. We also summarise more consistency tests for alternative analysis choices in Table~\ref{table:param_consistency_alt_analysis}. \resub{Finally, in Figs.\,\ref{fig:param_consistency_sigma8} and \ref{fig:param_consistency_OmegaM} we show the consistency between $\sigma_8$ and $\Omega_m$ inferred using various subsets of the data and alternative analysis choices. We note, that these parameters are individually only poorly constraint and the posteriors are highly non-Gaussian. Therefore we are unable to estimate the significance of the observed shifts. We note, however, that in all but three cases the mean $\sigma_8$ and $\Omega_m$ fall within the 68\% confidence region of our baseline analysis. The only exceptions arise when restricting the analysis to ecliptic latitudes larger than 30$^\circ$ for the Blue sample and considering only the NGC for the Blue or Green sample. In these cases the data volume is significantly reduced so that larger shifts are expected.}

\begin{deluxetable}{llCCCC}
    \tablehead{
    \nocolhead{} & \nocolhead{} & \colhead{$\Delta S_8$} & \colhead{$\Delta S_8^{\times}$} & \colhead{PTE}
    }
    \startdata
    \decimals
    \multirow{3}{*}{$TT$-only} & Blue & -0.035 (^{+0.027}_{-0.043})\tablenotemark{a} & -0.031(^{+0.025}_{-0.039})\tablenotemark{a} & -\\
    & Green & 0.012\pm0.010 & 0.010\pm0.009 & 0.24\\
    & Blue+Green & -0.014\pm0.010 & -0.011\pm0.009 & 0.22\\ \hline
    \multirow{3}{*}{$\beta_{\rm{ecliptic}}>30^{\circ}$} & Blue & 0.008(^{+0.033}_{-0.040})\tablenotemark{a} & 0.019(^{+0.031}_{-0.038})\tablenotemark{a} & -\\
    & Green & -0.0003\pm0.025 & -0.0004\pm0.021 & 0.99\\
    & Blue+Green & -0.005\pm0.020 & -0.003\pm0.017 & 0.86\\ \hline
    \multirow{3}{*}{$\beta_{\rm{ecliptic}}>30^{\circ}$ (restricted x-corr. $dN/dz$)}  & Blue & 0.016(^{+0.034}_{-0.046})\tablenotemark{a} & 0.026(^{+0.032}_{-0.043})\tablenotemark{a} & -\\
    & Green & 0.0142\pm0.025 & 0.0147\pm0.021 & 0.48\\
    & Blue+Green & 0.0101\pm0.022 & 0.0129\pm0.019 & 0.49\\ \hline
    \multirow{3}{*}{40\% Galactic mask} & Blue & -0.0001(^{+0.027}_{-0.034})\tablenotemark{a} & 0.006(^{+0.024}_{-0.031})\tablenotemark{a} & -\\
    & Green & -0.001\pm0.012 & 0.002\pm0.011 & 0.85\\
    & Blue+Green & -0.002\pm0.008 & 0.001\pm0.008 & 0.87\\ \hline
    \multirow{3}{*}{$\ell_{\rm{min}}^{gg}=150$, $\ell_{\rm{min}}^{\kappa g}=100$} & Blue & 0.019(^{+0.027}_{-0.034})\tablenotemark{a} & 0.019(^{+0.027}_{-0.033})\tablenotemark{a} & -\\
    & Green & 0.006\pm0.015 & 0.005\pm0.011 & 0.64\\
    & Blue+Green & 0.006\pm0.011 & 0.005\pm0.009 & 0.55\\ \hline
    \multirow{3}{*}{$\ell_{\rm{max}}=300$} & Blue & -0.002(^{+0.028}_{-0.041})\tablenotemark{a} & -0.003(^{+0.025}_{-0.036})\tablenotemark{a} & -\\
    & Green & -0.003\pm0.014 & -0.003\pm0.012 & 0.80\\
    & Blue+Green & -0.001\pm0.010 & -0.002\pm0.010 & 0.88\\ \hline
    \multirow{3}{*}{Higher order bias priors $\times 2$} & Blue & -0.033(^{+0.027}_{-0.077})\tablenotemark{a} & -0.031(^{+0.023}_{-0.070})\tablenotemark{a} & -\\
    & Green & -0.009\pm0.020 & -0.009\pm0.019 & 0.63\\
    & Blue+Green & -0.010\pm0.016 & -0.010\pm0.015 & 0.47\\ \hline
    \multirow{3}{*}{More restrictive $b(z)dN/dz$ scale cut} & Blue & 0.002(^{+0.025}_{-0.039})\tablenotemark{a} & 0.002(^{+0.022}_{-0.034})\tablenotemark{a} & -\\
    & Green & -0.004\pm0.005 & -0.004\pm0.005 & 0.46\\
    & Blue+Green & -0.002\pm0.002 & -0.002\pm0.003 & 0.52\\ \hline \hline
    \multirow{3}{*}{NGC - SGC} & Blue & -0.029\pm 0.063 & -0.039\pm 0.059 & 0.48\\
    & Green & 0.101\pm0.059 & 0.088\pm0.052 & 0.09\\
    & Blue+Green & 0.032\pm0.040 & 0.032\pm0.044 & 0.46
    \enddata
    \tablenotetext{a}{The posteriors on $S_8$ and $S_8^\times$ for analyses using only the Blue sample are not well approximated by Gaussians. Therefore, we were unable to reliably estimate the significance of the shift for these consistency tests. We nevertheless show here the shift in the parameter mean, which is in most cases small compared to the measurement errors shown in parentheses.}
    \caption{We perform a series of parameter level consistency tests detailed here. These tests were originally performed blind using random offsets on the cosmological parameters. We show the difference in the inferred $S_8$ and $S_8^\times$ for these analyses compared to our baseline analysis (and the difference between analyses using only the NGC and SGC respectively). Note that the uncertainty quoted is the uncertainty on the difference between the baseline analysis and the subset analysis estimated under the assumption that the posteriors are approximately Gaussian following the prescription from \cite{2020MNRAS.499.3410G}. In the case of the comparison of analyses performed on the NGC and SGC, the two analyses are fully independent and we estimate the consistency between the two data sets taking into account the full non-Gaussian posterior using the \textsc{tensiometer} package. We estimate the statistical likelihood of the shifts obtained assuming that we are effectively probing a single parameter, $S_8^{\times}$. We find all consistency tests to yield shifts consistent with expectations based on random fluctuations within $<2\sigma$.\label{table:param_consistency_subset}}
\end{deluxetable}

\begin{deluxetable}{llCCCCC}
    \tablehead{
    \nocolhead{} & \nocolhead{} & \colhead{$S_8$} & \colhead{$\Delta S_8$} & \colhead{$S_8^{\times}$} & \colhead{$\Delta S_8^{\times}$}
    }
    \startdata
    \decimals
    \multirow{3}{*}{CIB deprojection} & Blue & 0.833^{+0.026}_{-0.034} & 0.016 &  0.807^{+0.024}_{-0.031} & 0.020\\
    & Green & 0.789\pm 0.027 & 0.014 & 0.794\pm 0.024 & -0.015\\
    & Blue+Green & 0.800\pm 0.021 & 0.013 & 0.803\pm 0.019 & 0.013\\ \hline
    \multirow{3}{*}{Higher order biases centred on co-evolution} & Blue & 0.886^{+0.030}_{-0.070} &-0.037 & 0.860^{+0.028}_{-0.064} & -0.033\\
    & Green & 0.812^{+0.026}_{-0.032} & -0.009 & 0.818^{+0.024}_{-0.029} & -0.008\\
    & Blue+Green & 0.826^{+0.022}_{-0.026} & -0.013 & 0.829^{+0.020}_{-0.024} & -0.012\\ \hline
    \multirow{3}{*}{Different fid. cosmo. for $b(z)dN/dz$ } & Blue & 0.842^{+0.024}_{-0.038} & 0.007 & 0.819^{+0.021}_{-0.034} & 0.008\\
    & Green & 0.801\pm0.026 & 0.002 & 0.807\pm0.024 & 0.002\\
    & Blue+Green & 0.809\pm0.020 & 0.004 & 0.812\pm0.019 & 0.004\\ \hline
    \multirow{3}{*}{Independent set of simulations} & Blue & 0.847^{+0.025}_{-0.037} & 0.002 & 0.825^{+0.022}_{-0.033} & 0.002\\
    & Green & 0.803^{+0.025}_{-0.028} & 0.0004 & 0.809^{+0.022}_{-0.025} & 0.0003\\
    & Blue+Green & 0.813\pm 0.021 & 0.0001 & 0.816\pm 0.019 & 0.0001\\ \hline \hline
    \multirow{3}{*}{\texttt{HMCode}-only model} & Blue & 0.821\pm 0.024 & 0.028 & 0.806\pm 0.023 & 0.021\\
    & Green & 0.802\pm 0.025 & 0.001 & 0.808\pm 0.023 & 0.002\\
    & Blue+Green & 0.812\pm 0.019 & 0.002 & 0.814\pm 0.018 & 0.002\\ \hline
    \multirow{3}{*}{Fixed fid. higher order biases} & Blue & 0.869\pm 0.025 & -0.02 & 0.842\pm 0.023 & -0.015\\
    & Green & 0.807\pm 0.025 & -0.004 & 0.812\pm 0.022 & -0.003\\
    & Blue+Green & 0.826\pm 0.021 & -0.013 & 0.829\pm 0.018 & -0.012\\ \hline
    \multirow{3}{*}{No mag. bias} & Blue & 0.856^{+0.019}_{-0.047} & -0.007 & 0.833^{+0.017}_{-0.042} & -0.006\\
    & Green & 0.815^{+0.026}_{-0.029} & -0.012 & 0.823^{+0.024}_{-0.027} & -0.013\\
    & Blue+Green & 0.817\pm 0.021 & -0.004 & 0.823\pm 0.019 & -0.007\\
    \enddata
    \caption{Here we show further the parameter consistency tests not involving a subset of our baseline analysis. For our consistency test using a lensing reconstruction with explicit CIB deprojection, alternative analysis choices in regards to the higher order bias priors, and the fiducial cosmology assumed for the cross-correlation redshifts it is less straightforward to estimate the statistical significance of the observed parameter shifts. However, as can be seen here the impact of these alternative analysis choices is small compared to our uncertainties. We also show an analysis run with a covariance obtained from a independent set of 400 simulations. The stability of the results indicates that our covariance is sufficiently converged. Furthermore, we summarised analyses using simplified models, an analysis using only the \texttt{HMCode} matter power spectrum, an analysis fixing all higher order biases to the fiducial coevolution relations, and an analysis neglecting all magnification bias contributions. While these analyses are not necessarily expected to be consistent with out baseline results, they provide an indication of the importance of different model components. We find that the shifts induced by these simplified models are small compared to our uncertainties, but they also do not yield dramatic improvements in constraining power. \label{table:param_consistency_alt_analysis}}
\end{deluxetable}

\section{Alternative priors}\label{app:alt_priors}

\begin{deluxetable}{CC}
\tablehead{\colhead{Parameter}  &   \colhead{Prior}}
\startdata
\ln (10^{10}A_s)& [2,4]         \\ 
H_0             & [40,100]        \\ 
n_s             & \mathcal{N}(0.96,0.02)     \\ 
\Omega_bh^2     & \mathcal{N}(0.0223,0.0005) \\ 
\Omega_ch^2     & [0.005,0.99]
\enddata
\caption{The priors adopted from \cite{2024ApJ...962..112Q}. These are identical to those used in the lensing power spectrum analysis performed by the \textit{Planck} team~\citep{2020A+A...641A...8P}. For the galaxy nuisance parameters we adopt the same priors as in Table~\ref{tab:priors}.}
\label{table:qu_priors}
\end{deluxetable}

\begin{figure*}
    \centering
    \includegraphics[width=0.5\linewidth,trim=0.5cm 0.5cm 0.5cm 0.5cm, clip]{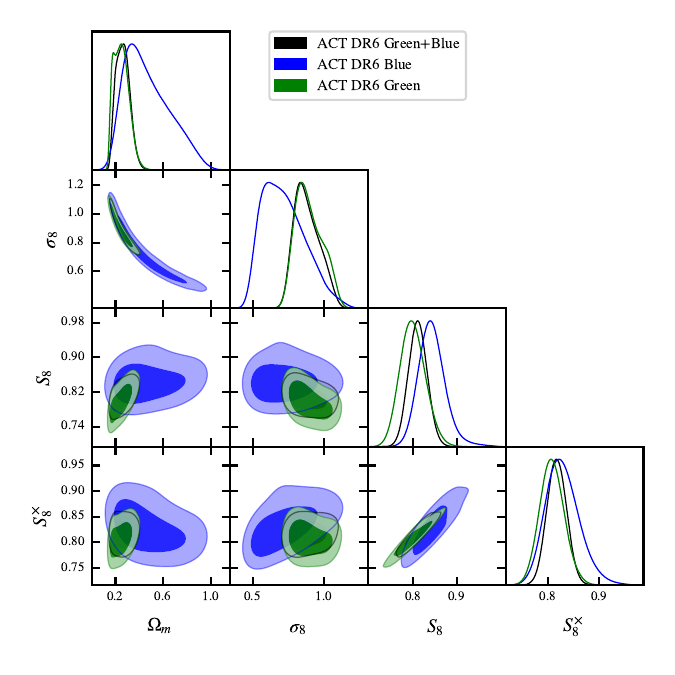}%
    \includegraphics[width=0.5\linewidth,trim=0.5cm 0.5cm 0.5cm 0.5cm, clip]{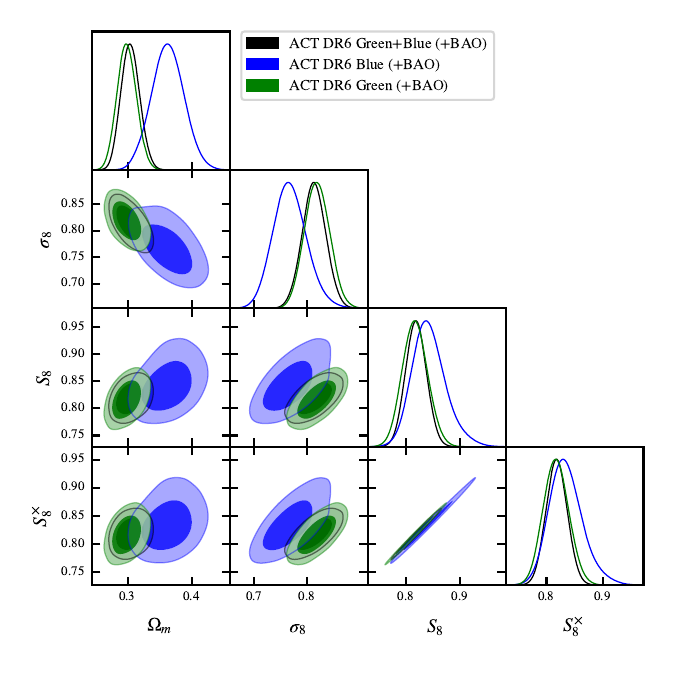}
    \caption{Parameter constraints from the cross-correlation of ACT DR6 lensing and unWISE galaxies using the priors from \cite{2024ApJ...962..112Q} (see Table~\ref{table:qu_priors}). The posteriors on $\Omega_m$ and $\sigma_8$ are significantly more non-Gaussian with this choice of priors compared to our baseline priors (Table~\ref{tab:priors}). The best constrained parameters $S_8$ and $S_8^\times$ are not strongly affected by this choice of priors, however. We find $S_8 = 0.810\pm 0.021$ from the combination of $C_\ell^{gg}$ and $C_\ell^{\kappa g}$ for both samples of unWISE galaxies (\textbf{left}). Adding BAO data (\textbf{right}) yields $\sigma_8=0.812\pm 0.023$, a slightly more significant degradation of constraining power compared to our baseline.}
    \label{fig:actXunWISE_corner_plots_auto_priors}
\end{figure*}

In this section we explore the impact of an alternative set of priors on cosmological parameters. In particular we adopt a set of priors compatible with \cite{2024ApJ...962..112Q}. The main difference between the priors adopted in this work (see Table~\ref{tab:priors}) and those from \cite{2024ApJ...962..112Q} is the fact that $H_0$ is left to vary with a uniform prior between $40\hun$ and $100\hun$. Additionally, \cite{2024ApJ...962..112Q} adopt priors on $n_s$ and $\Omega_b h^2$ rather than fixing those parameters to the mean values from \textit{Planck} \citep{2020A+A...641A...6P}. We summarise this alternative set of priors in Table~\ref{table:qu_priors}.

We find $S_8 = 0.810\pm 0.021$ from the combination of the galaxy auto-correlation and the cross-correlation measurements using both samples of unWISE galaxies, only a small degradation of constraining power compared to our baseline analysis (see Sec.~\ref{subsec:cosmo}). Adding BAO information we obtain $\sigma_8=0.812\pm 0.023$ which is a more significant degradation of constraining power compared to our baseline analysis with BAO. This is explained by the fact that BAO data provide constraints primarily in the $\Omega_m-H_0$ plane. In our baseline analysis these two parameter do not vary independently so that the addition of BAO information directly represents a constraint on $\Omega_m$ which breaks the degeneracy between $\sigma_8$ and $\Omega_m$ more effectively than in the case where $H_0$ is allowed to vary independently.

\begin{deluxetable}{lCCCC}
    \tablehead{
    \nocolhead{}    & \colhead{$\Omega_m$} & \colhead{$\sigma_8$} & \colhead{$S_8$} & \colhead{$S_8^{\times}$} }
    \startdata
    &\multicolumn{4}{c}{ACT DR6 $\times$ unWISE only}\\
    Blue & 0.47^{+0.12}_{-0.24} & 0.72^{+0.10}_{-0.19} & 0.843^{+0.026}_{-0.033} & 0.828^{+0.028}_{-0.037}\\
    Green & 0.249^{+0.044}_{-0.082} & 0.896^{+0.074}_{-0.12} & 0.799^{+0.027}_{-0.031} & 0.807\pm 0.024\\
    \textbf{Joint} & 0.261^{+0.049}_{-0.066} & 0.882^{+0.069}_{-0.11} & 0.810\pm 0.021 & 0.817\pm 0.019
    \\ \hline
    &\multicolumn{4}{c}{ACT DR6 $\times$ unWISE + BAO}\\
    Blue & 0.363\pm 0.025 & 0.768^{+0.029}_{-0.034} & 0.843^{+0.026}_{-0.034} & 0.835^{+0.025}_{-0.033}\\
    Green & 0.299\pm 0.015 & 0.819\pm 0.024 & 0.817\pm 0.023 & 0.817\pm 0.023\\
    \textbf{Joint} & 0.305\pm 0.014 & 0.813\pm 0.022 & 0.819\pm 0.019 & 0.818\pm 0.019
    \enddata
    \caption{The table summarises the constraints on cosmological parameters obtained from the cross-correlation of unWISE galaxies with ACT DR6 lensing reconstruction using the priors from \cite{2024ApJ...962..112Q} (see Table~\ref{tab:priors}). It should be compared to Table~\ref{table:results}. The constraining power on the best constrained parameters $S_8$ and $S_8^\times$ is similar to our baseline analysis, but the addition of BAO proves less powerful in constraining $\sigma_8$ with these alternative priors. \label{tab:results_auto_priors}}
\end{deluxetable}

As we argued in Sec.~\ref{subsec:inference_priors} the angular size of the sound horizon, $\theta_{\rm MC}$, which within a $\Lambda$CDM cosmology determines the combination $\Omega_m h^3$, is measured with great precision by \textit{Planck} \citep{2020A+A...641A...6P}. In the context of aiming to compare the amplitude of low redshift structure to expectations from fits to the primary CMB it is reasonable to adopt this constraint that helps break the degeneracy between distances and structure growth. Since we are here primarily interested in investigating the claimed `$S_8$ tension' we adopt this approach as our baseline, but will investigate other options that allow one to probe alternative cosmologies in future work. We also note that the comparison with other works that constrain $S_8$ without the addition of BAO is still meaningful as we showed above that our $S_8$ constraints are not significantly affected by this choice of priors.

\section{Consistency test of galaxy-galaxy cross-spectrum}\label{app:gg_cross_test}

As a further consistency test we compare the prediction for the cross-correlation between the Blue and Green sample of unWISE galaxies obtained from the model inference we presented in Sec.\,\ref{subsec:cosmo} with the observed cross-spectrum. This cross-spectrum contains potentially valuable information to constrain the magnification bias (particularly for the Green sample where the magnification kernel entirely overlaps with the Blue sample) and the high redshift tail of the Blue sample. Nevertheless, we chose to neglect this information in the model fit since the ratio of potential systematic contamination to signal is less favourable as compared to the auto-spectrum. While the signal is approximately a factor 1.5-2.5 smaller than the auto-correlation of the galaxy samples, observational systematics are expected to be of the same order as in the auto-correlation since they are likely common between both samples. We never performed any null or consistency test on the cross-spectrum and hence do not use it for cosmology analysis.

\begin{figure}
    \centering
    \includegraphics[width=0.55\linewidth]{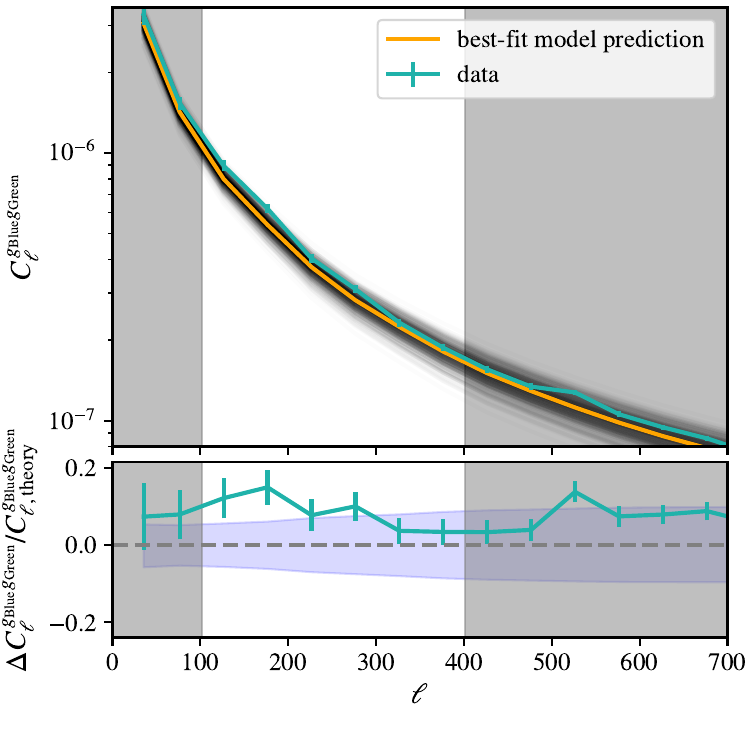}%
    \caption{The cross-correlation between the Blue and Green sample of galaxies is consistent with \resub{the posterior predictive distribution of} our model prediction from a joint fit to the Blue and Green sample from Sec.\,\ref{subsec:cosmo}. The grey lines show the spread in model predictions consistent with the posterior of the model fit. The best fit prediction is shown in orange, with the measured cross-spectrum shown in Blue. The grey bands indicate the range of scales not used for cosmological analysis. \resub{In the lower panel we show the fractional residuals between the best-fit model prediction and the measured cross-spectrum. The blue band shows the $1\sigma$ posterior on the model prediction. Note, that the amplitudes of different scales within the model posterior are significantly correlated.}}
    \label{fig:gg_cross_spec}
\end{figure}

Here we present a comparison of the model we constrain in Sec.\,\ref{subsec:cosmo} to the observed cross-spectrum as a consistency test (see Fig.\,\ref{fig:gg_cross_spec}). We find that the observed cross-spectrum is consistent with the \resub{posterior predictive distribution of the galaxy-galaxy cross-spectrum conditioned on the auto spectra of both galaxy samples and their cross-spectra with the ACT DR6 lensing reconstruction. This test which accounts for the data and model uncertainties yields a PTE of $0.18$. To compute the posterior predictive distribution we sample 500 random points from posterior of the joint fit to the Green and Blue auto- and cross-correlations (including BAO information; shown as grey lines in Fig.\,\ref{fig:gg_cross_spec})}. To compute the $\chi^2$ we add the covariance of these 500 model predictions (which has significant off-diagonal components) to an analytical approximation of the data covariance and compute the residual as the difference between the observed cross-spectrum and the best fit model prediction.

\section{Reanalysis of the cross-correlation of \textit{Planck} lensing and unWISE} \label{app:planck_reanalysis}
In this paper we present an analysis pipeline for the cross-correlation between CMB lensing reconstruction from ACT and unWISE galaxies. In the subsequent sections we discuss the differences to the previous analysis of the cross-correlation between \textit{Planck} lensing and unWISE presented by \cite{2021JCAP...12..028K} \citep[with spectra and redshift distributions measured in][]{2020JCAP...05..047K}. In Sec.\,\ref{app:planck_cosmo} we also show a reanalysis of the cross-correlation with \textit{Planck} based on the newest available lensing reconstruction \citep{2022JCAP...09..039C}.

\subsection{Summary of the Changes to the Analysis}

In Sec.~\ref{subsec:pk_model} we outlined the improvements we made to the model that was used for analysis of the cross-correlation of \textit{Planck} PR3 lensing \citep{2020A+A...641A...8P} with the unWISE galaxies presented in \cite{2020JCAP...05..047K}. This includes marginalisation over higher order biases, improved marginalisation over redshift uncertainties, and a different treatment of the fiducial cosmology dependence of the cross-correlation redshift estimates. Additionally, we have also introduced systematics weighting for the unWISE samples (see Sec.~\ref{subsec:sys_weights}) and we outline in Sec.~\ref{subsubsec:dndz} (with details provided in Appendix~\ref{app:spec_samples}) the additional data that are available to improve the cross-correlation redshift estimates for the unWISE samples. New and improved data are also available on the \textit{Planck} lensing side with the improved PR4 release \citep{2022JCAP...09..039C} briefly described in Sec.~\ref{subsec:planck_lensing}. Finally, it was noticed that previous cross-correlation work had neglected to include a Monte-Carlo correction that is required to achieve the correct normalisation of the lensing reconstruction; we briefly discuss this aspect in more detail in the subsequent section.

\begin{figure*}
    \centering
    \includegraphics[width=0.6\linewidth, trim=0cm 0.5cm 0cm 0cm, clip]{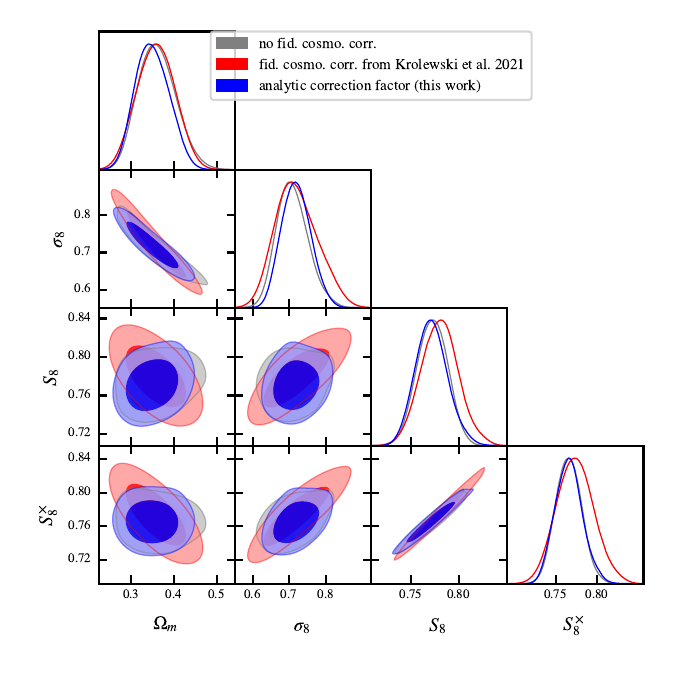}%
    \caption{Comparison of three methods for the fiducial cosmology correction on the Blue sample in \textit{Planck} $\times$ unWISE. Replacing the approximate fiducial cosmology correction in \cite{2021JCAP...12..028K} with the expressions in Sec.\,\ref{app:fid_cosmo_correction} rotates the $S_8$-$\Omega_m$ contours.}
    \label{fig:compare_fid_cosmo_correction}
\end{figure*}

The impact of the various changes and improvements is summarised in Table~\ref{table:changes_summary}. 
The Monte Carlo lensing norm correction is described in Appendix~\ref{sec:lensing_mc_norm} below, and the modelling
improvements are described in Section~\ref{sec:model}. The change in spectra from applying systematics weighting is shown in Fig.~\ref{fig:weights_impact}. The most significant driver
of the change in $S_8$ is the shift towards higher $C_\ell^{\kappa g}$ for the Blue
sample when applying the weights, while $C_\ell^{gg}$ is largely unchanged. In contrast, for the Green sample both $C_\ell^{\kappa g}$ and $C_\ell^{gg}$ increase, leading
to a smaller change in $S_8$. The change in redshift distribution
is shown in Fig.~\ref{fig:compare_old_new_dndz}; the impact on parameters
is dominated by the Blue sample, where the redshift distribution
changes much more ($0.74\sigma$) than for the Green sample ($0.14\sigma$). When adding the new data
to the cross-correlation redshifts, the Blue sample shifts to lower
redshift. At fixed parameters, this increases $C_\ell^{gg}$ (due to the larger
amplitude of structure at low redshift)
while decreasing $C_\ell^{\kappa g}$ (due to the smaller overlap with
the lensing kernel). To fit to the fixed data vector, the parameters
must change in the opposite direction, leading to a higher $S_8$.
The Green sample shifts to slightly higher redshift, although the dominant
change is likely from the narrower width of the new $dN/dz$, which increases
$C_{\ell}^{gg}$ more than $C_{\ell}^{\kappa g}$ at fixed parameters, which likewise leads to a smaller upwards shift in $S_8$.
The new redshift marginalisation is described in Section~\ref{subsec:dndz_marg}. Finally,
in Fig.\,\ref{fig:compare_fid_cosmo_correction} we show the impact of switching the approximate fiducial cosmology correction from \cite{2021JCAP...12..028K} to the formalism described in Sec.\,\ref{app:fid_cosmo_correction}. The impact is mostly a change in the $S_8$-$\Omega_m$ degeneracy direction.

\begin{deluxetable}{l c}
    \tablehead{
    \nocolhead{}    & \colhead{Impact on $S_8$}}
    \startdata
    Monte Carlo lensing norm correction& $+0.6\sigma$\\
    modelling improvements& $-0.5\sigma$\\
    Systematics weighting & $+0.4\sigma$\\
    Additional spectroscopic data & $+0.8\sigma$\\
    use of \textit{Planck} PR4 lensing reconstruction& $+0.2\sigma$\\ 
    PCA based $dN/dz$ marginalisation & $-0.2\sigma$ + $\sim$15\% wider posteriors \\
    fid. cosmo. correction & change in degeneracy directions\\ \hline
    \textbf{Total} & $\bf{+1.3 \sigma}$ + $\sim$$\bf{15\%}$ \textbf{wider posteriors}
    \enddata
    \caption{We make various improvements to the previous analysis of the cross-correlation between \textit{Planck} lensing and unWISE galaxies. This includes the exploitation of updated and additional data such as the PR4 lensing reconstruction from \textit{Planck} and additional spectroscopic data from eBOSS to constrain the sample redshifts. Additionally, we include systematics weighting for the unWISE galaxies, improvements to the modelling, and the Monte Carlo lensing norm correction for the \textit{Planck} lensing map. Here we summarise the impact of these improvements.\label{table:changes_summary}}
\end{deluxetable}

\subsection{Lensing Monte Carlo Norm Correction}
\label{sec:lensing_mc_norm}

A well known effect in CMB lensing reconstruction is a misnormalisation of the lensing reconstruction resulting from performing lensing reconstruction in the presence of a mask (see \cite{Sailer2023} for a pedagogical discussion of this effect and \cite{2013A+A...555A..37B} or \cite{2023JCAP...02..057C} for a more technical discussion). Typically CMB lensing auto-spectrum analyses have computed Monte Carlo corrections for this effect using Gaussian simulations \citep{2024ApJ...962..112Q,2022JCAP...09..039C,2020A+A...641A...8P}. The normalisation correction is obtained as the ratio of the cross-correlation between appropriately masked input lensing convergence with the lensing reconstruction to the auto-correlation of the known input convergence. For the case of the cross-correlation this implies
\beq
A^{\rm{MC}}_\ell = \frac{C_\ell^{\kappa_{\rm{in}, \kappa\rm{-mask}} \kappa_{\rm{in}, g\rm{-mask}}}}{C_\ell^{\hat{\kappa} \kappa_{\rm{in}, g\rm{-mask}}}}.
\eeq
Here $\hat{\kappa}$ is the masked CMB lensing reconstruction, $\kappa_{\rm{in}, \kappa\rm{-mask}}$ is the input lensing convergence masked with the lensing mask, and $\kappa_{\rm{in}, g\rm{-mask}}$ is the input lensing convergence masked using the galaxy mask. As one can see here the correction in principle depends on the region of overlap and thus cannot be universally computed, though we find in practice that the footprint dependence is small ($<1\%$). An unbiased estimate of the CMB lensing cross-correlation can then be obtained as 
\beq
C_\ell^{\hat{\kappa}_{\rm{MC}} g} = C_\ell^{\hat{\kappa} g} A^{\rm{MC}}_\ell.
\eeq

\begin{figure}
    \centering
    \includegraphics[width=0.5\linewidth]{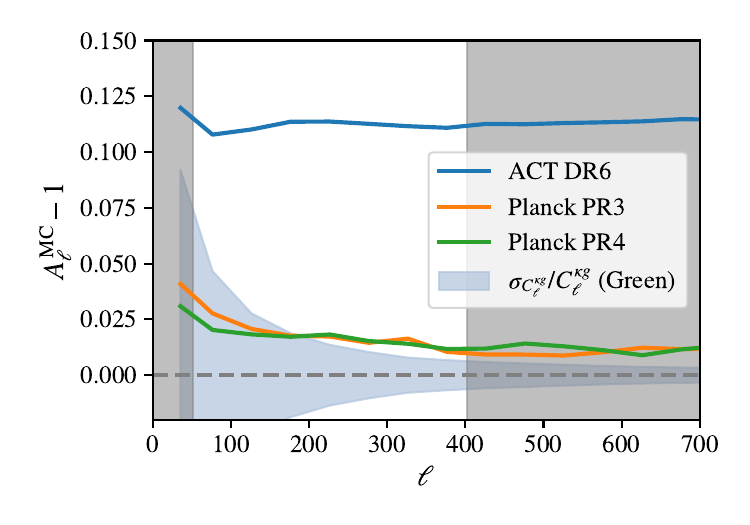}
    \caption{Shown here are the Monte Carlo norm corrections for the cross-correlation of the unWISE galaxies with \textit{Planck} PR4 \citep{2022JCAP...09..039C} and PR3 \citep{2020A+A...641A...8P} lensing reconstructions. We also show the equivalent norm correction for the ACT DR6 lensing reconstruction used in this work. That correction is much larger due to the application of Fourier space masking prior to lensing reconstruction that aims to reduce contamination from instrumental effects such as ground pick-up and features in the scan strategy. Moreover, we show the fractional $1\sigma$ errors on our $C_\ell^{\kappa g}$ (here shown for the Green galaxy sample) for comparison.}
    \label{fig:mc_norm_planck}
\end{figure}

We compute $A^{\rm{MC}}_\ell$ using 480 Gaussian simulations and corresponding lensing reconstructions provided by \cite{2022JCAP...09..039C}. We find that on the scales of interest for our analysis the norm correction results in a nearly scale invariant increase in the amplitude of $C_\ell^{\kappa g}$ by around 2\% (see Fig.\,\ref{fig:mc_norm_planck}). We note that \cite{2021JCAP...12..028K} used the \textit{Planck} PR3 lensing reconstruction \citep{2020A+A...641A...8P} which also exhibits very similar behaviour (see Fig.~\ref{fig:mc_norm_planck}). Given the errorbars on our cross-correlation measurement accounting for this effect is unavoidable and leads to a $\sim0.6\sigma$ increase in the inferred value of $S_8$ (see Table~\ref{table:changes_summary}). 

\begin{figure}
    \centering
    \includegraphics[width=\linewidth]{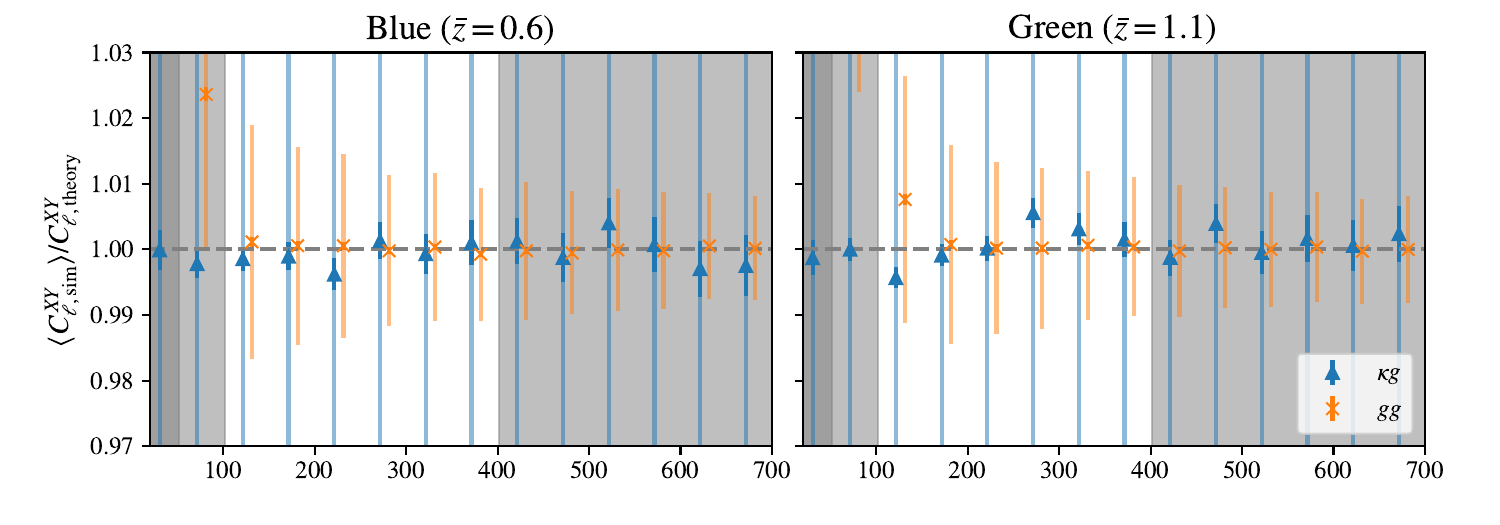}
    \caption{As with ACT we recover the input $C_\ell^{gg}$ and $C_\ell^{\kappa g}$ on \textit{Planck} simulations to better than 1\% on all scales. The lightly coloured errorbars indicate our measurement errors, while the dark errorbars show the error on the mean of 480 Gaussian simulations.}
    \label{fig:transfer_function_planck}
\end{figure}

We perform the same test of unbiased power spectrum recovery shown for ACT in Sec.~\ref{sec:sims} for \textit{Planck} and find recovery to within 1\% after application of the norm correction (see Fig.\,\ref{fig:transfer_function_planck}).

\subsection{Updated cosmological constraints from the cross-correlation of \textit{Planck} lensing and unWISE} \label{app:planck_cosmo}
\begin{figure}
    \centering
    \includegraphics[width=0.5\linewidth,trim=0.5cm 0.5cm 0.5cm 0.5cm, clip]{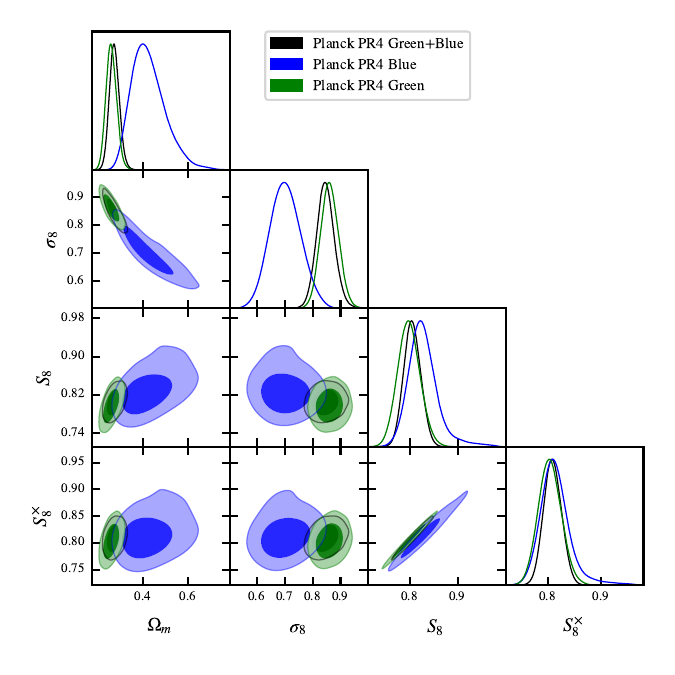}%
    \includegraphics[width=0.5\linewidth,trim=0.5cm 0.5cm 0.5cm 0.5cm, clip]{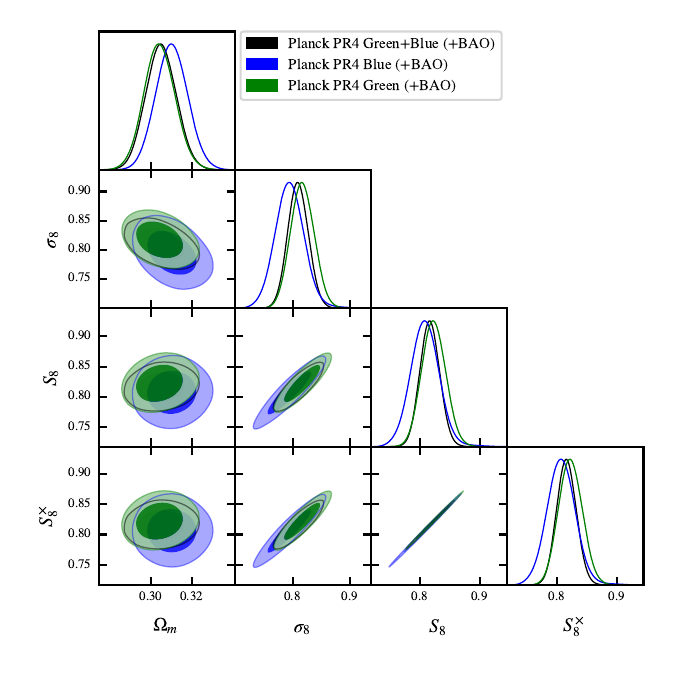}
    \caption{Updated parameter constraints from the cross-correlation of \textit{Planck} lensing and unWISE galaxies. From the combination of $C_\ell^{gg}$ and $C_\ell^{\kappa g}$ only for both samples of unWISE galaxies (\textbf{left}) we find $S_8 = 0.805\pm 0.018$. With additional information on from BAO (\textbf{right}) we obtain $\sigma_8=0.810\pm 0.018 $.}
    \label{fig:planckXunWISE_corner_plots}
\end{figure}

With the improvements laid out above we perform a reanalysis of the cross-correlation of \textit{Planck} lensing and unWISE galaxies. We adopt the priors and scale cuts consistent with our ACT analysis and estimate the covariance as descibed in Sec.~\ref{sec:covmat}. We find $S_8=0.805\pm 0.018$ from the combination of $C_\ell^{\kappa g}$ and $C_\ell^{gg}$ alone and $\sigma_8=0.810\pm0.018$ when adding BAO information. \resub{The parameter constraints from the re-analysis of the cross-correlation between \textit{Planck} lensing and unWISE are summarised in Table~\ref{table:results_planck} with posterior distributions shown in Fig.\,\ref{fig:planckXunWISE_corner_plots}.}
\begin{deluxetable}{lCCCCCCCC}
    \tablehead{
    \nocolhead{}    & \multicolumn{2}{c}{$\Omega_m$} & \multicolumn{2}{c}{$\sigma_8$} & \multicolumn{2}{c}{$S_8$} & \multicolumn{2}{c}{$S_8^{\times}$} }
    \startdata
    &\multicolumn{8}{c}{\textit{Planck} PR4 $\times$ unWISE only}\\
    Blue & 0.425^{+0.054}_{-0.088} & (0.390) & 0.703^{+0.052}_{-0.059} & (0.714) & 0.828^{+0.022}_{-0.033} & (0.814) & 0.815^{+0.021}_{-0.030} & (0.804)\\
    Green & 0.259^{+0.022}_{-0.025}  & (0.250) & 0.861\pm 0.032 & (0.865) & 0.798\pm 0.023 & (0.790) & 0.804\pm 0.022 & (0.797)\\
    \textbf{Joint} & 0.273^{+0.020}_{-0.024} & (0.279) & 0.846\pm 0.032 & (0.836) & 0.805\pm 0.018 & (0.807) & 0.809\pm 0.017 & (0.810)
    \\ \hline
    &\multicolumn{8}{c}{\textit{Planck} PR4 $\times$ unWISE + BAO}\\
    Blue & 0.3102\pm 0.0078 & (0.311) & 0.795\pm 0.026 & (0.793) & 0.808\pm 0.025 & (0.808) & 0.807\pm 0.025 & (0.806)\\
    Green & 0.3042\pm 0.0077 & (0.304) & 0.817\pm 0.021 & (0.814) & 0.823\pm 0.020 & (0.820) & 0.822\pm 0.020 & (0.820)\\
    \textbf{Joint} & 0.3050\pm 0.0075 & (0.305) & 0.810\pm 0.018 & (0.808) & 0.816\pm 0.017 & (0.816) & 0.816\pm 0.016 & (0.815)
    \enddata
    \caption{Here we summarise the one dimensional marginalise posteriors on the key cosmological parameters from our reanalysis of the cross-correlation between unWISE galaxies and the \textit{Planck} lensing reconstructions. We use the newest \textit{Planck} PR4 lensing maps \citep{2022JCAP...09..039C} and our improved model detailed in this paper. We provide best fit values in parentheses following the marginalised constraints. \label{table:results_planck}}
\end{deluxetable}

\section{Impact of galaxy systematics weighting}\label{app:sys_weights}

\begin{figure}
    \centering
    \includegraphics[height=0.95\textheight,trim=0cm 3cm 0cm 0cm, clip]{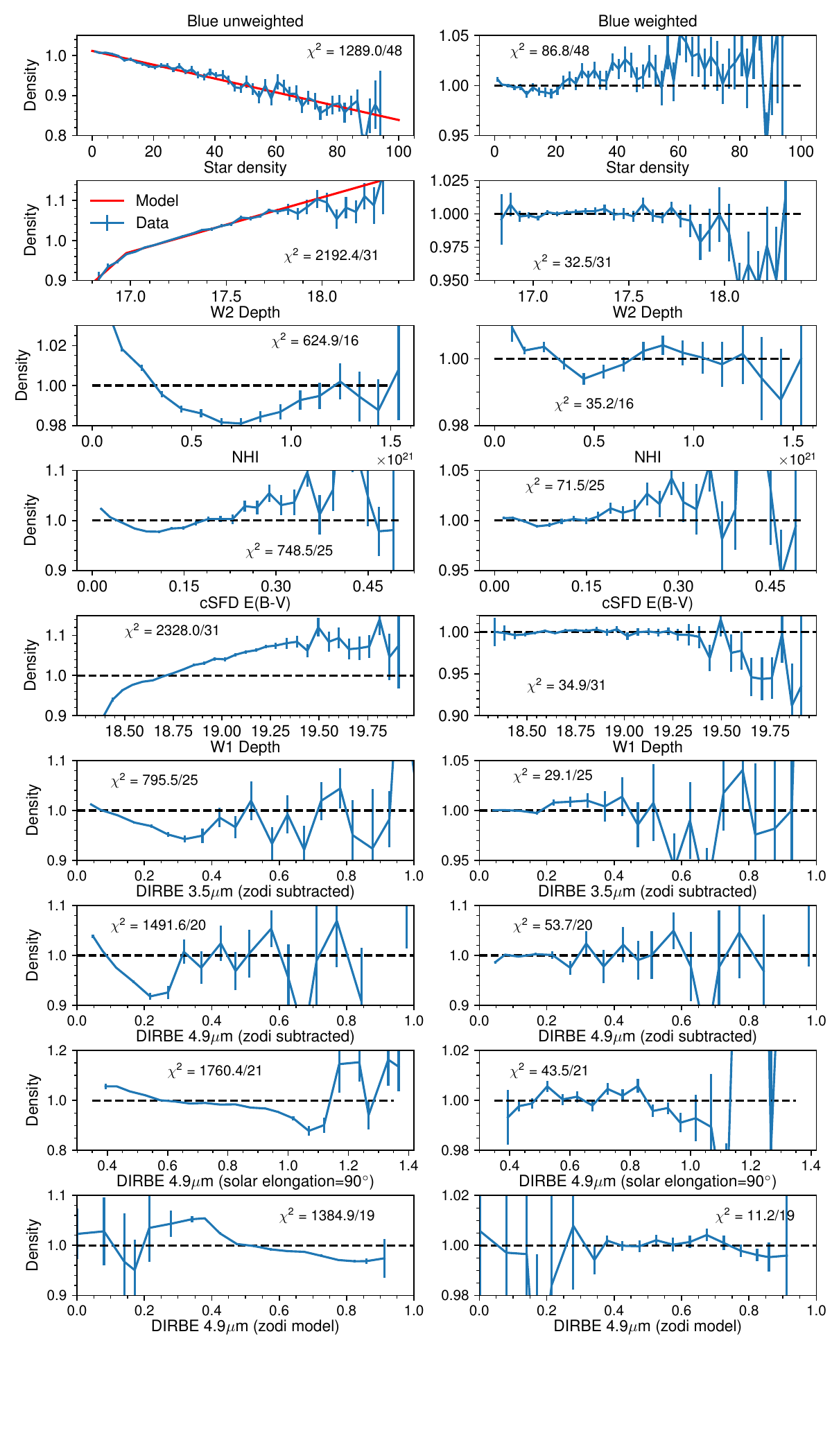}
    \caption{We show here the correlations of the unWISE number density from the Blue sample with the imaging systematics templates discussed in Sec.\,\ref{subsec:sys_weights}. Prior to applying any corrections the number density is significantly correlated with all templates. All correlations are reduced dramatically after applying corrections for the correlation with W2 depth and stellar density. Because many of the templates are correlated the improvement also extends to those templates for which we did not explicitly correct. To guard against overfitting and because the residual impact on the scales of interest is small we do not correct for further contamination even though some of the correlations are still significant even after the correction.}
    \label{fig:imaging_systematics_blue}
\end{figure}

\begin{figure}
    \centering
    \includegraphics[height=0.95\textheight]{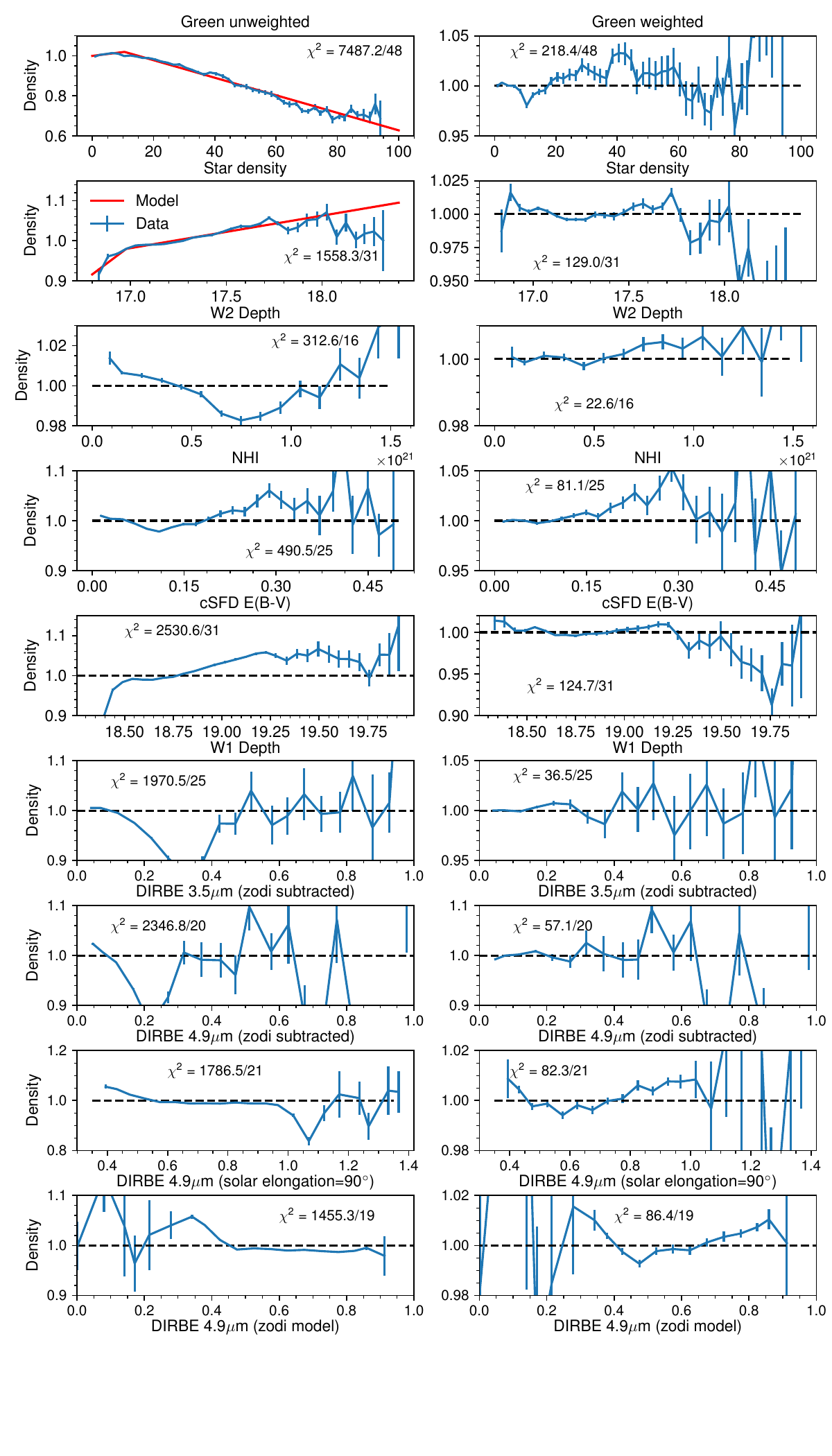}
    \vspace{-50pt}
    \caption{Similar to Fig.\,\ref{fig:imaging_systematics_blue} but here shown for the Green sample of unWISE galaxies. As with the blue sample we correct the data for correlations with W2 depth and stellar density, improving the correlations with all of the imaging templates used.}
    \label{fig:imaging_systematics_green}
\end{figure}

Figures~\ref{fig:imaging_systematics_blue} and \ref{fig:imaging_systematics_green} show the correlation of the unWISE number counts with various systematic tracers discussed in Sec.\,\ref{subsec:sys_weights} before and after correction for the trends with stellar density and W2 limiting magnitude (corrected correlations are shown as red lines in the relevant panels).

\begin{figure*}
    \centering
    \includegraphics[width=0.5\linewidth]{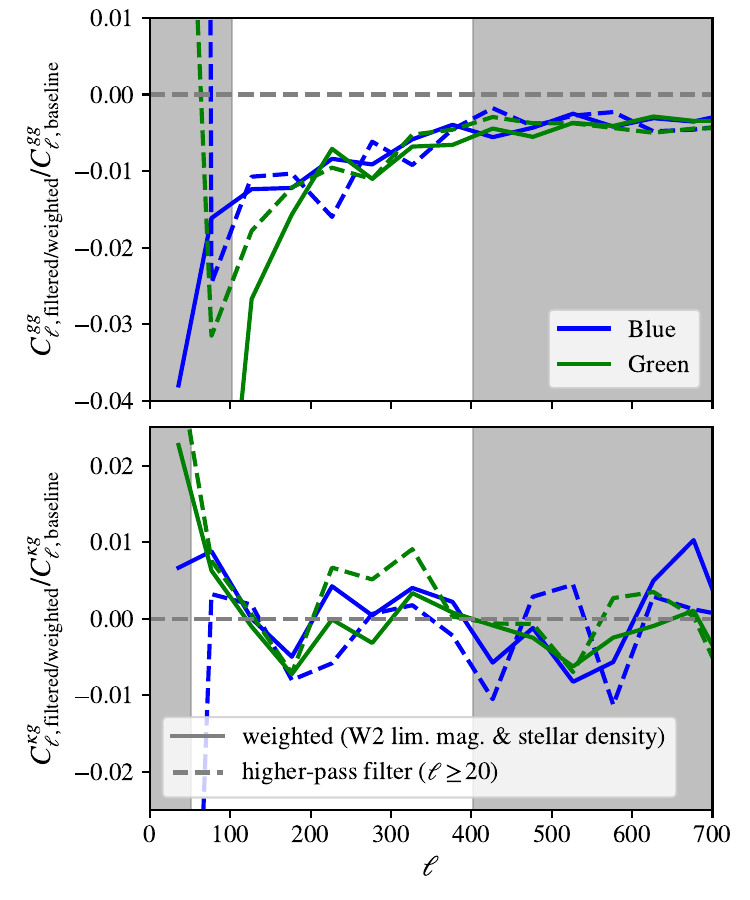}
    \caption{Ratio of galaxy-galaxy and galaxy-CMB lensing spectra after weighting against W2 $5\sigma$ limiting magnitude and Gaia stellar density (as described in Sec.\,\ref{subsec:sys_weights}) to the same spectra without such weighting. The correction for these observational effects reduces the power in the auto-correlation by up to 3\% on large scales, while it does not significantly impact the measured cross-correlation spectra. We also show for comparison the spectra used in \cite{2020JCAP...05..047K, 2021JCAP...12..028K}; these filtered our large scales ($\ell < 20$) in harmonic space as an alternative method for mitigating such observational effects. The difference between the spectra using the two methods is about 0.5\%. In this comparison we have corrected for the simulation derived transfer function described in Sec.\,\ref{sec:sims}.
    }
    \label{fig:weights_impact}
\end{figure*}

In this work, we include systematics weights which had not been applied previously, but we also no longer filter out the low-$\ell$ ($\ell < 20$) modes in the unWISE map \citep[in contrast to what was done in][]{2020JCAP...05..047K,2021JCAP...12..028K}. The large-scale filtering has a similar effect to systematics weighting, also reducing large-scale power by removing correlations between systematics and the true galaxy density. However, removing large scales in harmonic space complicates the use of the \textsc{MASTER} algorithm \cite{2002ApJ...567....2H} to obtain unbiased bandpowers through mode decoupling. Hence, we no longer adopt this method. Nevertheless, the net effect of replacing the low-$\ell$ filtering with the angular systematics weights is at most $0.5\%$ (Fig.~\ref{fig:weights_impact}).

\section{Spectroscopic samples used in cross-correlation redshifts}
\label{app:spec_samples}

\begin{figure*}
    \centering
    \includegraphics[width=\linewidth]{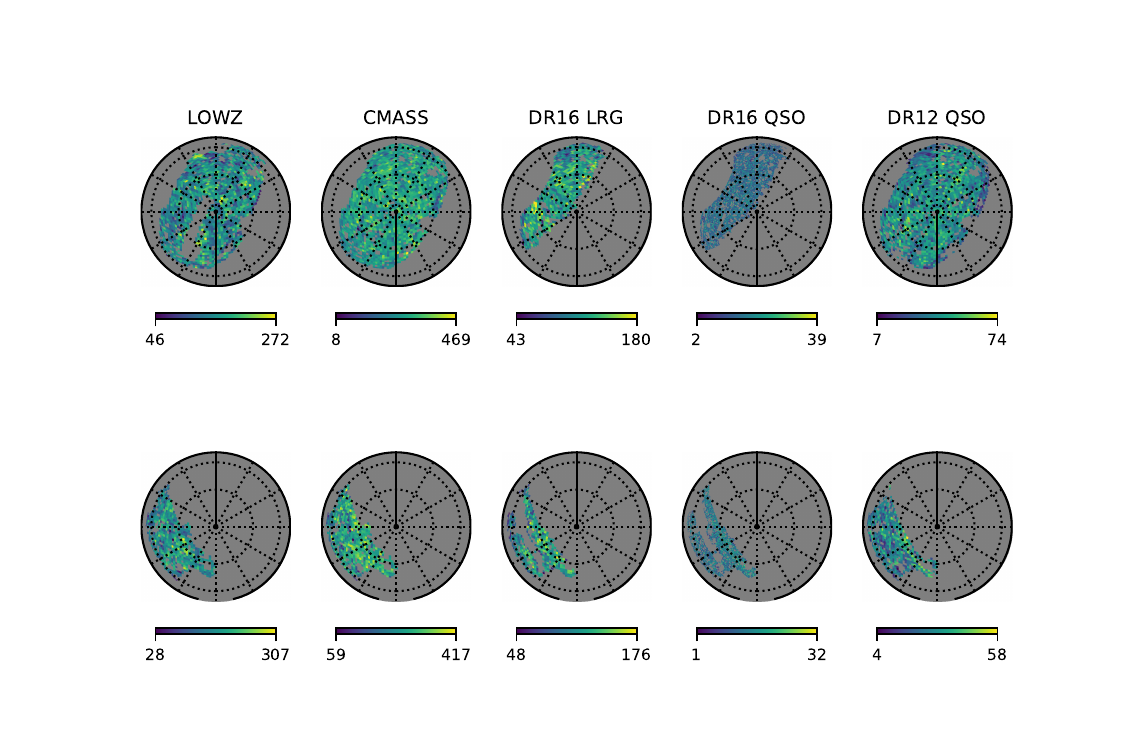}
    \caption{Maps of the spectroscopic samples used to determine $b(z)dN/dz$ of the unWISE galaxies. Top row is NGC and bottom row is SGC. Together the spectroscopic samples cover a representative subset of the unWISE footprint (as can also be seen in Fig.\,\ref{fig:abs_ecliptic_galactic}).}
    \label{fig:spec_skyplot}
\end{figure*}

\begin{figure}
    \centering
    \includegraphics[width=0.5\linewidth]{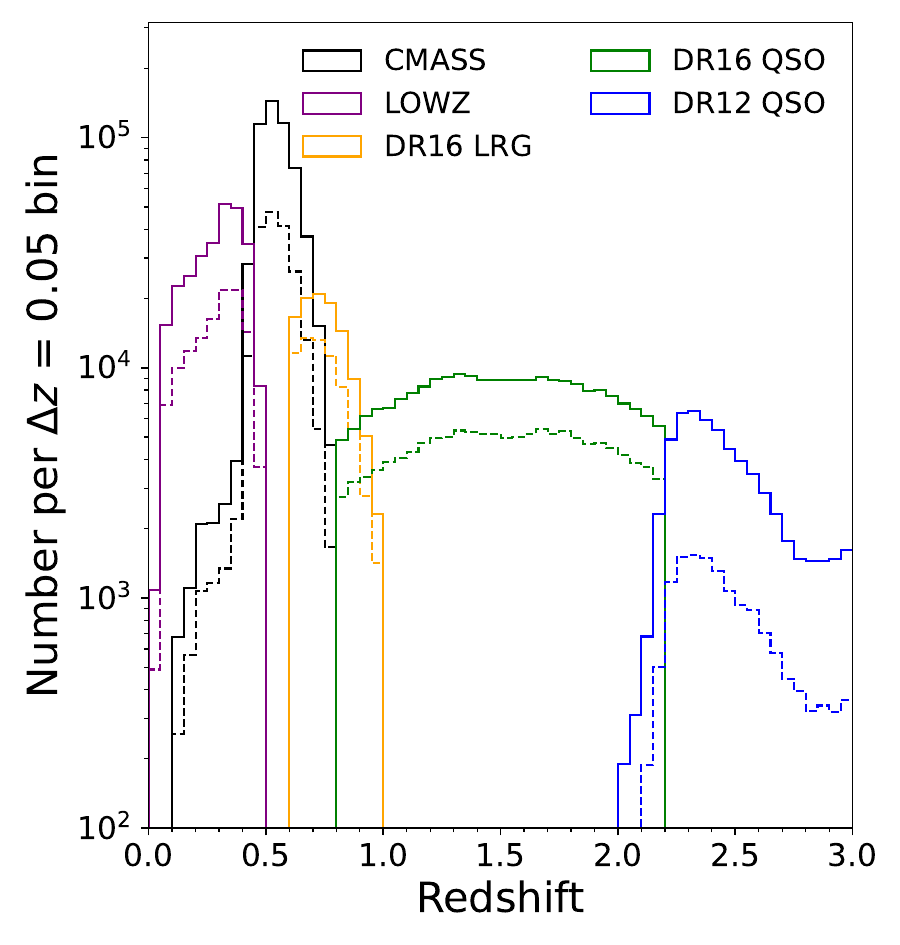}
    \caption{Here we show the redshift distribution of spectroscopic samples used to measure the cross-correlation redshifts for the unWISE samples (NGC and SGC are shown as solid and dashed lines respectively). The spectroscopic tracers used in this work span the full redshift range of the unWISE samples up to $z\sim3$.}
    \label{fig:spec_dndz}
\end{figure}

We follow the methodology described in Section 5.4 of \cite{2020JCAP...05..047K} with updated spectroscopic galaxy samples and revised magnification bias measurements for the spectroscopic samples. We split the spectroscopic samples into northern and southern galactic cap (NGC and SGC) regions, allowing for different spectroscopic biases between the hemispheres \citep{2017MNRAS.470.2617A,2017MNRAS.464.1168R} and to test their consistency. We summarise the spectroscopic samples used in Table~\ref{tab:spec_samples} and Figures~\ref{fig:spec_skyplot} and~\ref{fig:spec_dndz}.

\begin{table}
    \centering
        \begin{tabular}{c|ccccc}
        \multirow{2}{*}{Sample} & \multirow{2}{*}{$z_{\rm min}$} & \multirow{2}{*}{$z_{\rm max}$} & \multirow{2}{*}{$N$} & Jackknife & Area \\
        & & & & regions & (deg$^2$) \\
        \hline
         LOWZ NGC  & 0.0 & 0.5 & 273547 & 34 & 5661 \\
         LOWZ SGC  & 0.0 & 0.5 & 120585 & 13 & 2393 \\
         CMASS NGC  & 0.1 & 0.8 & 544308 & 37 & 6676 \\
         CMASS SGC  & 0.1 & 0.8 & 193899 & 13 & 2427 \\ 
        eBOSS DR16 LRG NGC  & 0.6 & 1.0  & 107500 & 63 & 2898 \\
        eBOSS DR16 LRG SGC  & 0.6 & 1.0  & 67316 & 38 & 1812 \\
        eBOSS DR16 QSO NGC  & 0.8 & 2.2  & 218209 & 61 & 2927 \\
        eBOSS DR16 QSO SGC  & 0.8 & 2.2  & 125499 & 38 & 1825 \\
         BOSS DR12 QSO NGC  & 2.0 & 4.0  & 69204 & 36 & 6569 \\
         BOSS DR12 QSO SGC  & 2.0 & 4.0  & 16590 & 12 & 2176 \\
        \end{tabular}
    \caption{Properties of the spectroscopic samples used for cross-correlation redshifts. Areas are defined using the overlap of the spectroscopic survey mask and the unWISE mask. \label{tab:spec_samples}}
\end{table}

\begin{table}
    \centering
        \begin{tabular}{cc|ccccc|ccccc|ccccc}
        & & \multicolumn{5}{c|}{LOWZ} &  \multicolumn{5}{c|}{CMASS} &  \multicolumn{5}{c}{eBOSS DR16 LRG}   \\
       $z_{\textrm{min}}$ & $z_{\textrm{max}}$ & $z_{\textrm{eff},ss}$ & $z_{\textrm{eff},sp}$ & $s$  & $b_{\textrm{NGC}}$ & $b_{\textrm{SGC}}$ & $z_{\textrm{eff},ss}$ & $z_{\textrm{eff},sp}$ & $s$ & $b_{\textrm{NGC}}$ & $b_{\textrm{SGC}}$ & $z_{\textrm{eff},ss}$ & $z_{\textrm{eff},sp}$ & $s$ & $b_{\textrm{NGC}}$ & $b_{\textrm{SGC}}$  \\
        \hline
        0.00 & 0.05 & 0.042 & --- & 1.975 & 1.326 & 1.244 & & & & & &   \\
        0.05 & 0.10 & 0.079 & 0.070 & 1.069 & 1.368 & 1.305 & & & & & &   \\
        0.10 & 0.15 & 0.124 & 0.123 & 0.912 & 1.518 & 1.491 & & & & & & & &   \\
        0.15 & 0.20 & 0.175 & 0.175 & 1.049 & 1.683 & 1.682 & 0.179 & 0.177 &  0.756 & 3.078 & 1.735 & & &   \\
        0.20 & 0.25 & 0.225 & 0.223 & 1.015 & 1.892 & 1.755 & 0.226 & 0.223 & 0.663 & 1.204 & 1.448 & & &   \\
        0.25 & 0.30 & 0.276 & 0.273 & 0.972 & 1.978 & 2.051 &  0.276 & 0.273 & 0.982 & 1.775 & 1.650 & & & &  \\
        0.30 & 0.35 & 0.327 & 0.325 & 0.923 & 1.984 & 2.062 & 0.324 & 0.323 & 1.121 & 1.765 & 1.798 & & &   \\
        0.35 & 0.40 & 0.372 & 0.369 & 1.014 & 2.051 & 2.077 &  0.382 & 0.371 & 1.161 & 2.273 & 2.078 & & &  \\
        0.40 & 0.45 & 0.420 & 0.419 & 1.179 & 2.211 & 2.109 & 0.439 & 0.429 & 0.972 & 2.055 & 2.048  & & &  \\
        0.45 & 0.50 & 0.460 & 0.463 & 1.511 & 2.541 & 2.187 & 0.479 & 0.475 & 0.937 & 2.065 & 2.000 & & &  \\
        0.50 & 0.55 & & & & & &  0.524 & 0.524 &  1.040 & 2.034 & 2.101 & & &   \\
        0.55 & 0.60 & & & & & & 0.572 & 0.573 &  1.148 & 2.136 & 2.073 \\
        0.60 & 0.65 & & & & & & 0.620 & 0.617 & 1.352 & 2.233 & 2.147 & 0.627 & 0.620 & 0.539 & 2.137 & 2.114 \\
        0.65 & 0.70 & & & & & & 0.669 & 0.672 & 1.553 & 2.400 & 2.236 & 0.675 & 0.675 & 0.595 & 2.094 & 2.107 \\
        0.70 & 0.75 & & & & & & 0.717 & 0.715 & 1.957 & 2.509 & 2.434 & 0.725 & 0.718 & 0.721 & 2.202 & 2.066 \\
        0.75 & 0.80 & & & & & & 0.765 & 0.772 & 2.239 & 2.666 & 2.918 & 0.773 & 0.777 & 0.798 & 2.363 & 2.333 \\
        0.80 & 0.85 & & & & & & & &  & &  & 0.821 & 0.825 & 0.884 & 2.458 & 2.514 \\
        0.85 & 0.90 & & & & & & & &  & &  & 0.871 & 0.868 & 0.890 & 2.572 & 2.400 \\
        0.90 & 0.95 & & & & & & & &   & &  & 0.919 & 0.922 & 0.949 & 2.517 & 2.405 \\
        0.95 & 1.00 & & & & & & & &  & &  & 0.969 & 0.971 & 0.947 & 2.211 & 2.382 \\
        \end{tabular}
    \caption{The table summarises the effective redshifts for the spectroscopic auto-correlation, $z_{\rm{eff}, ss}$, and the cross-correlation with Green unWISE sample $z_{\rm{eff}, sp}$. We also summarise the measured magnification ($s$) and clustering biases ($b_{\rm{NGC}}$ and $b_{\rm{SGC}}$ for the NGC and SGC respectively) for the spectroscopic samples. The effective redshifts and magnification bias are determined from the NGC only, whereas the clustering bias is measured separately for NGC and SGC. Estimates of the magnification biases on the SGC are consistent with the NGC but with increased uncertainties. Therefore, we use the magnification biases measured on the NGC throughout. We allow for a different clustering bias between the hemispheres. We do not report a value for $z_{\textrm{sp}}$ at $z < 0.05$, because there are zero unWISE-Green galaxies matched to COSMOS photometric redshifts in this redshift range.
    \label{tab:spec_magbias_zeff}}
\end{table}

\begin{table}
    \centering
        \begin{tabular}{cc|ccccc|ccc}
         & &   \multicolumn{5}{c|}{eBOSS DR16 QSO} &   \multicolumn{3}{c}{BOSS DR12 QSO} \\
       $z_{\textrm{min}}$ & $z_{\textrm{max}}$ & $z_{\textrm{eff},ss}$ & $z_{\textrm{eff},sp}$ & $s$  & $b_{\textrm{NGC}}$ & $b_{\textrm{SGC}}$ & $z_{\textrm{eff},ss}$ & $z_{\textrm{eff},sp}$ & $s$   \\
        \hline
        0.00 & 0.20 & & & & &   \\
        0.20 & 0.40 & & & & &   \\
        0.40 & 0.60 & & &  & & \\
        0.60 & 0.80 & & & & &   \\
        0.80 & 1.00 & 0.910 & 0.910 & 0.126 & 1.563 & 1.840 \\
        1.00 & 1.20 & 1.107 & 1.115 & 0.153 & 2.062 & 2.088 \\
        1.20 & 1.40 & 1.300 & 1.279 & 0.151 & 2.055 & 2.095 \\
        1.40 & 1.60 & 1.499 & 1.482 & 0.171 & 2.406 & 2.398 \\
        1.60 & 1.80 & 1.699 & 1.674 & 0.163 & 2.315 & 2.520 \\
        1.80 & 2.00 & 1.895 & 1.894 & 0.180 & 2.789 & 3.019 \\
        2.00 & 2.20 & 2.090 & 2.059 & 0.163 & 3.269 & 2.928 & 2.176 & 2.112 & 0.494 \\
        2.20 & 2.40 & & & & & &  2.307 & 2.302 & 0.213 \\
        2.40 & 2.60 & & & & & &  2.481 & 2.461 & 0.234 \\
        2.60 & 2.80 & &  & & & &  2.671 & 2.616 & 0.387 \\
        2.80 & 3.00 & &  & & & &  2.904 & 2.820 & 0.793 \\
        3.00 & 3.20 & & & & & &  3.095 & 3.150 & 0.821 \\
        3.20 & 3.40 & & & & & &  3.270 & --- & 1.047 \\
        \end{tabular}
    \caption{Same as Table~\ref{tab:spec_magbias_zeff}, but for the quasar samples. For the BOSS DR12 QSOs, we do not measure the spectroscopic bias, but rather use the fit from \cite{2017JCAP...07..017L}. 
    We do not report a value for $z_{\textrm{sp}}$ at $3.2 < z < 3.4$, because there are zero unWISE-Green galaxies matched to COSMOS photometric redshifts in this redshift range.
    \label{tab:spec_magbias_zeff2}}
\end{table}

We use both unWISE and spectroscopic sample sytematics weights in computing the Davis-Peebles \citep{1983ApJ...267..465D} estimator for $\bar{w}_{\textrm{sp}}(\theta)$. The unWISE weights are described in Section~\ref{subsec:sys_weights} above and the spectroscopic sample weights are the standard systematic weights for BOSS and eBOSS \citep{2016MNRAS.455.1553R,2015MNRAS.453.2779E}. 


The eBOSS LRGs are targeted at lower priority than eBOSS quasars. Unlike BOSS, which used multiple passes to eventually reach lower priority objects, eBOSS did not have enough passes to reach lower priority objects within the fiber radius of higher priority objects. Therefore, no LRGs were observed within the fiber scale ($62^{\prime \prime}$) of any quasar target. This poses a problem for comparing LRG-unWISE and quasar-unWISE cross-correlations in the region where both spectroscopic tracers overlap ($0.8 < z < 1.0$), since the LRGs and quasars will be anti-correlated on small scales due to fiber collisions. We therefore remove any unWISE galaxies and randoms within $62^{\prime \prime}$ of an eBOSS quasar target, as these unWISE galaxies will also be anti-correlated with the eBOSS LRGs due to their positions near the quasars.

We revise the magnification bias measurements presented in Appendix C of \cite{2020JCAP...05..047K} and update them for the eBOSS LRGs. Magnification bias of the spectroscopic samples can create spurious correlations between widely separately photometric and spectroscpic redshift bins, thus biasing the clustering $dN/dz$ in the tails; the spectroscopic magnification bias is relevant when the spectroscopic sample lies at higher redshift than the photometric sample.

First, \cite{2020JCAP...05..047K} did not correctly consider the impact of magnification bias on fixed-aperture flux measurements. Lensing magnification preserves surface brightness, brightening galaxies by making them larger on the sky. The impact of magnification on a fixed-aperture flux measurement therefore depends on the light profile: if it is spatially constant, then magnification has no effect, but in the more realistic case where it rises towards the galaxy's center, lensing magnification will brighten the galaxy, but not by as much as if the aperture
contained all of the galaxy's light. CMASS and LOWZ use four different measurements of magnitude: model magnitudes, consisting of the best-fit exponential or deVaucoleurs profile; cmodel magnitudes, a linear combination of the exponential and deVaucoleurs profiles; PSF magnitudes, which fit a PSF model to the source and therefore only consider flux within the PSF solid angle; and fib2 magnitudes, which only count flux within the $2^{\prime \prime}$ fibers. We approximate the model and cmodel magnitudes as measuring the full galaxy flux (therefore being maximally affected by lensing magnification). For PSF magnitudes and fib2 magnitudes, we magnify the galaxy profile (assuming an isotropic linear combination of deVaucoleurs and exponential profiles with effective radius given in the catalog) and truncate the fluxes at either $2^{\prime \prime}$ or the mean PSF size in a given band ($1.32^{\prime \prime}$ in $r$, $1.26^{\prime \prime}$ in $i$, $1.29^{\prime \prime}$ in $z$).

Second, in a spectroscopic survey, lensing magnification affects the redshift success rate, the probability with which a redshift can be estimated from a observed spectrum, which is lower for fainter galaxies. LOWZ galaxies are sufficiently bright that redshift failures and successes have very similar magnitude distributions \citep[Fig.\ 7 in ][]{2016MNRAS.455.1553R}, i.e.\ the faintest LOWZ galaxies are considerably brighter than the ``limiting magnitude'' of the spectrograph and therefore increasing or decreasing their flux by a small amount will not change the success rate. CMASS galaxies are fainter, and we estimate from Fig.\ 7 in \cite{2016MNRAS.455.1553R} that $\frac{dlnN}{di_{\textrm{fib2}}} = 0.32$ due to the change in the redshift success rate (by differencing the $i_{\textrm{fib2}}$ distribution at the faint end of CMASS galaxies, comparing $i_{\textrm{fib2}} = 21.3$ and 21.4). Since $i_{\textrm{fib2}}$ has a fixed aperture, we further correct for its reduced sensitivity to lensing magnification compared to total magnitudes. For eBOSS LRGs, we also correct for changes in redshift success rate using the measurement of redshift success against spectral signal-to-noise in Fig.\ 5 of \cite{2020MNRAS.498.2354R}. We translate signal-to-noise to magnitudes assuming background-limited sources and account for the finite aperture (fiber size) in which the spectral SNR is measured. For eBOSS quasars, we re-run the XDQSO quasar probability code to measure the response of the quasar selection to small magnifications \citep[the change in the redshift success rate with magnitude is negligible; see Fig.\ 4 of][]{2020MNRAS.498.2354R}. We also account for the reduced efficiency of star/galaxy separation at fainter magnitudes due to magnification \citep[where a point source can be demagnified and then spuriously considered an extended source][]{2002AJ....124.1810S}\footnote{\url{https://classic.sdss.org/dr7/products/general/stargalsep.html}}.

Third, we correctly account for the impact of lensing magnification at both the bright and faint ends of the sample (i.e.\ if the sample has a bright cut, de-magnified objects near the threshold are more likely to constitute the sample). This particularly matters for the eBOSS LRGs, which have a bright cut to prevent overlap with BOSS CMASS. We follow the method of \cite{2023MNRAS.523.3649E} and \cite{2023MNRAS.524.2195S}, in which the change in number counts is computed both when magnifying and de-magnifying the galaxies, and the total change is the difference of the two: $\delta N = \delta N_+ - \delta N_-$. We use a small finite step size of $\delta m = 0.02$ to compute magnification biases.

Magnification bias measurements are shown in Table~\ref{tab:spec_magbias_zeff}~and~\ref{tab:spec_magbias_zeff2}. These results are reasonably consistent with external estimates: \cite{2021MNRAS.504.1452V} find $s = 0.77$ for LOWZ at $z=0.35$ and $s = 1.05$ for CMASS at $z=0.65$; \citep{2023MNRAS.524.2195S} find $s=0.21$ for CMASS at $z=0.5$ and $s = 0.81$ for eBOSS LRG at $z=0.8$, and \cite{2022MNRAS.510.6150L} find $s \sim 1.2$ at $z=0.65$. 
The agreement between these measurements is hindered by differences in the exact samples used (i.e.\ redshift binning) and methodology adopted (measuring the slope of the luminosity function in a single band, or properly taking into account the complex selection function). During preparation of this paper, we compared results to \cite{2024MNRAS.527.1760W}, who also account for the full complexities of the selection function (with some methodological improvements over our results). We find very similar results for magnification bias of CMASS and LOWZ in redshift bins, agreeing to $\Delta s \sim 0.05$, well within our assumed systematic error of $\Delta s = 0.2$.

For quasars, \cite{2020MNRAS.499.2598W} finds $s \sim 0.1$--0.2 for quasars following the eBOSS quasar luminosity function with an absolute magnitude cutoff; \cite{2020JCAP...04..006L} and \cite{2016JCAP...02..051I} find $s=0.295$ for BOSS quasars; \cite{2003MNRAS.342..467M,2005MNRAS.359..741M} find $s=0.29$ for 2dF quasars.

We propagate errors on the magnification bias estimates into our uncertainty on the clustering redshift by assuming that magnification bias errors are dominated by a systematic error of $\Delta s = 0.2$. This is motivated by the level of agreement between our results and past results, and by \cite{2023MNRAS.523.3649E}, who find agreement to $\Delta s \sim 0.2$ between the approximate method that we use and more sophisticated methods based on image simulations. We find the magnification bias measurements are consistent between NGC and SGC, so we use NGC measurements due to its higher density and smaller uncertainties. We assume the magnification bias error is entirely correlated between NGC and SGC for each tracer.

We place the measurement in each bin at its effective redshift rather than the mean redshift of the bin. This is a very minor effect except for the $\Delta z =0.2$ quasar bins in the tail of the unWISE redshift distribution. We compute the effective redshift for the unWISE cross-correlations as
\begin{equation}
    z_{\textrm{eff}, sp} = 
    \frac{\sum_{p} \ z_{p,i} \ 1/\chi_{p,i}^2 \  dN_s/dz}{\sum_{p} \ 1/\chi_{p,i}^2 \ dN_s/dz}
\label{eqn:zeff_sp}
\end{equation}
and for the spectroscopic auto-correlations as
\begin{equation}\label{eqn:zeff_ss}
    z_{\textrm{eff}, ss} = \frac{\int \, dz \, z \, 1/\chi^2 dN_s/dz}
    {\int \, dz  \, 1/\chi^2 dN_s/dz}
\end{equation}
We can only obtain estimates of the redshifts of individual photometric galaxies, $z_{p,i}$, in the unWISE sample for the small number of objects that have counterparts in the 2 deg$^2$ COSMOS field. Because the number of objects is small we can not reliably estimate a smooth $dN_p/dz$ and therefore explicitly write the integral as a sum over the photometric sample in Eq.~\ref{eqn:zeff_sp}.

With an improved measurement of eBOSS quasar clustering, we switch from using the parameterised form of the quasar bias from \cite{2017JCAP...07..017L} to our own measurements of the quasar bias from its auto-correlation, similar to the method adopted for CMASS and LOWZ \citep[Fig.\,6 in][]{2020JCAP...05..047K}. 
We continue to use the \cite{2017JCAP...07..017L} bias function
for the DR12 QSOs, which have a noisier autocorrelation measurement.
Our quasar bias measurements are consistent with \cite{2017JCAP...07..017L} within the errorbars, but they allow for deviations from a smooth trend and take advantage of the larger sample size, especially compared to the measurement of \cite{2017JCAP...07..017L} which uses a preliminary release of eBOSS quasars (DR14). Spectroscopic biases are given in Table~\ref{tab:spec_magbias_zeff}~and~\ref{tab:spec_magbias_zeff2}.

As in \cite{2020JCAP...05..047K}, we verify that our cross-correlation redshifts are consistent between different spectroscopic tracers and between hemispheres. We find good agreement ($p > 0.05$) for all overlapping tracers with the same redshift binning ($\Delta z = 0.05$ for galaxies and $\Delta z = 0.20$ for quasars; see Table~\ref{tab:chi2_consistency}). Furthermore, since the eBOSS LRGs overlap with the eBOSS quasars in the redshift range $0.8  < z < 1.0$ we have an additional consistency check. We find that for both the Blue and Green unWISE samples, the quasar cross-correlation is slightly higher than the LRG cross-correlation (see Fig.~\ref{fig:compare_old_new_dndz}). To quantitatively compare the two, we create both a $\Delta z = 0.2$ bin for the LRGs and a $\Delta z = 0.05$ bin for the quasars. For Blue, in the single $\Delta z = 0.2$ bin, the quasars are higher by 1.77$\sigma$. In the four $\Delta z = 0.05$ bins, the quasars are higher by 1.25$\sigma$, 0.77$\sigma$, 2.35$\sigma$ and 1.65$\sigma$. For Green, in a single $\Delta z = 0.2$ bin, the quasars are higher by 1.06$\sigma$. Therefore, while a discrepancy certainly exists (especially for Blue), it is not large enough to be particularly concerning, especially given the large number of comparisons in Table~\ref{tab:chi2_consistency}. Future spectroscopic surveys such as DESI will significantly help the situation--both by improving the statistics and by offering multiple tracers for improved systematics testing.

\begin{table}
    \centering
        \begin{tabular}{cccc}
        Spec.\ sample 1 & Spec.\ sample 2 & Blue $\chi^2$/dof & Green $\chi^2$/dof \\
        \hline
        DR12 CMASS N & DR12 CMASS S & 7.4/14 & 17.7/14 \\
        DR12 CMASS N & DR12 LOWZ N & 5.6/7 & 6.6/7 \\
        DR12 CMASS N & DR12 LOWZ S & 6.2/7 & 5.1/7 \\
        DR12 CMASS N & DR16 LRG N & 3.6/4 & 2.5/4\\
        DR12 CMASS N & DR16 LRG S & 1.4/4 & 1.2/4 \\
        DR12 CMASS S & DR12 LOWZ N & 4.9/7 & 10.3/7 \\
        DR12 CMASS S & DR12 LOWZ S & 3.7/7 & 10.7/7 \\
        DR12 CMASS S & DR16 LRG N & 5.1/4 & 3.3/4\\
        DR12 CMASS S & DR16 LRG S & 3.6/4 & 1.4/4 \\
        DR12 LOWZ N & DR12 LOWZ S & 5.7/9 & 1.7/9 \\
        DR16 LRG N & DR16 LRG S & 2.9/8 & 8.5/8 \\
        DR16 QSO N & DR16 QSO S & 10.2/7 & 7.4/7 \\
        DR16 QSO N & DR12 QSO N & -- & 3.0/1 \\
        DR16 QSO N & DR12 QSO S & -- & 1.5/1 \\
         DR16 QSO S & DR12 QSO N & -- & 0.0/1 \\
        DR16 QSO S & DR12 QSO S & -- & 2.7/1 \\
        DR12 QSO N & DR12 QSO S & -- & 11.3/8 \\        
        \end{tabular}
    \caption{Consistency of cross-correlation estimates of $b(z)dN/dz$ for different spectroscopic tracers. We show the $\chi^2$ between the clustering redshift estimates obtained using different spectroscopic subsamples and the number of degrees of freedom, i.e. the number of overlapping redshift bins between the samples. The Blue sample of unWISE galaxies has no significant overlap with the high redshift BOSS DR12 quasars (compare Fig.\,\ref{fig:spec_dndz}). 
    \label{tab:chi2_consistency}}
\end{table}

\begin{figure}
    \centering
    \includegraphics[width=0.45\linewidth]{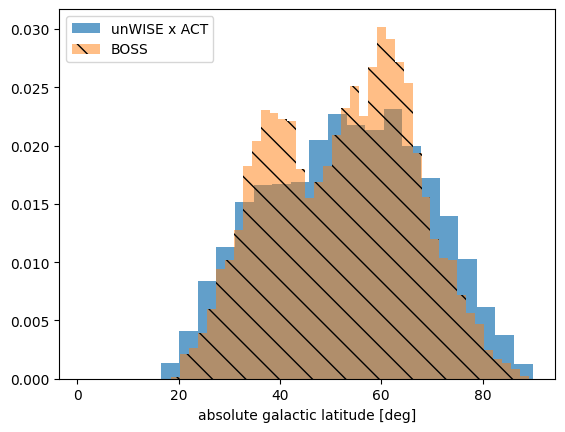}
    \includegraphics[width=0.45\linewidth]{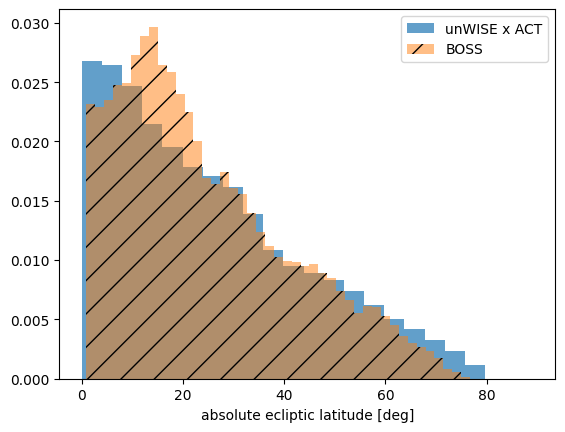}
    \caption{Distribution of absolute value of ecliptic and Galactic latitude for ACT-unWISE footprint that we measure $C_{\ell}^{gg}$ and $C_{\ell}^{\kappa g}$ on; and unWISE-BOSS footprint that we use for cross-correlation redshifts. The fact that the unWISE-BOSS footprint is a good match to the ACT-unWISE footprint means that our estimate of $b(z) dN/dz$ from the cross-correlation of unWISE with BOSS is likely to be representative of the bias weighted redshift distribution on the ACT-unWISE footprint.}
    \label{fig:abs_ecliptic_galactic}
\end{figure}

The unWISE sample exhibits some inhomogeneity, particularly in ecliptic latitude (see Sec.\,\ref{subsec:galaxy_sys}), likely due to the variable unWISE coverage depth. This inhomogeneity may also result in spatially varying $dN/dz$. Given that the $dN/dz$ is based on cross-correlations measured over only a fraction of the unWISE-ACT footprint (see Fig.~\ref{fig:spec_skyplot}), we desire that this area is  representative of the full footprint. In Fig.~\ref{fig:abs_ecliptic_galactic}, we show that the unWISE-ACT and unWISE-BOSS footprints match very well in (absolute value) ecliptic and Galactic latitude. The absolute value is most relevant here because the WISE depth is greatest at the poles and shallowest at the equator; likewise, Galactic contamination is small at both poles. Therefore, the sign of the latitude (hemisphere) matters much less than the distance from the equator. In Table~\ref{tab:abs_ecliptic_galactic}, we summarise the mean absolute ecliptic and Galactic latitude of the BOSS and eBOSS surveys. The combined spectroscopic regions are similar in mean absolute ecliptic and Galactic latitude to the unWISE-ACT footprint. Therefore, we expect our cross-correlation estimate of $b(z) dN/dz$ to be representative of the $b(z) dN/dz$ on the unWISE-ACT footprint. \cite{2023JCAP...07..044B} showed that as long as the redshift distribution assumed in the model matches the mean redshift distribution of the sample used, anisotropies in the $dN/dz$ across the survey footprint do not bias the inferred cosmology.

\begin{table}
    \centering
        \begin{tabular}{c|cc}
        Region & Galactic $\langle | b | \rangle$ ($^\circ$) & Ecliptic $\langle | \beta | \rangle$ ($^\circ$) \\
        \hline
        ACT-unWISE & 53.8 & 25.3 \\
        eBOSS N & 55.4 & 39.2 \\
        eBOSS S & 47.0 & 15.1 \\
        eBOSS N+S & 52.1 & 29.5 \\
        BOSS N & 55.2 & 29.5 \\
        BOSS S & 45.7 & 14.3 \\
        BOSS N+S & 52.9 & 25.8 
        \end{tabular}
    \caption{Mean absolute value of ecliptic and Galactic latitude for various footprints. The combined spectroscopic regions are similar in mean absolute ecliptic and Galactic latitude to the unWISE-ACT footprint.
    \label{tab:abs_ecliptic_galactic}}
\end{table}

\begin{figure}
    \centering
    \includegraphics[width=1.0\linewidth]{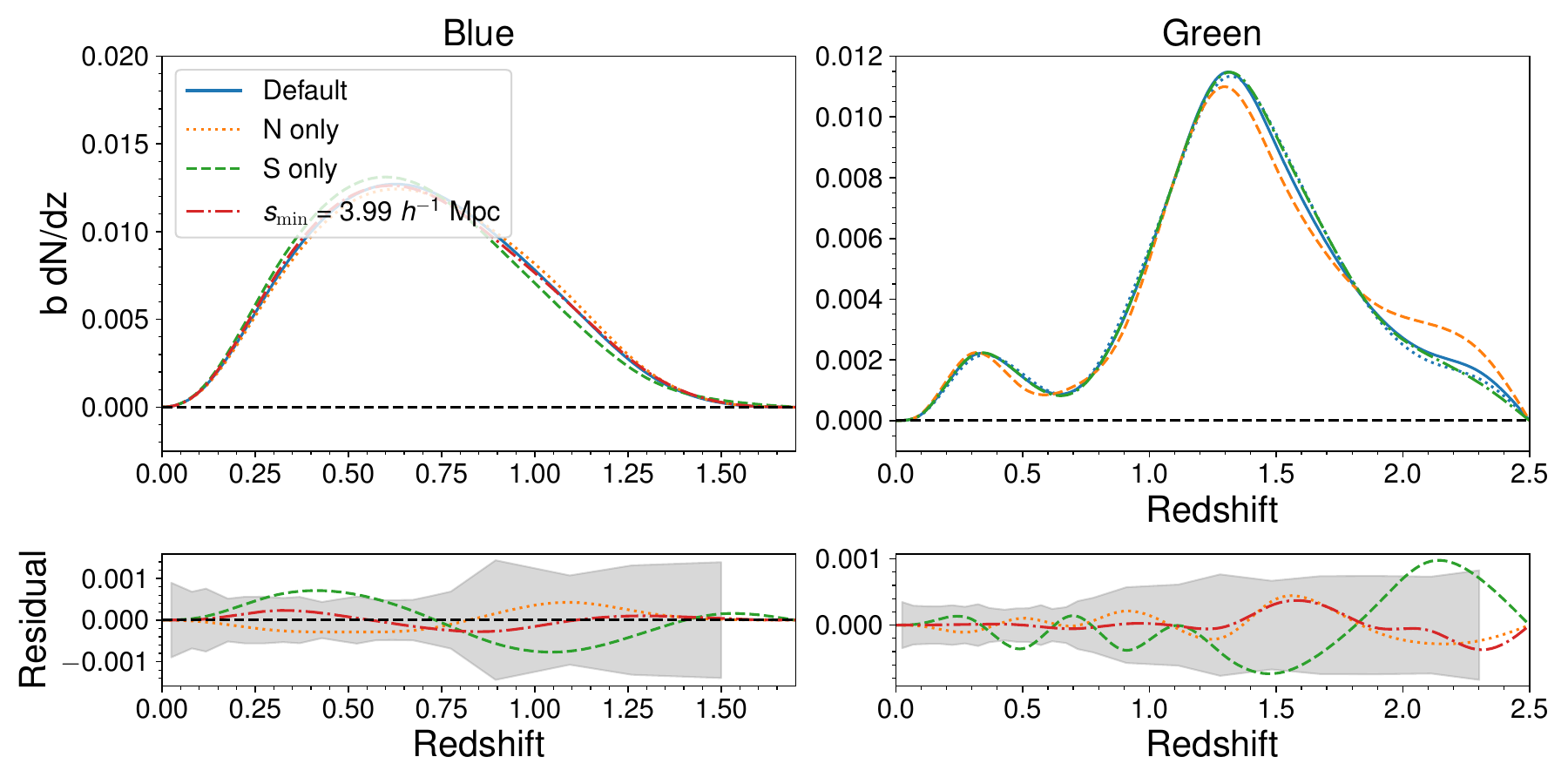}
    \caption{Comparison of $b(z)dN/dz$ used for different null tests: NGC only, SGC only, and increasing the minimum scale used in the clustering redshifts, with the residual (changed $b(z)dN/dz$ minus default) shown in bottom panels. The gray errorband shows the uncertainties on the data points used to fit the default $b(z) dN/dz$. The estimates are consistent within the uncertainties and we show in Sec.\,\ref{subsec:param_consistency_tests} that our cosmological inference is stable with respect to these choices.}
    \label{fig:dndz_null_tests}
\end{figure}

\begin{figure}
    \centering
    \includegraphics[width=1.0\linewidth]{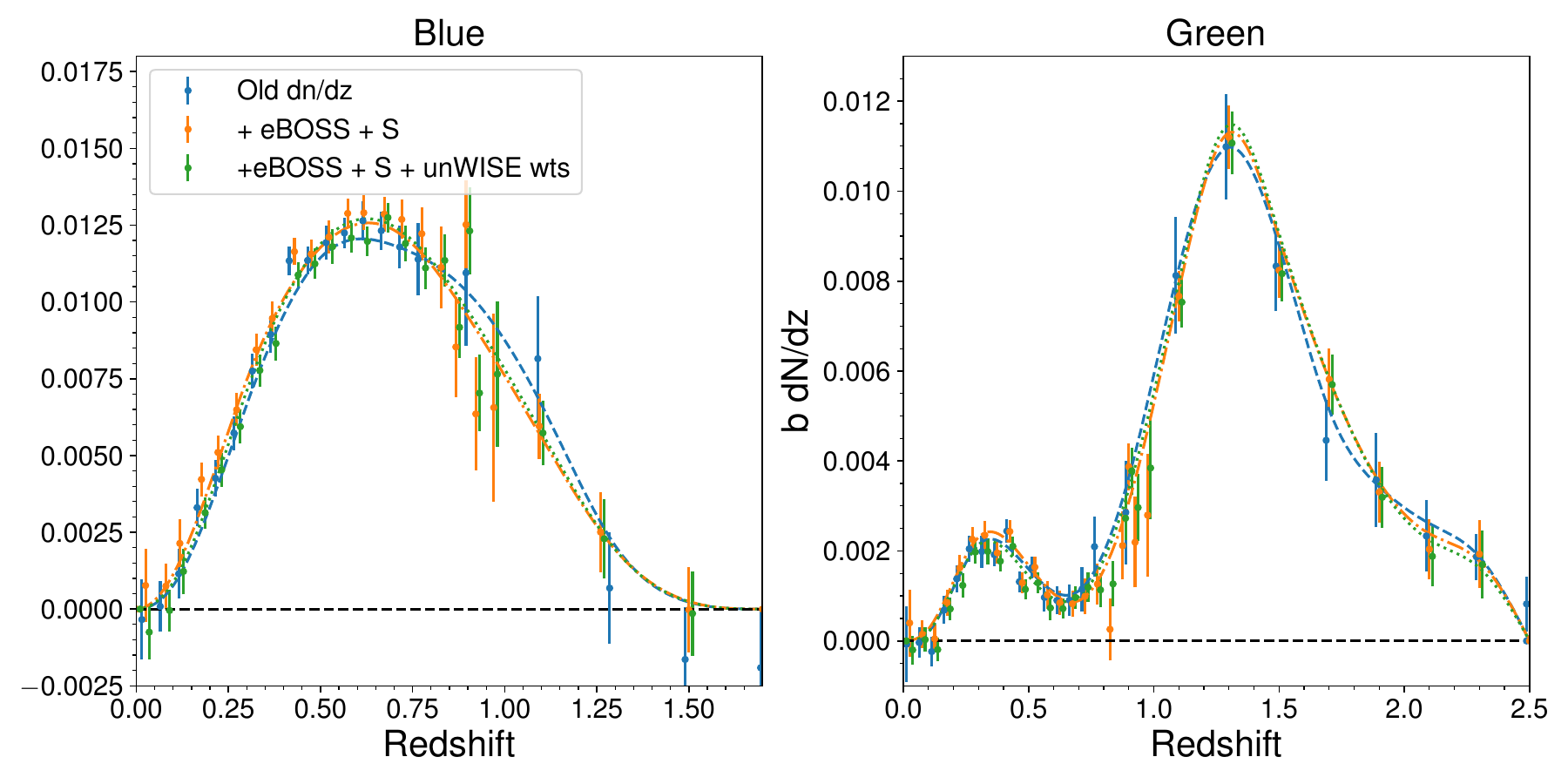}
    \caption{Comparison between the old cross-correlation $b(z)dN/dz$ \citep[from][]{2020JCAP...05..047K,2021JCAP...12..028K} and the updated $b(z)dN/dz$ used in this work. We show both the data points (i.e.\ cross-correlation measurement, averaged over scale, divided by spectroscopic sample bias) and the smooth best fits. The $b(z)dN/dz$ used in \cite{2021JCAP...12..028K} are shown in blue (with the smooth best fit shown as a dashed line). The orange data points and dashed line show the estimated $b dN/dz$ after adding the southern spectroscopic samples, the eBOSS LRGs, and replacing the eBOSS DR14 quasars with the DR16 quasars (this also includes the correction to the treatment of the spectroscopic magnification bias). Finally, the green data points and dotted line shows estimates which additionally include unWISE weights in the cross-correlation (final version used in paper). Points are offset in redshift for clarity.}
    \label{fig:compare_old_new_dndz}
\end{figure}

In Fig.~\ref{fig:dndz_null_tests}, we show the variations in $dN/dz$ considered as part of our null tests: restricting to the NGC; restricting to the SGC; and using a larger minimum scale cut in the clustering redshift measurement ($s_{\textrm{min}} = 3.99$ $h^{-1}$ Mpc rather than the default $s_{\textrm{min}} = 2.52$ $h^{-1}$ Mpc).
\resub{We show in Fig.~\ref{fig:param_consistency} the impact of these changes on $S_8^{\times}$; note that the NGC and SGC $S_8^{\times}$ measurements also use the appropriate $C_\ell^{\kappa g}$ and $C_{\ell}^{gg}$ measured only on the NGC or SGC respectively, so the observed $\sim 1\sigma$ shift is expected just due to the different sky areas.}

In Fig.~\ref{fig:compare_old_new_dndz}, we compare the new cross-correlation $b(z)dN/dz$ to the old ones, breaking the changes into two parts: the changes or additions to the dataset (adding SGC tracers, eBOSS LRGs, and exchanging the eBOSS DR14 quasars for the final eBOSS DR16 quasars), and a change in method (using unWISE weights in the cross-correlation as well as the spectroscopic weights). For Blue, the mean redshift changes from 0.714 to 0.690 when adding more data, and finally to 0.697 when adding the unWISE weights. This is a shift of 0.74$\sigma$ compared to the standard deviation of 0.023 as measured from 1000 samples of $dN/dz$, and this large shift is the dominant source for the upwards shift in $S_8$ in the Planck reanalysis. For Green, the mean redshift changes from 1.352 to 1.357 when adding new data, and to 1.355 when adding the unWISE weights (compared to an uncertainty of 0.022: a shift of 0.14$\sigma$). The new $dN/dz$ show clearly reduced uncertainty, especially in the eBOSS quasar redshift range $0.8 < z < 2.0$, where we have used almost four times as much area. Moreover, the new $dN/dz$ have improved redshift resolution at $0.8 < z < 1.0$ from the eBOSS LRG, which reduces the uncertainty on $dN/dz$ in this range although it is not apparent from the plot. It is also clear that most of the change in $dN/dz$ comes from the addition of the southern footprints of the spectroscopic tracers and the additional eBOSS data; adding unWISE weights makes little difference. Interestingly, the weights seem to make the largest difference at low redshift, where they suppress $dN/dz$, but make little difference at $z > 0.7$. This suggests that there may have been some small residual systematic power in the cross-correlation redshifts with BOSS CMASS and LOWZ. The changes between the old and new $dN/dz$ are consistent with statistical fluctuations related to adding more data. The changes in $dN/dz$ are largest at $z > 0.8$, where the statistical error is biggest.

\section{Detailed discussion of post-unblinding changes}\label{app:post-unblinding_changes}

As discussed in Sec.\,\ref{subsec:post_unblinding_changes} we made four minor corrections to the analysis after unblinding. This includes a modification to the treatment of the mask induced mode coupling of the shot noise, a small correction to the transfer function obtained from simulations, the inclusion of the cross-covariance between the two unWISE samples, and a bug-fix affecting the implementation of the higher order biases. The impact of these corrections partially offsets and they results in a net shift of $-0.1\sigma$ in $S_8$ compared to our initially unblinded results. Here we provide some more detail on these corrections.

In order to compare our theory predictions of $C_\ell^{gg}$ and $C_\ell^{\kappa g}$ with the observed spectra, we convolve them with a bandpower window which captures the effect of the approximate mode decoupling performed on the data. As discussed in Sec.~\ref{sec:tomography} the expectation value of the observed pseudo-$C_\ell$ is related to the true underlying $C_\ell$ through a mode coupling matrix $M_{\ell\ell'}$,
\beq
\av{\tilde{C}_\ell} = \sum_{\ell'} C_{\ell'} M_{\ell \ell'}.
\eeq
For the binned power spectra $\hat{C}_b$ this relation can only approximately be inverted. The decoupled observed power spectrum is then
\beq
\hat{C}^{\rm{decoupled}}_b = \sum_{b'} \widehat{M^{-1}}_{b b'} \hat{C}_{b'} \label{eq:approx_mode_decoupling}
\eeq
where $\widehat{M^{-1}}_{b b'}$ is the inverse of the binned mode coupling matrix obtained under the assumption that the power spectrum is piecewise constant across each bin\footnote{This is the default approximation implemented in \texttt{NaMaster} \citep{2019MNRAS.484.4127A}. An alternative approach suggested in \cite{2002ApJ...567....2H} is to assume the $\ell(\ell+1)C_\ell$ is constant instead.}. The expectation value of the decoupled power spectrum is therefore given in terms of the true underlying $C_\ell$ by
\beq
\av{\hat{C}^{\rm{decoupled}}_b} = \sum_{b'}\widehat{M^{-1}}_{b b'} \sum_{\ell} w_\ell^{b'} \sum_{\ell'} M_{\ell \ell'}  C_{\ell'}
\eeq
where $w_\ell^b$ are the weights associated with each multipole in bin $b$. In order to compare a given theory prediction with the observed and decoupled bandpowers, one thus needs to convolve it with a bandpower window.
\beq
F_{b, \ell'} = \sum_{b'}\widehat{M^{-1}}_{b b'} \sum_{\ell} w_\ell^{b'} M_{\ell \ell'}.
\eeq
For a purely white power spectrum no mode coupling is induced. This is because, as was pointed out in \cite{2002ApJ...567....2H}, $\sum_{\ell'} M_{\ell \ell'} = w_2$. So no mode decoupling is required on any white noise component of the signal. However, since the mode coupling matrix is computed in practice only to finite $\ell$, the operation of masking the noise and convolving the noise power spectrum with the finite mode coupling matrix are not equivalent ($\sum_{\ell'=0}^{\ell_{\rm{max}}} M_{\ell \ell'} \neq w_2$). Consequently, while the white noise present in the map is in fact scaled simply by $w_2$ the operation applied to the white noise component of the theory spectrum does not reflect this. In principle the mode coupling matrix is also inaccurate for the signal component of the spectrum for the same reason, however, because the signal falls off quickly with $\ell$ the effect is strongly suppressed. On our simulations we find that power spectrum is most accurately recovered if we instead adopt the following approach
\beq
C_b^{\rm{th},bpw\ conv.} = \sum_{b'}\widehat{M^{-1}}_{b b'} \sum_{\ell} w_\ell^{b'} 
 \left[\sum_{\ell'} \left(M_{\ell \ell'}  \left(C^{\rm{th}}_{\ell'} - N_{\rm{shot}}\right) + M_{0, \ell'} N_{\rm{shot}} \right)\right].
\eeq

The inaccuracy in our previous treatment, which did convolve the shot noise with the mode coupling matrix, due to this effect is very small ($<0.1\%$). In the process of correcting this effect, however, we noticed that because our Gaussian realisations of the lensing field are band limited to $\ell_{\rm{max}}^{\rm{Gaussian\ }\kappa}=5100<3*\textsc{nside}$, our correlated realisations of the galaxy field were not generated with the correct power at $\ell>\ell_{\rm{max}}^{\rm{Guassian\ }\kappa}$.  This was originally not taken into account when testing the power spectrum recovery and computing the transfer function (see Sec.~\ref{sec:sims}). Since the mode coupling matrix mixes even widely separated scales, the incorrect power at small scales also affects some scales within our analysis range at the $\sim0.5\%$ level. Given that the fractional error on $C_\ell^{gg}$ is of the order $2\%$ this explains the $\sim0.2\sigma$ shift in our results.

Additionally, it was noticed that the cross-covariance between the two redshift samples was erroneously omitted from the analysis. When corrected, this results in shift of $\sim-0.3\sigma$ in the value inferred for $S_8$ from the joint analysis of the Blue and Green sample ($\sim -0.17\sigma$ for $S_8^{\times}$) and a slight degradation of constraining power by about 5\%.

Finally, we noticed a small bug in the implementation of the bias evolution of the higher order biases. The fiducial bias evolution for $b_{2,L}$ was erroneously adopted for $b_{s,L}$ as well. Correcting this mistake required newly determining the priors on the higher order bias offsets from our $N$-body simulations. The prior mean on the offset for the second order Lagrangian bias for the Blue sample, $c_{2,L}^{\rm Blue}$, increased from $0.39$ to $0.55$ (approximately $0.3\sigma$) and since we set the width of the prior to be consistent with $c_{2,L}^{\rm Blue}=0$ this also required an increase of the width of the prior by $\sim30\%$. The prior mean for the shear bias parameter decreased from $c_{2,s}^{\rm Blue}=0.30$ to $c_{2,s}^{\rm Blue}=0.17$ with the width of the prior remaining basically unchanged. For the Green sample, the mean of the prior on $c_{2,L}^{\rm Green}$ also increased by about 30\% to 0.42 with the corresponding increase in width. The prior on $c_{2,s}^{\rm Green}$ increased in width by about 15\% while the mean increased by about 30\%. We reran the model verification on our $N$-body sims and recovered the input cosmology well within the requisite $1\sigma$ in terms of the covariance appropriate for the simulation volume. This indicates that the error introduce by this bug was at least partially absorbed by our higher order bias marginalisation.

Since the higher order terms are much more significant for the Blue sample, the impact on the value of $S_8$ inferred from the Blue sample alone lies at about $0.3\sigma$. On the other hand, our joint inference is unchanged apart from a small, $\sim5\%$, increase in the posterior width. This suggests that in the joint inference the combined data provide sufficient constraining power to probe the higher order biases beyond the information given by the prior.

\end{document}